\documentclass{article}

\usepackage[english]{babel}
\usepackage[a4paper,top=2cm,bottom=2cm,left=2.7cm,right=2.7cm]{geometry}
\usepackage{amsmath}
\usepackage{amssymb}
\usepackage{amsthm}
\usepackage{graphicx}
\usepackage[colorlinks=true, allcolors=black]{hyperref}
\usepackage{setspace}
\usepackage{authblk}
\usepackage{enumitem}
\usepackage{tabularx}
\usepackage{float}
\usepackage{booktabs}
\usepackage{multirow}
\usepackage{subcaption}
\usepackage{caption}
\usepackage[round, authoryear]{natbib}
\newcolumntype{L}{>{\raggedright\arraybackslash}X}

\usepackage{tikz}
\usetikzlibrary{positioning,arrows.meta,calc,fit,decorations.pathreplacing,patterns.meta,shapes.geometric}
\usepackage{algorithm}
\usepackage{algpseudocode}
\algrenewcommand\algorithmicrequire{\textbf{Requires}}
\algnewcommand\algorithmicsetup{\textbf{Setup}}
\algnewcommand\Setup{\item[\algorithmicsetup]}
\algrenewcommand\algorithmicensure{\textbf{Execution}}
\algnewcommand\algorithmicadjudicate{\textbf{Adjudication}}
\algnewcommand\Adjudicate{\item[\algorithmicadjudicate]}
\algnewcommand\algorithmicrefit{\textbf{Refit}}
\algnewcommand\Refit{\item[\algorithmicrefit]}
\makeatletter
\renewcommand{\fs@ruled}{%
  \def\@fs@cfont{\bfseries}%
  \let\@fs@capt\floatc@ruled
  \def\@fs@pre{\hrule height0.8pt depth0pt \kern4pt}%
  \def\@fs@post{\kern4pt\hrule height0.8pt\relax}%
  \def\@fs@mid{\kern3pt\hrule height0.4pt\kern3pt}%
  \let\@fs@iftopcapt\iftrue}
\makeatother
\floatstyle{ruled}
\restylefloat{algorithm}

\newcommand{\rungnum}[1]{\renewcommand{\thealgorithm}{#1}}

\theoremstyle{plain}
\newtheorem{proposition}{Proposition}
\newtheorem{corollary}{Corollary}
\theoremstyle{definition}
\newtheorem{definition}{Definition}
\theoremstyle{remark}
\newtheorem*{remark}{Remark}

\DeclareMathOperator*{\argmax}{arg\,max}

\DeclareMathOperator*{\mean}{mean}
\newcommand{\E}{\mathbb{E}}
\newcommand{\Mx}{\mathcal{M}}
\newcommand{\Rclass}{R_{\mathrm{class}}}
\newcommand{\Rinst}{R_{\mathrm{inst}}}

\providecommand{\keywords}[1]{\small\textbf{\textit{Keywords:}} #1}

\title{Better together?\\A performance theory of human–machine skill policies\\under generalization difficulty\\[30pt]}
\author[ ]{Riccardo Zanardelli}
\date{\small{Technical Report, v1.1\\[30pt]}}

\newcommand{\sqc}[1]{\textcolor{black}{#1}}

\begin{document}
\maketitle
\begin{spacing}{1}

\begin{abstract}
\noindent
Organizations that put a human and a machine on the same task decide, usually by default, which skill executes each instance and what happens to their work when both do. This paper gives an accounting of what that decision can attain in performance, under the squared rule. A skill's performance is a distribution over the instances of a task, which is how generalization difficulty enters. Between the mean of the skills' performances and the best of them on each instance lies a gross stake. It splits at the best single skill into a systematic component, identified from separate assessments and secured by committing to that skill, and a residual, the expected value of perfect information about which skill each instance favors, which only paired observation identifies and which classes of instances split again. The skills set the stake; the workflow decides the take, through two channels of augmentation. Routing directs work between the skills and combines their responses convexly or selects one; where the truth is binary it is bounded by the best of them on every instance, while a combination that reads the responses as joint evidence is not routing and is bounded instead by what they reveal together. Synergy lets the skills interact before a joint response is fixed; it is bounded by none of them, in either direction, and is measured by a lift against what the same skills would have returned independently, which must clear a break-even and can fall below one. Written as algorithms, the arrangements form a ladder that states what each must read, hold and record and, read in sequence, a procedure whose checks never give back what an earlier step secured when the class means are known, and give back little on a finite record once they require a margin. The accounting extends to rosters of many skills, where a skill is worth what it adds where it beats everything held, and to competing skills, where at parity what decides is how much a single failure may cost. Applied to the public data of a chest-radiograph experiment with 227 radiologists, whose original analysis asked whether showing them a model's prediction improved their performance, the accounting asks instead what was at stake and what each arrangement took of it. It finds the radiologists and the model close to parity and most of the stake residual; the assisted reading, no better on average in the original analysis, returned less than averaging the two responses; and recalibrating the model alone attains more than any deployable arrangement of the responses as reported. An organization can apply the accounting to its own skills and record, to learn what is at stake between them, which arrangement can collect it, and what evidence each step needs before it is taken; what these gains are worth once costs and the consequences of error are counted is the subject of future work.\\[20pt]
\end{abstract}

\noindent\keywords{human--AI complementarity; learning to defer; value of information; partial identification; automation and augmentation}
\clearpage

{\footnotesize\setcounter{tocdepth}{2}\tableofcontents}

\clearpage

\section{Introduction}\label{sec:introduction}

The deployment of artificial intelligence confronts organizations with a decision that is as old as the division of labor and newly pressing: whether the instances of a task should be executed by human skills, by machine skills, or by some combination of the two, and when by which. We call this the \emph{skill policy decision problem}, a choice with direct consequences for productivity, employment, safety and equity. It is demanding for reasons that compound. The performance of a skill is not a number but a distribution over the instances of a task, and the distributions of different skills differ in location, in spread and in how they co-occur on the same instance. Which skill prevails can change from one instance to the next, and what a workflow does with two responses can return more than either of them, or less than their average.

Faced with this complexity, a natural satisficing response \citep{simon1955behavioral} is to add human oversight: to keep a human in the loop, on the intuitive assumption that combining human and machine skills captures the benefits of automation while mitigating its risks. The assumption is embedded in regulation, most prominently in Article~14 of the European Union AI Act, in corporate governance and in public discourse. The evidence is less comfortable. When radiologists had access to an AI tool that alone exceeded most of their accuracy, providing its predictions did not improve their average performance, and the efficient allocation assigned cases to either the human or the machine rather than combining them \citep{agarwal2023combining}; a preregistered meta-analysis of 106 experiments finds human--AI combinations performing on average worse than the better of the two alone \citep{vaccaro2024combinations}. Neither result says that combining is a mistake. Both say that whether it helps depends on something the decision to combine does not fix.

This paper supplies an accounting of what that something is, stated in performance.\footnote{The earlier paper carried the archetypes to economic utility \citep{zanardelli2025navigating}; future work will carry the extended framework there, pricing the cost profiles reported here against performance and charging the consequences of error, and will examine the resulting trade-offs computationally. Every definition the present paper uses is stated here.} It does not ask how often combinations create or destroy value, which depends on costs and on the consequences of error. It asks what several skills available for the same task can attain before any workflow is designed, what each workflow can collect of it, and what it must already know to be run at all. The organizing distinction is between the \emph{stake}, which the skills set before any workflow exists, and the \emph{take}, which the workflow decides. The decomposition and its archetypes come from earlier work \citep{zanardelli2025navigating}. For a pair of skills, it split the gross stake into the systematic component and the residual and bounded both, read the mean, collaborative and superpower combinations of the two skills as the corner settings of routing and synergy, and ordered workflows into a ladder whose upper rungs it could bound but not run; through Monte Carlo simulation across levels of generalization difficulty, it then carried the analysis to bottom-line economic utility, finding that combining human and machine skills yields the most only where genuine augmentation is achieved, and destroys value where it is not. The present paper stays in performance and goes further in two directions. It extends the decomposition, to any number of skills and, within the residual, to the part a class-level rule can reach; and it traverses the ladder, defining the policies that occupy each rung, connecting the responses they act on to the performances the stake is defined on, and stating for each rung what it must hold, execute and record, including the routing on imperfect signals that the earlier model could only bound.

The tools are elementary. Linearity of expectation, Jensen's inequality, the value of information and the bounds of Fr\'echet and Hoeffding on joint distributions with given marginals carry almost every proposition, and none would surprise a reader of decision theory, ensemble learning or algorithm selection. The contribution is the accounting they support: an extended decomposition of what several skills put on the table into components that call for different capabilities and different evidence, refined where instances fall into classes; a bridge between the responses arrangements act on and the performances the stake is defined on; the separation of two channels of augmentation, one bounded by the skills and one not; and a hierarchy of the evidence each quantity requires, from separate assessments to paired observation.

Two scope conditions run through the paper. The bridge, and with it the ceiling on routing, holds where the correct response is an endpoint of a scalar response space, as in binary classification; it fails for multiclass responses and for continuous targets, where a combination of responses can beat every one of them (Section~\ref{sec:responses}). Performance is scored throughout by the squared rule, one minus the Brier score, which credits a response for the strength of the evidence it carries and charges it for stating that strength wrongly \citep{gneiting2007strictly}, and under which pooling the responses earns their disagreement over choosing among them at random. The linear rule, which takes performance to be the distance from the correct response itself and which field studies often report, appears as the special case in which that disagreement earns nothing: it is simpler, and where responses are final it loses nothing (Section~\ref{par:rule} and Appendix~\ref{app:linear}). Throughout, performance is separated from cost: arrangements are given a cost profile but not a price (Section~\ref{sec:costs}).

\subsection{Contributions}\label{sec:contributions}

The paper makes four contributions.

\emph{An accounting of the stake} (Sections~\ref{sec:anchors}--\ref{sec:insurance}, \ref{sec:magnitudes} and~\ref{sec:identification}), extending the decomposition introduced for a pair of skills by \citet{zanardelli2025navigating} to any number of skills and to classes. Between the mean of the skills' performances on each instance and the best of them lies a gross stake $G$, which splits at the best single skill into a systematic component $S$, identified from separate assessments and secured by an ex ante commitment, and a residual $R$, the expected value of learning which skill the instance favors, identified only from paired observation; where instances fall into classes, $R$ splits into a part a class-level rule reaches, $\Rclass$, and a part only an instance-level rule reaches, $\Rinst$. Engaging every skill without directing work is an insurance contract whose premium and coverage are stated, and which, against an assessment made on instances like those it will serve, pays only while the record is short; a hierarchy of evidence says which record identifies which quantity, and with what bias a finite record estimates it. At parity the stake is all residual and the insurance is free (Proposition~\ref{prop:parity}). Over rosters of many skills and many classes (Appendix~\ref{app:many}), a skill is worth what it adds where it beats everything already held, with diminishing returns under which a roster built greedily attains at least $1 - 1/e$ of the best, and a budget of paired instances is best spent on classes that are frequent, variable in their residual and cheap to pair.

\emph{A bridge between responses and performance, and two channels of augmentation} (Sections~\ref{sec:twochannels} and~\ref{sec:responses}--\ref{sec:reduced}). Where the correct response is a scalar endpoint, every convex combination of responses scores between the worst and the best of its members; under the squared rule it scores the combination of their performances plus the ambiguity term of ensemble theory, the disagreement between the responses, which vanishes under the linear rule. Routing, which directs work between the skills by class or by instance through assignment, ensembling or selection, is therefore bounded by the best skill on every instance, a bound that a rule pooling calibrated responses into one more extreme than any of them, as the Bayesian combination of independent evidence does, can exceed, and which is bounded instead by the score of the probability of the truth given the responses (Section~\ref{sec:responses}, Proposition~\ref{prop:evidence}); synergy, which lets the skills interact before a joint response is fixed, is bounded by none of them, and is measured by a lift, as a ratio or a difference, whose break-even is set by the informed baseline and by what pooling the responses attains.

\emph{The ladder of arrangements} (Section~\ref{sec:ladder}). From one unexamined skill to routing with synergy, each arrangement is written as an algorithm declaring what it reads, holds available, executes, requires and records. The algorithms give every arrangement a cost profile without a price, and expose that the arrangements engaging the fewest skills depend on the scarcest evidence; assignment with audit keeps that evidence growing at a chosen cost. Read in sequence, the arrangements give a procedure for climbing from an unexamined skill toward the whole stake, with a check at each step, made class by class on the classes as the policy reads them, that never gives back what the step before it secured when the class means are known, and on a finite record gives back little once it requires a margin (Section~\ref{sec:procedure}, Propositions~\ref{prop:checks} and~\ref{prop:margin}). What a policy takes is read component by component, exactly where it resolves its classes and by a stated crediting convention where its errors leave a class short of the incumbent (Corollary~\ref{cor:attribution}).

\emph{Field arrangements read through the theory} (Section~\ref{sec:field}). The chest-radiograph (CXR) experiment of \citet{agarwal2023combining} and the double-reading rules of mammography screening are placed on the ladder, including an outcome, a combination returning less than averaging the responses of the skills it engages, that a theory defined on performances alone cannot represent and a theory defined on protocols can. On the public data of the CXR experiment, reanalyzed here with estimands that differ from those of its authors, the stake and the take are measured: under the squared rule most of the stake is residual, on every subset of findings and after recalibration, and, since a radiologist reads whole films, only a third of the residual lies between films, within reach of a rule that assigns them; which skill leads depends on the findings, the rule and calibration; the arm that showed the radiologists the model's prediction returned less than pooling the two responses; and recalibrating the model alone attains more than any deployable arrangement of the responses as reported. A field experiment in patent drafting is read as a policy that moves the human skill it engages.

Section~\ref{sec:background} reviews the literatures the theory draws on, and Section~\ref{sec:primitives} defines performance, the skills and the skill policies through which they are deployed. Sections~\ref{sec:bounds} and~\ref{sec:ladder} set out the stake and traverse the arrangements, Section~\ref{sec:procedure} turns them into a procedure for climbing the ladder, Section~\ref{sec:field} reads field evidence through them and measures the stake and the take on the public data of one experiment, and Sections~\ref{sec:discussion} to~\ref{sec:conclusion} discuss, extend the theory to skills of equal capability in competition (Section~\ref{sec:contest}), qualify and conclude. The appendices map the features that govern the stake and the take (Appendix~\ref{app:factors}), state the linear rule as a special case (Appendix~\ref{app:linear}), prove the propositions on the stake (Appendix~\ref{app:proofs}), extend the accounting to many skills and many classes (Appendix~\ref{app:many}) and to skills in contest (Appendix~\ref{app:contest}), detail the records, the illustrative profiles and the CXR measurement by finding (Appendices~\ref{app:records}, \ref{app:profiles} and~\ref{app:cxr}), and close with a review map of the formulations and a reading of every proposition in plain words (Appendices~\ref{app:map} and~\ref{app:interpretations}).

\section{Background}\label{sec:background}

\subsection{Bounded humans, and bounded machines}\label{sec:bounded}

Herbert Simon argued that human rationality is bounded by limited cognitive capacity, incomplete information and time pressure, so that administrators faced with complex decisions do not optimize but \emph{satisfice} \citep{Simon1947AdministrativeBehavior, simon1955behavioral, simon1956rational, schwarz2022bounded}. The skill policy decision is a modern instance of that administrative behavior, with bounded rationality on both sides of the choice. Human skills are bounded by cognitive limits, fatigue and the inability to scale; machine skills by \emph{generalization difficulty} \citep{chollet2019measure}, the challenge of performing reliably when the conditions met at inference diverge from those met in training, which calls for the abstraction and analogy-making \citep{mitchell1993analogy} that remain hard to reproduce algorithmically. The set of automatable tasks expands as technology improves \citep{acemoglu2025simple}, but the frontier of machine capability is jagged rather than smooth in task difficulty: AI assistance improved the work of consultants inside it and worsened it outside, at seemingly similar levels of difficulty \citep{dell2026navigating}. If the frontier is jagged across tasks, it is natural to conjecture that it is jagged within a task as well, so that which skill prevails can change from one instance to the next. The theory below formalizes that dependence, and the formalization is what separates the systematic component of the stake from the residual.

\subsection{Allocating work between humans and machines}\label{sec:allocation}

\citet{fitts1951human} proposed comparing humans and machines on a list of capabilities and assigning each function to whichever was better at it; \citet{sheridan1978human} replaced the binary with a graded scale of supervisory control, from automation that offers no assistance, through automation that proposes and acts on approval, to automation that decides and acts alone, and \citet{parasuraman2000model} applied the scale to the stages of information processing. These frameworks describe where automation sits in a workflow, not what an arrangement can attain or what an organization must know before adopting one. \citet{puranam2021human} moves the question onto that ground, treating a human and an algorithm producing a decision together as an organization to be designed, observing that a division of labor without specialization is feasible because the outputs of decision tasks can be aggregated so that errors cancel, and anticipating a difficulty developed in Section~\ref{sec:records}: a serial division of labor censors what each party can learn. Economics frames the same choice as one of decision authority: \citet{agrawal2019exploring} separate the prediction a machine can supply from the judgment, the value placed on each outcome, that remains with a person, and \citet{athey2020allocation} ask when an organization should give a decision to a human or to an AI, given that who holds it changes how much effort the human puts into learning what each choice is worth. Both take who decides as the choice to be made; the present paper asks what each allocation can attain in performance before incentives enter.

\subsection{The limits of oversight}\label{sec:oversight}

The premise that adding a human makes an automated system safe has been contested since \citet{bainbridge1983ironies}: the operator left in place is given whatever could not be automated, an arbitrary collection of tasks for which little support was designed, and can follow the machine only if it decides in a way, and at a rate, that a person can follow. Operators retained as monitors lose the skills they would need to intervene \citep{rasmussen1983skills}; after an expert system failed, subjects who had supervised it were slower to decide and understood less of what had happened, while being more confident than subjects who had worked unaided \citep{endsley1995out}; and complacency and automation bias arise under divided attention, affect experts and novices alike, and resist training \citep{parasuraman1997humans, parasuraman2010complacency, romeo2026exploring}. Policy has nonetheless answered algorithmic shortcomings by inserting a human, the ``MABA-MABA trap'' of \citet{crootof2023humans}; \citet{green2022flaws} finds that forty-one policies prescribing human oversight of government algorithms rest on the uninterrogated assumption that people can oversee algorithmic decisions effectively, and that where they cannot, the policies legitimize the systems they were meant to check.

Two readings of this literature bear on what follows. First, oversight is itself work and can itself fail, so an exchange between skills is measured below rather than presumed to help. Second, the literature offers no way to tell whether a given combination of skills will fail before it is run. The meta-analysis of \citet{vaccaro2024combinations} suggests where to look: whether combinations gained or lost depended on whether humans or the machine performed better alone, a quantity that separate assessments of the skills can establish before any combination is built. How much of the stake that quantity accounts for, however, separate assessments cannot tell: the rest lies in the instances on which the weaker skill does better, which only observing both skills on the same instances reveals.

\subsection{Deciding per instance which skill executes}\label{sec:defer}

Where a criterion decides, for each instance, whether a machine answers or a person is consulted, the arrangement is what machine learning calls \emph{learning to defer}. \citet{madras2018predict} model it as a cascade in which the first model either predicts or passes the case on, the passing variable being a prediction of which party would incur the lower loss on the instance; only one of the two experts has parameters the designer can optimize, which is the position of an organization holding a human skill and a bought model, and consistent estimators followed \citep{mozannar2020consistent}. Its ancestor is learning with rejection, in which a classifier may abstain at a cost, with no model of who answers instead \citep{cortes2016learning}; and the model can be trained jointly with the deferral, to complement the human rather than to perform alone \citep{wilder2020learning}. In medical imaging, CoDoC learns from a predictive model's confidence score when to defer to the clinical workflow, in breast cancer screening and in tuberculosis triage on chest films, with gains over both the model and the clinicians alone in most of the settings tested \citep{dvijotham2023enhancing}; in the terms of this paper it assigns on a signal of the model's standing, and is bounded by the better of the two on every instance. The criterion need not be external to the skills: a language model carries an internal confidence signal that discriminates its own correctness and drives whether it answers or defers \citep{kumaran2026causal}, and large models can estimate the probability that they know an answer, although the calibration degrades on unfamiliar tasks \citep{kadavath2022know}. The same structure appears inside machine learning as dynamic selection, which distinguishes selecting one classifier for a sample from selecting an ensemble \citep{ko2008dynamic, cruz2018dynamic}; that distinction, choose one or combine several, is one this paper bounds.

\subsection{When combining beats choosing}\label{sec:combining}

Combining several responses into one pays under conditions that are well characterized. For an ensemble whose output is the weighted average of its members' outputs, \citet{krogh1994neural} show that under squared error the ensemble's error equals the weighted average of the members' errors less an ambiguity term, the spread of the members around the ensemble, so that disagreement pays against the \emph{average} member; the random forest rests on the same principle, its error falling with the strength of the trees and rising with their correlation \citep{breiman2001random}, and exact bias-variance-diversity decompositions extend to a wide range of losses \citep{wood2023unified}. A human and a machine are not exchangeable members of an ensemble: one is usually the more accurate, which gives the stake a systematic component that averaging forfeits, and the correlation between their performances is a property of the skills rather than a design choice.

The forecasting literature reaches a stronger result. For two unbiased forecasts combined under squared loss, the weights that minimize the variance of the combined error give a combination never worse than the better forecast, and strictly better unless that forecast encompasses the other \citep{bates1969combination, timmermann2006forecast}. That result concerns a continuous target, where the truth can lie between two forecasts. Where the truth is a class and the response a probability of it, no point between two responses is nearer to the truth than the nearer of the two, and Section~\ref{sec:responses} shows what survives. A linear pool of calibrated probability forecasts is itself uncalibrated, being underconfident, and invites recalibration, which may move it away from the base rate or toward it \citep{ranjan2010combining, lichtendahl2022extremizing}; since a recalibration can be applied to a single response as readily as to a combination, Section~\ref{sec:twochannels} credits it to no channel. Finally, the theory describes skills as \citet{breiman2001statistical} describes algorithms in his second culture, by how their responses score rather than how they are reached; two skills of similar accuracy are then a Rashomon pair, whose worth together lies almost entirely in the residual.

\subsection{Double reading, and a chest-radiograph experiment}\label{sec:doublereading}

The arrangement in which two skills execute every instance and their responses are combined is the standard of care in breast cancer screening across much of Europe, and its combination rules are named and distinguished. \citet{johnston1999two} set out three: recall if at least one of two independent readers suggests it; settle disagreement by consensus discussion; or settle it by a third reader. \citet{taylorphillips2018double}, evaluating 805,206 women and 7,055 detected cancers, report that $8.9\%$ of those cancers were detected by the second reader alone, and that the rule decides how many recalls it takes: $4.08\%$ after arbitration, against $6.19\%$ had women been recalled whenever either reader suggested it, and $4.76\%$ under single reading. The arrangement is being reconfigured with a machine among the readers: in the MASAI trial \citep{lang2023artificial, hernstrom2025screening, gommers2026interval} an AI risk score triaged each examination to single or double reading, deciding per instance \emph{how many} skills would execute; in the full cohort of more than $105{,}000$ women, detection rose from $5.0$ to $6.4$ per 1,000 without a significant rise in false positives and with $44\%$ less screen-reading workload, and the rate of interval cancers was non-inferior, $1.55$ against $1.76$ per 1,000; a nationwide implementation reports $6.7$ per 1,000 against $5.7$, with a non-inferior recall rate \citep{eisemann2025nationwide}.

The CXR experiment of \citet{agarwal2023combining}, whose data are released as Collab-CXR \citep{moehring2025dataset}, is the other field reference used throughout. Radiologists read chest films with and without the model's prediction in view and with and without clinical history the model could not see, 68 of them reading the same films both with and without the prediction. The model's area under the ROC curve exceeded that of $78\%$ of the participants; providing its predictions did not improve average performance, which the authors attribute to errors in updating; clinical history did improve performance; a logistic-regression benchmark combining the two reports would have ranked above $86\%$ of the radiologists on the same measure; and the delegation policy the authors derive sends each case to the model, to the radiologist, or rarely to the radiologist with the prediction in view, conditioning on the model's own predicted probability. Section~\ref{sec:field} reads each of these through the theory, and Section~\ref{sec:radiologydata} measures the CXR experiment on its released data.

\section{Skills, policies and performance}\label{sec:primitives}

The theory rests on three objects: the performance a skill realizes on an instance of a task, the distribution of that performance across the instances, and the policy that decides which skills execute and how what they produce becomes the response an organization delivers. This section introduces them in that order. Performance comes first, from its foundations in production (Sections~\ref{sec:performance} and~\ref{sec:instances}): whatever the method by which a response is determined, when one skill or several are involved, the response is scored against the correct one, when available, which yields a performance between zero, complete failure, and one, perfect execution. The skills are then described as distributions of performance (Section~\ref{sec:skills}), and the policy last (Section~\ref{sec:policies}), together with the two channels through which a policy can make several skills worth more than one: \emph{routing}, which directs work between the skills, and \emph{synergy}, which makes them interact (Section~\ref{sec:twochannels}). Everything the later sections derive or deploy is stated in terms of these two channels. The performance a skill realizes on an instance is stochastic, and the stochasticity is how the framework represents generalization difficulty, both \emph{global}, which shifts a skill's performance from one class of instances to another, and \emph{local}, which disperses it from one instance to the next within a class (Section~\ref{sec:instances}); everything downstream of it is a deterministic transformation. Section~\ref{sec:bounds} derives from these objects everything skill interaction can be worth in performance.

\subsection{Notational conventions}\label{sec:notation}

The symbols are defined where they first appear; this subsection collects the conventions that govern all of them.

Expectations $\E[\cdot]$ denote population quantities of the process generating instances and performances, taken over the instance space of the task at a given time, with the roster of skills fixed. Skills may evolve, humans through on-the-job learning and machines through retraining, so every quantity below is indexed implicitly by the date at which the profiles are taken. $\mu$, used for the mean performance of a skill within a class, is shorthand for an expectation within a class, a population quantity like $\E$, and not a statistic of a sample.

Organizations never observe population quantities; they hold finite records of past instances. A hat marks a quantity that stands in for one that cannot be had: a skill's attempt at the correct response, $\hat{r}_s$ in place of $r$; a prediction made from the signals, such as a class $\hat{k}$; or an estimate computed from a record, such as $\hat{\E}[\theta_s]$ of $\E[\theta_s]$ and $\hat{\mu}_{s,k}$ of $\mu_{s,k}$, each as good as the record is representative of the instances it is applied to. The bounds of Section~\ref{sec:bounds} are stated for the population, and no hatted quantity enters them; the arrangements of Section~\ref{sec:ladder} operate on records, and each hatted quantity they use is computed from a record they name.

$\mathcal{S}$ denotes the available set of a policy and the skills a statement concerns, and a skill is engaged on an instance when the policy commissions its execution (Definitions~\ref{def:available} and~\ref{def:engaged}); $i$ and $j$ name skills rather than index instances, and $s$ ranges over $\mathcal{S}$ when a statement holds for all of its members. In examples the pair is a human skill $H$ and a machine skill $M$. Two letters serve twice and are kept apart by their typeface: $\mathcal{S}$ is the available set and $S$ the systematic component of the stake, and $\Mx$ is the best performance on an instance, introduced in Section~\ref{sec:anchors}, and $M$ the machine skill. Maxima, minima and $\argmax$ range over $\mathcal{S}$ and sums over $s$ run over it, so changing the set changes every quantity. The exposition keeps $|\mathcal{S}| = 2$ wherever a larger set would obscure the argument; Appendix~\ref{app:many} relaxes the restriction and states the three properties that change with it.

\subsection{From production to performance}\label{sec:performance}

We build on a task-based model of production \citep{farboodi2021model} in which the quality of a firm's output is a strictly decreasing function of the squared distance between its production technique and the optimal one, so that the larger the gap, the lower the quality; the formulation is compatible with treatments in which automation deploys below a threshold of task complexity \citep{autor2025expertise}, labor holds a comparative advantage in newer or more complex tasks \citep{acemoglu2018race, acemoglu2025simple}, automation displaces workers from the tasks in which they held it \citep{acemoglu2022tasks}, and machines face generalization difficulty \citep{chollet2019measure}. We extend the notion of technique to models operated by the human mind on cognitive tasks, and abstract the distance between actual and optimal technique into a loss, which min-max normalization carries to a performance $\theta \in [0,1]$ on a common scale, one for a perfect match between skill and task and zero for complete failure. Performance is therefore not the response a skill produces but the distance of that response from the optimal one, and any measurement of it presupposes an instance whose optimal response eventually becomes knowable and a loss on which to grade it; the next subsection supplies both.

\subsection{Instances, responses and the scoring rule}\label{sec:instances}

A task is executed instance by instance. An instance belongs to a class $k \in \mathcal{K}$ with probability $p_k$, and carries a correct response $r$, the response that would have been right. A skill $s$ applied to the instance produces $\hat{r}_s$, its attempt at $r$, which is what a policy acts on; $\hat{r}$ denotes a response generically, whether one a skill produced or one a policy formed from several. Classes need not be anything more than a partition of the instances an organization can name: levels of difficulty, types of case, populations. They carry the distinction between two kinds of generalization difficulty. \emph{Global} difficulty differs across classes and moves the distribution of a skill's performance from one class to another; \emph{local} difficulty is the dispersion of performance within a class, the perturbation of task content from one instance to the next.

The correct response is established only when an instance is \emph{adjudicated}, which happens later and not for every instance. Performance is therefore realized at execution and measurable only at adjudication, through a scoring rule $\theta(\hat{r}, r) \in [0,1]$ the setting supplies, equal to one when the response is exactly right and falling as it departs from $r$. In probabilistic binary classification, the running example of this paper, the response is a probability $\hat{r} \in [0,1]$ and the truth is $r \in \{0,1\}$. What the results below use of the setting is exactly two properties, stated so that their scope is visible.

\begin{definition}[Endpoint truth, monotone rule]\label{def:endpoint}
The truth is an \emph{endpoint} if $r$ lies at an extreme of the response space, as $r \in \{0,1\}$ does for $\hat{r} \in [0,1]$. The rule is \emph{monotone} if $\theta(\hat{r}, r) = 1 - \phi(|\hat{r} - r|)$ for a non-decreasing $\phi$ with $\phi(0) = 0$ and $\phi(1) \leq 1$.
\end{definition}

Two members of the family recur. The \emph{squared} rule $\theta = 1 - (\hat{r} - r)^2$, one minus the Brier score, is the rule of this paper: it credits a response for the strength of the evidence it carries, and charges it where the response states more or less than that evidence supports, which is the response's \emph{calibration} (Section~\ref{par:calibration}). Section~\ref{par:rule} gives the full reasons for the choice, and its price, once the results they rest on are in place. The \emph{linear} rule $\theta = 1 - |\hat{r} - r|$, with $\phi$ the identity, recurs as a special case, the one under which the anchors of Section~\ref{sec:bounds} describe response-level arrangements exactly. The square renders the absolute value redundant, as does an endpoint truth, which fixes its sign; it is retained because the rule depends on the distance from the truth, not on the direction. The two rank responses identically on an instance and weight shortfalls differently across instances. Results that rest on the two properties alone hold for any classification task scored by any rule of the family; they do not hold for a continuous truth, where the correct response can lie between two responses, and Section~\ref{sec:responses} states which results require what.

\subsection{Skills as distributions}\label{sec:skills}

We write $\theta_s = \theta(\hat{r}_s, r)$ for the performance of skill $s$ on the instance at hand, $\theta_{s,k}$ for its distribution over instances of class $k$, with mean $\mu_{s,k} = \E[\theta_s \mid k]$, and $\E[\theta_s] = \sum_k p_k \mu_{s,k}$ for its marginal. A skill is thus described by a distribution per class, and a pair of skills by more than the two distributions: by how their performances co-occur on the same instance, their \emph{coupling}, which Section~\ref{sec:bounds} shows to carry part of the stake. Appendix~\ref{app:profiles} reports the illustrative skill profiles used in the figures: Beta distributions chosen for their support on $[0,1]$ and their flexibility, which illustrate the quantities and calibrate nothing.

\subsection{The skill policy}\label{sec:policies}

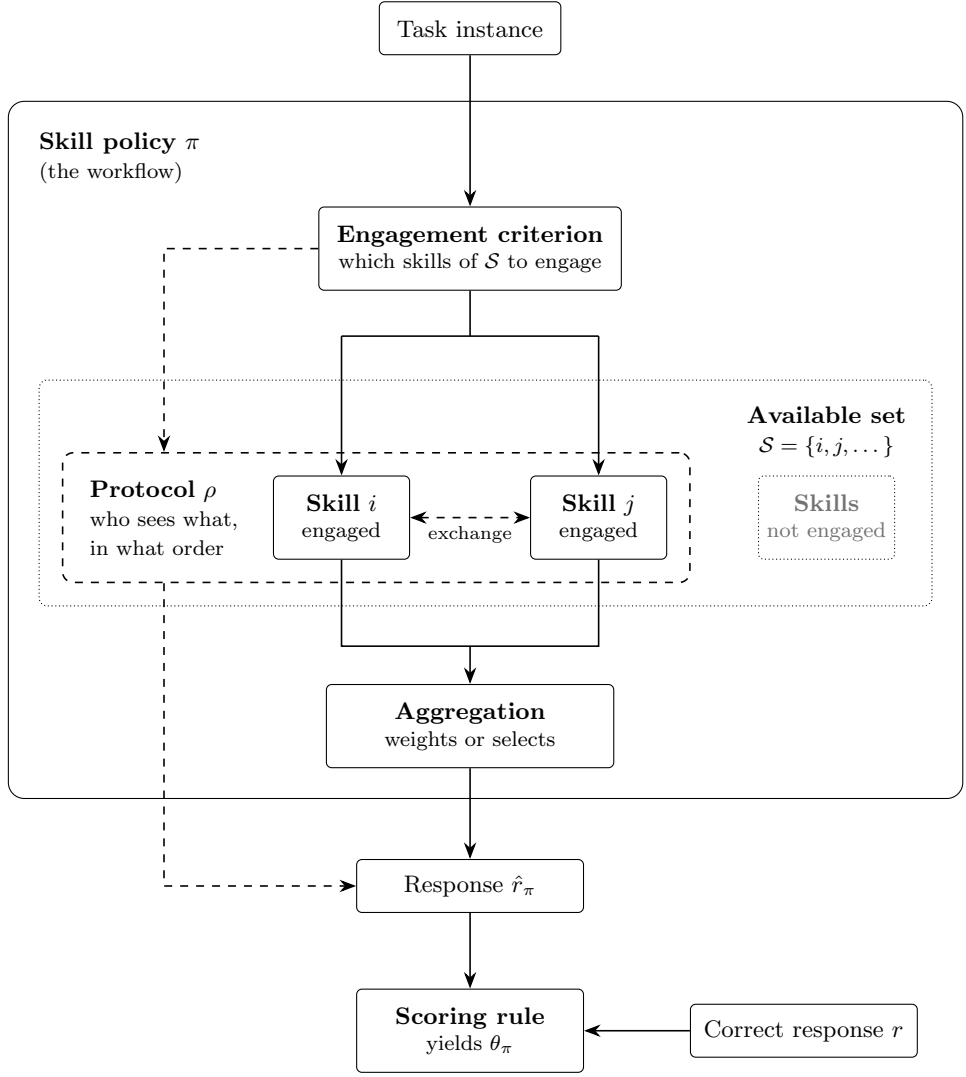
\begin{figure}[!htb]
\centering
\begin{tikzpicture}[
  font=\small, >={Stealth[length=2.2mm]},
  comp/.style={draw, rounded corners=2pt, align=center, minimum height=11mm, inner sep=3pt},
  plain/.style={draw, rounded corners=2pt, align=center, minimum height=7mm, inner sep=3pt},
  flow/.style={->, semithick},
  alt/.style={->, semithick, dashed},
  exch/.style={<->, semithick, dashed},
]
\node[plain, minimum width=24mm] (inst) at (0,0) {Task instance};
\node[comp, minimum width=40mm, below=20mm of inst] (crit) {\textbf{Engagement criterion}\\[-1pt] {\footnotesize which skills of $\mathcal{S}$ to engage}};
\node[comp, minimum width=18mm] (i) at ($(crit.south)+(-17mm,-30mm)$) {\textbf{Skill} $i$\\[-1pt] {\footnotesize engaged}};
\node[comp, minimum width=18mm] (j) at ($(crit.south)+(17mm,-30mm)$) {\textbf{Skill} $j$\\[-1pt] {\footnotesize engaged}};
\draw[exch] (i) -- node[below=0.6mm, font=\scriptsize, inner sep=1pt] {exchange} (j);
\node[align=left, left=4mm of i, inner sep=1.5pt] (plab) {\textbf{Protocol} $\rho$\\ {\footnotesize who sees what,}\\ {\footnotesize in what order}};
\node[draw, dashed, semithick, rounded corners=4pt, inner sep=3mm, fit=(plab)(i)(j)] (prot) {};
\node[comp, minimum width=18mm, right=12mm of j, densely dotted, text=black!55] (l) {\textbf{Skills}\\[-1pt] {\footnotesize not engaged}};
\node[align=center, above=2mm of l, inner sep=1pt] (slab) {\textbf{Available set}\\ {\footnotesize $\mathcal{S} = \{i, j, \dots\}$}};
\node[draw, densely dotted, rounded corners=3pt, inner sep=3mm, fit=(prot)(l)(slab)] (aset) {};
\node[comp, minimum width=38mm] (agg) at ($(crit.south |- i.south)+(0,-22mm)$) {\textbf{Aggregation}\\[-1pt] {\footnotesize weights or selects}};
\node[plain, minimum width=30mm, below=12mm of agg] (resp) {Response $\hat{r}_\pi$};
\draw[flow] (inst) -- (crit);
\coordinate (busA) at ($(crit.south)+(0,-6mm)$);
\draw[semithick] (crit.south) -- (busA);
\draw[semithick] (busA -| i.north) -- (busA -| j.north);
\draw[flow] (busA -| i.north) -- (i.north);
\draw[flow] (busA -| j.north) -- (j.north);
\coordinate (busB) at ($(agg.north)+(0,5mm)$);
\draw[semithick] (i.south) -- (i.south |- busB) -- (j.south |- busB) -- (j.south);
\draw[flow] (busB) -- (agg.north);
\draw[flow] (agg) -- (resp);
\coordinate (px) at (plab.north);
\draw[alt] (crit.west) -| (px |- prot.north);
\draw[alt] (px |- prot.south) |- (resp.west);
\node[anchor=south west, align=left, inner sep=0pt] (plabel) at ($(aset.west |- crit.north)+(0,3mm)$) {\textbf{Skill policy} $\pi$\\ {\footnotesize (the workflow)}};
\node[draw, rounded corners=6pt, inner sep=4mm, fit=(plabel)(crit)(aset)(agg)] (policy) {};
\node[comp, minimum width=30mm, below=10mm of resp] (score) {\textbf{Scoring rule}\\[-1pt] {\footnotesize yields $\theta_\pi$}};
\node[plain, minimum width=30mm, right=14mm of score] (truth) {Correct response $r$};
\draw[flow] (resp) -- (score);
\draw[flow] (truth) -- (score);
\end{tikzpicture}
\caption{A skill policy $\pi$, which an organization runs as a workflow. The policy holds the available set $\mathcal{S} = \{i, j, \dots\}$; on each instance the engagement criterion decides which of its skills to engage, that is, to execute, so a skill can be available and not engaged on the instance drawn. Without a protocol, the engaged skills run separately and their responses meet only in the aggregation, which weights them or selects one (solid connectors); the policy routes where the criterion or the aggregation directs work between the skills, by class or by instance, through assignment, ensembling or selection. With a protocol $\rho$ (dashed), the protocol surrounds the engaged skills, lets what one of them produces enter another's execution, fixes who is shown what and in what order, and itself returns the delivered response: the policy operates synergy, and routes as well where the engaged skills or their weights are set per instance. The scoring rule of the setting compares the delivered response $\hat{r}_\pi$ with the correct response $r$ to yield the performance $\theta_\pi$.}
\label{fig:policy}
\end{figure}

Between the skills an organization holds and the response it delivers stands a \emph{skill policy} $\pi$, which does two jobs for each instance, each through a distinct component: before execution it decides which skills execute, and after it forms the response that leaves the organization from what they produce. Its delivered response is $\hat{r}_\pi$ and its performance $\theta_\pi = \theta(\hat{r}_\pi, r)$. The paragraphs that follow introduce the elements of a policy in turn, and Figure~\ref{fig:policy} shows how they compose.

\begin{definition}[Available set]\label{def:available}
The \emph{available set} $\mathcal{S}$ of a policy is the set of skills it holds available, the only ones it may call on. Holding a skill available is a standing commitment, akin to a capital expenditure, incurred whether or not the skill is engaged (Section~\ref{sec:costs}).
\end{definition}

\begin{definition}[Engagement]\label{def:engaged}
A skill is \emph{engaged} on an instance when the policy commissions its execution on that instance. Engaging a skill incurs an execution cost, akin to an operating expenditure, on each instance on which it is engaged (Section~\ref{sec:costs}).
\end{definition}

\begin{definition}[Skill policy]\label{def:policy}
A \emph{skill policy} $\pi$ over an available set $\mathcal{S}$ consists of an engagement criterion (Definition~\ref{def:criterion}) and either an aggregation (Definition~\ref{def:aggregation}) or a protocol (Definition~\ref{def:protocol}). Its delivered response $\hat{r}_\pi$ is the response the aggregation forms or the protocol returns.
\end{definition}

Availability is a property of the policy; engagement is a property of the policy on an instance. A policy may therefore hold a skill available and engage it on some instances and not on others (Figure~\ref{fig:policy}). The stake of Section~\ref{sec:bounds} is defined over the available set, since it concerns what the skills a policy may call on could attain between them, whereas what a policy attains on an instance depends only on the skills it engages there.

\begin{definition}[Engagement criterion]\label{def:criterion}
An \emph{engagement criterion}, the \emph{criterion} for short, decides on each instance which skills of the available set are engaged, reading only what is observable before any of them executes.
\end{definition}

The criterion acts before execution, so it can read nothing any skill will produce on the instance at hand, and it decides engagement and nothing else. Any weighting of responses belongs to the aggregation, even where the weights are computed from the signals describing the instance before execution.

\begin{definition}[Aggregation]\label{def:aggregation}
An \emph{aggregation} forms the delivered response from the responses the engaged skills produced independently, as a convex combination
\begin{equation}\label{eq:aggregation}
\hat{r}_\pi \;=\; \sum_{s} w_s\, \hat{r}_s, \qquad w_s \geq 0, \quad \sum_s w_s = 1,
\end{equation}
with weights that may depend on anything observable when they are set, the responses included. Writing $w = (w_s)_s$, the delivered response is $\hat{r}_w$ and its score $\theta_w = \theta(\hat{r}_w, r)$. An aggregation whose weights are set once for the task, the same on every instance, is a \emph{fixed combination}; the fixed combination with equal weights $\bar{w} = (1/n, \dots, 1/n)$ over the $n$ engaged skills is \emph{pooling}, which delivers the mean response $\hat{r}_{\bar{w}} = \frac{1}{n}\sum_s \hat{r}_s$, with score $\theta_{\bar{w}}$.
\end{definition}

The aggregation acts after execution, on responses produced separately, none having seen another's, and may read everything the criterion could and the responses themselves. Where a single skill is engaged it returns that skill's response; where several are, it may \emph{weight} them, with fixed weights or weights set per class or per instance, or \emph{select} one of them once all are known. Selection returns one of its inputs unchanged whatever it conditions on, and is therefore an aggregation and not a protocol, provided that whoever selects does not execute the task on the instance (below). Because the delivered response is a convex combination of independent responses, it cannot score above the best of them on any instance when the truth is an endpoint (Proposition~\ref{prop:hull}).

\begin{definition}[Protocol]\label{def:protocol}
A \emph{protocol} $\rho$ is a procedure over the skills engaged on an instance, at least two, that prescribes which of them are shown what others have produced, whether responses, evidence or reasoning, in what order and before which responses are fixed, and that returns a joint response $\hat{r}_\rho$, which is then the delivered response, $\hat{r}_\pi = \hat{r}_\rho$.
\end{definition}

A protocol replaces the aggregation where the skills are to interact rather than only their responses. Under an aggregation the skills' outputs meet once every response is fixed; under a protocol what one skill produces enters another's execution before a joint response is fixed, so that $\hat{r}_\rho$ is not a function of the responses the skills would have produced independently. The order matters, since a skill shown another's response before committing can be drawn toward it, so two protocols that differ only in sequence are different protocols; where contributions are weighted, the weights are part of the protocol; and a protocol that shows no skill anything of another's work is an aggregation.

\begin{definition}[Exchange]\label{def:exchange}
An \emph{exchange} is the interaction a protocol runs: a step in which what one engaged skill has produced, its response, its evidence or its reasoning, is shown to another before that skill fixes its own response, which it may then revise or keep.
\end{definition}

An exchange may run one way, as when a reader sees a model's output, or both ways, as when two readers discuss a case; with more than two engaged skills, a protocol consists of the exchanges it prescribes, in the order it prescribes. The exchange is what happens between the skills and the protocol how it is arranged, so an exchange can occur without being designed, as when a skill is shown another's output with no rule for what to do with it (Section~\ref{sec:radiology}).
 
Who forms the delivered response decides between the two. A rule, or an agent, that reads the engaged skills' responses and the signals without executing the task on the instance aggregates, whatever it reads, and may weight the responses or select one. An agent that examines the instance itself after seeing their responses executes the task: it is an engaged skill shown what others produced before fixing its own response, the step is an exchange, and what it returns is a joint response, even where it coincides with one of theirs. A radiologist who reads a film with a model's prediction in view, or who chooses between that prediction and her own reading, is therefore in an exchange; a rule that recalls whenever either of two readers suggests it aggregates; and a third reader who settles a disagreement by reading the examination and the two reports is a third skill, engaged on the instances of disagreement and shown the others' work, which is synergy (Section~\ref{sec:mammography}).

The skill policy decision problem of the introduction is the choice of the available set, of the engagement criterion, and of an aggregation or a protocol. The problem can be posed at four resolutions, each a special case of the next, which the ladder of Section~\ref{sec:ladder} climbs: which single skill to commit to ex ante (rungs~0 and~1), how to combine several skills the same way on every instance (rungs~2 and~3), what to do with each class of instances (rung~4a), and what to do with each instance (rungs~4b to~5).

\subsection{The two channels of augmentation: routing and synergy}\label{sec:twochannels}

This subsection reads the components of Section~\ref{sec:policies} from a different angle, asking not what each of them does but what, together, they can make several skills worth.

\begin{definition}[Augmentation]\label{def:augmentation}
A policy operates \emph{augmentation} if it uses several skills of its available set in a way through which it can attain more than any fixed use of them, a commitment to one or a pooling with fixed weights: by directing work between them according to information about each instance, its class or the instance itself, read directly or inferred, or by letting them interact.
\end{definition}

Engaging several skills is not in itself augmentation. A policy that engages them all on every instance and pools their responses with fixed weights, as rung~2 of Section~\ref{sec:ladder} does, uses several skills without augmenting any of them, and what it buys is insurance against a wrong commitment (Section~\ref{sec:insurance}). Augmentation can also fail, since a policy can direct work the wrong way and a protocol can leave the skills worse off than they would have been alone.

Directing work between skills exploits the fact that the better skill is not the same everywhere, but changes from one class of instances to another or from one instance to the next, and involves skills that do not interact. A policy can direct work before execution, sending each instance to the skill that will do it best and engaging no other; or after it, engaging several skills independently on the same instance and forming the delivered response from their responses by the method of its choice, weighting them or selecting one of them once all are known. Letting skills interact means engaging several of them on the same instance and letting them work on one another's output, so that the joint response is something none of them would have produced alone. The first is routing, done by the engagement criterion and the aggregation; the second is synergy, done by a protocol.

\begin{definition}[Routing]\label{def:routing}
A policy \emph{routes} if it directs work between the skills according to information about each instance, whether about its class or about the instance itself, and whether read directly or inferred from what is observable, as when a class or the leading skill is predicted from signals: its criterion engages different skills on different instances, or the weight its aggregation or protocol gives each skill's contribution differs across instances.
\end{definition}

\begin{definition}[Synergy]\label{def:synergy}
A policy operates \emph{synergy} if it forms its delivered response by a protocol.
\end{definition}

\begin{figure}[!htb]
\centering
\begin{tikzpicture}[
  font=\small, >={Stealth[length=1.8mm]},
  bx/.style={draw, rounded corners=2pt, align=center, minimum height=9mm, inner sep=2pt, font=\footnotesize, fill=white},
  sk/.style={draw, rounded corners=2pt, align=center, minimum height=7mm, minimum width=1.2cm, inner sep=2pt, font=\footnotesize, fill=white},
  skm/.style={sk, densely dotted, text=black!55},
  skd/.style={sk, dashed},
  mflow/.style={->, semithick, dashed},
  ttl/.style={font=\small\bfseries},
  flow/.style={->, semithick},
  alt/.style={->, semithick, dashed},
  exch/.style={<->, semithick, dashed},
  info/.style={->, semithick, densely dotted},
]

\def\W{6.60} \def\H{6.40}
\begin{scope}[shift={(0,2*\H+1.2)}]
\node[ttl, anchor=west] at (0,1.68) {One skill engaged on every instance};
\node[bx, dashed, minimum width=4.6cm, minimum height=1.25cm] (c1) at (2.9,0.65) {\textbf{Engagement criterion}\\ the incumbent, or\\ the assessed leader};
\node[sk] (i1) at (6.6,0.65) {Skill $i$};
\node (r1) at (8.6,0.65) {$\hat{r}_\pi = \hat{r}_i$};
\draw[flow] (c1) -- (i1);
\draw[flow] (i1) -- (r1);
\node[font=\footnotesize, align=left, anchor=west] at (9.6,0.65) {nothing to augment:\\ by ignorance or by\\ ex ante commitment};
\end{scope}
\node[font=\small\bfseries, anchor=south west] at (0,2*\H+0.05) {Several skills engaged};
\draw[thin] (0,0) rectangle (2*\W,2*\H);
\draw[thin] (\W,0) -- (\W,2*\H);
\draw[thin] (0,\H) -- (2*\W,\H);
\begin{scope}[shift={(0.00,0.00)}]
\node[ttl] at (3.30,5.95) {Neither channel};
\node[bx, minimum width=4.6cm] (c) at (3.55,4.85) {\textbf{Engagement criterion}\\ constant};
\node[sk] (i) at (2.40,3.05) {Skill $i$};
\node[sk] (j) at (4.70,3.05) {Skill $j$};
\draw[flow] (c.south -| i.north) -- (i.north);
\draw[flow] (c.south -| j.north) -- (j.north);
\node[font=\scriptsize, fill=white, inner sep=1pt] at ($(c.south -| i.north)!0.5!(j.north)$) {and/or};
\node[bx, minimum width=4.6cm] (a) at (3.55,1.55) {\textbf{Aggregation}\\ fixed weights};
\draw[flow] (i.south) -- (i.south |- a.north);
\draw[flow] (j.south) -- (j.south |- a.north);
\node (r) at (3.55,0.35) {$\hat{r}_\pi$};
\draw[flow] (a.south) -- (r.north);
\end{scope}
\begin{scope}[shift={(6.60,0.00)}]
\node[ttl] at (3.30,5.95) {Routing only};
\node[bx, minimum width=4.6cm] (c) at (3.55,4.85) {\textbf{Engagement criterion}\\ reads the class or the instance};
\draw[info] (0.35,4.85) node[left=-1pt, font=\scriptsize, align=center, anchor=south west] {} -- (c.west);
\node[font=\scriptsize, align=center] at (0.62,5.25) {signals};
\node[skm] (k) at (5.30,3.05) {Skill $\cdots$};
\node[sk] (i) at (1.80,3.05) {Skill $i$};
\node[sk] (j) at (3.55,3.05) {Skill $j$};
\draw[flow] (c.south -| i.north) -- (i.north);
\draw[flow] (c.south -| j.north) -- (j.north);
\node[font=\scriptsize, fill=white, inner sep=1pt] at ($(c.south -| i.north)!0.5!(j.north)$) {and/or};
\node[bx, minimum width=4.6cm] (a) at (3.55,1.55) {\textbf{Aggregation}\\ weights or selects\\ per class or instance};
\draw[flow] (i.south) -- (i.south |- a.north);
\draw[flow] (j.south) -- (j.south |- a.north);
\node (r) at (3.55,0.35) {$\hat{r}_\pi$};
\draw[flow] (a.south) -- (r.north);
\draw[info] (0.35,1.55) -- (a.west);
\node[font=\scriptsize, align=center] at (0.62,1.95) {signals};
\end{scope}
\begin{scope}[shift={(0.00,6.40)}]
\node[ttl] at (3.30,5.95) {Synergy only};
\node[bx, minimum width=4.6cm] (c) at (3.55,4.85) {\textbf{Engagement criterion}\\ constant};
\node[sk] (i) at (2.40,3.05) {Skill $i$};
\node[sk] (j) at (4.70,3.05) {Skill $j$};
\draw[flow] (c.south -| i.north) -- (i.north);
\draw[flow] (c.south -| j.north) -- (j.north);
\draw[exch] (i) -- (j);
\node[draw, semithick, rounded corners=3pt, inner xsep=2.5mm, inner ysep=2.5mm, fit=(i)(j)] (p) {};
\node[font=\scriptsize, align=center, anchor=north] (pl) at (p.south) {\textbf{Protocol} $\rho$: exchanges among\\ the engaged skills};
\node (r) at (3.55,0.45) {$\hat{r}_\pi = \hat{r}_\rho$};
\draw[flow] (pl.south -| r.north) -- (r.north);
\end{scope}
\begin{scope}[shift={(6.60,6.40)}]
\node[ttl] at (3.30,5.95) {Routing and synergy};
\node[bx, minimum width=4.6cm] (c) at (3.55,4.85) {\textbf{Engagement criterion}\\ reads the class or the instance};
\draw[info] (0.35,4.85) node[left=-1pt, font=\scriptsize, align=center, anchor=south west] {} -- (c.west);
\node[font=\scriptsize, align=center] at (0.62,5.25) {signals};
\node[skm] (k) at (5.30,3.05) {Skill $\cdots$};
\node[sk] (i) at (1.80,3.05) {Skill $i$};
\node[sk] (j) at (3.55,3.05) {Skill $j$};
\draw[flow] (c.south -| i.north) -- (i.north);
\draw[flow] (c.south -| j.north) -- (j.north);
\draw[exch] (i) -- (j);
\node[draw, semithick, rounded corners=3pt, inner xsep=2.5mm, inner ysep=2.5mm, fit=(i)(j)] (p) {};
\node[font=\scriptsize, align=center, anchor=north] (pl) at (p.south) {\textbf{Protocol} $\rho$: exchanges among\\ the engaged skills, weights\\ set per class or instance};
\node (r) at (2.67,0.45) {$\hat{r}_\pi = \hat{r}_\rho$};
\draw[flow] (pl.south -| r.north) -- (r.north);
\end{scope}
\node[font=\footnotesize] at (0.5*\W,-0.35) {no};
\node[font=\footnotesize] at (1.5*\W,-0.35) {yes};
\node[font=\small, align=center] at (\W,-0.95) {\textbf{Routing}: does the policy direct work between the skills, by class or by instance?};
\node[font=\footnotesize, rotate=90] at (-0.35,0.5*\H) {no};
\node[font=\footnotesize, rotate=90] at (-0.35,1.5*\H) {yes};
\node[font=\small, align=center, rotate=90] at (-0.95,\H) {\textbf{Synergy}: does the policy run a protocol?};
\end{tikzpicture}
\caption{Routing and synergy as flows. Above: a policy that engages one skill on every instance, by ignorance or by an ex ante commitment, has nothing to augment, and its criterion makes no choice among skills at instance level. Below: augmentation requires at least two skills available and the use of at least one of them based on a criterion that runs at instance level. A policy routes when information about the class or the instance, read or inferred from signals, directs which skills are engaged or how their contributions are weighted; it operates synergy when a protocol $\rho$ (frame) surrounds the engaged skills and returns the joint response after the exchanges among them. Where no synergy, the criterion may engage one of the skills or several (and/or), so that the criterion can implement assignment, ensembling and selection. Without routing the criterion is constant; without synergy the skills run separately and an aggregation forms the response. The quadrant of neither holds the fixed use of several skills, pooling with fixed weights, which buys insurance against a wrong commitment rather than augmentation. Routing without a protocol is bounded by the best response on every instance where the truth is an endpoint; a protocol is bounded by none of the skills, and is measured by its lift.}
\label{fig:augmentation}
\end{figure}
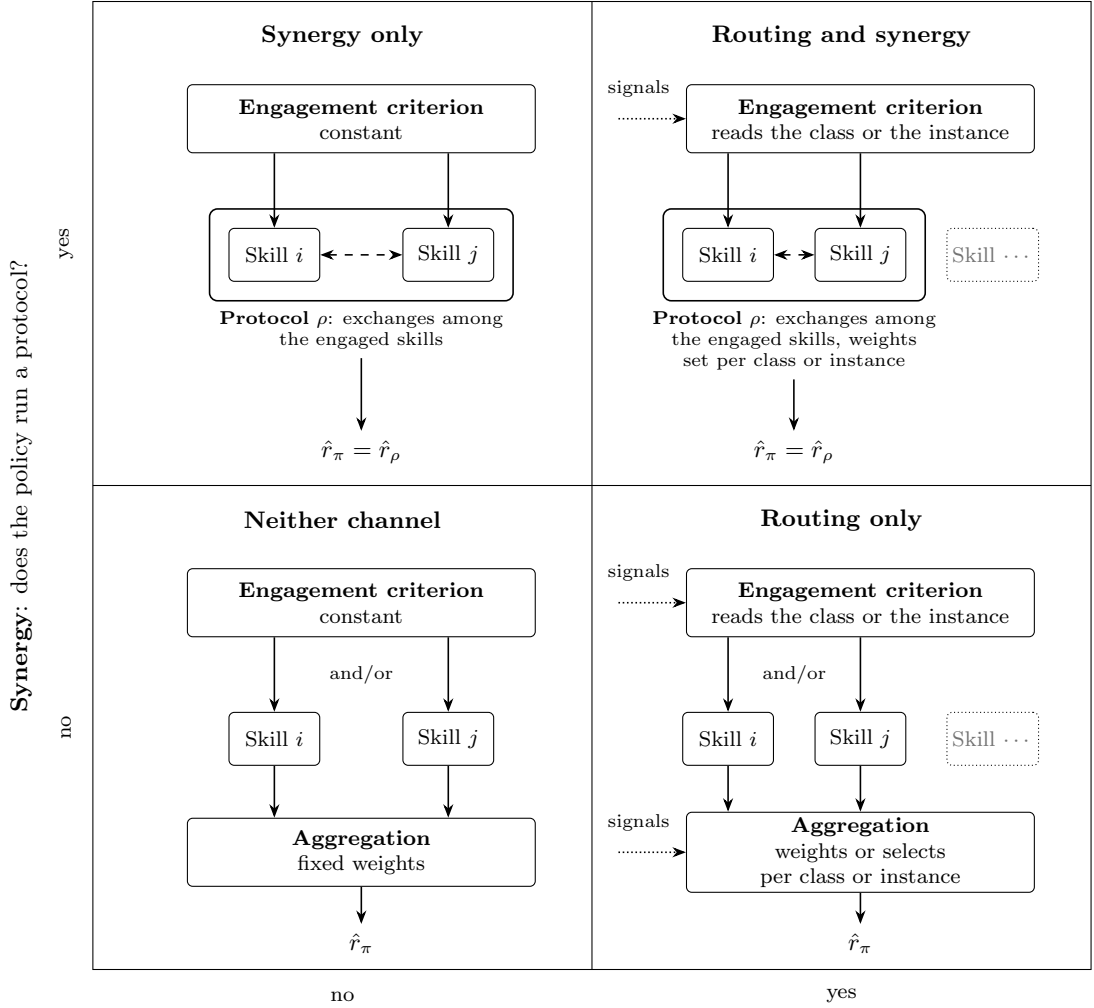

Augmentation is therefore the operation of at least one of the two channels. The two channels are orthogonal in their definition. Whether a protocol is run places, in principle, no restriction on which skills the engagement criterion engages or on how the contributions of the engaged skills are weighted, with the one exception that a protocol needs at least two engaged skills to work with, which rules out assignment. A policy may therefore operate either channel, both or neither (Figure~\ref{fig:augmentation}). A policy whose criterion engages the same skills on every instance and whose aggregation treats every instance alike operates neither: it commits to one skill, or pools several with fixed weights. A protocol run under a constant criterion, with no weights set per instance, is synergy without routing. A protocol whose engaged skills or weights are set per class or per instance is both, routing with synergy, with the weights part of the protocol.

Routing without a protocol is deployed entirely through the engagement criterion and aggregation, and returns a convex combination of responses produced independently, so what it can collect is bounded by the best of them on every instance where the truth is an endpoint (Proposition~\ref{prop:hull}), and how the work is divided between the skills defines its implementations. Synergy, in contrast, is bounded by none of the skills, only by the scale of performance itself: its joint response can exceed every independent response, up to a perfect execution, or fall below all of them, and what it adds can only be measured.

\begin{definition}[Assignment]\label{def:assignment}
A policy that routes without a protocol operates by \emph{assignment} if its engagement criterion engages a single skill $\hat{s}$ on each instance, chosen from what it reads before execution, and its aggregation returns that skill's response, $\hat{r}_\pi = \hat{r}_{\hat{s}}$, with no weights to set.
\end{definition}

\begin{definition}[Ensembling]\label{def:ensembling}
A policy that routes without a protocol operates by \emph{ensembling} if its engagement criterion engages two or more skills of the available set, not necessarily all of them, and its aggregation weights their responses, $\hat{r}_\pi = \sum_s \hat{w}_s \hat{r}_s$, with weights set per class or per instance from what is observable when they are set. Choosing which skills an ensemble engages and giving a zero weight to the responses of some engaged skills deliver the identical response; the two mechanisms differ in cost. A skill the criterion does not engage saves its execution cost, whereas a zero weight discards a response that has already been paid for, so that the same response is obtained at a higher operating cost (Section~\ref{sec:costs}).
\end{definition}

Pooling has the form of an ensemble, $\hat{r}_{\bar{w}} = \frac{1}{n}\sum_s \hat{r}_s$, but its weights are fixed and equal on every instance (Definition~\ref{def:aggregation}): it reads nothing, does not route, and is the information-free limit of ensembling rather than one of its cases. What it earns over a single skill chosen at random it owes to the scoring rule, not to information about the skills (Section~\ref{sec:insurance}).

\begin{definition}[Selection]\label{def:selection}
A policy that routes without a protocol operates by \emph{selection} if its engagement criterion engages two or more skills of the available set, not necessarily all of them, and its aggregation returns one of their responses unchanged, $\hat{r}_\pi = \hat{r}_{\hat{s}}$, with $\hat{s}$ chosen once all the engaged skills' responses are known and possibly from the responses themselves.
\end{definition}

The three differ in what the aggregation has to do once the engaged skills have executed. Under assignment it has nothing to do: a single response exists and is returned, and no weights are needed. Under ensembling it weights the responses, and under selection it returns one of them unchanged. Assignment executes one skill and chooses it before execution: it can condition only on signals and records, and a wrong choice leaves no other response to fall back on. Ensembling and selection execute several skills, and their choice acts on responses already produced. Ensembling may fix its weights before execution, from signals, and apply them afterwards, hedging between responses where it cannot tell which is better; selection chooses only once the responses are known, and can read them. Both produce, among the skills they engage, the paired record that assignment needs and cannot generate. With a pair, ensembling and selection engage both skills and only assignment leaves one unengaged; with more than two skills, any of the three may leave skills unengaged, and the criterion's choice of which to engage then directs work before execution, in addition to whatever the aggregation does after it.

On the ladder of Section~\ref{sec:ladder}, assignment is rungs 4a to 4c, conditioned on the class, on each skill's standing, or on both; ensembling is rung 4d and selection rung 4e; and the ceilings of Section~\ref{sec:ceilings} close the families from above, assignment by the oracles on the class and on the instance, ensembling and selection by the oracle weighting, whose best choice is the winner of the instance. The rungs below 4 are the cases that operate no routing: rungs 0 and 1 commit to one skill, and rung 2 pools both with fixed weights.

Synergy is deployed through the criterion and a protocol, and what it adds cannot be read from the joint response alone. It is measured against what the same skills would have delivered had their independent responses been aggregated, which requires those responses to be produced and retained on the instances used to measure it; Section~\ref{sec:channels} defines the measure, the \emph{lift}, which can fall below one. Because the order of disclosure is part of the protocol, the lift belongs to a protocol and not to a pair of skills: the same skills under two sequences can deliver two different lifts. Rung~3 of the ladder is synergy without routing, and rung~5 is routing with synergy.

The channels are properties of the policy's design and can be read from it before anything is run; a policy is classified by the channels it operates and not by what it recovered, so a protocol whose exchange changed nothing is synergy with a lift of one, and selection is routing without synergy whatever it conditioned on. For the same reason, operations that can be applied to a single skill's response as readily as to a combination are credited to no channel. Sharpening, recalibration (Section~\ref{par:calibration}) and extremizing by a fixed factor, which the forecasting literature develops for pooled probabilities \citep{ranjan2010combining, lichtendahl2022extremizing}, are of this kind; a combination whose push depends on how the responses agree is not (Section~\ref{sec:responses}). Throughout, a response is taken as the skill returned it, and an aggregation is compared with what its members would have returned untransformed, so that what is credited is the policy and not an operation every policy could undergo.

\subsection{Arrangements and costs}\label{sec:costs}

A skill policy as an organization runs it is a \emph{workflow}. An \emph{arrangement} is a class of policies sharing a structure: the available set, what the criterion reads, whether the response is formed by an aggregation or a protocol and, for an aggregation, which implementation of routing it is. The rungs of Section~\ref{sec:ladder} are arrangements, written $i$ for a single skill, $ij$ for both engaged with a fixed aggregation, $i,j \mid \cdot$ for assignment or selection conditioned on what follows the bar, $ij(w \mid \cdot)$ for weights set on it, and with a trailing ${+}$ for a protocol. The scoring rule belongs to the setting and no policy can change it; where the literature speaks of a combination rule, as in double reading, it is an aggregation in these terms, or a protocol where the readers see one another's work.

Definitions~\ref{def:available} and~\ref{def:engaged} attach a cost to each side of the distinction. Holding skill $s$ available is a standing commitment, akin to a capital expenditure, written $\kappa_s$, so that a policy over $\mathcal{S}$ carries $\kappa_\pi = \sum_{s \in \mathcal{S}} \kappa_s$ whatever the volume. Engaging it is an operating expenditure, $c_s$ per execution, with $c_\rho$ for the exchanges of a protocol, so that a policy's execution cost on an instance is $c_\pi = \sum_{s \text{ engaged}} c_s$, plus $c_\rho$ where it runs a protocol. Every arrangement therefore has a \emph{cost profile}, $\kappa_\pi$ and $\E[c_\pi]$, which follows from its design as directly as its channels do and is reported for every rung (Table~\ref{tab:costs}). The profile is not priced against performance, which would require an output function, the volume over which standing costs are spread, and the consequences of poor executions; every result is stated in performance and holds whatever those are, and pricing is left to future work.

\section{Stake and take}\label{sec:bounds}

Section~\ref{sec:primitives} defined what a policy is and the two channels through which it can make several skills worth more than one. This section asks what that can amount to in performance, before any particular policy is run, and its organizing distinction is the one between the stake and the take: \emph{the skills set the stake; the workflow decides the take}. The stake comes first: the anchors between which it lies and its decomposition into a systematic component and a residual (Section~\ref{sec:anchors}), then the refinement of the residual by classes (Section~\ref{sec:classes}), all properties of the skills on a task at a date, fixed before any policy exists. Policies act on responses rather than on performances, and Section~\ref{sec:responses} connects the two registers, which coincide under the linear rule and, under the squared rule this paper uses, differ by the disagreement between the responses. The take follows, channel by channel: policies that operate neither channel, which buy insurance rather than augmentation (Section~\ref{sec:insurance}); routing, bounded by the best response on every instance, whose take climbs the components of the stake in order and can be read against them one by one (Section~\ref{sec:routing}); synergy, bounded by none of the skills and characterized instead by the lift it must clear (Section~\ref{sec:channels}); and the two channels in a reduced form that locates their corner cases (Section~\ref{sec:reduced}). A worked example on four instances shows every quantity at once (Section~\ref{sec:worked}). The section closes with the magnitudes on illustrative profiles (Section~\ref{sec:magnitudes}) and with the evidence that identifies each quantity before deployment (Section~\ref{sec:identification}). Everything in it follows from the marginal and joint distributions of the skills' performances over the instance space, and holds whatever the cost profile of a policy (Section~\ref{sec:costs}) and whatever the consequences of error.

\subsection{The stake, and its decomposition}\label{sec:anchors}

On each instance, write
\[
m \;=\; \mean_{s \in \mathcal{S}} \theta_s, \qquad \Mx \;=\; \max_{s \in \mathcal{S}} \theta_s, \qquad \Delta \;=\; \Mx - m \;\geq\; 0,
\]
for the \emph{within-instance} mean, maximum and gap (see Figure~\ref{fig:stake}a). Across instances, the skills have expected performances $\E[\theta_s]$; write
\[
b \;=\; \max_{s \in \mathcal{S}} \E[\theta_s]
\]
for the \emph{informed baseline}, the expected performance of an ex ante commitment to the skill that is better on average, and call that skill the \emph{leader} $L$ (and the other the \emph{underdog} $U$, when $|\mathcal{S}| = 2$), so that $b = \E[\theta_L]$. Two anchors delimit what information about the skills can be worth to an organization that holds them. $\E[m]$ is the performance of engaging the skills in ignorance, handing each instance to one of them chosen blindly, with no idea of which to prefer; averaging their responses instead attains it plus the disagreement between them (Section~\ref{sec:insurance}). $\E[\Mx]$ is the performance of routing omniscience, every instance given to the skill that turns out to be better on it. These anchors are ordered, $\E[m] \leq b \leq \E[\Mx]$ (Propositions~\ref{prop:S} and~\ref{prop:R}), and the gap between the outer two splits at the middle one (Figure~\ref{fig:stake}b). The quantities this delimits are the following.

\begin{definition}[Gross stake]\label{def:G}
The \emph{gross stake} of an available set is $G = \E[\Mx] - \E[m]$, equivalently $G = \E[\Delta]$: what the heterogeneity of the skills puts on the table, everything that knowing their joint realization on each instance can be worth to an organization that holds them, against knowing nothing.
\end{definition}

\begin{definition}[Systematic component]\label{def:S}
The \emph{systematic component} is $S = b - \E[m]$, the portion of the gross stake secured by an ex ante commitment to the leader skill.
\end{definition}

\begin{definition}[Residual]\label{def:R}
The \emph{residual} is $R = \E[\Mx] - b$, the portion left above the informed baseline. It cannot be reached by an ex ante commitment, however well informed, but only by a policy that lets information about each instance, its class or the instance itself, discriminate in how instances are treated.
\end{definition}

By construction the two components exhaust the stake:
\begin{equation}\label{eq:gsr}
\underbrace{\E[\Delta]}_{G} \;=\; \E[\Mx] - \E[m] \;=\; \underbrace{b - \E[m]}_{\substack{S \text{, systematic} \\ \text{(task level)}}} \;+\; \underbrace{\E[\Mx] - b}_{\substack{R \text{, residual} \\ \text{(instance level)}}} .
\end{equation}

The two components are conditional on different information: $S$ on information available at the level of the task, which is only as reliable as the separate assessments of each skill over the space of adjudicated instances. $R$, in contrast, is conditional on information about how the skills' performances are realized jointly on each instance, which has no existence at the level of the task.

The stake is measured from the mean of the skills' performances rather than from the worst of them, $\E[\min_s \theta_s]$, for two reasons. The first is that the worst is not what ignorance attains. Reaching it on every instance requires knowing which skill fails there, the same information that reaching $\E[\Mx]$ requires, put to the opposite use. A policy that engages one skill on each instance, chosen independently of the performances, attains a mixture of the skills' expected performances and, with no reason to prefer any of them, their mean: $\E[m]$ is the point at which information about the skills is worth nothing, and the stake is what that information can be worth against it. Measured from the worst, the stake would also count the loss that a perverse use of information can inflict, which is not something the skills put on the table: for a pair, $\E[\Mx] - \E[\min_s \theta_s] = 2G$, and half of it is that loss (Proposition~\ref{prop:routing}). The second is that the worst, like the best, depends on how the performances are realized jointly, while the mean depends on the marginal distributions alone (Proposition~\ref{prop:anchor}). Measured from the mean, $S$ depends only on the separate assessments of each skill, as the previous paragraph requires, and everything that depends on the joint realization is confined to $R$; measured from the worst, part of that dependence would move into the systematic component. The mean anchor does depend on the roster in a way the other anchors do not. Adding a skill that leads on no instance and is not the best on average leaves $\E[\Mx]$, $b$ and $R$ unchanged and moves $\E[m]$ toward its own mean, so that a skill weaker than the average raises $S$ and $G$ by the same amount. This is not an artefact: holding a weak skill without knowing which skill is weak makes that knowledge worth more. It does make $S$ and $G$ incomparable across rosters of different composition, whereas $b$, $\E[\Mx]$ and $R$ compare directly (Appendix~\ref{app:change}).

The mean anchor is the reference from which the stake is measured, not a floor on what a policy can attain. It is what assigning instances to the skills at random attains, the least that engaging them without information secures, and policies that use information badly can fall below it: a commitment to a weaker skill attains less on average, $b - 2S$ for a pair (Proposition~\ref{prop:S}); routing that directs work the wrong way can fall to $\E[\min_s \theta_s]$, which for a pair is $\E[m] - G$ (Proposition~\ref{prop:routing}); and a protocol whose lift falls below one can return less than any of its skills on some instances, bounded below by nothing but the scale itself. Measuring the stake from $\E[m]$ therefore assumes nothing about how well organizations use their skills. It fixes the point at which information about the skills begins to be worth something, and a policy's take can be negative as readily as positive.

$S$ therefore raises the benchmark against which any combination is measured. An arrangement that engages both skills but cannot direct work between them is judged against $b$ rather than against $\E[m]$, and enters the comparison $S$ behind before its workflow has done anything, less whatever the disagreement between the responses repays when they are pooled; routing is necessary only for $R$. The human-machine arrangements bracket the single-skill benchmark from both sides, and which side a given arrangement lands on is decided by what its workflow does with the skills, not by the decision to combine them. \emph{The skills set the stake; the workflow decides the take.} In the terms of Section~\ref{sec:twochannels}, augmentation is how a policy takes: routing collects part of the stake, and synergy can add to it or subtract from it.

\begin{figure}[!htb]
\centering
\input{stake_schema.tex}
\caption{The anchors and the decomposition of the stake, as defined in Equation~\ref{eq:gsr}, for an available set of any size, drawn with two skills. (a) On one instance: $m$ is the mean of the skills' performances, $\Mx$ the best of them, and the gap $\Delta = \Mx - m$. (b) Over the instance space: the distributions of skills $i$ (dotted) and $j$ (dash-dotted), of the mean anchor $m$ (dashed) and of the routed anchor $\Mx$ (solid, hatched), drawn for two independent Beta-distributed skills, independence being what lets the distributions of $m$ and $\Mx$ follow from the skills' own (Appendix~\ref{app:factors}); below, their expectations magnified, with $S$ between $\E[m]$ and $b$ and $R$ between $b$ and $\E[\Mx]$.}
\label{fig:stake}
\end{figure}

\paragraph{Why the decomposition matters.} $G$ alone says little about what to do: it sizes the prize without saying which instrument reaches it. Two skill pairs with the same gross stake can differ entirely in how it divides between the part an instance-agnostic commitment secures and the part only instance-aware routing reaches, and only the split tells an organization whether what it needs is a better assessment or a routing capability. Figures~\ref{fig:gsr_grid} and~\ref{fig:gsr_equal_means} show the two components responding to different features of the skills: the gap between their means fixes $S$, while the shapes of their distributions move $R$, so that a stake can be mostly systematic, where the means are far apart and the distributions tight, or wholly residual, where the means coincide (Section~\ref{sec:magnitudes}). The distinction echoes the classical question of selecting versus combining forecasts \citep{bates1969combination, timmermann2006forecast} and, in econometric terms, the gap between the best uniform policy and the first-best individualized assignment \citep{manski2004statistical, kitagawa2018should, athey2021policy}, whose ancestor is the sorting on comparative advantage of \citet{roy1951some}. Written as $\E[\max_s \theta_s] - \max_s \E[\theta_s]$, the residual is the expected value of perfect information \citep{howard1966information, raiffa1961applied}, with the skills as the acts and the instance as the state: the value of learning, before committing, which skill this instance favors. Section~\ref{sec:adjacent} develops these correspondences.

The following propositions establish the properties the decomposition rests on. They make no assumption about the form of the distributions or about independence between the skills, and their proofs are collected in Appendix~\ref{app:proofs}.

\begin{proposition}[Closed form of the systematic component]\label{prop:S}
Let $n = |\mathcal{S}| \geq 2$ and write $d_s = b - \E[\theta_s] \geq 0$ for the expected shortfall of skill $s$ from the informed baseline, so that $d_L = 0$ for the leader $L$. Then
\[
S \;=\; b - \E[m] \;=\; \frac{1}{n} \sum_{s \in \mathcal{S}} d_s \;\geq\; 0,
\]
the mean expected shortfall of the skills from the leader. A commitment to skill $s$ falls short of $b$ by $d_s$, and a commitment to the weakest skill by $\max_s d_s$, which lies between $\frac{n}{n-1}\,S$ and $n S$; the mean anchor falls short of $b$ by $S$ whichever skill leads. For a pair, $S = \tfrac{1}{2}\big|\E[\theta_i] - \E[\theta_j]\big|$, and a commitment to the weaker skill falls short by $2S$.
\end{proposition}

Propositions~\ref{prop:gap} and~\ref{prop:S} are the same operation applied at two levels: to the two performances on an instance, and to the two averages across instances. A combination that engages both skills without directing work between them halves the cost of having committed to the wrong one, and that is the whole of what it obtains in return for forfeiting $S$ unconditionally. Section~\ref{sec:insurance} prices the trade.

\begin{proposition}[The gap and the gross stake]\label{prop:gap}
On every instance, with $n = |\mathcal{S}|$, the gap is the mean shortfall of the skills from the best of them,
\[
\Delta \;=\; \Mx - m \;=\; \frac{1}{n} \sum_{s \in \mathcal{S}} (\Mx - \theta_s),
\]
which for a pair is $\tfrac{1}{2}|\theta_i - \theta_j|$. In expectation,
\[
G \;=\; \E[\Delta] \;=\; \frac{1}{n} \sum_{s \in \mathcal{S}} \E[\Mx - \theta_s], \qquad \E[\Mx - \theta_s] \;=\; R + d_s,
\]
so that each skill's expected shortfall from the leader of each instance splits into the residual, common to every skill, and its own shortfall $d_s$ from the informed baseline (Proposition~\ref{prop:S}). For a pair, $G = \tfrac{1}{2}\E|\theta_i - \theta_j|$.
\end{proposition}

The identity holds instance by instance and uses no property of the joint distribution. It states the cost of averaging: a combination anchored on $m$ gives up exactly half of whatever separates the two skills on that instance, whichever is better and by however much.

\begin{proposition}[Decomposition and factorization of the stake]\label{prop:R}
$R = \E[\Mx] - b \geq 0$ and $G = S + R$. Moreover, with $\theta_U = \max_{s \neq L} \theta_s$, the other skill's performance for a pair,
\begin{equation}\label{eq:Rpos}
R \;=\; \E\big[(\theta_U - \theta_L)^{+}\big] \;=\; \Pr(\theta_U > \theta_L)\;\E\big[\theta_U - \theta_L \,\big|\, \theta_U > \theta_L\big],
\end{equation}
the second expression holding whenever $\Pr(\theta_U > \theta_L) > 0$.
\end{proposition}

The inequality has a direct reading: taking the better skill on each instance cannot do worse than taking the better skill on average and using it throughout, since selecting per instance includes the option of always selecting the same skill. The factorization gives the size of $R$ a further reading. It is an average over all instances, with zero recorded wherever the underdog does not win, which is what makes it comparable with $b$, itself an average over all instances. It factors into two properties of the pair. The \emph{overlap} of the two distributions fixes how often the underdog wins: the more the densities coincide, the more often the identity of the better skill changes from one instance to the next. Their \emph{dispersion} fixes by how much it wins when it does: two skills that are individually consistent differ little on any instance, whichever is ahead. Both are needed, since $R$ is their product. Overlap without dispersion gives reversals that are frequent but trivial, dispersion without overlap gives reversals that are large but rare, and either leaves little worth routing for. Both fall together as one skill pulls ahead, which is why $R$ can be small when the two skills are far apart on average.

\begin{proposition}[When the residual vanishes]\label{prop:R0}
$R = 0$ if and only if $\Pr(\theta_U > \theta_L) = 0$: the residual is positive if and only if the identity of the better skill changes with positive probability. For a pair, where ties have probability zero, this reads $\Pr(\theta_i > \theta_j) \in \{0,1\}$.
\end{proposition}

$R$ can be zero while $S$ is arbitrarily large. A machine that outperforms the human on every instance yields $R = 0$ however large the gap between them, so the distance between two skills is not by itself informative about what a combination can win from them.

\begin{proposition}[Bounds on the stake]\label{prop:bound}
For performance in $[0,1]$ and an available set of $n = |\mathcal{S}|$ skills,
\[
0 \;\leq\; R \;\leq\; G \;\leq\; 1 - \frac{1}{n},
\]
which for a pair is $G \leq \tfrac{1}{2}$. The upper bound is attained, with $S = 0$ and $R = G$, when on every instance one skill is perfect and the others fail, each skill being the perfect one on a share $1/n$ of instances.
\end{proposition}

Beyond the means, a bound that holds for any distributions with given pairwise moments limits the residual (Proposition~\ref{prop:twomoment}, Appendix~\ref{app:factors}).

The extreme is the case of perfectly complementary skills, each failing exactly where the other succeeds: the whole stake is residual, and an ex ante commitment secures none of it.

\begin{proposition}[Instance-wise ordering of the anchors]\label{prop:disp}
On every instance, $\min_s \theta_s \leq m \leq \Mx$ and $\Mx \geq \theta_s$ for every $s$. The routed anchor therefore weakly dominates every single skill instance by instance, while the mean anchor lies strictly between the extremes whenever they differ.
\end{proposition}

Two consequences follow. No realization of $\Mx$ is worse than the corresponding realization of any skill alone, so routing at its bound cannot increase the frequency of poor outcomes relative to a single skill, and by Proposition~\ref{prop:R} it does not reduce the mean either. The mean anchor, bounded on both sides by the performances, cannot attain the better skill's best realizations and cannot fall to the weaker skill's worst: the reduction in dispersion it obtains is paid for on both tails, and the shortfall $S$ is its cost in the mean.

\subsection{Classes: the residual refined}\label{sec:classes}

Both $\E[m]$ and $\E[\Mx]$ are expectations over instances, so both can be taken class by class and reassembled. Write $b_k = \max_s \mu_{s,k}$ for the best mean within class $k$, attained by the class \emph{incumbent} $\bar{s}(k) = \argmax_s \mu_{s,k}$, and $G_k$, $S_k$ and $R_k$ for the decomposition of Equation~\eqref{eq:gsr} computed within class $k$. Let $L$ be the skill an ex ante commitment selects, the leader of Section~\ref{sec:anchors}, so that $b = \E[\theta_{L}] = \sum_k p_k \mu_{L,k}$.

\begin{proposition}[Class decomposition of the residual]\label{prop:class} (i) $R = \Rclass + \Rinst$, with
\[
\Rclass \;=\; \sum_k p_k b_k - b \;=\; \sum_k p_k\big(b_k - \mu_{L,k}\big) \;\geq\; 0, \qquad \Rinst \;=\; \sum_k p_k\big(\E[\Mx \mid k] - b_k\big) \;=\; \sum_k p_k R_k \;\geq\; 0 .
\]
(ii) $\Rclass = 0$ if and only if $L$ attains $b_k$ in every class of positive probability. (iii) The class-level decompositions aggregate as $\sum_k p_k G_k = G$ and $\sum_k p_k S_k = S + \Rclass$.
\end{proposition}

\begin{proof}
(i) $\E[\Mx] = \sum_k p_k \E[\Mx \mid k] = \sum_k p_k b_k + \sum_k p_k (\E[\Mx \mid k] - b_k)$; subtract $b$. Each term of $\Rclass$ is non-negative since $b_k$ is a maximum over quantities including $\mu_{L,k}$; each $R_k$ is non-negative by Proposition~\ref{prop:R} applied within the class. (ii) follows from the non-negativity of the terms. (iii) $\E[m]$ and $\E[\Mx]$ are linear in the class mixture, so $\sum_k p_k G_k = G$; and $\sum_k p_k S_k = \sum_k p_k b_k - \E[m] = (b + \Rclass) - \E[m]$.
\end{proof}

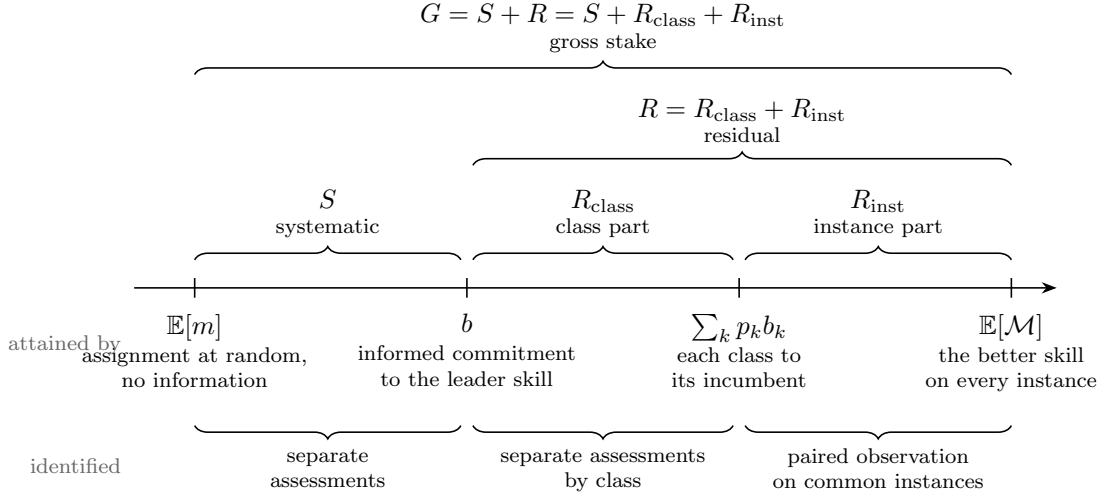
\begin{figure}[!htb]
\centering
\begin{tikzpicture}[
  font=\small,
  >={Stealth[length=2mm]},
  tick/.style={semithick},
  brace/.style={decorate, decoration={brace, amplitude=5pt, raise=2pt}, semithick},
  mbrace/.style={decorate, decoration={brace, mirror, amplitude=5pt, raise=2pt}, semithick},
  lab/.style={font=\footnotesize, align=center},
]
\def\pm{0.6} \def\pb{4.2} \def\pc{7.8} \def\pM{11.4}
\draw[->, semithick] (-0.2,0) -- (12.0,0);
\foreach \p in {\pm,\pb,\pc,\pM} \draw[tick] (\p,0.13) -- (\p,-0.13);
\node[lab, anchor=north] at (\pm,-0.22) {\normalsize $\E[m]$\\[1pt] assignment at random,\\ no information};
\node[lab, anchor=north] at (\pb,-0.22) {\normalsize $b$\\[1pt] informed commitment\\ to the leader skill};
\node[lab, anchor=north] at (\pc,-0.22) {\normalsize $\sum_k p_k b_k$\\[1pt] each class to\\ its incumbent};
\node[lab, anchor=north] at (\pM,-0.22) {\normalsize $\E[\Mx]$\\[1pt] the better skill\\ on every instance};
\draw[brace] (\pm,0.3) -- (\pb-0.08,0.3) node[midway, above=7pt, lab] {\normalsize $S$\\ systematic};
\draw[brace] (\pb+0.08,0.3) -- (\pc-0.08,0.3) node[midway, above=7pt, lab] {\normalsize $\Rclass$\\ class part};
\draw[brace] (\pc+0.08,0.3) -- (\pM,0.3) node[midway, above=7pt, lab] {\normalsize $\Rinst$\\ instance part};
\draw[brace] (\pb+0.08,1.55) -- (\pM,1.55) node[midway, above=7pt, lab] {\normalsize $R = \Rclass + \Rinst$\\ residual};
\draw[brace] (\pm,2.75) -- (\pM,2.75) node[midway, above=7pt, lab] {\normalsize $G = S + R = S + \Rclass + \Rinst$\\ gross stake};
\draw[mbrace] (\pm,-1.75) -- (\pb-0.08,-1.75) node[midway, below=7pt, lab] {separate\\ assessments};
\draw[mbrace] (\pb+0.08,-1.75) -- (\pc-0.08,-1.75) node[midway, below=7pt, lab] {separate assessments\\ by class};
\draw[mbrace] (\pc+0.08,-1.75) -- (\pM,-1.75) node[midway, below=7pt, lab] {paired observation\\ on common instances};
\node[lab, anchor=east, text=black!60] at (-0.25,-2.35) {identified};
\node[lab, anchor=east, text=black!60] at (-0.25,-0.75) {attained by};
\end{tikzpicture}
\caption{The full decomposition of the stake, along expected performance. Between the mean anchor $\E[m]$ and the routed anchor $\E[\Mx]$, the gross stake $G$ splits at the informed baseline $b$ into the systematic component $S$ and the residual $R$ (Equation~\ref{eq:gsr}); where instances fall into classes, the residual splits again, at $\sum_k p_k b_k$, into the part $\Rclass$ that sending each class to its incumbent reaches and the part $\Rinst$ that only a policy reading the instance itself can (Proposition~\ref{prop:class}). Below the axis, what attains each anchor and what evidence identifies each component. Positions are schematic.}
\label{fig:decomposition}
\end{figure}

In decision-analytic terms, $\Rclass$ is the expected value of partial perfect information about the class \citep{felli1998sensitivity}, and $\Rinst$ the rest of the expected value of perfect information. Nothing about an individual instance enters $\Rclass$: it is reachable by a policy that sends each instance to the incumbent of its class, an incumbency determined from marginals computed within each class (Figure~\ref{fig:decomposition}). The second term is what remains once that switch has been made. On instances of class $k$ some skill other than the incumbent realizes the higher performance, and the class label cannot say which, so collecting $\Rinst$ requires changing skill within a class and is reachable only by a policy conditioning on the instance. Part (iii) states the same fact from the other side: what the task-level view calls residual is partly systematic within classes, and a finer partition moves stake from the residual into class-level systematic components, up to the limit in which classes are single instances and $\Rinst$ vanishes. The residual is therefore not a property of the available set alone but of that set together with what the policy is permitted to read, and $R/G$ computed at a commitment is an upper bound on what survives once the class is known. Section~\ref{sec:worked} works through these quantities, and those of the subsections that follow, on four instances.

\subsection{Scoring a combination of responses}\label{sec:responses}

The stake is defined on performances. Arrangements act on responses: they return one response, or a combination of several, and it is that returned response which is scored. When a radiologist reads a probability of pathology of $90\%$ and a model reads $10\%$, the workflow that averages them delivers $50\%$, and what matters is how that $50\%$ scores against the truth, not the average of how the two readings score. The two registers coincide under a condition that the endpoint property makes exact.

\begin{proposition}[The hull bound]\label{prop:hull}
Let the truth be an endpoint and the rule monotone (Definition~\ref{def:endpoint}). For any weights $w_s \geq 0$ with $\sum_s w_s = 1$, which may depend on anything, including the responses and the truth, let $\hat{r}_w = \sum_s w_s \hat{r}_s$ and $\theta_w = \theta(\hat{r}_w, r)$, the performance of the weighted response; the subscript is the weight vector, not a skill or the signal $W$. Then, on every instance,
\[
\min_s \theta_s \;\leq\; \theta_w \;\leq\; \max_s \theta_s .
\]
Moreover: (a) under the linear rule, $\theta_w = \sum_s w_s \theta_s$; (b) if $\phi$ is convex, $\theta_w \geq \sum_s w_s \theta_s$, and if $\phi$ is concave, $\theta_w \leq \sum_s w_s \theta_s$; (c) under the squared rule,
\begin{equation}\label{eq:ambiguity}
\theta_w \;=\; \sum_s w_s \theta_s \;+\; \sum_s w_s\,(\hat{r}_s - \hat{r}_w)^2 .
\end{equation}
\end{proposition}

\begin{proof}
Write $e_s = |\hat{r}_s - r|$. With $r \in \{0,1\}$ and responses in $[0,1]$, $e_s = \hat{r}_s$ if $r = 0$ and $e_s = 1 - \hat{r}_s$ if $r = 1$: on each instance the distance is affine in the response, with slope $\pm 1$ common to all skills. Hence $|\hat{r}_w - r| = \sum_s w_s e_s$, which lies between $\min_s e_s$ and $\max_s e_s$, and the monotonicity of the rule gives the bound. (a) and (b) follow from $\theta_w = 1 - \phi(\sum_s w_s e_s)$ and Jensen's inequality. For (c), $\sum_s w_s e_s^2 - (\sum_s w_s e_s)^2 = \sum_s w_s (e_s - \bar{e})^2$ with $\bar{e} = \sum_s w_s e_s$, and $e_s - \bar{e} = \pm(\hat{r}_s - \hat{r}_w)$ with the sign common to all $s$.
\end{proof}

Three readings follow, and they carry the rest of the paper.

The \emph{ceiling} holds for every rule of the family: no convex combination of responses, however its weights were chosen, scores above the best response it combines. This is the property that bounds the routing channel (Proposition~\ref{prop:routing} below). It rests on the endpoint property, and it fails for a continuous truth, where a point between two forecasts can be closer to the target than either, which is the setting of the forecast-combination results recalled in Section~\ref{sec:combining}. It also fails for multiclass responses, where the correct response is a vertex of a simplex rather than an endpoint of an interval: with truth $(1,0,0)$, the responses $(0,1,0)$ and $(0,0,1)$ are each at squared distance $2$ from it, while their average $(0,\frac{1}{2},\frac{1}{2})$ is at $1.5$, and so scores above both under any rule decreasing in that distance. The ceiling is a result for binary truth, and for rules affine in the response, such as the absolute distance summed over coordinates, which against a vertex is twice one minus the probability given to the correct class.

The \emph{floor} of combining without design rests on the curvature of the rule. Pooling engages the whole available set, $n = |\mathcal{S}|$, and its score $\theta_{\bar{w}}$, the score of the mean response (Definition~\ref{def:aggregation}), is to be read against $m$, the mean of the scores. Under the squared rule, Equation~\eqref{eq:ambiguity} is the ambiguity decomposition of \citet{krogh1994neural}: averaging responses earns, over averaging performances, a term measuring the disagreement between the responses, their variance $\frac{1}{n}\sum_s (\hat{r}_s - \hat{r}_{\bar{w}})^2$ around the mean response $\hat{r}_{\bar{w}}$, which for a pair is $\tfrac{1}{4}(\hat{r}_i - \hat{r}_j)^2$ and does not depend on the truth. Write the dividend that pooling earns in expectation, which under the squared rule is the expected variance,
\begin{equation}\label{eq:Lambda}
\Lambda \;=\; \E[\theta_{\bar{w}}] - \E[m] \;=\; \E\Big[\tfrac{1}{n}\textstyle\sum_s (\hat{r}_s - \hat{r}_{\bar{w}})^2\Big] \;\geq\; 0 \quad\text{under the squared rule},
\end{equation}
with $\hat{r}_{\bar{w}}$ and $\theta_{\bar{w}}$ the response and the score of pooling (Definition~\ref{def:aggregation}); for a pair, \sqc{$\Lambda = \tfrac{1}{4}\E[(\hat{r}_i - \hat{r}_j)^2]$}. Unlike $\E[m]$, \sqc{$\Lambda$} depends on how the responses co-occur on the same instance, and separate assessments do not identify it. The gain is a gain over the average member, bounded above by the best one. Under the linear rule, $\Lambda = 0$: the mean of the responses scores exactly the mean of the performances, and every statement made on the performance anchor $m$ is also a statement about the arrangement that averages responses.

\begin{corollary}[Fixed combinations against the informed baseline]\label{cor:fixed}
Under the squared rule, a combination with fixed weights $w$ attains $\sum_s w_s \E[\theta_s] + \E[\sum_s w_s (\hat{r}_s - \hat{r}_w)^2]$, which can exceed $b$; under any rule of the family it never exceeds $\E[\Mx]$. Under the linear rule the second term vanishes, and the combination attains $\sum_s w_s \E[\theta_s] \leq b$, with equality only if all weight sits on skills whose mean is $b$: no fixed convex combination of responses then beats the informed commitment.
\end{corollary}

\begin{proof}
Take expectations in Proposition~\ref{prop:hull}(c) and~(a), and apply the bound. For an example of the first statement, let two skills return opposite certainties, $\hat{r}_i \in \{0,1\}$ and $\hat{r}_j = 1 - \hat{r}_i$, each right on half the instances: each has $\E[\theta_s] = 0.5 = b$, while the average response $0.5$ scores $0.75$ on every instance.
\end{proof}

Any gain a fixed combination makes over the better single skill is therefore a gain it owes to the curvature of the rule, paid for by disagreement between the responses, and not a gain of augmentation in the sense of Section~\ref{sec:twochannels}; under the linear rule there is none. The \emph{exclusion} of Section~\ref{sec:twochannels} is what keeps the ceiling general. A rule that takes independent responses and returns a point outside their convex hull escapes the bound, and not always through an operation a single response could undergo. A pool of log-odds scaled by a fixed factor above one moves every pooled response toward its nearer endpoint, which a recalibration of a single skill could do alike, raising performance where the pooled response was on the correct side and lowering it where it was not. A rule whose push depends on the responses jointly does more: the Bayesian combination of two calibrated and conditionally independent probabilities, with even prior odds, returns about $0.94$ when both report $0.8$, using their agreement as evidence, and can attain more than $\E[\Mx]$, as the example below shows. The ceilings of this paper are ceilings on convex aggregation and selection. Rules of the second kind are neither routing nor synergy in its terms, a third way of combining skills that the ladder does not contain \citep{kerrigan2021combining, steyvers2022bayesian}; Proposition~\ref{prop:evidence} below gives their ceiling, and they are otherwise left outside the comparisons.

\paragraph{A combination outside the hull, by example.} Let one instance in ten be positive, and let two skills read each instance independently given the truth, each flagging a positive instance with probability $0.9$ and a negative one with probability $0.2$. Each reports the share of positive instances among those it flags, $1/3$, or among those it does not, $1/73$, and so is calibrated; each attains $0.9301$, which is both $b$ and $\E[m]$. Table~\ref{tab:outside} lists what each arrangement returns. Pooling attains $0.9380$, and routing omniscience, which on every instance takes the better of the two responses, $\E[\Mx] = 0.9509$. Whatever a rung does with the two responses, it returns a point between them, and where both skills flag, that point is $1/3$. Read as independent evidence, two flags move the odds of a positive from $1$ to $9$ before reading to $9$ to $4$, and the combination returns $9/13 \approx 0.692$. It is a logistic regression of the truth on the two log-odds, with both coefficients equal to one and intercept $\ln 9$, the prior log-odds against a positive. It attains $0.9571$, above $\E[\Mx]$. It gains where both skills flag a positive instance, and pays where both flag a negative one, or only one flags a positive one, where routing omniscience pays nothing. No exchange takes place and no skill is chosen on any instance: the combination is neither synergy nor routing, and $\E[\Mx]$, which bounds every rung, does not bound it. Section~\ref{sec:radiologydata} measures such a combination on the Collab-CXR data.
 
\begin{table}[!htb]
\centering
\footnotesize
\caption{A combination outside the hull. Two calibrated skills, independent given the truth, on instances one in ten of which is positive; each flags a positive instance with probability $0.9$ and a negative one with probability $0.2$. For each truth and number of flags: the share of instances, the two responses, the pooled and the combined response, and the scores under the squared rule, the first of them the mean of the two skills' scores. The last row is the expectation}
\label{tab:outside}
\setlength{\tabcolsep}{4.5pt}
\begin{tabular}{c c c c c c c c c c}
\toprule
 & & & & & & \multicolumn{4}{c}{Score} \\
\cmidrule(lr){7-10}
$r$ & Flags & Share & Responses & Pooled & Combined & Each skill & Pooling & $\Mx$ & Combination \\
\midrule
1 & 2 & 0.081 & 0.333, 0.333 & 0.333 & 0.692 & 0.556 & 0.556 & 0.556 & 0.905 \\
1 & 1 & 0.018 & 0.333, 0.014 & 0.174 & 0.059 & 0.291 & 0.317 & 0.556 & 0.114 \\
1 & 0 & 0.001 & 0.014, 0.014 & 0.014 & 0.002 & 0.027 & 0.027 & 0.027 & 0.003 \\
0 & 2 & 0.036 & 0.333, 0.333 & 0.333 & 0.692 & 0.889 & 0.889 & 0.889 & 0.521 \\
0 & 1 & 0.288 & 0.333, 0.014 & 0.174 & 0.059 & 0.944 & 0.970 & 1.000 & 0.997 \\
0 & 0 & 0.576 & 0.014, 0.014 & 0.014 & 0.002 & 1.000 & 1.000 & 1.000 & 1.000 \\
\midrule
 & & 1 & & & & 0.9301 & 0.9380 & 0.9509 & 0.9571 \\
\bottomrule
\end{tabular}
\end{table}

\begin{proposition}[The evidence ceiling]\label{prop:evidence}
Let the truth be binary, and let $\mathcal{I}$ be what is observable once the engaged skills have responded: the signals $Z$ and $W$ and the responses $\{\hat{r}_s\}_s$. Write $\hat{r}_{\mathcal{E}} = \Pr(r = 1 \mid \mathcal{I})$ for the probability of a positive given all of it, and $\theta_{\mathcal{E}} = \theta(\hat{r}_{\mathcal{E}}, r)$. Under the squared rule, every response $\hat{r}$ formed from $\mathcal{I}$, by weights that read it, by selection or by any other function, attains
\[
\E[\theta(\hat{r}, r)] \;=\; \E[\theta_{\mathcal{E}}] - \E\big[(\hat{r} - \hat{r}_{\mathcal{E}})^2\big] \;\leq\; \E[\theta_{\mathcal{E}}] \;=\; 1 - \E\big[\hat{r}_{\mathcal{E}}\,(1 - \hat{r}_{\mathcal{E}})\big].
\]
Every rung of routing whose criterion and aggregation read only what is observable therefore attains at most $\min\{\E[\Mx], \E[\theta_{\mathcal{E}}]\}$, and a combination that reads the responses as evidence at most $\E[\theta_{\mathcal{E}}]$. Neither ceiling bounds the other.
\end{proposition}
 
\begin{proof}
Given $\mathcal{I}$, both $\hat{r}$ and $\hat{r}_{\mathcal{E}}$ are fixed and $\E[r \mid \mathcal{I}] = \hat{r}_{\mathcal{E}}$, so that $\E[(\hat{r} - r)^2 \mid \mathcal{I}] = (\hat{r} - \hat{r}_{\mathcal{E}})^2 + \hat{r}_{\mathcal{E}}(1 - \hat{r}_{\mathcal{E}})$; taking expectations gives the identity. A rung of routing forms its response from $\mathcal{I}$, and is bounded by $\E[\Mx]$ by Proposition~\ref{prop:hull}. In the example above the two flags are all that the responses carry, so that $\hat{r}_{\mathcal{E}}$ is the combined response and $\E[\theta_{\mathcal{E}}] = 0.9571 > \E[\Mx] = 0.9509$. In the example of Corollary~\ref{cor:fixed} the two responses are opposite certainties; if which of them is right does not depend on what they say, $\hat{r}_{\mathcal{E}} = \frac{1}{2}$ on every instance, so that $\E[\theta_{\mathcal{E}}] = 0.75 < \E[\Mx] = 1$.
\end{proof}
 
A combination that reads the responses as evidence is therefore bounded, though not by the skills: its ceiling is what the responses and the signals reveal together about the truth, which paired observation identifies as it identifies $\E[\Mx]$. The argument holds under any proper rule, with the expected score of $\hat{r}_{\mathcal{E}}$ in place of its squared form; under the linear rule, which is not proper, the ceiling is attained by the hard response that states the more likely outcome. The two ceilings count different things. $\E[\Mx]$ credits routing with knowing, on every instance, which response is nearer the truth, which nothing observable need reveal; $\E[\theta_{\mathcal{E}}]$ credits a combination with reading perfectly what the responses do reveal, and the agreement of responses formed independently can reveal more than either states. Which is higher depends on the skills: responses that err independently given the truth, as in the example, lift the second above the first; responses each right where the other is wrong, with nothing in what they say to tell which, leave the second at the base rate while the first is perfect. A fitted combination estimates $\E[\theta_{\mathcal{E}}]$ from below, since it attains at most the ceiling; on the Collab-CXR data the logistic combination of the two probabilities attains $0.9674$, below $\E[\Mx] = 0.9699$ (Section~\ref{sec:radiologydata}).
 
\paragraph{Why the squared rule.}\label{par:rule} The squared rule is the paper's rule for three reasons. It credits a response for the strength of the evidence it carries: averaged over instances a skill cannot tell apart, it ranks a response that keeps that strength above one rounded toward certainty, which the linear rule need not do \citep[Sec.~4.1]{gneiting2007strictly}. It gives combining its due: pooling the responses attains the mean anchor plus their disagreement, \sqc{$\E[m] + \Lambda$}, and so improves on choosing among the skills at random, as the literature on combining forecasts finds (Section~\ref{sec:combining}). And it is the standard score for reported probabilities, one minus the Brier score. It is also a score on each instance, which the decomposition needs. The metrics more usual in classification, the areas under the ROC and precision-recall curves, and precision, recall and F1 once a threshold turns probabilities into decisions, are computed over sets of instances, and the comparison of the skills instance by instance on which the stake rests cannot be written in them; accuracy, which is a score on each instance, is the linear rule applied to hard responses. On a binary truth the score also has a level that no skill sets. A reporter that reads nothing and states on every instance the share $\bar{r} = \E[r]$ of positive ones attains $1 - \bar{r}(1 - \bar{r})$, one that states each class's share $\bar{r}_k = \E[r \mid k]$ attains $1 - \sum_k p_k \bar{r}_k(1 - \bar{r}_k)$, and one that always states zero attains $1 - \bar{r}$. Where positive instances are rare all three are close to one, so that a performance close to one need not be a skill's. The first is the reference of the decomposition of the Brier score into uncertainty, resolution and miscalibration \citep{murphy1973new}, and the reporter of class shares the same reference within each class: a skill beats it exactly when what its responses resolve exceeds what their miscalibration costs (Section~\ref{par:calibration}). The stake is made of differences between skills on the same instances and is read on its own terms; the reference says whether the level at which those differences sit is worth having, and sets the scale on which their size is read. These advantages have a price, however. The two registers no longer coincide: the score of a combination of responses exceeds the combination of their scores by the disagreement between them (Equation~\eqref{eq:ambiguity}), so that what pooling attains, the premium of the insurance contract and the break-even of synergy depend on how the responses co-occur on the same instance, and so on paired observation, not on the marginals alone. The linear rule is the special case $\Lambda = 0$, and it has a territory of its own: it coincides with the squared rule on every instance when responses are hard, it is the right valuation where a graded response is itself an action whose errors cost in proportion to their size, and where the skills have only been assessed separately it prices pooling, insurance and synergy from the marginals. Appendix~\ref{app:linear} states the results in that special case and delimits where each rule fits. The closed forms of the rest of the paper are stated under the squared rule, with \sqc{$\Lambda$} marking where the response register departs from the performance register.

\paragraph{Calibration.}\label{par:calibration} The squared rule separates what a skill knows from how it states it. Write $\bar{r}_{s,k}(p) = \Pr(r = 1 \mid \hat{r}_s = p,\, k)$ for the share of positive instances among those of class $k$ to which skill $s$ responds $p$. The skill is \emph{calibrated} in the class if $\bar{r}_{s,k}(p) = p$ for every response it gives: of the instances it puts at $0.9$, nine in ten are positive. Under the squared rule, in every class,
\begin{equation}\label{eq:murphy}
\mu_{s,k} \;=\; 1 - \bar{r}_k(1 - \bar{r}_k) + \mathrm{Res}_{s,k} - \mathrm{Mis}_{s,k},
\end{equation}
with
\[
\mathrm{Res}_{s,k} = \E\big[(\bar{r}_{s,k}(\hat{r}_s) - \bar{r}_k)^2 \mid k\big], \qquad
\mathrm{Mis}_{s,k} = \E\big[(\hat{r}_s - \bar{r}_{s,k}(\hat{r}_s))^2 \mid k\big] \;\geq\; 0
\]
\citep{murphy1973new}. The first term is the uncertainty of the class, a property of the task. The \emph{resolution} $\mathrm{Res}_{s,k}$ is what the skill knows: how far the groups of instances its responses tell apart differ in their share of positives. The \emph{miscalibration} $\mathrm{Mis}_{s,k}$ is what its way of stating that knowledge costs: the squared distance between what the skill says and what its responses turn out to mean, zero exactly when the skill is calibrated. The reporter of class shares has neither, and so a skill beats it in class $k$ exactly when $\mathrm{Res}_{s,k} > \mathrm{Mis}_{s,k}$. Calibration is not ranking: a skill whose responses order the instances well, positives above negatives, can still be badly calibrated, and the area under the ROC curve, which reads only the order, does not see it. A \emph{recalibration} replaces each response $\hat{r}_s$ by $\bar{r}_{s,k}(\hat{r}_s)$, estimated in each class from the skill's own record. With the shares known it removes $\mathrm{Mis}_{s,k}$ and leaves $\mathrm{Res}_{s,k}$ as it is; an estimate that does not decrease in $\hat{r}_s$, such as isotonic regression gives, also leaves the order of the responses. It changes the skill, not the arrangement the skill takes part in, and the paper credits it to no channel (Section~\ref{sec:twochannels}). Under the linear rule the decomposition does not hold and calibration earns nothing, since the response that maximizes expected performance is the more likely outcome and not its probability (Appendix~\ref{app:linear}).

\subsection{Without augmentation: commitment, pooling and insurance}\label{sec:insurance}

The first policies to read against the stake are those that operate neither channel (Section~\ref{sec:twochannels}): a commitment to one skill, which secures $S$ when the assessment behind it is right, and a pooling of several with fixed weights, which secures neither $S$ nor $R$ and earns, under the squared rule, only the disagreement between their responses. The comparison between them turns on that assessment. An ex ante commitment is only as good as it is, and the assessment is made on instances already worked while the commitment is used on instances yet to come. Pooling's defining property, that it does not depend on which skill is better, makes it an insurance against an assessment that does not survive. Under the squared rule the pooled response scores more than the mean of the performances, by the disagreement \sqc{$\Lambda$} between the responses (Equation~\eqref{eq:Lambda}). Since the pooled response never scores above the best skill on an instance (Proposition~\ref{prop:hull}), the disagreement on every instance is at most $\Delta$, and \sqc{$\Lambda \leq G$}; equivalently, \sqc{$\Lambda - S \leq R$}, so that pooling beats the commitment by at most the residual, and not at all where one skill leads on every instance. A numeric worked example is provided in Section~\ref{sec:worked}.

\begin{proposition}[The insurance contract]\label{prop:insurance}
Let $n = |\mathcal{S}|$, and let the ex ante assessment name the leader $L$ with probability $p_a$ and each other skill $s$ with probability $q_s$, so that $\sum_{s \neq L} q_s = 1 - p_a$. With the shortfalls $d_s = b - \E[\theta_s]$ of Proposition~\ref{prop:S}, the commitment attains $b$ when the assessment is right and $b - d_s$ when it names $s$, and in expectation $b - D$, where $D = \sum_{s \neq L} q_s d_s$; its outcomes spread over $[\,b - \max_s d_s,\ b\,]$. Pooling the responses with equal weights attains \sqc{$\E[\theta_{\bar{w}}] = b - S + \Lambda$} whatever the assessment. Engaging every skill on every instance therefore costs, relative to the commitment, an expected premium of \sqc{$S - \Lambda - D$} and removes the exposure to the assessment; it is preferable in expectation if and only if \sqc{$D > S - \Lambda$}. If a wrong assessment names each other skill with equal probability and \sqc{$S > 0$}, $D = (1 - p_a)\,\frac{n}{n-1}\,S$, and the condition becomes \sqc{$p_a < \frac{1}{n} + \frac{n-1}{n}\frac{\Lambda}{S}$}. For a pair, the commitment attains $b - 2S(1 - p_a)$ in expectation, with a spread of $2S$, the premium is \sqc{$S(2p_a - 1) - \Lambda$}, and the condition is \sqc{$p_a < \tfrac{1}{2} + \tfrac{\Lambda}{2S}$}.
\end{proposition}

\begin{proof}
The commitment's expected performance is $p_a b + \sum_{s \neq L} q_s (b - d_s) = b - D$. Pooling attains \sqc{$\E[m] + \Lambda$} by Equation~\eqref{eq:Lambda}, and $\E[m] = b - S$ by Proposition~\ref{prop:S}, so the premium is \sqc{$(b - D) - (b - S + \Lambda) = S - \Lambda - D$}. Under equal probabilities, $q_s = (1 - p_a)/(n - 1)$ and $\sum_{s \neq L} d_s = nS$, since $d_L = 0$, giving $D = (1 - p_a)\frac{n}{n-1} S$; then \sqc{$D > S - \Lambda$} if and only if \sqc{$p_a < \frac{1}{n} + \frac{n-1}{n}\frac{\Lambda}{S}$}. For a pair, $n = 2$ and the single other skill has $d_s = 2S$, so $D = 2S(1 - p_a)$, the premium is \sqc{$S(2p_a - 1) - \Lambda$}, and the spread $\max_s d_s$ is $2S$.
\end{proof}

The contract is an insurance in the strict sense: for an organization whose assessment is usually right it lowers the expected outcome, unless the responses disagree by more than $S$, and what it buys is the removal of the spread between being right and being wrong. Whether the premium is worth paying depends on how confident the organization is in an assessment it cannot continuously revalidate, which is part of what makes the skill policy decision a satisficing problem rather than an optimization. One further feature makes the contract better than the proposition states: across instances as well as across states, pooling narrows dispersion on both tails (Propositions~\ref{prop:disp} and~\ref{prop:hull}), which in a setting that punishes the lower tail can avoid the consequences that a threshold on the lower tail would impose on both skills. Two features make it worse. The premium is stated in performance; in its cost profile the contract also commits a second execution on every instance and a second skill held available (Table~\ref{tab:costs}), whose price is left to future work. And the insurance covers only the risk that the assessment is wrong about which skill is better; it does nothing to collect $R$.
 
Against an assessment drawn from instances like those the commitment will serve, the probability of a wrong assessment is not free: it is high only where the skills are close, and there a wrong commitment costs little.
 
\begin{corollary}[The insurance contract under sampling]\label{cor:insurance}
Let $|\mathcal{S}| = 2$, and let the assessment name the skill with the higher estimated mean, the difference between the two estimates being normal with mean $\E[\theta_L] - \E[\theta_U] = 2S$ and standard deviation $\tau$. Then $1 - p_a = \Phi(-2S/\tau)$ and
\[
D \;=\; 2S\,\Phi(-2S/\tau) \;\leq\; \min\{S,\ \kappa\,\tau\}, \qquad \kappa \;=\; \max_{x \geq 0}\, x\,\Phi(-x) \;\approx\; 0.170,
\]
the maximum reached where $2S \approx 0.75\,\tau$. With each skill assessed on $N$ instances, $\tau = \sigma/\sqrt{N}$, where $\sigma^2$ is $\mathrm{Var}(\theta_H) + \mathrm{Var}(\theta_M)$ for separate assessments and $\mathrm{Var}(\theta_H - \theta_M)$ for paired ones. Pooling is then preferable in expectation only if $S - \Lambda < \kappa\,\sigma/\sqrt{N}$, and whatever the assessment if $\Lambda > S$. Without normality, for performances in $[0,1]$, $D \leq 2S\,e^{-2NS^2} \leq e^{-1/2}/\sqrt{N}$.
\end{corollary}
 
\begin{proof}
The assessment is wrong when the estimated difference falls below zero, which has probability $\Phi(-2S/\tau)$, and for a pair $D = 2S(1 - p_a)$ (Proposition~\ref{prop:insurance}). Since $\Phi(-x) \leq \frac{1}{2}$ for $x \geq 0$, $D \leq S$; with $x = 2S/\tau$, $D = \tau\,x\,\Phi(-x) \leq \kappa\tau$, and the first-order condition $\Phi(-x) = x\,\phi(x)$ gives $x \approx 0.752$ and $\kappa \approx 0.170$. Pooling is preferable if and only if $D > S - \Lambda$ (Proposition~\ref{prop:insurance}), which requires $\kappa\tau > S - \Lambda$, and holds for every $D \geq 0$ if $S < \Lambda$. Without normality, the estimated difference is a mean of $N$ independent terms in $[-1, 1]$ with mean $2S$, pairing the instances of the two records in any order where the assessments are separate, so that $1 - p_a \leq e^{-2NS^2}$ by the inequality of \citet{hoeffding1963probability}; the function $2S\,e^{-2NS^2}$ is largest at $S = 1/(2\sqrt{N})$, where it equals $e^{-1/2}/\sqrt{N}$.
\end{proof}
 
Against sampling error, then, the insurance pays only while the record is short: $D$ is largest where the skills are close enough for the assessment to err and far enough apart for the error to matter, and it falls as $1/\sqrt{N}$ whatever $S$. What keeps $1 - p_a$ from shrinking with the record is not its size but its relevance: an assessment made in another period, on another caseload or on a vendor's benchmark errs with a probability that no record of that kind reduces, and that is the case for which Proposition~\ref{prop:insurance} leaves $p_a$ free. The same function bounds what a check of the procedure gives back on a finite record (Proposition~\ref{prop:margin}). On the Collab-CXR data the disagreement exceeds the systematic component, $\Lambda = 0.0101$ against $S = 0.0018$ as reported and $0.0049$ against $0.0035$ recalibrated, so that pooling attains more than the commitment whatever the assessment, by an amount that after recalibration the data do not establish (Section~\ref{sec:radiologydata}).

\paragraph{At parity.} When the skills are equally capable on average, the question of which to commit to has no answer in expectation, and the commitment becomes a bet without an edge.
 
\begin{proposition}[Parity]\label{prop:parity}
Let $|\mathcal{S}| = 2$ and $\E[\theta_H] = \E[\theta_M]$. Then (i) $S = 0$ and $b = \E[m]$, and a commitment to either skill attains $b$; (ii) $R = G = \tfrac{1}{2}\,\E|\theta_H - \theta_M|$; (iii) $R = 0$ if and only if $\theta_H = \theta_M$ almost surely, which with identical distributions is the comonotone coupling; (iv) under the squared rule, the premium of pooling over a commitment is \sqc{$-\Lambda \leq 0$}.
\end{proposition}
 
\begin{proof}
(i) $S$ is half the gap between the expected performances (Proposition~\ref{prop:S}). (ii) By Proposition~\ref{prop:R}, $R = \E[(\theta_U - \theta_L)^{+}]$, whichever skill is labeled the leader; since $\E[(\theta_H - \theta_M)^{+}] - \E[(\theta_M - \theta_H)^{+}] = \E[\theta_H - \theta_M] = 0$ and the two terms sum to $\E|\theta_H - \theta_M|$, each equals half of it, and $G = S + R = R$. (iii) $\E|\theta_H - \theta_M| = 0$ if and only if the two coincide almost surely; with a common distribution $F$, the comonotone coupling sets both to $F^{-1}(V)$. (iv) With equal expectations every shortfall $d_s$ of Proposition~\ref{prop:insurance} is zero, so $D = 0$ and the premium \sqc{$S - \Lambda - D$} is \sqc{$-\Lambda$}.
\end{proof}
 
Neither commitment is wrong on average, but each is wrong on the instances where the other skill leads, and the expected shortfall from the best on every instance is half the mean distance between the two performances: it grows with their dispersion and vanishes when the two fail on the same instances. The insurance against a wrong commitment is free at parity, since pooling earns the disagreement between the responses and gives up nothing in mean, before the cost of the second execution. What a wrong commitment can cost is bounded in performance, by one on any instance; what it costs the organization need not be, and Section~\ref{sec:contest} returns to the case in which a single failure can cost without limit.

\begin{proposition}[Pooling against randomizing]\label{prop:randomize}
A policy that engages on each instance one skill drawn uniformly at random from the available set, independently of the instance, attains $\E[m]$ with one execution per instance; pooling the responses attains \sqc{$\E[m] + \Lambda$} with $n$. The randomized performance is also more dispersed than the mean of the performances:
\[
\mathrm{Var}(\theta_{\mathrm{rand}}) - \mathrm{Var}(m) \;=\; \E\Big[\tfrac{1}{n}\textstyle\sum_s (\theta_s - m)^2\Big] \;\geq\; 0,
\]
which for a pair is $\tfrac{1}{4}\E[(\theta_i - \theta_j)^2]$.
\end{proposition}

\begin{proof}
Conditionally on the instance, the policy returns each $\theta_s$ with probability $1/n$, so its conditional mean is $m$ and its conditional variance $\frac{1}{n}\sum_s (\theta_s - m)^2$. Taking expectations gives $\E[m]$, and the law of total variance gives the difference; pooling attains \sqc{$\E[m] + \Lambda$} by Equation~\eqref{eq:Lambda}. For a pair, $\theta_s - m = \pm\frac{1}{2}(\theta_i - \theta_j)$.
\end{proof}

What the second execution buys over randomizing, then, is the disagreement between the responses, \sqc{$\Lambda$} in expectation; lower dispersion across instances; and the paired record, which randomizing never produces because it never executes two skills on the same instance. Under the linear rule the first of these vanishes, and randomizing secures the same expectation as pooling at a fraction of the execution cost. The two arrangements also differ in what they can achieve against the informed commitment. Random assignment never beats it, since it attains $\E[m] = b - S \leq b$ whatever the skills; pooling beats it whenever the disagreement it earns exceeds the systematic component, \sqc{$\Lambda > S$}, which separate assessments cannot settle, since they identify $S$ but not $\Lambda$.

The mean anchor nevertheless remains the reference from which the take is measured, for the reason it is the reference of the stake: it is a property of the skills, fixed by their marginals and identified from separate assessments, whereas \sqc{$\Lambda$} is a property of the rule and of how the responses are arranged, identified only from paired observation. Moving the reference to pooling would make the stake depend on the rule, and would turn $S$ into \sqc{$S - \Lambda$}, which is negative whenever the responses disagree by more than $S$. Measured from $\E[m]$, the take of any arrangement has three sources. Combining responses earns a \emph{curvature dividend}, the disagreement it averages, without any information about the skills: it is all that pooling collects, \sqc{$\Lambda$}, and part of what an ensemble collects. Routing collects part of the stake, bounded by $G$. Synergy adds to or subtracts from either, bounded by nothing. Assignment and selection return a single response and earn no dividend. The dividend is not augmentation in the sense of Section~\ref{sec:twochannels}: it is paid for by executing every skill on every instance, and it is collected equally by an arrangement that knows nothing about them. Nor is all of it complementarity. Part of it reflects responses that err on different instances; part is the correction of each skill's own overconfidence, which averaging pulls back toward the middle: a reduction of the miscalibration of Equation~\eqref{eq:murphy}, which a recalibration of each skill would also deliver, raising $b$ instead. Like recalibration, that part is credited to no channel.

\subsection{Routing and the climb}\label{sec:routing}

Applied to the responses a policy without a protocol combines, the hull bound fixes what routing can take.

\begin{proposition}[The routing range]\label{prop:routing}
Under the conditions of Proposition~\ref{prop:hull}, every arrangement that forms its delivered response by an aggregation, and so every routing arrangement without a protocol, attains, on every instance, a performance between $\min_s \theta_s$ and $\max_s \theta_s$, and in expectation a performance in $[\E[\min_s \theta_s],\ \E[\Mx]]$, whatever it conditions on. The upper end is attained by placing all weight on the winner $\dot{s} = \argmax_s \theta_s$ of each instance, the lower end by placing it on the laggard. For a pair, $\E[\min_s \theta_s] = \E[m] - G$, so the range is $[\E[m] - G,\ \E[m] + G]$.
\end{proposition}

\begin{proof}
The first statement is Proposition~\ref{prop:hull} applied to independently produced responses. Weights concentrated on $\dot{s}$ return $\hat{r}_{\dot{s}}$ and score $\Mx$. For a pair, $\min + \max = 2\,\mean$ on every instance, so $\E[\min_s\theta_s] = 2\E[m] - \E[\Mx] = \E[m] - G$.
\end{proof}

For a pair the routing range is symmetric about the mean anchor: a policy directing work the wrong way costs exactly what one directing it the right way earns. A policy carrying no information about the instance but fitted to the marginals collapses onto the better skill and returns $b$, a commitment in routing's clothing; from there routing's headroom is exactly $R$, of which a policy reading only the class reaches at most $\Rclass$. A criterion that uses what it reads as well as it can never falls below $b$ and never loses by reading more; ensembling exceeds assignment on the same information only by the disagreement between the responses it weights, and under the linear rule never (Propositions~\ref{prop:information} and~\ref{prop:ensembling}, Appendix~\ref{app:factors}).

\paragraph{What a policy takes, in plain words.} The skills put a fixed amount on the table, the stake, and it comes in the three components presented in Sections~\ref{sec:anchors} and~\ref{sec:classes}: $S$, what an organization gains by knowing which skill is better on average and using it throughout, $\Rclass$, what it adds by knowing which skill is better for each kind of instance, and $\Rinst$, what it adds by knowing which is better on each instance. The components are climbed in order: a policy that holds a component in full holds every component below it (Proposition~\ref{prop:nested}). Not everything a combination returns is stake, though: averaging responses earns their disagreement, a protocol adds or subtracts its own lift, and neither says which skill should do the work. What a policy takes can therefore be read component by component, with one qualification. Where a policy does without error what a step of the climb prescribes, committing to the skill better on average, sending a class to its incumbent or an instance to the skill that leads on it, the reading is exact: its take is a known sum of components. Where it errs, because the criterion that reads the class or the instance is wrong on some instances, its take is still exact in total, but it is a blend of right and wrong decisions, and how much of the shortfall falls on one component rather than another is no longer fixed by the decomposition: it depends on how the shortfall is credited. What follows states the order of the components, the reading, the condition under which it is exact, and the two natural ways of crediting a take short of it.

The decomposition is an ordering of what each level of information reaches, and the levels are nested: the ceiling of each contains the ones below it.

\begin{proposition}[The components are reached in order]\label{prop:nested}
(a) Let an assignment engage on each instance a skill $\hat{s}(k)$ determined by its class. It attains
\[
\E[\theta_{\hat{s}}] \;=\; b + \Rclass - \textstyle\sum_k p_k \big(b_k - \mu_{\hat{s}(k),k}\big) \;\leq\; b + \Rclass,
\]
with equality if and only if $\hat{s}(k)$ is an incumbent of every class of positive probability. An assignment that attains $b + \Rclass$ therefore attains, in every class, at least $\mu_{L,k}$, what the commitment attains there, and its take over the mean anchor is $S + \Rclass$. (b) Let an arrangement return on every instance a performance $\theta_\pi \leq \Mx$, as every arrangement returning one skill's response does (Proposition~\ref{prop:disp}) and every aggregation does (Proposition~\ref{prop:routing}). It attains $\E[\Mx] = b + \Rclass + \Rinst$ if and only if $\theta_\pi = \Mx$ almost surely, and then $\theta_\pi \geq \theta_{\bar{s}(k)}$ and $\theta_\pi \geq \theta_{L}$ almost surely: instance by instance, at least what the assignment of (a) and the commitment return, and its take is $S + \Rclass + \Rinst = G$. (c) Short of the ceilings the order is not guaranteed: the assignment of (a) attains $b + \sum_k p_k (\mu_{\hat{s}(k),k} - \mu_{L,k})$, below $b$ whenever what it loses on classes where it engages a skill weaker than $L$ exceeds what it gains where it engages a stronger one.
\end{proposition}

\begin{proof}
(a) On an instance of class $k$ the assignment engages $\hat{s}(k)$, so $\E[\theta_{\hat{s}}] = \sum_k p_k \mu_{\hat{s}(k),k}$; adding and subtracting $\sum_k p_k b_k = b + \Rclass$ (Proposition~\ref{prop:class}) gives the expression. Each term $b_k - \mu_{\hat{s}(k),k}$ is non-negative, so equality holds if and only if every term of positive probability vanishes, that is, $\mu_{\hat{s}(k),k} = b_k$. Then $\mu_{\hat{s}(k),k} = b_k \geq \mu_{L,k}$ in every class, and $\E[\theta_{\hat{s}}] - \E[m] = (b - \E[m]) + \Rclass = S + \Rclass$. (b) $\E[\Mx] - \E[\theta_\pi] = \E[\Mx - \theta_\pi]$ has a non-negative integrand, so it vanishes if and only if $\theta_\pi = \Mx$ almost surely; then $\theta_\pi \geq \theta_s$ for every $s$ (Proposition~\ref{prop:disp}), in particular for $\bar{s}(k)$ and $L$. $\E[\Mx] = b + R = b + \Rclass + \Rinst$ by Propositions~\ref{prop:R} and~\ref{prop:class}, and subtracting $\E[m] = b - S$ gives the take. (c) Since $b = \sum_k p_k \mu_{L,k}$, the attainment of (a) is $b + \sum_k p_k (\mu_{\hat{s}(k),k} - \mu_{L,k})$, whose terms are positive, zero or negative as $\hat{s}(k)$ is stronger than, equal to or weaker than $L$ in class $k$.
\end{proof}

The components are therefore not separate prizes to be pursued in any order. Collecting $\Rclass$ in full requires the incumbent of every class, including the classes where the commitment's skill is already the incumbent, and so never does worse than the commitment in any class; collecting $\Rinst$ in full requires the skill that leads on every instance, and so never does worse, on any instance, than the class assignment that collects $S + \Rclass$. What the proposition does not make safe is the climb between ceilings: a class assignment that knows some incumbents and misjudges others can end below $b$, as rung~4a does below a threshold accuracy (Proposition~\ref{prop:classpred}), and an arrangement reading the instance can fall to $\E[\min_s \theta_s]$ (Proposition~\ref{prop:routing}). Part (a) concerns assignment: an ensemble whose weights read only the class can exceed $b + \Rclass$ under the squared rule, by the disagreement it earns (Proposition~\ref{prop:ensembling}), which is a curvature dividend and not stake.

More generally, the three sources of a take named in Section~\ref{sec:insurance} separate on every instance, and the part that is stake can be credited to its components in the order in which Proposition~\ref{prop:nested} nests their ceilings.

\begin{corollary}[What a policy takes, component by component]\label{cor:attribution}
Let a policy engage the skills with weights $w_s \geq 0$, $\sum_s w_s = 1$, one skill carrying all the weight under assignment, selection and random assignment, and deliver the aggregated response, scored $\theta_w$, or the joint response of a protocol run on the engaged skills. Write $\tilde{\theta}_\pi = \sum_s w_s \theta_s$ for the weighted mean of the engaged performances and $a_k = \E[\tilde{\theta}_\pi \mid k]$ for its mean in class $k$. Then
\[
\E[\theta_\pi] - \E[m] \;=\; T_S + T_{\mathrm{class}} + T_{\mathrm{inst}} \;+\; \sqc{\E[\theta_w - \tilde{\theta}_\pi]} \;+\; \E[\theta_\pi - \theta_w],
\]
where, with $C = \sum_k p_k \big(\min(a_k, b_k) - \E[m \mid k]\big)$,
\[
T_S \;=\; \min(C, S) \;\leq\; S, \qquad 0 \;\leq\; T_{\mathrm{class}} \;=\; (C - S)^{+} \;\leq\; \Rclass, \qquad 0 \;\leq\; T_{\mathrm{inst}} \;=\; \textstyle\sum_k p_k (a_k - b_k)^{+} \;\leq\; \Rinst .
\]
The fourth term is the curvature dividend, \sqc{$\E[\sum_s w_s (\hat{r}_s - \hat{r}_w)^2]$} under the squared rule with an endpoint truth (Proposition~\ref{prop:hull}), and zero under the linear rule and wherever one response is returned unchanged; the fifth is the lift of the protocol in difference (Definition~\ref{def:lift}), and zero without a protocol.
\end{corollary}

\begin{proof}
On every instance $\theta_\pi = \tilde{\theta}_\pi + (\theta_w - \tilde{\theta}_\pi) + (\theta_\pi - \theta_w)$, and $\E[\tilde{\theta}_\pi] = \sum_k p_k a_k$. In each class $a_k = \min(a_k, b_k) + (a_k - b_k)^{+}$; averaging over the classes and subtracting $\E[m] = \sum_k p_k \E[m \mid k]$ gives $\E[\tilde{\theta}_\pi] - \E[m] = C + T_{\mathrm{inst}}$, and $C = T_S + T_{\mathrm{class}}$. Since $\tilde{\theta}_\pi \leq \Mx$ on every instance, $a_k \leq \E[\Mx \mid k] = b_k + R_k$, so $T_{\mathrm{inst}} \leq \sum_k p_k R_k = \Rinst$; and $C \leq \sum_k p_k (b_k - \E[m \mid k]) = S + \Rclass$ by Proposition~\ref{prop:class}(iii), so $T_{\mathrm{class}} \leq \Rclass$. The dividend is Proposition~\ref{prop:hull}(c), and vanishes by (a) under the linear rule.
\end{proof}

\paragraph{When the split is exact.} The total take is fixed by the corollary; its split is fixed only at points of the climb. Say that a policy \emph{resolves} class $k$ if at the level of the class it attains exactly what the commitment or the incumbent attains there, $\min(a_k, b_k) \in \{\mu_{L,k},\, b_k\}$. If it resolves every class of positive probability, it takes
\[
T_S \;=\; S, \qquad T_{\mathrm{class}} \;=\; \textstyle\sum_{k \in \mathcal{K}_\pi} p_k \big(b_k - \mu_{L,k}\big), \qquad T_{\mathrm{inst}} \;=\; \textstyle\sum_k p_k (a_k - b_k)^{+},
\]
with $\mathcal{K}_\pi$ the classes it sends to their incumbent: all of $S$, the terms of $\Rclass$ of exactly those classes, and whatever it gains above the incumbents, however imperfectly it reads the instance. Each class sits on a point of the climb and contributes a known combination of components, so no convention is needed, and the two rules below agree. A class is unresolved when errors leave its attainment strictly between the two points, or below both. Its take is then a blend: a class criterion that names the class correctly with probability $a$ sends a share $1 - a$ of the class's instances to the wrong incumbent, and an instance criterion that misjudges which skill leads on the instance can fall below the incumbent's mean on a whole class. The decomposition fixes how much such a class takes, but not which component its shortfall belongs to.

\paragraph{Crediting a take short of the climb.} The corollary credits in order, the convention that matches Proposition~\ref{prop:nested}: a class-level take counts toward $\Rclass$ only beyond $S$, what a commitment secures, so that a full $S$ reads as doing at least as well as the commitment. Assignment on a predicted class shows what the convention decides. With two classes whose incumbents differ and accuracy $a$, its take at the level of the class is $C = (2a - 1)(S + \Rclass)$ under any convention (Proposition~\ref{prop:classpred}); credited in order it is $\min(C, S)$ and the rest, so that $S$ fills first and is complete from the accuracy $a^{*}$ at which the policy reaches $b$. Other conventions split the same total differently, and none changes a conclusion; Appendix~\ref{app:crediting} compares one.

\paragraph{What the record must show.} Whether a policy resolves its classes, and the shares it then takes, need only the class-wise means $a_k$, $b_k$ and $\E[m \mid k]$, which an audited record supplies (Proposition~\ref{prop:audit}). A step of the climb taken without error moves the take by a known component; a step taken with error moves it by a known amount, whose split is a matter of accounting. Section~\ref{sec:worked} reads the arrangements of a worked example this way (Figure~\ref{fig:attribution}), and Section~\ref{sec:procedure} turns the reading into the checks of a procedure.

\subsection{Synergy: what a protocol must achieve}\label{sec:channels}

A protocol is bounded by none of the skills (Section~\ref{sec:twochannels}), so synergy has no counterpart to Proposition~\ref{prop:routing}: nothing bounds it at $\E[\Mx]$, because the responses that bound $\E[\Mx]$ are not the ones it returns, and nothing guarantees it either, since an exchange can pull a right response toward a wrong one. What can be stated in advance is how it is measured and the lift it must clear.

\begin{definition}[Lift]\label{def:lift}
Let a policy operate synergy through a protocol $\rho$, and let the responses its engaged skills would have produced independently be retained. The \emph{lift} of the protocol on an instance is
\begin{equation}\label{eq:gamma}
\gamma \;=\; \frac{\theta(\hat{r}_\rho, r)}{\theta_w},
\end{equation}
the ratio between what the protocol attained and what the same skills would have attained had their independent responses been aggregated, $\theta_w$ of Proposition~\ref{prop:hull}, with equal weights where the protocol weights none. We call $\theta_w$ the \emph{mechanical baseline}: the score of a response that a fixed rule computes from the skills' independent responses, without any exchange between them, so that the lift isolates what the interaction adds to what the same responses would have yielded without it. The distinction is that between the mathematical and the behavioral aggregation of expert judgments \citep{clemen1999combining}. Over a caseload, the lift is summarized by the ratio of means
\begin{equation}\label{eq:gammapi}
\gamma_\pi \;=\; \frac{\E[\theta(\hat{r}_\rho, r)]}{\E[\theta_w]},
\end{equation}
and, separately, by the share of instances on which $\gamma < 1$. The ratio of means, rather than the mean of the instance lifts, is the summary that recovers expected performance, $\E[\theta(\hat{r}_\rho, r)] = \gamma_\pi\,\E[\theta_w]$, and so can be compared with a break-even; it is the mean of the instance lifts weighted by $\theta_w$, so that instances on which the aggregated response performs poorly, and the ratio is unstable, count for less. Where $\theta_w$ is small the ratio is unstable, and the lift is also reported as a difference, $\delta = \theta(\hat{r}_\rho, r) - \theta_w$ on an instance and $\delta_\pi = \E[\theta(\hat{r}_\rho, r)] - \E[\theta_w]$ over a caseload, which is positive exactly when $\gamma_\pi > 1$.
\end{definition}

\begin{proposition}[Break-even of synergy]\label{prop:breakeven}
A synergy protocol whose mechanical baseline is $\theta_w$, with $\E[\theta_w] > 0$, attains $\gamma_\pi \E[\theta_w]$, and matches the informed baseline if and only if
\begin{equation}\label{eq:gammastar}
\gamma_\pi \;\geq\; \gamma^{*} \;=\; \frac{b}{\E[\theta_w]} .
\end{equation}
Equivalently, it matches the informed baseline if and only if $\delta_\pi \geq \delta^{*} = b - \E[\theta_w]$. With equal weights, \sqc{$\E[\theta_{\bar{w}}] = b - S + \Lambda$}, so that \sqc{$\gamma^{*} = b/(b - S + \Lambda)$} and \sqc{$\delta^{*} = S - \Lambda$}: the bar rises with $S$ and falls with the disagreement between the responses. Under the linear rule, $\Lambda = 0$, and $\gamma^{*} = b/(b - S)$ and $\delta^{*} = S$ follow from the marginals alone.
\end{proposition}

\begin{proof}
The first statement is Equation~\eqref{eq:gammapi} rearranged. With equal weights, \sqc{$\E[\theta_{\bar{w}}] = \E[m] + \Lambda$} by Equation~\eqref{eq:Lambda}, and $\E[m] = b - S$ by Proposition~\ref{prop:S}; the additive form subtracts $\E[\theta_w]$ from both sides instead of dividing.
\end{proof}

The break-even is the least synergy a protocol needs to match the best single skill; the bar is higher where the skills are further apart on average and lower where their responses disagree more on the instance. Because \sqc{$\Lambda$} depends on how the responses co-occur, the bar is known before the protocol is designed only from the skills observed on common instances; under the linear rule the marginals would suffice.

\subsection{The two channels in reduced form}\label{sec:reduced}

A reduced form locates the two channels at their corners. It represents the lift as a uniform multiplier applied to a performance anchor, which is the right abstraction for establishing that a synergy channel exists and what it must clear.

\begin{proposition}[The archetypes under the squared rule]\label{prop:archetypes}
The archetypes of \citet{zanardelli2025navigating} extend to the squared rule as follows. When the lift is represented as a uniform multiplier $\gamma \geq 1$ applied to the pooled response or to the routed anchor, and the ceiling at one does not bind, $\gamma\,\theta_{\bar{w}} \leq 1$ and $\gamma\,\Mx \leq 1$ almost surely (Proposition~\ref{prop:ceiling} gives the shortfall when it does), the three archetypes
\begin{equation}\label{eq:archetypes}
\sqc{\theta_{HM} = \theta_{\bar{w}}, \qquad \theta_{HM{+}} = \min(1, \gamma\, \theta_{\bar{w}})}, \qquad \theta_{HM{+}{+}} = \min(1, \gamma\, \Mx)
\end{equation}
attain in expectation
\[
\sqc{\E[\theta_{HM}] = b - S + \Lambda}, \qquad \sqc{\E[\theta_{HM{+}}] = \gamma\,(b - S + \Lambda)}, \qquad \E[\theta_{HM{+}{+}}] = \gamma\,(b + R),
\]
and so, relative to $b$, respectively \sqc{$-(S - \Lambda)$}, \sqc{$(\gamma - 1)b - \gamma(S - \Lambda)$}, and $(\gamma - 1)b + \gamma R$. With $\Lambda = 0$, the linear rule, these are the archetypes and attainments of the earlier paper, defined there on performances, $\theta_{HM} = m$ and $\theta_{HM{+}} = \min(1, \gamma\, m)$; $HM{+}{+}$, which returns a single response on each instance, is the same under both rules.
\end{proposition}

\begin{proof}
By Equation~\eqref{eq:Lambda} and Propositions~\ref{prop:S} and~\ref{prop:R}, \sqc{$\E[\theta_{\bar{w}}] = b - S + \Lambda$} and $\E[\Mx] = b + R$; without the ceiling, the lift multiplies each expectation by $\gamma$, which gives the attainments, and subtracting $b$ gives the rest.
\end{proof}

The reduced form separates the two channels in the simplest terms, and its three consequences survive the refinement. The non-augmenting combination $HM$ loses \sqc{$S - \Lambda$} at every level of the lift, since it carries none: it forfeits in performance the advantage a single-skill policy obtains at no cost, and recovers only the disagreement between the responses; under the linear rule it loses exactly $S$. The strongly augmenting $HM{+}{+}$ carries no $S$ term, because selecting the better skill on each instance also selects it on average, so its gain is $R \geq 0$ at $\gamma = 1$ and grows thereafter; on performance it has no break-even. The collaborating $HM{+}$ has one, at the $\gamma^{*}$ of Proposition~\ref{prop:breakeven}. The three archetypes are the corner settings of the two channels: $HM$ operates neither, $HM{+}$ synergy without routing, $HM{+}{+}$ perfect routing with synergy on top. They bound the arrangements rather than traverse them, which is the task of Section~\ref{sec:ladder}.

The refinement concerns the lift: a reduced form applies one multiplier to every instance, whereas an exchange reconciles responses that disagree differently from one instance to the next and falls below one where a right response is talked toward a wrong one, which is why Equation~\eqref{eq:gamma} measures the lift from what a protocol did on a caseload rather than positing it. The reduced form also carries an artefact the measured form does not: the ceiling at one. Applied instance by instance, the multiplier must be clipped at one, and the clipping lowers expected performance below what clipping the anchor would suggest, by more for the routed anchor than for the pooled response (Proposition~\ref{prop:ceiling}, Appendix~\ref{app:factors}).

With binary performance the reduced form adds only on the instances on which the skills disagree, which makes it a weak representation of synergy for tasks decided by hard responses (Proposition~\ref{prop:binarylift}, Appendix~\ref{app:factors}).

\subsection{A worked example}\label{sec:worked}

Four instances, two in each class, are enough to see every quantity of this section at once (Table~\ref{tab:worked}). Each skill responds with a probability that the instance is positive, and its performance is the squared rule's score of that response: skill $i$, responding $0.4$ on a positive instance, scores \sqc{$1 - 0.6^2 = 0.64$}.

\paragraph{The stake.} The skills' means are \sqc{$0.865$} for $i$ and \sqc{$0.825$} for $j$, so \sqc{$b = 0.865$}, \sqc{$\E[m] = 0.845$} and \sqc{$S = 0.020$}. The best skill on each instance averages \sqc{$\E[\Mx] = 0.955$}, so \sqc{$G = 0.110$} and \sqc{$R = 0.090$}: most of the stake is residual, although $i$ is better on average. Skill $j$ leads class~1, with \sqc{$0.875$} against \sqc{$0.775$}, and skill $i$ leads class~2, with \sqc{$0.955$} against \sqc{$0.775$}, so \sqc{$\sum_k p_k b_k = 0.915$}, \sqc{$\Rclass = 0.050$} and \sqc{$\Rinst = 0.040$}.

\paragraph{The arrangements, in ascending order.} Assigning each instance to a skill at random attains the mean anchor, and can never beat the commitment to the leader, since $\E[m] = b - S \leq b$. Pooling the responses with equal weights can, because their score exceeds the mean of the performances wherever they disagree (Equation~\eqref{eq:Lambda}): on the first instance $\hat{r}_i = 0.4$ and $\hat{r}_j = 1.0$ pool to $\hat{r}_{\bar{w}} = 0.7$, which scores \sqc{$0.91$} against a mean performance of \sqc{$0.82$}. Over the four instances the pooled response averages \sqc{$\E[\theta_{\bar{w}}] = 0.8925$}, above $b$, because the disagreement it earns, \sqc{$\Lambda = 0.0475$}, exceeds $S$. Sending each class to its incumbent attains $\sum_k p_k b_k$, collecting $\Rclass$ on top of $S$.

A protocol makes the skills interact before a joint response is fixed (Definition~\ref{def:protocol}). Take a consensus conference: each skill first reads the instance alone and records its response, which is retained; the two then show each other the evidence behind their responses; and they agree on one joint probability, given in the column $\hat{r}_\rho$. The joint response is not a function of the two retained responses, since it depends on what the exchange revealed. On the first and third instances the skill that saw the decisive evidence persuades the other, and the pair settles on $0.9$; on the second, where the skills responded $0.3$ and $0.5$, the exchange traces the doubt to an artefact and settles on $0.2$, which scores \sqc{$0.96$}, above the better skill's \sqc{$0.91$}; on the fourth, where they agreed on $0.3$, the discussion raises a doubt neither had held and moves them to $0.5$, which scores \sqc{$0.75$}, below both. Over the four instances the protocol attains \sqc{$0.9225$}: its lift over pooling is \sqc{$\gamma_\pi = 1.034$}, or \sqc{$\delta_\pi = 0.030$}, against a break-even of \sqc{$\gamma^{*} = b/\E[\theta_{\bar{w}}] = 0.969$}, which pooling's disagreement has pushed below one. Synergy is bounded by neither skill, in either direction.

Routing each instance to its winner, were the winner known, would attain $\E[\Mx]$ and collect the whole of $G$. The protocol, though it beats the best skill on one instance, falls short of it over the four: on these instances an exchange is worth more than knowing the class, and less than knowing the instance.

\begin{figure}[!htb]
\centering
\includegraphics[width=\linewidth]{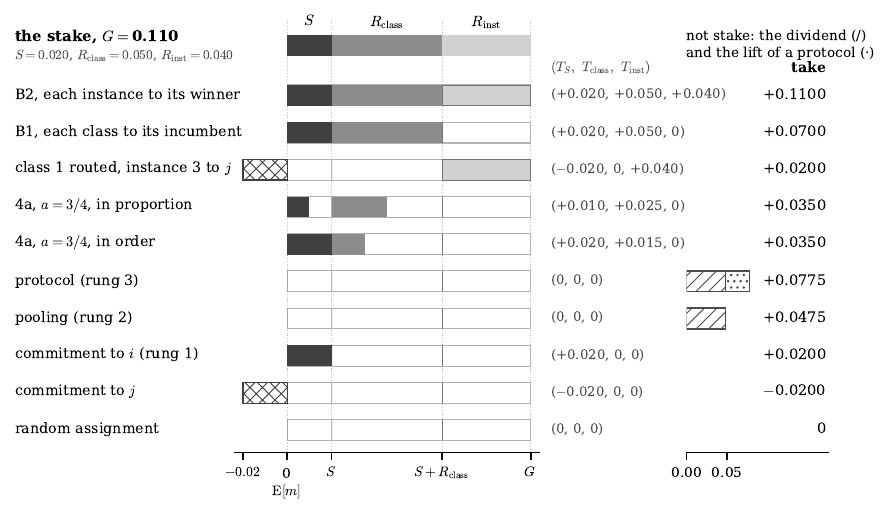}
\caption{What each arrangement of the worked example takes, component by component (Corollary~\ref{cor:attribution}): two skills, two classes and four instances, scored under the squared rule (Table~\ref{tab:worked}). At the top, the stake and its three components. Each row is a policy. In the left panel the lanes are the components, each filled from its start with the part the policy takes, and a cross-hatched bar left of $\E[m]$ marks a share below zero; the triple gives the shares. The right panel shows what the policy returns that is not stake, the dividend of combining responses (hatched) and the lift of the consensus protocol (dotted), and the take, $\E[\theta_\pi] - \E[m]$. Assignment on a class predicted correctly with probability $a = 3/4$ resolves neither class and is shown under both crediting rules (Figure~\ref{fig:attribution_sweep}). The class-1 router resolves class~1 but not class~2: it takes the whole of $\Rinst$, falls $S$ below the mean anchor at the level of the class, and ends at $b$.}
\label{fig:attribution}
\end{figure}

\paragraph{What each arrangement takes.} Corollary~\ref{cor:attribution} reads these arrangements against the stake (Figure~\ref{fig:attribution}). The commitment to $i$, assignment on the true class and routing each instance to its winner resolve every class, and take exactly $(T_S, T_{\mathrm{class}}, T_{\mathrm{inst}}) = $ \sqc{$(0.020, 0, 0)$}, \sqc{$(0.020, 0.050, 0)$} and \sqc{$(0.020, 0.050, 0.040)$}. Random assignment and pooling decide nothing and take nothing at the level of the class, so there is nothing to split: pooling collects only the dividend, \sqc{$0.0475$}, and the consensus protocol adds its lift, \sqc{$0.030$}, to it. The commitment to $j$ is the mirror of the commitment to $i$, and takes \sqc{$(-0.020, 0, 0)$} under either rule. Assignment on a class predicted correctly three times in four resolves neither class: its take at the level of the class, \sqc{$0.035$}, is credited \sqc{$(0.020, 0.015)$} in order, and \sqc{$(0.010, 0.025)$} in proportion under the convention of Appendix~\ref{app:crediting}, the values that Figure~\ref{fig:attribution_sweep} traces at $a = \frac{3}{4}$. A policy that leaves a class unresolved by reading the instance shows why the qualification matters: one that routes both instances of class~1 to their winner and sends the third instance to $j$ attains \sqc{$0.955$} in class~1 and \sqc{$0.775$} in class~2, and so $b$ overall. It resolves class~1 and takes the whole of \sqc{$\Rinst = 0.040$} there, but its error leaves class~2 below both points, \sqc{$0.090$} under its mean anchor, so that at the level of the class it takes \sqc{$-0.020$} in all, and credited in order it takes \sqc{$(-0.020, 0, 0.040)$}: a policy that reads the instance has given back at the level of the class what a commitment would have held.

\begin{table}[!htb]
\centering\footnotesize
\caption{A worked example: four instances, two classes, two skills and one protocol. Responses are probabilities that the instance is positive; performances are their scores under the squared rule, $\theta = 1 - (\hat{r} - r)^2$}
\label{tab:worked}
\setlength{\tabcolsep}{4.5pt}
\begin{tabular}{ccc cccc cccccc}
\toprule
 & & & \multicolumn{4}{c}{Responses} & \multicolumn{6}{c}{Performances} \\
\cmidrule(lr){4-7}\cmidrule(lr){8-13}
Instance & Class $k$ & $r$ & $\hat{r}_i$ & $\hat{r}_j$ & $\hat{r}_{\bar{w}}$ & $\hat{r}_\rho$ & $\theta_i$ & $\theta_j$ & $m$ & $\Mx$ & $\theta_{\bar{w}}$ & $\theta(\hat{r}_\rho, r)$ \\
\midrule
1 & 1 & 1 & 0.4 & 1.0 & 0.7 & 0.9 & \sqc{0.64} & \sqc{1.00} & \sqc{0.82} & \sqc{1.00} & \sqc{0.91} & \sqc{0.99} \\
2 & 1 & 0 & 0.3 & 0.5 & 0.4 & 0.2 & \sqc{0.91} & \sqc{0.75} & \sqc{0.83} & \sqc{0.91} & \sqc{0.84} & \sqc{0.96} \\
\multicolumn{7}{l}{\emph{Mean over class 1}} & \sqc{0.775} & \sqc{0.875} & \sqc{0.825} & \sqc{0.955} & \sqc{0.875} & \sqc{0.975} \\
\addlinespace
3 & 2 & 1 & 1.0 & 0.4 & 0.7 & 0.9 & \sqc{1.00} & \sqc{0.64} & \sqc{0.82} & \sqc{1.00} & \sqc{0.91} & \sqc{0.99} \\
4 & 2 & 0 & 0.3 & 0.3 & 0.3 & 0.5 & \sqc{0.91} & \sqc{0.91} & \sqc{0.91} & \sqc{0.91} & \sqc{0.91} & \sqc{0.75} \\
\multicolumn{7}{l}{\emph{Mean over class 2}} & \sqc{0.955} & \sqc{0.775} & \sqc{0.865} & \sqc{0.955} & \sqc{0.910} & \sqc{0.870} \\
\midrule
\multicolumn{7}{l}{Mean over all instances} & \sqc{0.865} & \sqc{0.825} & \sqc{0.845} & \sqc{0.955} & \sqc{0.8925} & \sqc{0.9225} \\
\bottomrule
\end{tabular}
\end{table}

\subsection{Magnitudes}\label{sec:magnitudes}

\begin{figure}[!htb]
\centering
\includegraphics[width=\linewidth]{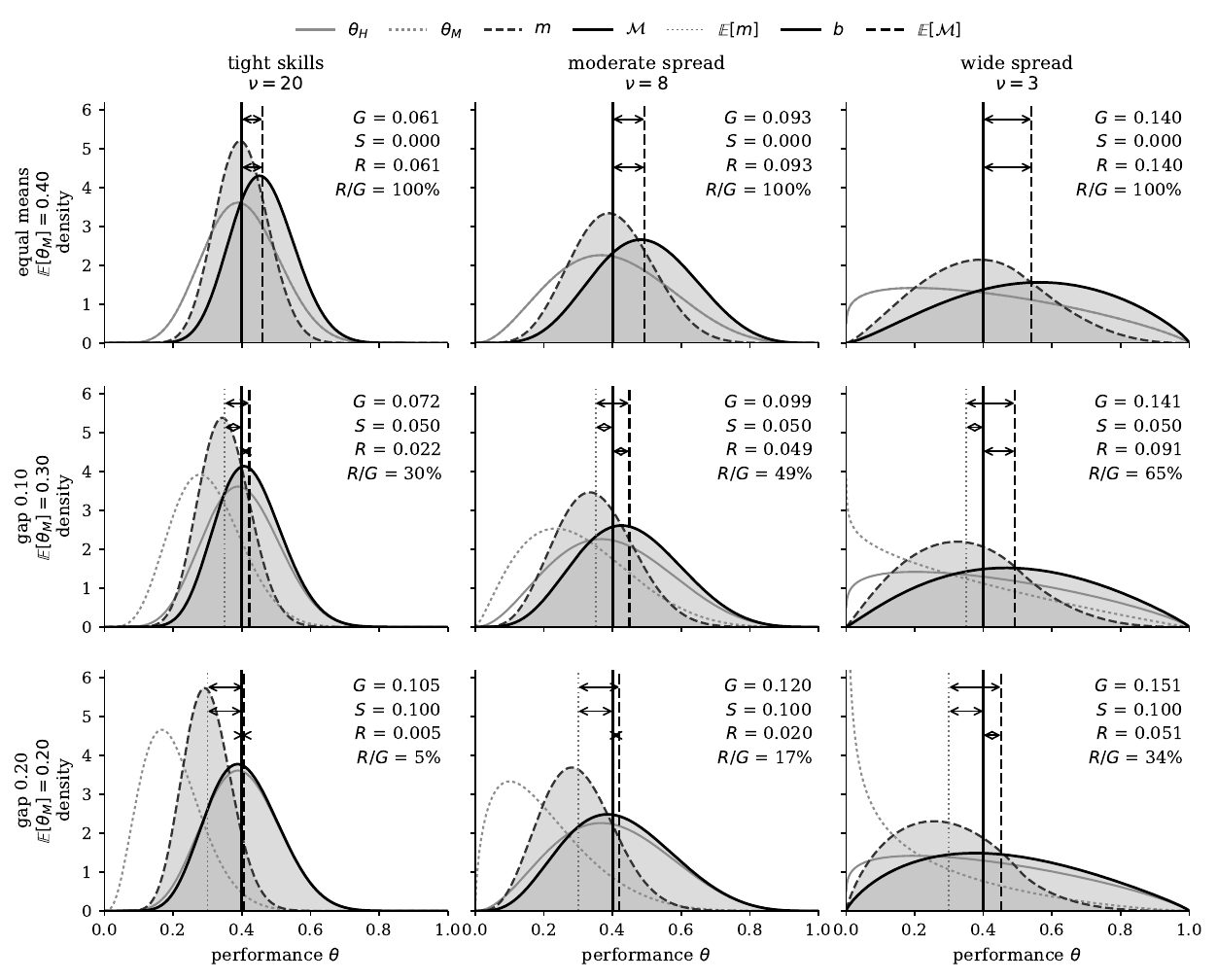}
\caption{The decomposition of Equation~\eqref{eq:gsr} on a grid of synthetic skill pairs, both skills Beta-distributed and independent on each instance, parameterized by their mean and a concentration $\nu$, with $\alpha = \E[\theta]\,\nu$ and $\beta = (1-\E[\theta])\,\nu$; $\E[\theta_H] = 0.40$ in every panel, so the informed baseline is $b = 0.40$ throughout. Rows set the gap $\E[\theta_H] - \E[\theta_M]$ and therefore $S = \mathrm{gap}/2$ (Proposition~\ref{prop:S}); columns set the concentration, larger $\nu$ giving tighter distributions. gray curves are the exact densities of the two skills ($\theta_H$ solid, $\theta_M$ dotted), and the shaded densities those of the anchors, $m$ with a dashed outline and $\Mx$ with a solid one; vertical lines mark $\E[m]$ (dotted), $b$ (solid) and $\E[\Mx]$ (dashed), and the arrows and the column on the right give $G$, $S$ and $R$ with the routable share $R/G$}
\label{fig:gsr_grid}
\end{figure}

Figure~\ref{fig:gsr_grid} takes the decomposition onto a grid of synthetic skill pairs in which its two drivers are varied independently. Rows fix the gap between the means, which by Proposition~\ref{prop:S} sets $S$ at exactly half the gap; columns vary the concentration of both distributions, which widens or narrows them without moving their means. The human is held at $\E[\theta_H] = 0.40$ and the machine is moved only downward from it, so $b = 0.40$ in every cell. Nothing is lost by fixing the direction: $\Mx$, $m$ and $b$ are symmetric functions of the two performances, so $G$, $S$ and $R$ depend on the pair only through $|\E[\theta_H] - \E[\theta_M]|$ and the unordered pair of distributions.

\begin{figure}[!htb]
    \centering
    \includegraphics[width=\textwidth]{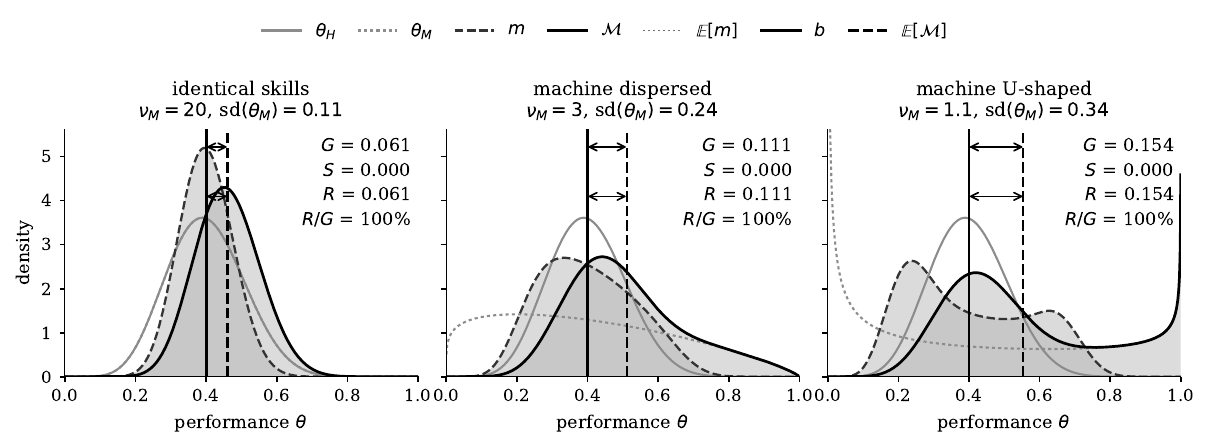}
    \caption{Equal means, three shapes. Both skills carry $\E[\theta] = 0.40$, so $\E[m]$ and $b$ coincide and $S = 0$; only the machine's density changes, from identical to the human's, through dispersed, to U-shaped. Reshaping moves $\E[\Mx]$ and raises $G$ and $R$ together, from $0.061$ to $0.154$, leaving the routable share at $100\%$. The invariance is not a property of the Beta family: with equal means the shapes cancel in the ratio whatever they are. Conventions as in Figure~\ref{fig:gsr_grid}; the machine's density in the third panel is unbounded at both ends}
    \label{fig:gsr_equal_means}
\end{figure}

Read as a whole, the grid shows the two drivers pulling in opposite directions. Dispersion is what creates routable value: across any row, widening the skills leaves $S$ untouched and multiplies $R$, from $0.005$ to $0.051$ in the bottom row. A gap between the means converts that value into the other component: down any column, $S$ rises by construction while $R$ collapses, from $0.061$ to $0.005$ in the tight column. With equal means the whole stake is routable, $R/G = 100\%$ whatever the shapes, since $S$ vanishes and the shapes cancel in the ratio; an ex ante commitment secures none of it. Figure~\ref{fig:gsr_equal_means} carries this further, holding both means at $0.40$ while the machine is reshaped from identical to the human, through dispersed, to U-shaped: $G$ and $R$ climb together from $0.061$ to $0.154$ and the share stays at $100\%$. The U-shaped machine is worse than the human on most instances and yet carries the largest $R$ of the three. Unreliability with occasional excellence is exactly what routing monetizes, and two track records reporting the same $0.40$ would conceal the whole difference. At the opposite corner, with a wide gap and tight skills, only $5\%$ of the stake is routable, and an organization there should commit to the better skill and stop. What the grid does not show is $G$ varying much: it stays between $0.061$ and $0.151$ across the nine panels while its composition changes completely. The size of the prize is the least informative thing about it.

Table~\ref{tab:gsr} reports the decomposition for the two illustrative profiles of Appendix~\ref{app:profiles}, a human and a machine at three levels of difficulty and two dates between which the human learns and the machine does not. In Profile A the machine holds a comparative advantage at low difficulty and the human at high; in Profile B the machine leads at every level. Within each difficulty level the pattern of Figure~\ref{fig:gsr_grid} reappears. In Profile A at the initial date, the residual is largest at high difficulty, where the two densities overlap most, and accounts for $70\%$ of the stake there against $12\%$ at low difficulty, where the machine leads on $85\%$ of instances. Proposition~\ref{prop:R} decomposes the high-difficulty figure exactly: the underdog, the machine, wins $40.5\%$ of instances by an average of $0.192$, which gives $R = 0.078$; at the final date, with the human improved, it wins $21.5\%$ by $0.158$, and $R$ falls to $0.034$. The ordering of $R/G$ across difficulty levels then inverts, from largest at high difficulty to largest at medium: the level at which routing is worth most is not fixed by the task but is wherever learning has last left the two skills level.

\begin{table}[!htb]
\centering\footnotesize
\setlength{\tabcolsep}{2.6pt}
\caption{The stake profile of the illustrative skill pairs of Appendix~\ref{app:profiles}, per difficulty level and for the task as a whole, taking the three levels as equiprobable classes. At the level of the task, the residual splits into the part a class-level rule reaches, $\Rclass$, and the rest, $\Rinst = R - \Rclass$ (Proposition~\ref{prop:class}); within a level the whole residual is instance-level. \sqc{$\E[\theta_{\bar{w}}]$} is what pooling the two responses attains under the squared rule, taking the profiles' distributions as performances under that rule, with the two responses on the same side of the truth, and \sqc{$\gamma^{*} = b/\E[\theta_{\bar{w}}]$} the break-even lift of Proposition~\ref{prop:breakeven}. The last column relaxes an assumption of the profiles: $R_{\mathrm{cop}}$ is the residual with the two performances coupled by a Gaussian copula of parameter $0.5$ instead of independent (Proposition~\ref{prop:coupling}). Expectations of the maximum are computed by numerical integration, performances being independent on each instance by construction of the profiles}
\label{tab:gsr}
\begin{tabular}{llrrrrrrrrrrr}
\toprule
 & $d$ & $\E[\theta_H]$ & $\E[\theta_M]$ & $b$ & $G$ & $S$ & $R$ & $\Rclass$ & $R/G$ & \sqc{$\E[\theta_{\bar{w}}]$} & \sqc{$\gamma^{*}$} & $R_{\mathrm{cop}}$ \\
\midrule
\multicolumn{13}{l}{\emph{Profile A}, $t=0$} \\
 & Low & 0.600 & 0.833 & 0.833 (M) & 0.132 & 0.117 & 0.015 & --- & 12\% & \sqc{0.740} & \sqc{1.126} & 0.005 \\
 & Med & 0.500 & 0.667 & 0.667 (M) & 0.118 & 0.083 & 0.035 & --- & 30\% & \sqc{0.598} & \sqc{1.114} & 0.015 \\
 & High & 0.400 & 0.333 & 0.400 (H) & 0.111 & 0.033 & 0.078 & --- & 70\% & \sqc{0.375} & \sqc{1.065} & 0.047 \\
 & \emph{Task} & 0.500 & 0.611 & 0.611 (M) & 0.121 & 0.056 & 0.065 & 0.022 & 54\% & \sqc{0.571} & \sqc{1.070} & 0.044 \\[3pt]
\multicolumn{13}{l}{\emph{Profile A}, $t=T$} \\
 & Low & 0.750 & 0.833 & 0.833 (M) & 0.077 & 0.042 & 0.035 & --- & 46\% & \sqc{0.803} & \sqc{1.037} & 0.018 \\
 & Med & 0.625 & 0.667 & 0.667 (M) & 0.098 & 0.021 & 0.078 & --- & 79\% & \sqc{0.658} & \sqc{1.014} & 0.050 \\
 & High & 0.545 & 0.333 & 0.545 (H) & 0.140 & 0.106 & 0.034 & --- & 24\% & \sqc{0.455} & \sqc{1.200} & 0.013 \\
 & \emph{Task} & 0.640 & 0.611 & 0.640 (H) & 0.105 & 0.015 & 0.090 & 0.042 & 86\% & \sqc{0.638} & \sqc{1.003} & 0.068 \\[3pt]
\multicolumn{13}{l}{\emph{Profile B}, $t=0$} \\
 & Low & 0.600 & 0.833 & 0.833 (M) & 0.132 & 0.117 & 0.015 & --- & 12\% & \sqc{0.740} & \sqc{1.126} & 0.005 \\
 & Med & 0.500 & 0.762 & 0.762 (M) & 0.140 & 0.131 & 0.009 & --- & 7\% & \sqc{0.651} & \sqc{1.170} & 0.002 \\
 & High & 0.400 & 0.600 & 0.600 (M) & 0.132 & 0.100 & 0.032 & --- & 24\% & \sqc{0.514} & \sqc{1.168} & 0.013 \\
 & \emph{Task} & 0.500 & 0.732 & 0.732 (M) & 0.135 & 0.116 & 0.019 & 0.000 & 14\% & \sqc{0.635} & \sqc{1.152} & 0.007 \\[3pt]
\multicolumn{13}{l}{\emph{Profile B}, $t=T$} \\
 & Low & 0.750 & 0.833 & 0.833 (M) & 0.077 & 0.042 & 0.035 & --- & 46\% & \sqc{0.803} & \sqc{1.037} & 0.018 \\
 & Med & 0.625 & 0.762 & 0.762 (M) & 0.099 & 0.068 & 0.031 & --- & 31\% & \sqc{0.707} & \sqc{1.078} & 0.013 \\
 & High & 0.545 & 0.600 & 0.600 (M) & 0.101 & 0.027 & 0.074 & --- & 73\% & \sqc{0.583} & \sqc{1.030} & 0.047 \\
 & \emph{Task} & 0.640 & 0.732 & 0.732 (M) & 0.092 & 0.046 & 0.047 & 0.000 & 50\% & \sqc{0.697} & \sqc{1.049} & 0.026 \\[3pt]

\bottomrule
\end{tabular}
\end{table}

Taking the difficulty levels as classes shows what Proposition~\ref{prop:class} adds. In Profile A at the initial date, the task-level commitment goes to the machine, whose mean over the whole task is higher, and the residual of $0.065$ contains $0.022$ that a class rule sending high-difficulty instances to the human would collect without looking at any instance. At the final date the task-level commitment switches to the human, while the machine remains incumbent at low and medium difficulty, and $\Rclass$ nearly doubles, to $0.042$ of a residual of $0.090$. In Profile B the machine is incumbent at every level, $\Rclass = 0$, and the whole residual can be reached only instance by instance. The class structure pays only where incumbency changes across classes.

\begin{proposition}[Coupling]\label{prop:coupling}
For a pair with given marginal distributions in each class, a joint distribution that is larger in the concordance order (a larger joint distribution function at every point, the marginals fixed; \citealp{nelsen2006introduction}), as a Gaussian copula with a larger parameter is, leaves $\E[m]$, $b$, $S$ and $\Rclass$ unchanged and does not raise $\E[\Mx]$, $G$, $R$ or $\Rinst$.
\end{proposition}

\begin{proof}
The first four depend on the marginals alone (Propositions~\ref{prop:anchor} and~\ref{prop:class}). Within each class, $\E[\max(\theta_i, \theta_j)] = \E[\theta_i] + \E[\theta_j] - \E[\min(\theta_i, \theta_j)]$, and $\E[\min(\theta_i, \theta_j)]$ does not fall as the joint distribution rises in the concordance order, $\min$ being supermodular; Gaussian copulas are ordered in concordance by their parameter \citep{nelsen2006introduction}. $G$, $R$ and $\Rinst$ differ from $\E[\Mx]$ by quantities fixed by the marginals.
\end{proof}

The profiles draw the two performances independently on each instance, and independence is not a neutral choice. Human and machine errors are often positively correlated, both failing on the same hard instances, and for given marginals a more concordant coupling leaves $b$, $S$ and $\Rclass$ unchanged and does not raise $\E[\Mx]$, $G$ or $R$ (Proposition~\ref{prop:coupling}). The column $R_{\mathrm{cop}}$ of Table~\ref{tab:gsr} recomputes the residual with the performances coupled by a Gaussian copula of parameter $0.5$: at high difficulty in Profile~A at the initial date it falls from $0.078$ to $0.047$, and at the level of the task from $0.065$ to $0.044$, while $S$ and $\Rclass$ stay where they were. The independent profiles are therefore a generous case. Complementary skills, whose errors are negatively correlated, would carry more than they show, and which case holds is a property of the pair that only paired observation reveals. Assistance can itself raise the concordance: people shown a model's advice converge on it \citep{fugener2021will}, which by Proposition~\ref{prop:coupling} lowers the residual the pair carries.

The features that govern these magnitudes, the coordinates that reduce them to a few, and a protocol for exploring them numerically are set out in Appendix~\ref{app:factors}.

\subsection{Identification before deployment}\label{sec:identification}

Equation~\eqref{eq:gsr} and Proposition~\ref{prop:class} separate the identification problem into questions requiring different evidence. What the marginals alone determine comes first; the evidence then comes in four levels (Table~\ref{tab:evidence}). The levels concern the studies an organization runs; Appendix~\ref{app:layers} orders the same quantities by what must be known of the skills to establish them, and shows that pairwise moments already bound the residual before its full law is known.

\begin{proposition}[What the marginals determine]\label{prop:anchor}
$\E[m] = \mean_s \E[\theta_s]$ and $b = \max_s \E[\theta_s]$ are determined by the marginal distributions of the performances, and so is $S = b - \E[m]$. $\E[\Mx]$ is not determined by the means $\{\E[\theta_s]\}$, nor by the marginal distributions of the $\theta_s$, and neither are $G$ and $R$. Given the marginal distribution functions $F_s(x) = \Pr(\theta_s \leq x)$, with quantile functions $F_s^{-1}$, they are bounded: with $V$ uniform on $[0,1]$,
\[
\E\big[\max_s F_s^{-1}(V)\big] - b \;\leq\; R \;\leq\; \min_v \Big\{ v + \textstyle\sum_s \E\big[(\theta_s - v)^{+}\big] \Big\} - b,
\]
the lower bound attained when the performances are comonotone, and the upper, for a pair, when they are countermonotone.
\end{proposition}

The mean anchor is therefore recoverable from separate evaluations of the skills, and the routed anchor only bounded. The two bounds serve differently. The lower one sets every skill at the same quantile of its own distribution on each instance, $\theta_s = F_s^{-1}(V)$, so that the skills are good and bad together; it is positive only where the leader's distribution does not dominate every other skill's, their quantile functions crossing, so that separate assessments can establish that a residual exists only in that case; the upper one is always available, and caps what any routing could collect. In it, $v$ is a threshold on the scale of performance: on every instance the best performance is at most $v$ plus what each skill scores above $v$, with equality where exactly one skill exceeds it, so that every $v$ gives a bound and the minimum over $v$ gives the tightest. For continuous distributions the minimizing threshold leaves on average exactly one skill above it, $\sum_s \Pr(\theta_s > v) = 1$, and the bound is attained by the coupling under which, on every instance, exactly one skill exceeds that threshold: the skills take turns at being good and are never good together. For Profile~A at high difficulty and the initial date (Table~\ref{tab:gsr}), the marginals alone place $R$ between $0$ and $0.125$, against $0.078$ under independence. The problem is that of the joint distribution of potential outcomes in program evaluation, where the marginals, and with them the means and the best uniform policy, are identified from the separate arms of an experiment, while features of the distribution of gains are at best bounded \citep{manski1997mixing, heckman1997making, fan2010sharp}. This asymmetry is the reason why the quantities of Equation~\eqref{eq:gsr} cannot all be recovered from separate track records, and it runs through the rest of the paper.

\paragraph{Level 1: separate assessments.} Each skill is assessed on the task on its own. The assessments identify the marginals, and with them $b$, $S$ and the commitment; the break-even $\gamma^{*}$ needs more under the squared rule, since what pooling attains depends on how the responses co-occur. They can come from separate periods, separate caseloads, even separate organizations, though a vendor's benchmark is measured on instances that are not the organization's own and is not the record the estimate $\hat{\E}[\theta_s]$ requires.

\paragraph{Level 2: separate assessments by class.} The same assessments, sorted by class, identify the incumbencies and $\Rclass$, still without any instance worked twice. They need only the class of each assessed instance to be recorded.

\paragraph{Level 3: paired assessment.} A two-arm trial in which the same instances are executed by each skill alone identifies $\E[\Mx]$, and with it $G$, $R$ and $\Rinst$, together with the disagreement \sqc{$\Lambda$} that pooling earns and with it the break-even $\gamma^{*}$. That pairing cannot be relaxed: separate track records, vendor benchmarks and published accuracy figures estimate $b$ and $S$ but cannot reach $R$ at all, and two pairs with identical marginal distributions can have $R = 0$ or $R$ at its maximum depending only on their coupling (Proposition~\ref{prop:anchor}). Since both arms return responses that must be scored, the trial can run only on instances whose correct response eventually becomes knowable (Section~\ref{sec:performance}).

Nothing at levels~1 to~3 depends on a workflow: they measure the stake, not what any workflow takes of it. What the paired trial delivers is a ceiling rather than a target: $R$ vanishes when one skill wins almost every instance and grows with how often the other takes over (Propositions~\ref{prop:R} and~\ref{prop:R0}), while a workflow decides on a signal it can see beforehand and so collects only part of it, and one with nothing to decide on collects none, however large $R$ may be. This is the distinction \citet{hemmer2025complementarity} draw between the complementarity potential of a human--AI pair and the effect a team actually realizes.

\paragraph{Level 4: a third arm.} A third arm, in which the same instances are executed by the workflow itself, identifies the take: the share of $R$ a routing arrangement collects, or the lift $\gamma_\pi$ a synergy protocol achieves, which by Definition~\ref{def:lift} requires the independent responses to have been retained alongside the joint one. Such designs are neither hypothetical nor rare: the CXR experiment of \citet{agarwal2023combining} implemented one, reading the same films under different conditions, and the preregistered inclusion criteria of \citet{vaccaro2024combinations} required all 106 experiments they synthesize to report the performance of the human alone, the AI alone, and the human--AI system.

\paragraph{What a finite record adds.} Each of these quantities is estimated with error, and not symmetrically. The informed baseline is estimated as the larger of the estimated means, $\hat{b} = \max_s \hat{\E}[\theta_s]$, which is biased upward, since the maximum of noisy estimates tends to exceed the estimate of the maximum: the winner's curse of selecting the apparent leader, or the optimizer's curse of decision analysis \citep{smith2006optimizer}. The estimate $\hat{S}$ inherits the upward bias, and $\hat{R} = \hat{\E}[\Mx] - \hat{b}$ the downward one, since $\hat{\E}[\Mx]$ is unbiased. Where a skill's response to the same instance varies from one execution to the next, as for a model sampled at a positive temperature or a reader on a second reading, $\Mx$ is the best of single executions, and part of $\Rinst$ is that variability, which no criterion reading the instance can predict; repeated executions on a subset of instances separate it from the part a criterion can reach. The bias is largest where the skills are closest: for two skills with equal means and Beta performances of concentration $20$, the bias of $\hat{R}$ is $-0.008$ with $50$ paired instances and $-0.003$ with $500$. Splitting the record removes the selection bias. With $\hat{\E}_1$ and $\hat{\E}_2$ the means over two disjoint parts of the record, the leader is chosen on the first and its mean estimated on the second,
\[
\hat{b}_{\mathrm{split}} = \hat{\E}_2\big[\theta_{i^{*}_1}\big], \qquad i^{*}_1 = \argmax_s \hat{\E}_1[\theta_s],
\]
which is unbiased for the mean of the skill it selects, and so falls short of $b$ when that skill is not the leader: it errs, if anything, downward, and in the case of equal means above it is unbiased. Cross-fitting exchanges the roles of the parts, or rotates them over several folds, and averages the resulting estimates, so that the whole record serves both to choose and to estimate. Computed from $\hat{b}_{\mathrm{split}}$, $\hat{S}$ errs, if anything, downward and $\hat{R}$ upward: the split reverses the directions of their biases rather than removing them. Precision is set by the spread of $\Mx - \theta_L$: in Profile~A at high difficulty and the initial date, about $180$ paired instances give $\hat{R}$ a standard error of $0.01$, against a residual of $0.078$.

\begin{table}[!htb]
\centering\footnotesize
\caption{The levels of evidence and the layers of knowledge. Each level of evidence is a kind of record (Appendix~\ref{app:records}); each layer is what is known of the skills (Appendix~\ref{app:layers}). The two do not coincide: paired observation reaches two layers at once, the third arm reaches none, since what it measures belongs to the workflow, and a record of finite size is not a further level but the error with which every level is read}
\label{tab:evidence}
\setlength{\tabcolsep}{4pt}
\begin{tabular}{>{\raggedright\arraybackslash}p{2.9cm} >{\raggedright\arraybackslash}p{3.4cm} >{\raggedright\arraybackslash}p{2.6cm} >{\raggedright\arraybackslash}p{4.4cm}}
\toprule
Level of evidence & Record & Layer & What it identifies \\
\midrule
Separate assessments & $\mathcal{R}_s$ for every $s$ & 1: the means, and the marginal distributions & $b$, $S$ and the commitment $i^{*}$; bounds on $R$ (Proposition~\ref{prop:anchor}) \\[3pt]
Separate assessments by class & $\mathcal{R}_{s,k}$ for every $s$ and $k$ & 1, by class & the incumbents $\bar{s}(k)$ and $\Rclass$ \\[3pt]
Paired assessment & $\mathcal{R}_{\mathcal{S}}$, with the responses retained & 2: pairwise moments; 3: the joint law & \sqc{$\Lambda$} and $\gamma^{*}$; $\E[\Mx]$, and with it $G$, $R$ and $\Rinst$ \\[3pt]
A third arm & the workflow's output on the instances of $\mathcal{R}_{\mathcal{S}}$, with the independent responses retained & none: the take is a property of the workflow & the take of a routing arrangement; the lift $\gamma_\pi$ of a protocol \\[3pt]
\midrule
A finite record & any of the above & 4: finite records & each of the above with error; $\hat{b}$ biased upward, $\hat{R}$ downward \\
\bottomrule
\end{tabular}
\end{table}

\section{The ladder: traversing the arrangements}\label{sec:ladder}

Section~\ref{sec:bounds} fixed what the skills put on the table and what each channel can take of it. Between those bounds lies everything an organization can actually deploy. This section enumerates it. It asks, rung by rung, which policies an organization could run, what each attains, what each holds available and executes, and what each requires to have been recorded before it can be run at all. What it delivers is a structural inventory: the attainment and the cost profile of every position, the second reported without being priced against the first, which is left to future work. The purpose is to establish the option space over which a skill policy decision is taken, so that the alternatives can be compared rather than collapsed into a binary choice between autonomous and supervised operation.

\subsection{What is known, and when}\label{sec:signals}

An arrangement decides before the outcome of the instance is known, so what it can condition on is limited to what exists at that moment. A \emph{signal} is any observable quantity describing the instance before execution, on which assignment or weighting can be conditioned. Two are distinguished throughout. $Z$ is informative about the class of the instance, at quality $q_z$, written $Z(q_z)$ or $Z_q$. $W$ is informative about each skill's standing on the individual instance, at quality $q_w$, written $W(q_w)$ or $W_q$; a skill's own confidence in its response, where it is available and informative \citep{kadavath2022know, kumaran2026causal}, is such a signal about that skill. Combined with what adjudication supplies, the true class for $Z$ and the correct response for $W$, from which the vector of true performances follows, signals can be used to fit predictive models. They can also be used to choose which policy applies to an instance rather than to set the parameters within one, which we call \emph{gating}; since the choice among policies turns on what they execute and hold available, and so on their cost profiles, gating is left to future work.

Three symbols denote the skill a policy sends an instance to. $\hat{s}$ is the skill a fitted policy selects: its criterion's choice where it is made from the signals before execution, as in assignment, or its aggregation's where it is made from the responses after it, as in selection. $\bar{s}(k)$ is the incumbent of class $k$, which requires a class to be named, the true one or a predicted $\hat{k}$. And $\dot{s} = \argmax_s \theta_s$ is the winner of the instance, as distinct from the leader $L$, which is chosen ex ante; it requires the outcome. They differ only in how the skill was determined: the bar averages over a class, the dot is realized on the instance, the hat is predicted from the signals. Where a policy weights rather than selects, the weights belong to its aggregation (Section~\ref{sec:policies}) and may be set from the same signals; $\hat{w}_s$ denotes the weight given to skill $s$ on the instance, hatted because it is predicted and the weighting that would have been best is not available when it must be set. As throughout, the hat marks what stands in for a quantity that cannot be had; $\widehat{\Pr}(\cdot)$ denotes a fitted estimate of a probability, and $\hat{\E}[\cdot \mid \cdot\,]$ a fitted estimate of a conditional expectation, a function of what is conditioned on, evaluated on the instance. $p_k$ denotes the probability of class $k$, a property of the caseload no policy alters.

Everything a policy estimates comes from \emph{records} of past instances, and records differ in what they hold. $\mathcal{R}_s$ holds instances executed by skill $s$, scored; $\mathcal{R}_{s,k}$ is its part of class $k$; $\mathcal{R}^{Z}$ holds instances with their class and their $Z$, whoever executed them; and $\mathcal{R}_{\mathcal{S}}$ holds instances executed by \emph{every} skill in $\mathcal{S}$, with superscripts naming the signals and responses retained. The family $\{\mathcal{R}_s\}$ and the single record $\mathcal{R}_{\mathcal{S}}$ are different objects: the second can be empty while every member of the first is large, since an organization that has never engaged two skills on the same instance has no comparison to draw however long it has been working. Appendix~\ref{app:records} sets the records out. Figure~\ref{fig:when} places every quantity in time.

\begin{figure}[H]
\centering
\small
\setlength{\tabcolsep}{4pt}

\begin{tabular}{@{}p{0.30\textwidth}p{0.30\textwidth}p{0.30\textwidth}@{}}
\centering\bfseries Setup & \centering\bfseries Execution
  & \centering\arraybackslash\bfseries Adjudication \\[2pt]
\end{tabular}%
\par\vspace{-2pt}
\rule{0.95\textwidth}{0.4pt}%
\par\vspace{-9pt}
\makebox[0.95\textwidth]{%
  \hspace{0.145\textwidth}\rule{0.4pt}{7pt}%
  \hfill\rule{0.4pt}{7pt}%
  \hspace{0.155\textwidth}\hfill\rule{0.4pt}{7pt}%
  \hspace{0.135\textwidth}}
\par\vspace{2pt}
\begin{tabular}{@{}p{0.30\textwidth}p{0.30\textwidth}p{0.30\textwidth}@{}}
\centering\itshape\footnotesize once, on the records
  & \centering\itshape\footnotesize every instance
  & \centering\arraybackslash\itshape\footnotesize later, some instances \\
\end{tabular}

\vspace{12pt}

\begin{tabular}{@{}>{\raggedright\arraybackslash}p{0.46\textwidth}
                  >{\raggedright\arraybackslash}p{0.46\textwidth}@{}}

\textbf{Carried by the instance} &
\textbf{Arrives at adjudication} \\[2pt]

$Z_q$\quad observations bearing on its class &
$r$\quad the correct response \\
$W_q$\quad observations bearing on each skill's standing &
$k$\quad the class it really was \\
 & $\theta_s$\quad what each skill actually achieved \\[14pt]

\textbf{Derived from the records, at Setup} &
\textbf{Available only then} \\[2pt]

$\hat{\E}[\theta_s]$\quad from $\mathcal{R}_s$, hence $i^{*}$ &
$\bar{s}(k)$\quad incumbent of the true class \\
$\hat{\mu}_{s,k}$\quad from $\mathcal{R}_{s,k}$, hence $\bar{s}(\cdot)$ &
$\dot{s} = \argmax_s \theta_s$\quad the winner \\
$\widehat{\Pr}(k \mid Z_q)$\quad from $\mathcal{R}^{Z}$ & \\
$\hat{\E}[\theta_s \mid \cdot\,]$\quad from $\mathcal{R}_{\mathcal{S}}$ & \\[18pt]

\textbf{Predicted on the instance} &
\textbf{And the records grow} \\[2pt]

$\hat{k}$\quad its class &
$\mathcal{R}_s$, $\mathcal{R}_{s,k}$\quad if one skill executed \\
$\hat{s}$\quad the skill that will do best &
$\mathcal{R}_{\mathcal{S}}$\quad only if every skill executed \\
$\hat{w}_s$\quad the weight each skill's response carries &
$\mathcal{R}^{Z}$\quad whoever executed \\[14pt]

\textbf{Produced by executing} &
\textbf{Measured only here} \\[2pt]

$\hat{r}_s$\quad what each engaged skill answered &
$\theta_s$, $\theta$\quad performance, by $\theta(\hat{r}, r)$ \\
$\hat{r}$\quad what the policy returns, by $\rho$, by $\hat{w}$, or unchanged &
$\gamma$\quad what a protocol $\rho$ was worth \\

\end{tabular}

\caption{What exists when. A policy reads what the instance carries and what the records have already yielded, and returns a response; everything on the right arrives later and on some instances only. The records a policy leaves behind are not the records it needs: $\mathcal{R}_{\mathcal{S}}$ grows only where every skill executed, so an arrangement that commissions one consumes that record and replenishes none of it}
\label{fig:when}
\end{figure}

\subsection{Reading the ladder}\label{sec:reading}

Table~\ref{tab:spectrum} enumerates the arrangements, and Table~\ref{tab:ladder} summarizes what each rung pays and buys in performance. The rungs are read from the bottom up, each adding one decision to the one below. The \emph{ladder orders arrangements by how much they decide, not by what they attain}: a rung can sit high and still attain less than one below it, as Section~\ref{sec:ladderfig} shows on the illustrative profiles. Two rungs are special. Rung~2 does not dominate rung~1 in expectation unless the responses disagree by more than $S$: it pays \sqc{$S - \Lambda$} to buy the insurance of Proposition~\ref{prop:insurance}. Rung~3 pays the same and recovers it only if its protocol lifts performance past $\gamma^{*}$.

\emph{Attains} is the performance of the policy, in expectation over the instance space and per instance. It does not settle what a policy is worth, which depends also on what it executes and holds available and on the consequences of its errors, and is left to future work. The \emph{Rout.} and \emph{Syn.} columns record which channel the policy operates, read from its design (Definitions~\ref{def:routing} and~\ref{def:synergy}); \emph{Arch.} records its exact correspondence with an archetype of Equation~\eqref{eq:archetypes}; and \emph{Depl.} whether it is deployable, meaning that everything it conditions on is observable when it acts: before execution for the engagement criterion, and before adjudication for the aggregation. Attainment is stated under the squared rule; under the linear rule, $\Lambda = 0$ and the entries built on a mechanical combination reduce to the performance anchors (Proposition~\ref{prop:hull}).

Table~\ref{tab:costs} gives the cost profile of each arrangement (Section~\ref{sec:costs}). Two regularities stand out. Availability separates the bottom of the ladder from the rest: rungs~0 and~1 hold one skill, every arrangement from rung~2 upward holds the whole available set, and assignment on the class is the one exception, holding only the skills that lead some class. Above that line the arrangements differ in what they execute: assignment executes one skill per instance, ensembling and selection execute every skill whatever the weights or the choice, and a protocol adds the cost of its exchanges. The profile records what each position commits and consumes; it does not say whether the commitment repays itself, which depends on the value of what is attained (Section~\ref{sec:costs}).

\begin{table}[!htb]
\centering
\footnotesize
\caption{The spectrum of skill policies: the ceilings of each routing family above, the ladder below, its rungs read from the bottom up. Sums over $s$ run over the available set; attainment is an expectation over the instance space, per instance}
\label{tab:spectrum}
\setlength{\tabcolsep}{3pt}\begin{tabular}{c l l c c c c}
\toprule
Rung & Skill policy & Attains & Rout. & Syn. & Arch. & Depl. \\
\midrule
\multicolumn{7}{l}{\textit{Ceilings on what each routing family can attain (Section~\ref{sec:ceilings})}} \\[2pt]
B3& $ij(w \mid \mathrm{inst})$ & $\E[\Mx]$ & Y & N & $HM{+}{+}_{\gamma=1}$ & N \\
B2& $i,j \mid \mathrm{inst}$ & $\E[\Mx]$ & Y & N & --- & N \\
B1& $i,j \mid \mathrm{class}$ & $\sum_k p_k \, \mu_{\bar{s}(k),k} = b + \Rclass$ & Y$^\dagger$ & N & --- & N \\
\midrule
\multicolumn{7}{l}{\textit{The ladder}} \\[2pt]
5 & $ij(w \mid Z_q, W_q){+}$ & $\gamma_\pi\, \E[\theta(\textstyle\sum_s \hat{w}_s \hat{r}_s, r)]$ & Y & Y &  ---  & Y \\[3pt]
\cmidrule(lr){1-1}
4e & $i,j \mid Z_q, W_q, \{\hat{r}_s\}$  & $\E[\theta_{\hat{s}}]$, $\hat{s}$ read on the responses & Y & N &  ---  & Y \\
4d & $ij(w \mid Z_q, W_q)$ & $\E[\theta(\textstyle\sum_s \hat{w}_s \hat{r}_s, r)]$ & Y & N &  ---  & Y \\
4c+a & $i,j \mid Z_q, W_q$, audited & as 4c & Y & N & --- & Y \\
4c & $i,j \mid Z_q, W_q$ & no closed form & Y & N & --- & Y \\
4b & $i,j \mid W_q$ & no closed form & Y & N & --- & Y \\
4a & $i,j \mid Z_q$ & $\sum_k p_k \sum_{\hat{k}} \Pr(\hat{k} \mid k)\, \mu_{\bar{s}(\hat{k}),k}$ & Y$^\dagger$ & N & --- & Y \\[3pt]
\cmidrule(lr){1-1}
3 & $ij{+}$ & \sqc{$\gamma_\pi\, \E[\theta_{\bar{w}}]$} & N & Y & $HM{+}$ & Y \\[3pt]
\cmidrule(lr){1-1}
2 & $ij$ & \sqc{$\E[\theta_{\bar{w}}] = b - S + \Lambda$} & N & N & $HM$ & Y \\
2r & $i,j \mid \varnothing$ & $\E[m] = b - S$ & N & N & --- & Y \\[3pt]
\cmidrule(lr){1-1}
1 & $i^{*} = \argmax_s \hat{\E}[\theta_s]$ & $b$ & N & N & $H$ or $M$ & Y \\[3pt]
\cmidrule(lr){1-1}
0 & $i$ & $\E[\theta_i]$ & N & N & $H$ or $M$ & Y \\
\bottomrule
\addlinespace[3pt]
\multicolumn{7}{p{0.92\textwidth}}{\footnotesize
Attainment under the squared rule. At rungs 2 and 3, \sqc{$\E[\theta_{\bar{w}}]$} is the mean response scored, which exceeds the mean of the performances by the disagreement \sqc{$\Lambda$} and equals it under the linear rule (Proposition~\ref{prop:hull}); at rungs 3 and 5 the entry is the measured lift times the attainment of the combination the protocol modifies (Equation~\ref{eq:gammapi}). Row 2r, random assignment, conditioned on nothing ($\varnothing$), is not a rung but the arrangement that attains the mean anchor exactly, under any rule; under the squared rule it falls short of pooling by \sqc{$\Lambda$}, the curvature dividend (Proposition~\ref{prop:randomize}). In the class-assignment row the two subscripts of $\mu_{\bar{s}(\hat{k}),k}$ differ on purpose: the criterion chooses the skill through $\hat{k}$ and the instance chooses the distribution through $k$, so a misclassified instance is executed by the incumbent of a class it does not belong to and scored on the class it does. $\dagger$ marks routing at class level, where every instance of a class is assigned alike. $HM{+}{+}$ corresponds to B3 only at $\gamma = 1$; with $\gamma > 1$ it is the reduced-form upper bound of rung~5.}
\end{tabular}
\end{table}

\begin{table}[!htb]
\centering
\footnotesize
\setlength{\tabcolsep}{3pt}
\caption{What each rung pays and buys in performance, with its anchors in the literature}
\label{tab:ladder}
\begin{tabularx}{\textwidth}{@{}c >{\raggedright\arraybackslash\hsize=0.8\hsize}X >{\raggedright\arraybackslash\hsize=1.1\hsize}X >{\raggedright\arraybackslash\hsize=1.1\hsize}X@{}}
\toprule
Rung & Configuration & Pays and buys & Anchors \\
\midrule
5 & \textbf{Combining information, not just responses}: routing, then an exchange in which each skill can revise its response in light of the other's & Synergy on top of rung~4's routing; the lift is measured, is bounded by none of the skills, and can fall below one & Second opinions between skills; centaur mode \citep{randazzo2025cyborgs, dell2026navigating} \\ \addlinespace[3pt]
4 & \textbf{Directing work given signals}: a policy conditioned on the class or on the instance assigns, weights or selects & Collects $R$ in proportion to the value of the signal, up to $\Rclass$ on a class signal and up to all of it at perfect routing; requires paired evidence, which assignment stops producing & Function allocation per instance \citep{parasuraman2000model}; learning to defer \citep{madras2018predict, mozannar2020consistent}; dynamic selection \citep{cruz2018dynamic} \\ \addlinespace[3pt]
3 & \textbf{Together against uncertainty}: independent responses, then an exchange before the joint response is fixed & Inherits rung~2's insurance at cost \sqc{$S - \Lambda$}; recovers it only past $\gamma^{*}$; lift measured against the pre-exchange mean response & Consensus double reading \citep{johnston1999two}; the organizational loop \citep{herrmann2023keeping} \\ \addlinespace[3pt]
2 & \textbf{Putting both skills at work}: independent responses averaged, no routing & Pays \sqc{$S - \Lambda$}; caps the shortfall at \sqc{$S - \Lambda$} against $2S$ for a wrong commitment (Proposition~\ref{prop:insurance}); narrows dispersion on both tails; the default landing place of oversight by mandate & Supervisory-control failure \citep{parasuraman2010complacency, green2022flaws}; the human in the average \citep{crootof2023humans} \\ \addlinespace[3pt]
1 & \textbf{One decision takes all}: commit ex ante from separate assessments & Secures $S$ for the assessment alone; a wrong commitment forfeits $2S$ & Comparative advantage \citep{autor2003skill, acemoglu2025simple} \\ \addlinespace[3pt]
0 & \textbf{Nothing to decide} & No assessment of the skill in use, and no alternative considered & The status quo \\
\bottomrule
\end{tabularx}
\end{table}

\begin{table}[!htb]
\centering
\footnotesize
\caption{The cost profile of each arrangement: the standing cost of what it holds available, incurred whatever the volume, and the expected execution cost of what it engages on an instance. Sums over $s$ run over the available set; $\bar{s}(\mathcal{K})$ is the set of skills that lead at least one class}
\label{tab:costs}
\setlength{\tabcolsep}{4pt}
\renewcommand{\arraystretch}{1.3}
\begin{tabular}{c l l l}
\toprule
Rung & Skill policy & Standing cost $\kappa_\pi$ & Expected execution cost $\E[c_\pi]$ \\
\midrule
\multicolumn{4}{l}{\textit{Ceilings (Section~\ref{sec:ceilings})}} \\[2pt]
B3 & $ij(w \mid \mathrm{inst})$ & $\sum_s \kappa_s$ & $\sum_s c_s$ \\
B2 & $i,j \mid \mathrm{inst}$ & $\sum_s \kappa_s$ & $\E[c_{\dot{s}}]$ \\
B1 & $i,j \mid \mathrm{class}$ & $\sum_{s \in \bar{s}(\mathcal{K})} \kappa_s$ & $\sum_k p_k\, c_{\bar{s}(k)}$ \\
\midrule
\multicolumn{4}{l}{\textit{The ladder}} \\[2pt]
5 & $ij(w \mid Z_q, W_q){+}$ & $\sum_s \kappa_s$ & $\sum_s c_s + c_\rho$ \\
4e & $i,j \mid Z_q, W_q, \{\hat{r}_s\}$ & $\sum_s \kappa_s$ & $\sum_s c_s$ \\
4d & $ij(w \mid Z_q, W_q)$ & $\sum_s \kappa_s$ & $\sum_s c_s$ \\
4c+a & $i,j \mid Z_q, W_q$, audited & $\sum_s \kappa_s$ & $\E[c_{\hat{s}}] + \E[\epsilon \sum_{s \neq \hat{s}} c_s]$ \\
4c & $i,j \mid Z_q, W_q$ & $\sum_s \kappa_s$ & $\E[c_{\hat{s}}]$ \\
4b & $i,j \mid W_q$ & $\sum_s \kappa_s$ & $\E[c_{\hat{s}}]$ \\
4a & $i,j \mid Z_q$ & $\sum_{s \in \bar{s}(\mathcal{K})} \kappa_s$ & $\sum_k p_k \sum_{\hat{k}} \Pr(\hat{k} \mid k)\, c_{\bar{s}(\hat{k})}$ \\
3 & $ij{+}$ & $\sum_s \kappa_s$ & $\sum_s c_s + c_\rho$ \\
2 & $ij$ & $\sum_s \kappa_s$ & $\sum_s c_s$ \\
2r & $i,j \mid \varnothing$ & $\sum_s \kappa_s$ & $\frac{1}{n}\sum_s c_s$ \\
1 & $i^{*}$ & $\kappa_{i^{*}}$ & $c_{i^{*}}$ \\
0 & $i$ & $\kappa_i$ & $c_i$ \\
\bottomrule
\end{tabular}
\end{table}

The arrangements also connect to the human factors tradition. The allocation of functions across stages and at graded levels within each \citep{sheridan1978human, parasuraman2000model} makes the position of the operator a documentable property of a workflow, so a workflow can be placed on the ladder before any instance is worked. Treating the human-machine policy as the midpoint of a scale running from full human to full automation is, on that reading, an oversimplification: as \citet{shneiderman2020human} argues, human and machine control develop along separate dimensions, and the archetypes make the point in performance terms, with $HM$ where the midpoint intuition leads and $HM{+}$ and $HM{+}{+}$ in the region the two-dimensional view opens up. The ladder traverses that region.

\subsection{Conventions}\label{sec:conventions}

\begin{figure}[!htb]
\centering
\begin{subfigure}[t]{0.48\textwidth}\centering
\begin{tikzpicture}[font=\footnotesize, >={Stealth[length=1.8mm]},
  bx/.style={draw, rounded corners=2pt, align=center, minimum height=8mm, inner sep=2pt},
  off/.style={bx, densely dotted, text=black!50},
  flow/.style={->, semithick}, offflow/.style={->, semithick, densely dotted, black!50},
  exch/.style={<->, semithick, dashed}]
\path[use as bounding box] (-37mm,5mm) rectangle (37mm,-38mm);
\node[bx, minimum width=48mm] (c) at (0,0) {\textbf{Engagement criterion}\\ engages the incumbent $i$};
\node[bx, minimum width=15mm] (i) at (-13mm,-14mm) {Skill $i$};
\draw[flow] (c.south -| i.north) -- (i.north);
\node[bx, minimum width=48mm] (a) at (0,-26mm) {\textbf{Aggregation}\\ returns $\hat{r}_i$};
\draw[flow] (i.south) -- (i.south |- a.north);
\node (r) at (0,-35mm) {$\hat{r}_\pi$};
\draw[flow] (a.south) -- (r.north);
\end{tikzpicture}
\caption{Rung 0: $i$}
\end{subfigure}\hfill
\begin{subfigure}[t]{0.48\textwidth}\centering
\begin{tikzpicture}[font=\footnotesize, >={Stealth[length=1.8mm]},
  bx/.style={draw, rounded corners=2pt, align=center, minimum height=8mm, inner sep=2pt},
  off/.style={bx, densely dotted, text=black!50},
  flow/.style={->, semithick}, offflow/.style={->, semithick, densely dotted, black!50},
  exch/.style={<->, semithick, dashed}]
\path[use as bounding box] (-37mm,5mm) rectangle (37mm,-38mm);
\node[bx, minimum width=48mm] (c) at (0,0) {\textbf{Engagement criterion}\\ commits to $i^{*}$ from the records};
\node[bx, minimum width=15mm] (i) at (-13mm,-14mm) {Skill $i$};
\node[off, minimum width=15mm] (j) at (13mm,-14mm) {Skill $j$};
\draw[flow] (c.south -| i.north) -- (i.north);
\node[bx, minimum width=48mm] (a) at (0,-26mm) {\textbf{Aggregation}\\ returns $\hat{r}_{i^{*}}$};
\draw[flow] (i.south) -- (i.south |- a.north);
\node (r) at (0,-35mm) {$\hat{r}_\pi$};
\draw[flow] (a.south) -- (r.north);
\end{tikzpicture}
\caption{Rung 1: $i^{*}$}
\end{subfigure}\\[2mm]
\begin{subfigure}[t]{0.48\textwidth}\centering
\begin{tikzpicture}[font=\footnotesize, >={Stealth[length=1.8mm]},
  bx/.style={draw, rounded corners=2pt, align=center, minimum height=8mm, inner sep=2pt},
  off/.style={bx, densely dotted, text=black!50},
  flow/.style={->, semithick}, offflow/.style={->, semithick, densely dotted, black!50},
  exch/.style={<->, semithick, dashed}]
\path[use as bounding box] (-37mm,5mm) rectangle (37mm,-38mm);
\node[bx, minimum width=48mm] (c) at (0,0) {\textbf{Engagement criterion}\\ engages both};
\node[bx, minimum width=15mm] (i) at (-13mm,-14mm) {Skill $i$};
\node[bx, minimum width=15mm] (j) at (13mm,-14mm) {Skill $j$};
\draw[flow] (c.south -| i.north) -- (i.north);
\draw[flow] (c.south -| j.north) -- (j.north);
\node[bx, minimum width=48mm] (a) at (0,-26mm) {\textbf{Aggregation}\\ fixed weights $\tfrac{1}{2}, \tfrac{1}{2}$};
\draw[flow] (i.south) -- (i.south |- a.north);
\draw[flow] (j.south) -- (j.south |- a.north);
\node (r) at (0,-35mm) {$\hat{r}_\pi$};
\draw[flow] (a.south) -- (r.north);
\end{tikzpicture}
\caption{Rung 2: $ij$}
\end{subfigure}\hfill
\begin{subfigure}[t]{0.48\textwidth}\centering
\begin{tikzpicture}[font=\footnotesize, >={Stealth[length=1.8mm]},
  bx/.style={draw, rounded corners=2pt, align=center, minimum height=8mm, inner sep=2pt},
  off/.style={bx, densely dotted, text=black!50},
  flow/.style={->, semithick}, offflow/.style={->, semithick, densely dotted, black!50},
  exch/.style={<->, semithick, dashed}]
\path[use as bounding box] (-37mm,5mm) rectangle (37mm,-38mm);
\node[bx, minimum width=48mm] (c) at (0,0) {\textbf{Engagement criterion}\\ engages both};
\node[bx, minimum width=15mm] (i) at (-13mm,-14mm) {Skill $i$};
\node[bx, minimum width=15mm] (j) at (13mm,-14mm) {Skill $j$};
\draw[flow] (c.south -| i.north) -- (i.north);
\draw[flow] (c.south -| j.north) -- (j.north);
\draw[exch] (i) -- (j);
\node[inner sep=1pt, align=center, left=2mm of i] (rl) {$\rho$};
\node[draw, dashed, semithick, rounded corners=3pt, inner sep=2mm, fit=(rl)(i)(j)] (p) {};
\node (r) at (0,-35mm) {$\hat{r}_\pi = \hat{r}_\rho$};
\draw[->, semithick, dashed] (p.south -| r.north) -- (r.north);
\end{tikzpicture}
\caption{Rung 3: $ij{+}$}
\end{subfigure}\\[2mm]
\caption{The lower rungs as skill policies (Section~\ref{sec:policies}). Rungs 0 to 2 operate no channel: the criterion engages the same skills on every instance and the aggregation treats every instance alike. Rung~3 adds a protocol $\rho$ (dashed), which surrounds the engaged skills and returns the delivered response. Dotted boxes are skills available and not engaged. Random assignment (Algorithm~\ref{alg:2r}) is not drawn: its criterion engages one skill per instance, as at rung~1, but draws it at random instead of committing to $i^{*}$.}
\label{fig:rungs03}
\end{figure}
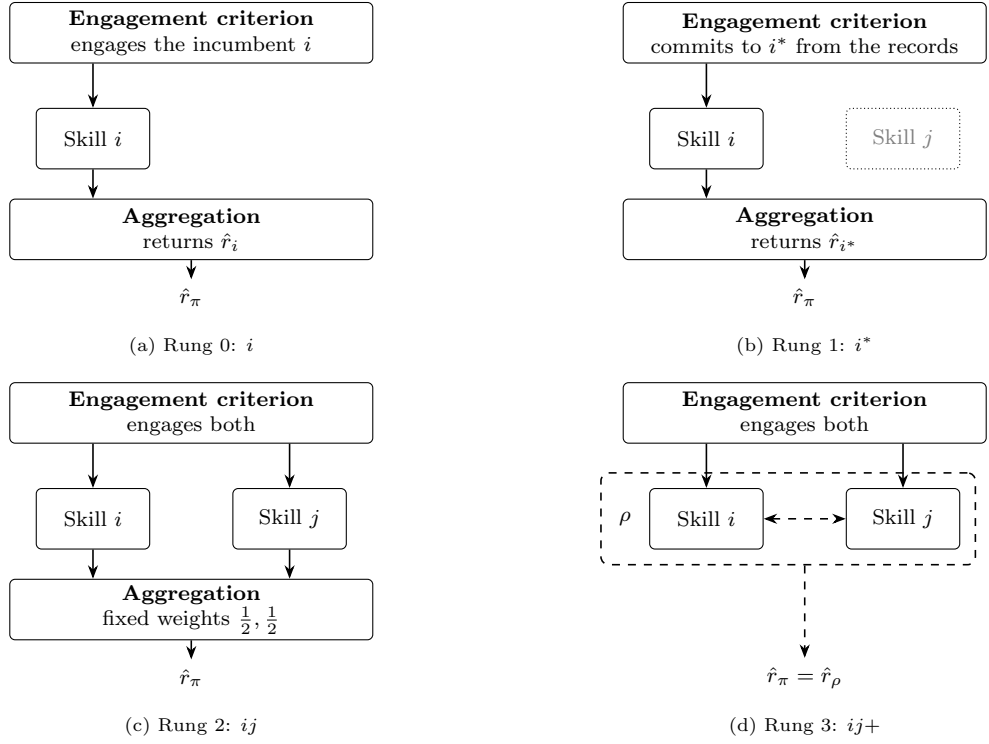

\begin{figure}[p]
\centering
\begin{subfigure}[t]{0.48\textwidth}\centering
\begin{tikzpicture}[font=\footnotesize, >={Stealth[length=1.8mm]},
  bx/.style={draw, rounded corners=2pt, align=center, minimum height=8mm, inner sep=2pt},
  off/.style={bx, densely dotted, text=black!50},
  flow/.style={->, semithick}, offflow/.style={->, semithick, densely dotted, black!50},
  exch/.style={<->, semithick, dashed}]
\path[use as bounding box] (-37mm,5mm) rectangle (37mm,-38mm);
\node[bx, minimum width=48mm] (c) at (0,0) {\textbf{Engagement criterion}\\ reads $Z_q$, engages $\bar{s}(\hat{k})$};
\node[bx, minimum width=15mm] (i) at (-13mm,-14mm) {Skill $i$};
\node[off, minimum width=15mm] (j) at (13mm,-14mm) {Skill $j$};
\draw[flow] (c.south -| i.north) -- (i.north);
\draw[offflow] (c.south -| j.north) -- (j.north);
\node[bx, minimum width=48mm] (a) at (0,-26mm) {\textbf{Aggregation}\\ returns the engaged response};
\draw[flow] (i.south) -- (i.south |- a.north);
\draw[offflow] (j.south) -- (j.south |- a.north);
\node (r) at (0,-35mm) {$\hat{r}_\pi$};
\draw[flow] (a.south) -- (r.north);
\end{tikzpicture}
\caption{Rung 4a: $i,j \mid Z_q$}
\end{subfigure}\hfill
\begin{subfigure}[t]{0.48\textwidth}\centering
\begin{tikzpicture}[font=\footnotesize, >={Stealth[length=1.8mm]},
  bx/.style={draw, rounded corners=2pt, align=center, minimum height=8mm, inner sep=2pt},
  off/.style={bx, densely dotted, text=black!50},
  flow/.style={->, semithick}, offflow/.style={->, semithick, densely dotted, black!50},
  exch/.style={<->, semithick, dashed}]
\path[use as bounding box] (-37mm,5mm) rectangle (37mm,-38mm);
\node[bx, minimum width=48mm] (c) at (0,0) {\textbf{Engagement criterion}\\ reads $W_q$, engages $\hat{s}$};
\node[bx, minimum width=15mm] (i) at (-13mm,-14mm) {Skill $i$};
\node[off, minimum width=15mm] (j) at (13mm,-14mm) {Skill $j$};
\draw[flow] (c.south -| i.north) -- (i.north);
\draw[offflow] (c.south -| j.north) -- (j.north);
\node[bx, minimum width=48mm] (a) at (0,-26mm) {\textbf{Aggregation}\\ returns the engaged response};
\draw[flow] (i.south) -- (i.south |- a.north);
\draw[offflow] (j.south) -- (j.south |- a.north);
\node (r) at (0,-35mm) {$\hat{r}_\pi$};
\draw[flow] (a.south) -- (r.north);
\end{tikzpicture}
\caption{Rung 4b: $i,j \mid W_q$}
\end{subfigure}\\[2mm]
\begin{subfigure}[t]{0.48\textwidth}\centering
\begin{tikzpicture}[font=\footnotesize, >={Stealth[length=1.8mm]},
  bx/.style={draw, rounded corners=2pt, align=center, minimum height=8mm, inner sep=2pt},
  off/.style={bx, densely dotted, text=black!50},
  flow/.style={->, semithick}, offflow/.style={->, semithick, densely dotted, black!50},
  exch/.style={<->, semithick, dashed}]
\path[use as bounding box] (-37mm,5mm) rectangle (37mm,-38mm);
\node[bx, minimum width=48mm] (c) at (0,0) {\textbf{Engagement criterion}\\ reads $Z_q, W_q$, engages $\hat{s}$};
\node[bx, minimum width=15mm] (i) at (-13mm,-14mm) {Skill $i$};
\node[off, minimum width=15mm] (j) at (13mm,-14mm) {Skill $j$};
\draw[flow] (c.south -| i.north) -- (i.north);
\draw[offflow] (c.south -| j.north) -- (j.north);
\node[bx, minimum width=48mm] (a) at (0,-26mm) {\textbf{Aggregation}\\ returns the engaged response};
\draw[flow] (i.south) -- (i.south |- a.north);
\draw[offflow] (j.south) -- (j.south |- a.north);
\node (r) at (0,-35mm) {$\hat{r}_\pi$};
\draw[flow] (a.south) -- (r.north);
\end{tikzpicture}
\caption{Rung 4c: $i,j \mid Z_q, W_q$}
\end{subfigure}\hfill
\begin{subfigure}[t]{0.48\textwidth}\centering
\begin{tikzpicture}[font=\footnotesize, >={Stealth[length=1.8mm]},
  bx/.style={draw, rounded corners=2pt, align=center, minimum height=8mm, inner sep=2pt},
  off/.style={bx, densely dotted, text=black!50},
  flow/.style={->, semithick}, offflow/.style={->, semithick, densely dotted, black!50},
  exch/.style={<->, semithick, dashed}]
\path[use as bounding box] (-37mm,5mm) rectangle (37mm,-38mm);
\node[bx, minimum width=48mm] (c) at (0,0) {\textbf{Engagement criterion}\\ engages both};
\node[bx, minimum width=15mm] (i) at (-13mm,-14mm) {Skill $i$};
\node[bx, minimum width=15mm] (j) at (13mm,-14mm) {Skill $j$};
\draw[flow] (c.south -| i.north) -- (i.north);
\draw[flow] (c.south -| j.north) -- (j.north);
\node[bx, minimum width=48mm] (a) at (0,-26mm) {\textbf{Aggregation}\\ weights $\hat{w}$ set from $Z_q, W_q$};
\draw[flow] (i.south) -- (i.south |- a.north);
\draw[flow] (j.south) -- (j.south |- a.north);
\node (r) at (0,-35mm) {$\hat{r}_\pi$};
\draw[flow] (a.south) -- (r.north);
\end{tikzpicture}
\caption{Rung 4d: $ij(w \mid Z_q, W_q)$}
\end{subfigure}\\[2mm]
\begin{subfigure}[t]{0.48\textwidth}\centering
\begin{tikzpicture}[font=\footnotesize, >={Stealth[length=1.8mm]},
  bx/.style={draw, rounded corners=2pt, align=center, minimum height=8mm, inner sep=2pt},
  off/.style={bx, densely dotted, text=black!50},
  flow/.style={->, semithick}, offflow/.style={->, semithick, densely dotted, black!50},
  exch/.style={<->, semithick, dashed}]
\path[use as bounding box] (-37mm,5mm) rectangle (37mm,-38mm);
\node[bx, minimum width=48mm] (c) at (0,0) {\textbf{Engagement criterion}\\ engages both};
\node[bx, minimum width=15mm] (i) at (-13mm,-14mm) {Skill $i$};
\node[bx, minimum width=15mm] (j) at (13mm,-14mm) {Skill $j$};
\draw[flow] (c.south -| i.north) -- (i.north);
\draw[flow] (c.south -| j.north) -- (j.north);
\node[bx, minimum width=48mm] (a) at (0,-26mm) {\textbf{Aggregation}\\ selects $\hat{s}$ from $Z_q, W_q, \{\hat{r}_s\}$};
\draw[flow] (i.south) -- (i.south |- a.north);
\draw[flow] (j.south) -- (j.south |- a.north);
\node (r) at (0,-35mm) {$\hat{r}_\pi$};
\draw[flow] (a.south) -- (r.north);
\end{tikzpicture}
\caption{Rung 4e: $i,j \mid Z_q, W_q, \{\hat{r}_s\}$}
\end{subfigure}\hfill
\begin{subfigure}[t]{0.48\textwidth}\centering
\begin{tikzpicture}[font=\footnotesize, >={Stealth[length=1.8mm]},
  bx/.style={draw, rounded corners=2pt, align=center, minimum height=8mm, inner sep=2pt},
  off/.style={bx, densely dotted, text=black!50},
  flow/.style={->, semithick}, offflow/.style={->, semithick, densely dotted, black!50},
  exch/.style={<->, semithick, dashed}]
\path[use as bounding box] (-37mm,5mm) rectangle (37mm,-38mm);
\node[bx, minimum width=48mm] (c) at (0,0) {\textbf{Engagement criterion}\\ engages both};
\node[bx, minimum width=15mm] (i) at (-13mm,-14mm) {Skill $i$};
\node[bx, minimum width=15mm] (j) at (13mm,-14mm) {Skill $j$};
\draw[flow] (c.south -| i.north) -- (i.north);
\draw[flow] (c.south -| j.north) -- (j.north);
\draw[exch] (i) -- (j);
\node[inner sep=1pt, align=center, left=2mm of i] (rl) {$\rho$ with\\ $\hat{w}(Z_q, W_q)$};
\node[draw, dashed, semithick, rounded corners=3pt, inner sep=2mm, fit=(rl)(i)(j)] (p) {};
\node (r) at (0,-35mm) {$\hat{r}_\pi = \hat{r}_\rho$};
\draw[->, semithick, dashed] (p.south -| r.north) -- (r.north);
\end{tikzpicture}
\caption{Rung 5: $ij(w \mid Z_q, W_q){+}$}
\end{subfigure}\\[2mm]
\caption{Rungs 4 and 5 as skill policies. Under assignment (4a to 4c) the criterion engages one skill per instance from what it reads, $i$ on the instance drawn and $j$ on others (dotted); under ensembling (4d) and selection (4e) it engages both, and the aggregation weights their responses or selects one. Rung~5 runs a protocol $\rho$ (dashed) whose weights are set per instance, and so routes and operates synergy at once. Assignment with audit (4c+a, Algorithm~\ref{alg:audit}) is not drawn: it is 4c with the other skill also engaged on a random fraction of instances, its response retained for the record and not returned.}
\label{fig:rungs45}
\end{figure}
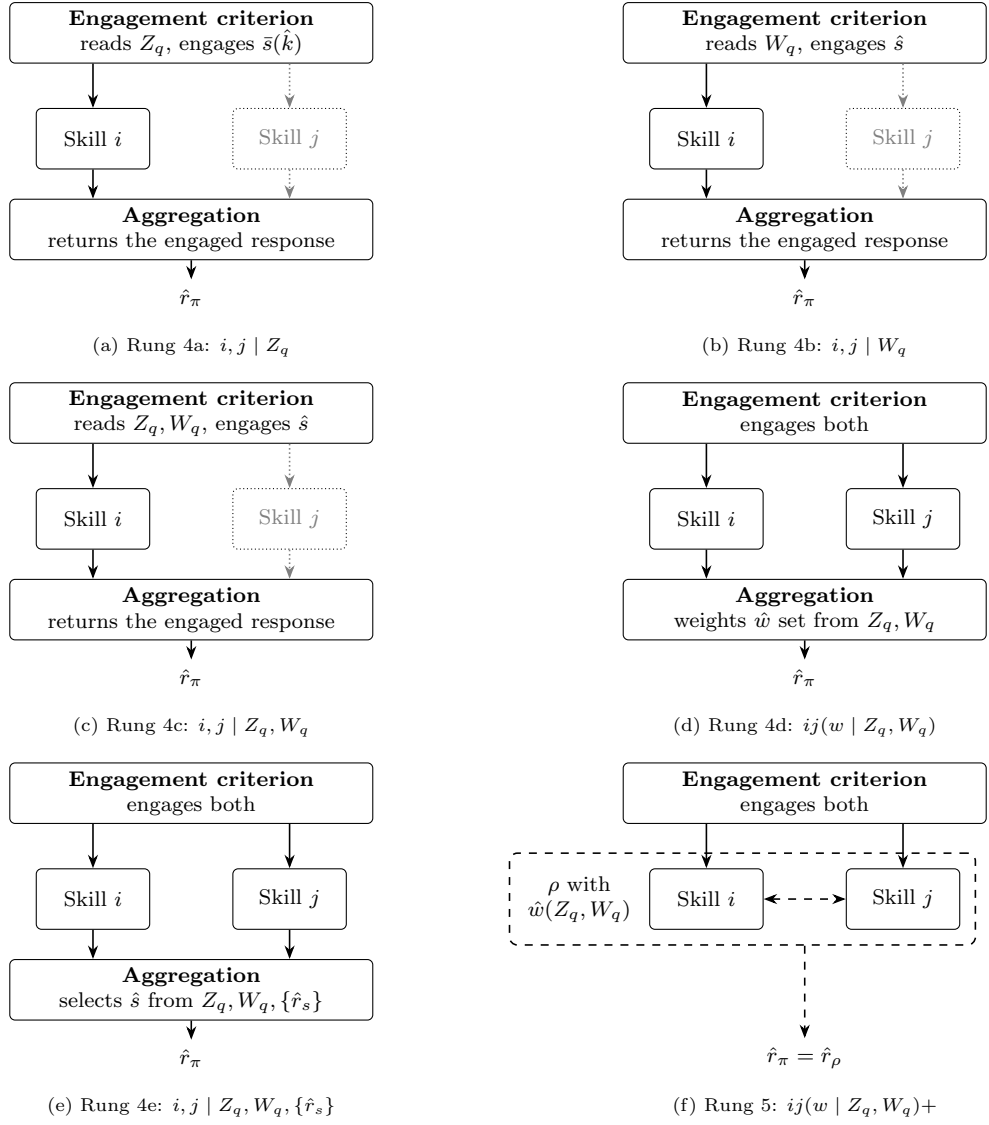

\begin{figure}[!htb]
\centering
\begin{subfigure}[t]{0.48\textwidth}\centering
\begin{tikzpicture}[font=\footnotesize, >={Stealth[length=1.8mm]},
  bx/.style={draw, rounded corners=2pt, align=center, minimum height=8mm, inner sep=2pt},
  off/.style={bx, densely dotted, text=black!50},
  flow/.style={->, semithick}, offflow/.style={->, semithick, densely dotted, black!50},
  exch/.style={<->, semithick, dashed}]
\path[use as bounding box] (-37mm,5mm) rectangle (37mm,-38mm);
\node[bx, minimum width=48mm] (c) at (0,0) {\textbf{Engagement criterion}\\ oracle class $k$, engages $\bar{s}(k)$};
\node[bx, minimum width=15mm] (i) at (-13mm,-14mm) {Skill $i$};
\node[off, minimum width=15mm] (j) at (13mm,-14mm) {Skill $j$};
\draw[flow] (c.south -| i.north) -- (i.north);
\draw[offflow] (c.south -| j.north) -- (j.north);
\node[bx, minimum width=48mm] (a) at (0,-26mm) {\textbf{Aggregation}\\ returns the engaged response};
\draw[flow] (i.south) -- (i.south |- a.north);
\draw[offflow] (j.south) -- (j.south |- a.north);
\node (r) at (0,-35mm) {$\hat{r}_\pi$};
\draw[flow] (a.south) -- (r.north);
\end{tikzpicture}
\caption{B1: $i,j \mid \mathrm{class}$}
\end{subfigure}\hfill
\begin{subfigure}[t]{0.48\textwidth}\centering
\begin{tikzpicture}[font=\footnotesize, >={Stealth[length=1.8mm]},
  bx/.style={draw, rounded corners=2pt, align=center, minimum height=8mm, inner sep=2pt},
  off/.style={bx, densely dotted, text=black!50},
  flow/.style={->, semithick}, offflow/.style={->, semithick, densely dotted, black!50},
  exch/.style={<->, semithick, dashed}]
\path[use as bounding box] (-37mm,5mm) rectangle (37mm,-38mm);
\node[bx, minimum width=48mm] (c) at (0,0) {\textbf{Engagement criterion}\\ oracle outcome, engages $\dot{s}$};
\node[bx, minimum width=15mm] (i) at (-13mm,-14mm) {Skill $i$};
\node[off, minimum width=15mm] (j) at (13mm,-14mm) {Skill $j$};
\draw[flow] (c.south -| i.north) -- (i.north);
\draw[offflow] (c.south -| j.north) -- (j.north);
\node[bx, minimum width=48mm] (a) at (0,-26mm) {\textbf{Aggregation}\\ returns the engaged response};
\draw[flow] (i.south) -- (i.south |- a.north);
\draw[offflow] (j.south) -- (j.south |- a.north);
\node (r) at (0,-35mm) {$\hat{r}_\pi$};
\draw[flow] (a.south) -- (r.north);
\end{tikzpicture}
\caption{B2: $i,j \mid \mathrm{inst}$}
\end{subfigure}\\[2mm]
\begin{subfigure}[t]{0.48\textwidth}\centering
\begin{tikzpicture}[font=\footnotesize, >={Stealth[length=1.8mm]},
  bx/.style={draw, rounded corners=2pt, align=center, minimum height=8mm, inner sep=2pt},
  off/.style={bx, densely dotted, text=black!50},
  flow/.style={->, semithick}, offflow/.style={->, semithick, densely dotted, black!50},
  exch/.style={<->, semithick, dashed}]
\path[use as bounding box] (-37mm,5mm) rectangle (37mm,-38mm);
\node[bx, minimum width=48mm] (c) at (0,0) {\textbf{Engagement criterion}\\ engages both};
\node[bx, minimum width=15mm] (i) at (-13mm,-14mm) {Skill $i$};
\node[bx, minimum width=15mm] (j) at (13mm,-14mm) {Skill $j$};
\draw[flow] (c.south -| i.north) -- (i.north);
\draw[flow] (c.south -| j.north) -- (j.north);
\node[bx, minimum width=48mm] (a) at (0,-26mm) {\textbf{Aggregation}\\ oracle weights, all on $\dot{s}$};
\draw[flow] (i.south) -- (i.south |- a.north);
\draw[flow] (j.south) -- (j.south |- a.north);
\node (r) at (0,-35mm) {$\hat{r}_\pi$};
\draw[flow] (a.south) -- (r.north);
\end{tikzpicture}
\caption{B3: $ij(w \mid \mathrm{inst})$}
\end{subfigure}\hfill
\caption{The ceilings of the routing families as skill policies (Section~\ref{sec:ceilings}). Each receives from an oracle what no workflow can have: the true class (B1), the winner of the instance (B2), or the weighting that would have been best (B3), which on an endpoint truth places all weight on the winner. B1 and B2 close assignment, B3 ensembling.}
\label{fig:ceilingspanels}
\end{figure}
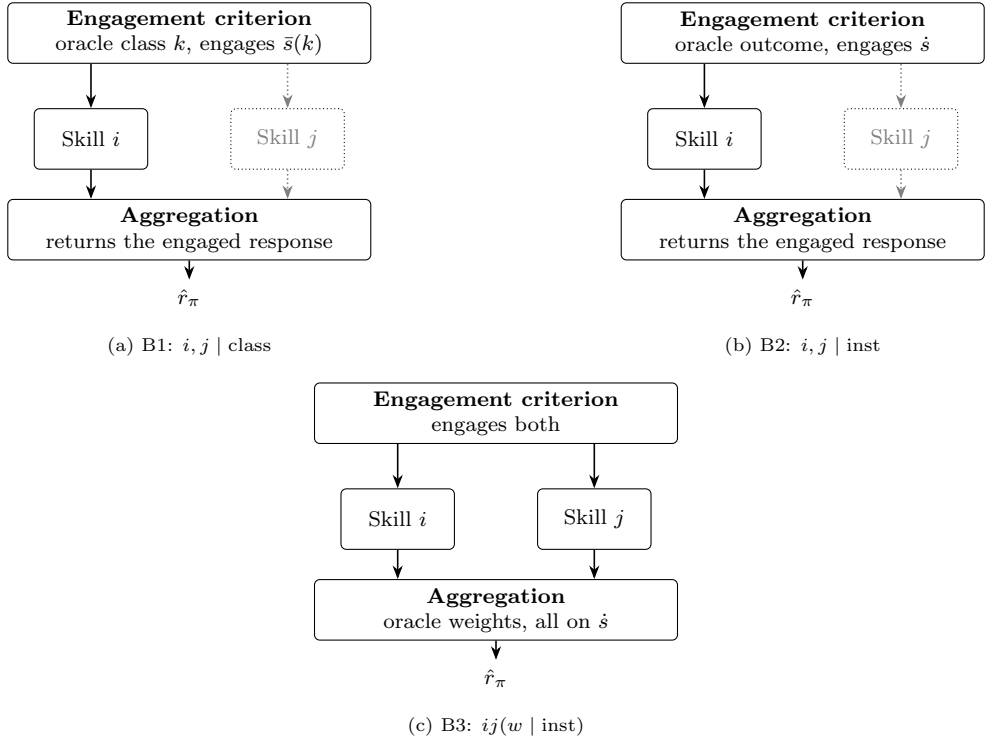

Each arrangement is described in plain terms and as an algorithm, by what the workflow does with an instance, what it consumes, and what must be known before it can be run at all. All algorithms are in four blocks and a coda. \textbf{Requires} states what must already exist for the policy to be run. \textbf{Setup} runs once for the task and its outputs are held fixed thereafter; it is where standing costs are incurred. \textbf{Execution} runs for each arriving instance: it decides, commissions execution and returns a response, and it is where execution costs are incurred. The comments on each algorithm state both, so that the cost profile of Table~\ref{tab:costs} can be read off the steps. \textbf{Adjudication} also runs per instance, but later and only for instances whose truth arrives: it is where performance is observed and records are updated. \textbf{Refit} states what the policy can re-estimate from what adjudication added. Parenthetical steps are discretionary: the operating policy runs without them and nothing in its own setup consumes what they produce. They are not optional for evaluating what the policy attains: an organization that never scores its responses can execute any policy in Table~\ref{tab:spectrum} and can compare none of them, which is the position rung~0 describes. Adjudication is kept separate from execution because an organization that intends to refit the criterion it operates needs the correct response to come from somewhere the arrangement does not reach (Section~\ref{sec:records}). Each algorithm follows the paragraph that describes its arrangement, and Figures~\ref{fig:rungs03}, \ref{fig:rungs45} and~\ref{fig:ceilingspanels} draw the arrangements as skill policies in the notation of Section~\ref{sec:policies}; Appendix~\ref{app:records} sets out the records and the notation they use.

The CXR experiment of \citet{agarwal2023combining} supplies the setting followed through the rungs, with a radiologist and a model reading chest films and the probability of a pathology as the response.

\subsection{Rungs 0 to 3: one skill, or both without routing}\label{sec:rungs03}

\paragraph{Rung 0: one skill, unexamined.} A policy is in force whether or not anyone chose it. A single skill executes every instance, no signal is consulted, and no instance is treated differently from any other (Algorithm~\ref{alg:0}). This is the status quo of an organization that has a way of doing the work and has not asked whether it is the right one: the radiologist reads every film, or the model does, and no alternative is under evaluation. It attains the marginal performance of the incumbent skill and needs nothing: it can be operated by an organization that knows nothing about its own skills, including which of them is best.

What rung~0 forgoes is the information on which any alternative would be chosen. An organization can record nothing about its executions, or score and record what it executes, so that when an alternative is considered the evidence for rung~1 already exists; testing candidate skills on its own instances, rather than relying on a vendor's benchmark, is an investment of the same kind.

\begin{algorithm}[H]
\rungnum{0}
\caption{$i$: one skill, unexamined}
\label{alg:0}
\begin{algorithmic}
\Require nothing
\Setup
\State make skill $i$ available \Comment{standing cost $\kappa_i$}
\Ensure
\State $\hat{r}_i \gets$ skill $i$ executes the instance \Comment{execution cost $c_i$}
\State \Return $\hat{r} \gets \hat{r}_i$
\Adjudicate
\State (observe $r$)
\State (compute $\theta_i$)
\State (append the instance, with $\hat{r}_i$ and $r$, to $\mathcal{R}_i$)
\Refit
\State ---
\end{algorithmic}
\end{algorithm}

\paragraph{Rung 1: one skill, assessed.} One skill is still engaged on all instances; what changes is how that skill is chosen. The organization commits ex ante to the skill that performs best on average according to its assessment, $i^{*} = \argmax_s \hat{\E}[\theta_s]$ (Algorithm~\ref{alg:1}); where the assessment is correct, $i^{*}$ is the leader $L$ of Section~\ref{sec:classes}, and the rung attains $b$. This is the comparative-advantage logic of \citet{autor2003skill} and \citet{acemoglu2025simple} applied once and at the level of the task, and it is no trivial step, since estimating performance over a representative instance space is not the standard for every skill and task an organization manages. The estimates $\hat{\E}[\theta_s]$ are computed over records $\mathcal{R}_s$, and since each record concerns one skill alone, the rung can be reached without ever observing two skills on the same instance: it is the first step above rung~0 that separate assessments support. In the CXR setting, all films go to the radiologists or all to the model, regardless of the characteristics of any film and with no interaction between the two. Two cautions attach to it. The assessment must be correct, and it must remain correct: it is computed on instances already worked, and nothing in the policy detects that the caseload has moved away from them. A wrong commitment forfeits $2S$ (Proposition~\ref{prop:S}).

\begin{algorithm}[H]
\rungnum{1}
\caption{$i^{*}$: one skill, assessed}
\label{alg:1}
\begin{algorithmic}
\Require $\mathcal{R}_s \;\forall\, s \in \mathcal{S}$ \Comment{assessments only; the skills need not be available}
\Setup
\State $i^{*} \gets \argmax_{s \in \mathcal{S}} \hat{\E}[\theta_s]$
\State make skill $i^{*}$ available \Comment{standing cost $\kappa_{i^{*}}$}
\Ensure
\State $\hat{r}_{i^{*}} \gets$ skill $i^{*}$ executes the instance \Comment{execution cost $c_{i^{*}}$}
\State \Return $\hat{r} \gets \hat{r}_{i^{*}}$
\Adjudicate
\State observe $r$
\State compute $\theta_{i^{*}}$
\State append the instance, with $\hat{r}_{i^{*}}$ and $r$, to $\mathcal{R}_{i^{*}}$ \Comment{one record grows; the setup consumed all of them}
\Refit
\State recompute $\hat{\E}[\theta_{i^{*}}]$
\end{algorithmic}
\end{algorithm}

\paragraph{Rung 2: both skills, no routing.} Both skills execute every instance and their responses are averaged, under a constant criterion and with fixed weights (Algorithm~\ref{alg:2}). The workflow decides nothing: every instance is treated alike and both skills are consumed on all of them. It attains \sqc{$\E[\theta_{\bar{w}}] = b - S + \Lambda$}, below $b$ unless the responses disagree by more than $S$, while engaging both skills on every instance. It needs nothing, which distinguishes it from every other multi-skill policy in the table except random assignment: no assessment, no incumbencies, no estimated criterion. If the radiologist reads $90\%$ and the model $10\%$, the report delivers $50\%$. The outcome differs from either reading, doubt in place of good or bad news, but the average is a lossy compression and the workflow keeps only the compressed value: neither the patient nor the clinic can tell whether the $50\%$ conceals a violent disagreement, $90\%$ against $10\%$, or a shared uncertainty, $51\%$ against $49\%$, though the two call for different responses. The averaging is itself a rule, but a blind one: a human in the average rather than in the loop \citep{crootof2023humans}.

The combination is what makes this rung an insurance rather than a waste: it forfeits \sqc{$S - \Lambda$} with certainty in exchange for never forfeiting $2S$ (Proposition~\ref{prop:insurance}). It is also where oversight is most exposed to the failure modes of Section~\ref{sec:oversight}, since an organization that combines skills without a specific design lands here, and the expectation that the combination improves on either skill alone is, on performance, mistaken unless the responses disagree by more than $S$. A variant fixes unequal weights fitted on a record of instances both skills executed; Corollary~\ref{cor:fixed} prices it: it beats rung~1 only by the disagreement it earns, and under the linear rule it cannot; and estimating the weights has its own cost, which is why equal weights so often beat estimated ones in forecast combination \citep{smith2009simple}.

\begin{algorithm}[H]
\rungnum{2}
\caption{$ij$: both skills, no routing (pooling)}
\label{alg:2}
\begin{algorithmic}
\Require nothing
\Setup
\State make every skill in $\mathcal{S}$ available \Comment{standing cost $\sum_s \kappa_s$}
\Ensure
\State $\{\hat{r}_s\} \gets$ each $s \in \mathcal{S}$ executes the instance \Comment{execution cost $\sum_s c_s$}
\State \Return $\hat{r} \gets \hat{r}_{\bar{w}} = \mean_{s \in \mathcal{S}} \hat{r}_s$ \Comment{fixed, equal weights}
\Adjudicate
\State (observe $r$)
\State (compute $\theta_{\bar{w}} = \theta(\hat{r}_{\bar{w}}, r)$, and $\theta_s \;\forall\, s \in \mathcal{S}$)
\State (compute the disagreement \sqc{$\frac{1}{n}\sum_s (\hat{r}_s - \hat{r}_{\bar{w}})^2$}) \Comment{under the squared rule, \sqc{$\theta_{\bar{w}} - m$}}
\State (append the instance, with every $\hat{r}_s$, $Z_q$, $W_q$ and $r$, to $\mathcal{R}^{Z,W,\hat{r}}_{\mathcal{S}}$) \Comment{what the rung 4 fits require}
\Refit
\State (\sqc{$\hat{\Lambda} \gets$} the mean disagreement over the instances adjudicated so far) \Comment{nothing the policy uses is refitted; \sqc{$\hat{\Lambda}$} prices the insurance and the break-even}
\end{algorithmic}
\end{algorithm}

\paragraph{Random assignment: $i,j \mid \varnothing$.} Each instance goes to one skill drawn at random, independently of the instance (Algorithm~\ref{alg:2r}). The policy holds every skill available but executes one per instance, a fraction $1/n$ of rung~2's execution cost, and attains the mean anchor exactly, $\E[m] = b - S$, under any rule (Proposition~\ref{prop:randomize}). It is not a rung, since it decides nothing and combines nothing, but the reference from which every take is measured. Against rung~2 it gives up the disagreement \sqc{$\Lambda$} and the paired record, and keeps the insurance: since the draw does not depend on the assessment, it is exposed to no wrong commitment, and its outcomes are only more dispersed than pooling's.

\begin{algorithm}[H]
\rungnum{2r}
\caption{$i,j \mid \varnothing$: random assignment}
\label{alg:2r}
\begin{algorithmic}
\Require nothing
\Setup
\State make every skill in $\mathcal{S}$ available \Comment{standing cost $\sum_s \kappa_s$}
\Ensure
\State $\hat{s} \gets$ a skill drawn uniformly from $\mathcal{S}$, independently of the instance
\State $\hat{r}_{\hat{s}} \gets$ skill $\hat{s}$ executes the instance \Comment{execution cost $c_{\hat{s}}$; $\frac{1}{n}\sum_s c_s$ in expectation}
\State \Return $\hat{r} \gets \hat{r}_{\hat{s}}$
\Adjudicate
\State (observe $r$)
\State (compute $\theta_{\hat{s}}$)
\State (append the instance, with $\hat{r}_{\hat{s}}$ and $r$, to $\mathcal{R}_{\hat{s}}$) \Comment{each skill's record grows; no instance is paired}
\Refit
\State --- \Comment{nothing is refitted}
\end{algorithmic}
\end{algorithm}

\paragraph{Rung 3: both skills, with synergy.} Both skills execute every instance independently, and the workflow then runs a protocol in which what each produced enters the other's execution before a joint response is returned (Algorithm~\ref{alg:3}). Consensus double reading \citep{johnston1999two} is this rung in clinical form: two readers work the same study without sight of each other's findings, which is rung~2 up to the point where the reports are compared, and where they conflict they confer and produce one report that neither had written. The distinction between rungs~2 and~3 is therefore not in how many skills execute but in whether the responses may change each other before the case is closed, and a lift above the bare average can come only from acting on the disagreement the average discards. The arrangement attains \sqc{$\gamma_\pi \E[\theta_{\bar{w}}]$}, engages both skills and the exchange on every instance, and needs $\gamma$, the one quantity in the table that no measurement taken before deployment supplies: it is observable only once the protocol has run and its joint responses have been scored against the independent responses retained before the exchange. What can be known before the protocol is designed, from the skills observed on common instances, is the break-even \sqc{$\gamma^{*} = b/\E[\theta_{\bar{w}}]$}, the burden of proof a proposed protocol carries before any of it is built.

\begin{algorithm}[H]
\rungnum{3}
\caption{$ij{+}$: both skills, with synergy}
\label{alg:3}
\begin{algorithmic}
\Require a protocol $\rho$ under which the skills see something of one another's responses before a joint one is fixed \Comment{a protocol that shows nothing is rung 2}
\Setup
\State make every skill in $\mathcal{S}$ available \Comment{standing cost $\sum_s \kappa_s$}
\Ensure
\State $\{\hat{r}_s\} \gets$ each $s \in \mathcal{S}$ executes the instance independently \Comment{retained; execution cost $\sum_s c_s$}
\State $\hat{r}_\rho \gets \rho\bigl(\mathcal{S}, \{\hat{r}_s\}\bigr)$ \Comment{the exchange, at cost $c_\rho$; the skills may revise}
\State \Return $\hat{r} \gets \hat{r}_\rho$
\Adjudicate
\State (observe $r$)
\State (compute $\theta_s \;\forall\, s \in \mathcal{S}$, and $\theta = \theta(\hat{r}_\rho, r)$)
\State ($\gamma \gets \theta \,/\, \theta\bigl(\mean_s \hat{r}_s, r\bigr)$) \Comment{the lift on this instance, against the pre-exchange mean}
\State (append the instance, with every $\hat{r}_s$, $\hat{r}_\rho$, $Z_q$, $W_q$ and $r$, to $\mathcal{R}^{Z,W,\hat{r}}_{\mathcal{S}}$)
\Refit
\State ($\gamma_\pi \gets$ the ratio of the means of $\theta(\hat{r}_\rho, r)$ and $\theta(\mean_s \hat{r}_s, r)$ over the instances adjudicated so far)
\end{algorithmic}
\end{algorithm}

\subsection{Rung 4: routing}\label{sec:rung4}

At this rung more than one skill is available and the workflow decides, by class or by instance, what to do with them. Most of the spectrum's complexity is here, because the routing channel admits three implementations (Definitions~\ref{def:assignment}--\ref{def:selection}). Under \emph{assignment} one skill executes and the problem is to determine which. Under \emph{ensembling} every skill executes and the problem is how to weight the responses. Under \emph{selection} every skill executes and one response is returned unchanged, chosen rather than combined. Selection resembles assignment in what it returns and ensembling with all weight but one set to zero in what it executes; it is neither, because the three differ both in how the returning skill is determined and in how many skills are executed before the response is returned. Every implementation lies within the routing range of Proposition~\ref{prop:routing}: between the no-information point $\E[m]$ and the perfect-routing point $\E[\Mx]$ when what directs its work is informative, down to $\E[m] - G$ when it points the wrong way, and at $b$ when it reads nothing about the instance and collapses onto the better skill.

Each implementation can be deployed in several ways, and only a subset is available to any organization. A criterion or a weighting conditioned on a signal requires that signal to have been collected alongside the responses and the correct response on the adjudicated instances used to fit it, and to be available again on each arriving instance. Selecting which skill should execute is the estimation problem that learning-to-defer and algorithmic-triage methods address \citep{madras2018predict, mozannar2020consistent}, with $R$ as the ceiling no router can exceed, and the criterion is itself an estimate that new uncertainty can strain.

\paragraph{Assignment on the class: $i,j \mid Z_q$.} A criterion estimates the class of the instance, and the instance goes to the incumbent of that class (Algorithm~\ref{alg:4a}). Every instance of a class is treated alike, so the policy is only as fine as the class structure: within a class it has nothing left to distinguish. It executes one skill. It needs the class-wise means, which give the incumbencies, and a fitted posterior over classes given the signal; the first requires the skills to have been observed by class, the second a record of instances with their class and signal, and the two need no instance in common. Its ceiling is $i,j \mid \mathrm{class}$ (Algorithm~\ref{alg:B1}), which attains $b + \Rclass$; a criterion that misclassifies collects less. Although the criterion is evaluated per instance, what it produces is a partition, every instance of a class going to the same skill, so the arrangement partitions the instances into groups with a skill attached to each, which is the decomposition by instances of Section~\ref{sec:defer} and not a decomposition into sub-tasks. It can be written down as a standing rule, encoded in a workflow or an organizational chart, and applied without per-instance evaluation wherever the class is observable at intake. It is a criterion reading what kind of instance it faces, not how each candidate skill stands on it. Trusting the predicted class however unreliable the prediction, it attains $\E[m]$ at chance accuracy and exceeds $b$ only above a threshold accuracy (Proposition~\ref{prop:classpred}, Appendix~\ref{app:factors}).

\begin{algorithm}[H]
\rungnum{4a}
\caption{$i,j \mid Z_q$: assignment based on class prediction}
\label{alg:4a}
\begin{algorithmic}
\Require $\mathcal{R}_{s,k} \;\forall\, s,k$, and $\mathcal{R}^{Z}$ \Comment{two records, needing no instance in common}
\Setup
\State $\hat{\mu}_{s,k} \gets \mean$ of $\theta$ over $\mathcal{R}_{s,k}$, $\;\forall\, s,k$
\State $\bar{s}(k) \gets \argmax_{s \in \mathcal{S}} \hat{\mu}_{s,k} \;\forall\, k$ \Comment{the incumbency table}
\State make every incumbent available \Comment{standing cost $\sum_{s \in \bar{s}(\mathcal{K})} \kappa_s$}
\State fit $\widehat{\Pr}(k \mid Z_q)$ on $\mathcal{R}^{Z}$
\Ensure
\State $\hat{k} \gets \argmax_k \widehat{\Pr}(k \mid Z_q)$ on the instance
\State $\hat{r}_{\bar{s}(\hat{k})} \gets$ skill $\bar{s}(\hat{k})$ executes the instance \Comment{execution cost $c_{\bar{s}(\hat{k})}$}
\State \Return $\hat{r} \gets \hat{r}_{\bar{s}(\hat{k})}$
\Adjudicate
\State observe $r$ and $k$
\State compute $\theta_{\bar{s}(\hat{k})}$
\State append the instance, with $Z_q$ and its class $k$, to $\mathcal{R}^{Z}$, and, with $\hat{r}_{\bar{s}(\hat{k})}$ and $r$, to $\mathcal{R}_{\bar{s}(\hat{k}),k}$
\Refit
\State refit $\widehat{\Pr}(k \mid Z_q)$ on the updated $\mathcal{R}^{Z}$
\State recompute $\hat{\mu}_{\bar{s}(\hat{k}),k}$, hence $\bar{s}(k)$
\end{algorithmic}
\end{algorithm}

\paragraph{Assignment on the standing: $i,j \mid W_q$.} An incumbency is settled by the means and does not change from one instance to the next; which skill wins is settled on the instance and changes constantly, and a skill that holds no incumbency anywhere can still lead on a substantial share of instances. That gap is $\Rinst$. A model estimates from $W$ alone each skill's expected performance on the instance, and the instance goes to the skill expected to do best (Algorithm~\ref{alg:4b}), the criterion of Proposition~\ref{prop:information}; predicting which skill will lead is a surrogate, which can rank a skill that wins often by little above one that loses rarely by much. It executes one skill, unless $W$ is itself a skill's output, as the model's probability is in the CXR experiment, in which case that skill executes on every instance as well. Fitting the model free of selection requires past instances on which every skill executed: separate records carrying $W$ would serve only where each skill has been observed across the whole range of $W$, which a deployed assignment, observing each skill only where it sends it, does not supply. Hence the record $\mathcal{R}^{W}_{\mathcal{S}}$. The more consequential limitation is not in the criterion but in what the arrangement does to the record behind it. Once deployed, the policy executes one skill per instance and identifies no winner, so the record it consumes stops growing on the day it starts. The criterion can only be as good as a period of operation the policy has replaced. The delegation rule derived by \citet{agarwal2023combining}, which conditions on the model's own predicted probability, is of this kind.

\begin{algorithm}[H]
\rungnum{4b}
\caption{$i,j \mid W_q$: assignment based on standing}
\label{alg:4b}
\begin{algorithmic}
\Require $\mathcal{R}_{\mathcal{S}}^{W}$
\Setup
\State make every skill in $\mathcal{S}$ available \Comment{standing cost $\sum_s \kappa_s$}
\State fit $\hat{\E}[\theta_s \mid W_q]$ on $\mathcal{R}_{\mathcal{S}}^{W}$
\Ensure
\State $\hat{s} \gets \argmax_{s \in \mathcal{S}} \hat{\E}[\theta_s \mid W_q]$ on the instance
\State $\hat{r}_{\hat{s}} \gets$ skill $\hat{s}$ executes the instance \Comment{execution cost $c_{\hat{s}}$}
\State \Return $\hat{r} \gets \hat{r}_{\hat{s}}$
\Adjudicate
\State observe $r$
\State compute $\theta_{\hat{s}}$
\State append the instance, with $\hat{r}_{\hat{s}}$ and $r$, to $\mathcal{R}_{\hat{s}}$ \Comment{not to $\mathcal{R}_{\mathcal{S}}$: only one skill executed, so the winner is unidentified}
\Refit
\State --- \Comment{the fit needs jointly executed instances, and this policy executes one skill}
\end{algorithmic}
\end{algorithm}

\paragraph{Assignment based on class and standing prediction: $i,j \mid Z_q, W_q$.} The two signals are combined (Algorithm~\ref{alg:4c}). Since they are informative about different objects, the posterior conditions on more than either alone supports, and conditioning on more cannot hurt in the population, though it can in a finite record. Its ceiling is $i,j \mid \mathrm{inst}$, the ceiling of the rung. It inherits the record problem of assignment on the standing unchanged; Algorithm~\ref{alg:audit} writes the variant that executes the other skills on a fraction of the instances to keep the joint record growing, and Section~\ref{sec:records} prices it.

\begin{algorithm}[H]
\rungnum{4c}
\caption{$i,j \mid Z_q, W_q$: assignment based on class and standing prediction}
\label{alg:4c}
\begin{algorithmic}
\Require $\mathcal{R}_{\mathcal{S}}^{Z,W}$
\Setup
\State make every skill in $\mathcal{S}$ available \Comment{standing cost $\sum_s \kappa_s$}
\State fit $\hat{\E}[\theta_s \mid Z_q, W_q]$ on $\mathcal{R}_{\mathcal{S}}^{Z,W}$
\Ensure
\State $\hat{s} \gets \argmax_{s \in \mathcal{S}} \hat{\E}[\theta_s \mid Z_q, W_q]$ on the instance
\State $\hat{r}_{\hat{s}} \gets$ skill $\hat{s}$ executes the instance \Comment{execution cost $c_{\hat{s}}$}
\State \Return $\hat{r} \gets \hat{r}_{\hat{s}}$
\Adjudicate
\State observe $r$
\State compute $\theta_{\hat{s}}$
\State append the instance, with $\hat{r}_{\hat{s}}$ and $r$, to $\mathcal{R}_{\hat{s}}$ \Comment{as in 4b}
\Refit
\State --- \Comment{the fit needs jointly executed instances, and this policy executes one skill}
\end{algorithmic}
\end{algorithm}

\begin{algorithm}[H]
\rungnum{4c+a}
\caption{$i,j \mid Z_q, W_q$ with audit: assignment that keeps the joint record growing}
\label{alg:audit}
\begin{algorithmic}
\Require $\mathcal{R}_{\mathcal{S}}^{Z,W}$; an audit probability $\epsilon(Z_q, W_q) \geq \underline{\epsilon} > 0$
\Setup
\State make every skill in $\mathcal{S}$ available \Comment{standing cost $\sum_s \kappa_s$}
\State fit $\hat{\E}[\theta_s \mid Z_q, W_q]$ on $\mathcal{R}_{\mathcal{S}}^{Z,W}$, weighting each audited instance by $1/\epsilon$
\Ensure
\State $\hat{s} \gets \argmax_{s \in \mathcal{S}} \hat{\E}[\theta_s \mid Z_q, W_q]$ on the instance
\State $\hat{r}_{\hat{s}} \gets$ skill $\hat{s}$ executes the instance \Comment{execution cost $c_{\hat{s}}$}
\State draw $A \sim \mathrm{Bernoulli}(\epsilon(Z_q, W_q))$
\If{$A = 1$}
\State $\hat{r}_s \gets$ skill $s$ executes the instance, for every $s \neq \hat{s}$ \Comment{execution cost $\sum_{s \neq \hat{s}} c_s$; kept, not delivered}
\EndIf
\State \Return $\hat{r} \gets \hat{r}_{\hat{s}}$
\Adjudicate
\State observe $r$
\State compute $\theta_{\hat{s}}$, and every $\theta_s$ if $A = 1$
\State append the instance, with $\hat{r}_{\hat{s}}$ and $r$, to $\mathcal{R}_{\hat{s}}$; if $A = 1$, append it, with every $\hat{r}_s$, $Z_q$, $W_q$, $r$ and the weight $1/\epsilon(Z_q, W_q)$, to $\mathcal{R}_{\mathcal{S}}^{Z,W}$
\Refit
\State refit $\hat{\E}[\theta_s \mid Z_q, W_q]$ on the updated $\mathcal{R}_{\mathcal{S}}^{Z,W}$
\end{algorithmic}
\end{algorithm}

\paragraph{Ensembling based on class and standing prediction: $ij(w \mid Z_q, W_q)$.} Both skills execute every instance and the responses are combined with weights set from the signals: the weights under which the combined response has scored best on past instances carrying the same signals (Algorithm~\ref{alg:4d}), the comparison of Proposition~\ref{prop:ensembling}. Setting them requires the responses to have been retained, not only the performances, since what is scored is the combined response. It executes both skills whatever the weights say, and incurs both executions: a zero weight does not undo an execution already commissioned. In performance it spans the continuum between equal combination and assignment. In the CXR setting a reading rule may give the model $25\%$ of the weight for one population of patients and $75\%$ for another, reflecting a different incidence of the pathology and the resulting scarcity of training examples, or condition on the patient's clinical history. Whether the interior of the continuum is worth occupying is settled by the hull bound and not by the curvature of the rule: the returned response is a convex combination, so on every instance it scores at most what the winner scored, and the best weighting, had it been known, places all its mass on the winner (Proposition~\ref{prop:routing}). Soft weighting is weakly dominated by selecting the winner, were the winner known. On the same information it exceeds assignment only by the disagreement it earns, and under the linear rule never (Proposition~\ref{prop:ensembling}). An interior weighting can beat the fitted assignment it softens, on instances where the assignment would have picked the laggard, and under a curved rule it earns the ambiguity term of Proposition~\ref{prop:hull} as well; it never beats the winner. The arrangement earns its place from what it insures against: the failure of the model governing the assignment.

\begin{algorithm}[H]
\rungnum{4d}
\caption{$ij(w \mid Z_q, W_q)$: ensembling based on class and standing prediction}
\label{alg:4d}
\begin{algorithmic}
\Require $\mathcal{R}_{\mathcal{S}}^{Z,W,\hat{r}}$
\Setup
\State make every skill in $\mathcal{S}$ available \Comment{standing cost $\sum_s \kappa_s$}
\State fit the weights $\hat{w}(Z_q, W_q) = \argmax_{w \geq 0,\, \sum_s w_s = 1} \hat{\E}\big[\theta(\textstyle\sum_s w_s \hat{r}_s, r) \mid Z_q, W_q\big]$ on $\mathcal{R}_{\mathcal{S}}^{Z,W,\hat{r}}$ \Comment{scored on retained responses}
\Ensure
\State $\hat{w} \gets \hat{w}(Z_q, W_q)$ on the instance
\State $\{\hat{r}_s\} \gets$ each $s \in \mathcal{S}$ executes the instance \Comment{execution cost $\sum_s c_s$, whatever the weights}
\State \Return $\hat{r} \gets \sum_{s \in \mathcal{S}} \hat{w}_s\, \hat{r}_s$
\Adjudicate
\State observe $r$
\State compute $\theta_s \;\forall\, s \in \mathcal{S}$, and $\theta = \theta(\hat{r}, r)$
\State append the instance, with every $\hat{r}_s$, $Z_q$, $W_q$ and $r$, to $\mathcal{R}_{\mathcal{S}}^{Z,W,\hat{r}}$
\Refit
\State refit $\hat{w}(Z_q, W_q)$ on the updated $\mathcal{R}_{\mathcal{S}}^{Z,W,\hat{r}}$
\end{algorithmic}
\end{algorithm}

\paragraph{Selection on both and the responses: $i,j \mid Z_q, W_q, \{\hat{r}_s\}$.} All skills are engaged and one of the responses is returned (Algorithm~\ref{alg:4e}). It is the only arrangement in the table that can condition on the responses themselves, which do not exist until the skills have executed; disagreement between them, for instance, is visible to it and to no arrangement before it. It needs a rule for choosing, estimated from instances on which the responses were observed jointly with the signals and adjudicated. Selection returns one of the responses it commissioned, unchanged: it is routing whose aggregation reads an unusually rich set of inputs, and not synergy, although it can attain more than averaging the same responses. The constraint of executing every available skill can be relaxed to a designated subset without changing its nature. Its record requirement is self-limiting in a way the others are not. Fitting the rule needs instances on which several skills executed, and this arrangement produces them, while an assignment does not; an organization that moves from here to an assignment loses the evidence that would let it revalidate the rule it moved away from. The structure is that of the selective labels problem \citep{lakkaraju2017selective}, with one difference: the unobserved performances are not outcomes the world never produced but executions the organization declined to commission, which it could commission at any time by asking for a second reading. With more than two skills, ensembling and selection apply alike to any subset of two or more that a criterion engages, ensembling's weights running over the engaged skills only.

\begin{algorithm}[H]
\rungnum{4e}
\caption{$i,j \mid Z_q, W_q, \{\hat{r}_s\}$: selection after execution}
\label{alg:4e}
\begin{algorithmic}
\Require $\mathcal{R}_{\mathcal{S}}^{Z,W,\hat{r}}$
\Setup
\State make every skill in $\mathcal{S}$ available \Comment{standing cost $\sum_s \kappa_s$}
\State fit $\hat{\E}[\theta_s \mid \{\hat{r}_{s'}\}, Z_q, W_q]$ on $\mathcal{R}_{\mathcal{S}}^{Z,W,\hat{r}}$ \Comment{conditions on the responses}
\Ensure
\State $\{\hat{r}_s\} \gets$ each $s \in \mathcal{S}$ executes the instance \Comment{execution cost $\sum_s c_s$}
\State $\hat{s} \gets \argmax_{s \in \mathcal{S}} \hat{\E}[\theta_s \mid \{\hat{r}_{s'}\}, Z_q, W_q]$ on the instance \Comment{the responses now exist}
\State \Return $\hat{r} \gets \hat{r}_{\hat{s}}$
\Adjudicate
\State observe $r$
\State compute $\theta_s \;\forall\, s \in \mathcal{S}$ \Comment{$\theta = \theta_{\hat{s}}$}
\State append the instance, with every $\hat{r}_s$, $Z_q$, $W_q$ and $r$, to $\mathcal{R}_{\mathcal{S}}^{Z,W,\hat{r}}$
\Refit
\State refit $\hat{\E}[\theta_s \mid \{\hat{r}_{s'}\}, Z_q, W_q]$ on the updated $\mathcal{R}_{\mathcal{S}}^{Z,W,\hat{r}}$
\end{algorithmic}
\end{algorithm}

\subsection{The ceiling of each family}\label{sec:ceilings}

Proposition~\ref{prop:routing} bounds routing as a whole. The families of routing just traversed are closed from above by three arrangements, each of which conditions on something a workflow cannot have, and is written here as receiving it from an oracle, so that what makes it unreachable is stated in its \textbf{Requires} line rather than buried in the steps.

$i,j \mid \mathrm{class}$ (Algorithm~\ref{alg:B1}) closes assignment on the class. It sends each instance to the incumbent of its true class and attains $\sum_k p_k \mu_{\bar{s}(k),k} = b + \Rclass$ by Proposition~\ref{prop:class}, engaging one skill per instance. It attains at least $b$. $i,j \mid \mathrm{inst}$ (Algorithm~\ref{alg:B2}) closes assignment generally: it executes whichever skill turns out best on the instance and attains $\E[\Mx] = b + R$ with one execution per instance. $ij(w \mid \mathrm{inst})$ (Algorithm~\ref{alg:B3}) closes weighted ensembling, and selection with it, since the best response to select after execution is the winner's. The last two attain the same performance on every instance and differ by an entire execution: since the truth is an endpoint, the best weighting places all its mass on the winner whatever the monotone rule (Proposition~\ref{prop:routing}), so the ensemble returns that skill's response having executed every skill rather than one. In performance they are the same arrangement; what separates them is whether the winner is chosen before the executions, by the engagement criterion, or after them, by the aggregation. Their cost profiles record the difference: the two hold the same skills available, and the second executes every skill where the first executes one, at an expected extra execution cost of $\sum_s c_s - \E[c_{\dot{s}}]$ per instance, which is the whole of what executing selectively saves (Table~\ref{tab:costs}). The archetype $HM{+}{+}$ at $\gamma = 1$ is $ij(w \mid \mathrm{inst})$; with $\gamma > 1$ it is the reduced-form upper bound of routing with synergy.

The three differ in what their oracle supplies, and therefore in whether anything remains to be learned. The oracle of $i,j \mid \mathrm{class}$ supplies the class alone, so the correct response still arrives later and the arrangement still has something to observe and records to extend. The oracles of the other two supply the performances themselves, which is everything adjudication would have brought, so their last two blocks are empty. That emptiness is the mark of a bound: every deployable arrangement on the ladder has something to observe and something to refresh. Synergy has no ceiling of this kind below perfect performance, since no oracle over the independent responses bounds what an exchange returns.

\begin{algorithm}[H]
\rungnum{B1}
\caption{$i,j \mid \mathrm{class}$: the upper bound of class assignment}
\label{alg:B1}
\begin{algorithmic}
\Require $\mathcal{R}_{s,k} \;\forall\, s,k$, and an oracle supplying $k$ for the instance at hand
\Setup
\State $\hat{\mu}_{s,k} \gets \mean$ of $\theta$ over $\mathcal{R}_{s,k}$, $\;\forall\, s,k$
\State $\bar{s}(k) \gets \argmax_{s \in \mathcal{S}} \hat{\mu}_{s,k} \;\forall\, k$
\State make every incumbent available \Comment{standing cost $\sum_{s \in \bar{s}(\mathcal{K})} \kappa_s$}
\Ensure
\State $\hat{r}_{\bar{s}(k)} \gets$ skill $\bar{s}(k)$ executes the instance \Comment{$k$ from the oracle; execution cost $c_{\bar{s}(k)}$}
\State \Return $\hat{r} \gets \hat{r}_{\bar{s}(k)}$
\Adjudicate
\State observe $r$
\State compute $\theta_{\bar{s}(k)}$
\State append the instance, with $\hat{r}_{\bar{s}(k)}$ and $r$, to $\mathcal{R}_{\bar{s}(k),k}$
\Refit
\State recompute $\hat{\mu}_{\bar{s}(k),k}$, hence $\bar{s}(k)$
\end{algorithmic}
\end{algorithm}

\begin{algorithm}[H]
\rungnum{B2}
\caption{$i,j \mid \mathrm{inst}$: the upper bound of assignment}
\label{alg:B2}
\begin{algorithmic}
\Require an oracle supplying $\theta_s \;\forall\, s \in \mathcal{S}$ for the instance at hand
\Setup
\State make every skill in $\mathcal{S}$ available \Comment{standing cost $\sum_s \kappa_s$}
\Ensure
\State $\dot{s} \gets \argmax_{s \in \mathcal{S}} \theta_s$ \Comment{from the oracle}
\State $\hat{r}_{\dot{s}} \gets$ skill $\dot{s}$ executes the instance \Comment{execution cost $c_{\dot{s}}$}
\State \Return $\hat{r} \gets \hat{r}_{\dot{s}}$
\Adjudicate
\State --- \Comment{the oracle has already supplied what adjudication would bring}
\Refit
\State --- \Comment{nothing is refitted}
\end{algorithmic}
\end{algorithm}

\begin{algorithm}[H]
\rungnum{B3}
\caption{$ij(w \mid \mathrm{inst})$: the upper bound of ensembling and selection}
\label{alg:B3}
\begin{algorithmic}
\Require an oracle supplying $r$, hence $\theta_s \;\forall\, s \in \mathcal{S}$, for the instance at hand
\Setup
\State make every skill in $\mathcal{S}$ available \Comment{standing cost $\sum_s \kappa_s$}
\Ensure
\State $\{\hat{r}_s\} \gets$ each $s \in \mathcal{S}$ executes the instance \Comment{execution cost $\sum_s c_s$}
\State $w \gets \argmax_{w} \theta\bigl(\sum_s w_s \hat{r}_s, r\bigr)$ \Comment{from the oracle; under any monotone rule on an endpoint truth this places all mass on $\dot{s}$ (Proposition~\ref{prop:routing})}
\State $\{\hat{r}_s\} \gets$ each $s \in \mathcal{S}$ executes the instance \Comment{execution cost $\sum_s c_s$}
\State \Return $\hat{r} \gets \sum_{s \in \mathcal{S}} w_s\, \hat{r}_s$
\Adjudicate
\State --- \Comment{the oracle has already supplied what adjudication would bring}
\Refit
\State --- \Comment{nothing is refitted}
\end{algorithmic}
\end{algorithm}

\subsection{Rung 5: routing, with synergy}\label{sec:rung5}

This rung operates both channels, and it is the only place in the table where they meet. Synergy composes only with arrangements that engage several skills on the same instance, since a protocol needs several of them to work with, which rules out assignment; and selection returns one of the responses it commissioned, which is routing. What remains is the weighted ensemble with an exchange between the skills before the joint response is fixed, the weights governing how much each is heard rather than how the independent responses are averaged (Algorithm~\ref{alg:5}). An example: the radiologist reads $90\%$; the model responds $10\%$ and attaches similar films from its training set; the radiologist revises her reading to $50\%$; the weights the workflow assigns then yield an aggregate of $30\%$. One skill corrected its response on information the other actively provided, which goes beyond routing, since nothing in either original response was selected or weighted into the result as it stood. This is the division of labor described as \emph{centaurs}, operators who direct the machine to selected sub-tasks and reserve their own judgment for what it cannot supply \citep{randazzo2025cyborgs}. The arrangement engages every skill and runs the exchange, needs the weights, retains the independent responses, and measures the lift against the weighted combination those responses would have produced. Whether it attains more than the rungs below is a separate question, since the ladder orders arrangements by how they decide and not by what they attain.

\begin{algorithm}[H]
\rungnum{5}
\caption{$ij(w \mid Z_q, W_q){+}$: both channels}
\label{alg:5}
\begin{algorithmic}
\Require $\mathcal{R}_{\mathcal{S}}^{Z,W,\hat{r}}$, and a protocol $\rho$ producing a joint response from several, not one of them
\Setup
\State make every skill in $\mathcal{S}$ available \Comment{standing cost $\sum_s \kappa_s$}
\State fit the weights $\hat{w}(Z_q, W_q) = \argmax_{w \geq 0,\, \sum_s w_s = 1} \hat{\E}\big[\theta(\textstyle\sum_s w_s \hat{r}_s, r) \mid Z_q, W_q\big]$ on $\mathcal{R}_{\mathcal{S}}^{Z,W,\hat{r}}$ \Comment{scored on retained responses}
\Ensure
\State $\hat{w} \gets \hat{w}(Z_q, W_q)$ on the instance
\State $\{\hat{r}_s\} \gets$ each $s \in \mathcal{S}$ executes the instance \Comment{execution cost $\sum_s c_s$}
\State $\hat{r}_\rho \gets \rho\bigl(\mathcal{S}, \{\hat{r}_s\}, \hat{w}\bigr)$ \Comment{the exchange, at cost $c_\rho$; the independent responses are retained}
\State \Return $\hat{r} \gets \hat{r}_\rho$
\Adjudicate
\State observe $r$
\State compute $\theta_s \;\forall\, s \in \mathcal{S}$, and $\theta = \theta(\hat{r}_\rho, r)$
\State ($\gamma \gets \theta \,/\, \theta\bigl(\sum_s \hat{w}_s \hat{r}_s, r\bigr)$) \Comment{the lift on this instance, over the weighted combination}
\State append the instance, with every $\hat{r}_s$, $Z_q$, $W_q$ and $r$, to $\mathcal{R}_{\mathcal{S}}^{Z,W,\hat{r}}$
\Refit
\State refit $\hat{w}(Z_q, W_q)$ on the updated $\mathcal{R}_{\mathcal{S}}^{Z,W,\hat{r}}$
\State ($\gamma_\pi \gets$ the ratio of the means of $\theta(\hat{r}_\rho, r)$ and $\theta\bigl(\sum_s \hat{w}_s \hat{r}_s, r\bigr)$ over the instances adjudicated so far)
\end{algorithmic}
\end{algorithm}

\subsection{The ladder against the decomposition}\label{sec:ladderfig}

\begin{figure}[!htb]
    \centering
    \includegraphics[width=\linewidth]{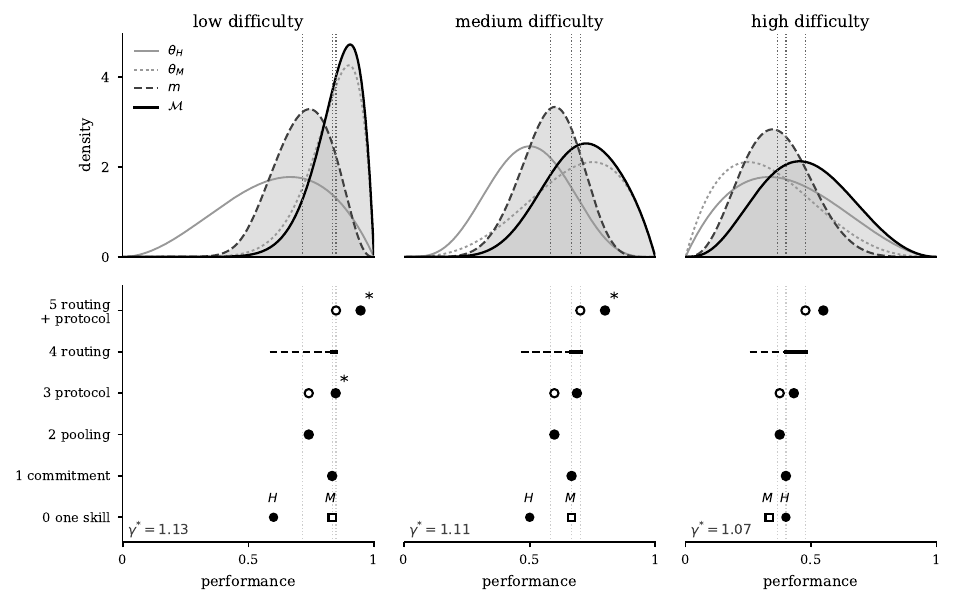}
    \caption{The ladder against the decomposition, illustrative Profile A at the initial date, one column per level of difficulty. Above, the exact densities of the two skills in gray ($\theta_H$ solid, $\theta_M$ dotted) and of the within-instance anchors, $m$ with a dashed outline and $\Mx$ with a solid one, with dotted lines at $\E[m]$, $b$ and $\E[\Mx]$; below, the performance each rung of Table~\ref{tab:ladder} can attain, rung~0 showing both skills and rung~2 the pooled response scored under the squared rule, above $\E[m]$ by the disagreement \sqc{$\Lambda$}; random assignment (Algorithm~\ref{alg:2r}) is not drawn, since it sits on the dotted line at $\E[m]$. Hollow marks carry no lift; solid marks apply a uniform lift $\gamma = 1.15$ in the reduced form of Proposition~\ref{prop:archetypes}, computed by clipping each instance and averaging afterwards, with an asterisk where the ceiling clips more than $1\%$ of instances (Proposition~\ref{prop:ceiling}). Rung~4 is drawn solid from $b$ to perfect routing and dashed down to $\E[m] - G$, where routing the wrong way on every instance would leave it (Proposition~\ref{prop:routing}). The ladder orders arrangements by how much of the two channels they operate, not by what they attain. $\gamma^{*}$ is reported in each column footer}
    \label{fig:ladder_grid}
\end{figure}

Figure~\ref{fig:ladder_grid} places the rungs on the performance axis for the illustrative Profile A at the initial date, one column per level of difficulty, above the skill densities and anchors that set the stake. What each rung can collect is fixed by the skills. At low difficulty $b$ sits within $0.015$ of $\E[\Mx]$, since the machine leads on $85\%$ of instances, so routing has almost nothing left to find and the stake is overwhelmingly systematic. At high difficulty the densities overlap almost completely, and the residual accounts for $70\%$ of the stake. Rung~2 sits \sqc{$S - \Lambda$} below rung~1, but rung~1 is drawn at its best: a wrong assessment returns $2S$ less, and rung~2 sits just above the midpoint, the disagreement between the responses repaying part of what it forgoes to be protected against the second. Rung~4 is drawn solid from $b$, because a criterion that uses what it reads never falls below it (Proposition~\ref{prop:information}). With no lift, rung~5 coincides with the top of rung~4 and rung~3 with rung~2, so the lift is all that separates each pair. With a uniform lift of $1.15$ in the reduced form, rung~3 passes the informed baseline at every level, where $\gamma^{*}$ is \sqc{$1.13$}, \sqc{$1.11$} and \sqc{$1.07$}, at low difficulty only narrowly, the ceiling already clipping \sqc{$10\%$} of the pooled responses there and $48\%$ of instances at the routed anchor (Proposition~\ref{prop:ceiling}). What the climb is worth is set by the skills: the same rungs span $0.111$ of performance at high difficulty and leave only $0.015$ above the informed baseline at low, where a fixed protocol must clear \sqc{$\gamma^{*} = 1.13$} while routing at its bound has no threshold, starting as it does from the routed anchor. A well-tuned rung~3 can overtake a poorly routed rung~4.

\begin{figure}[!htb]
\centering
\includegraphics[width=\linewidth]{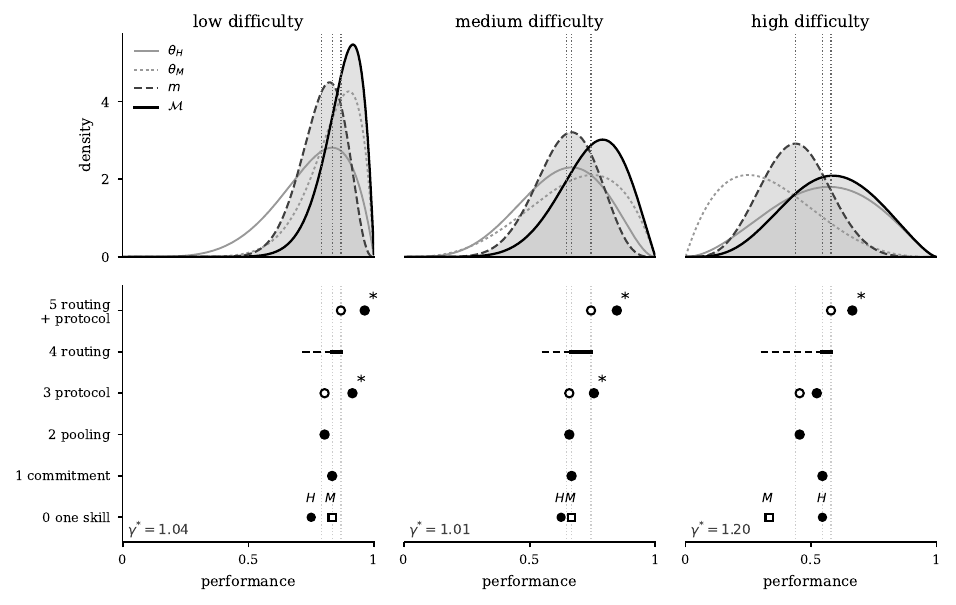}
\caption{The ladder against the decomposition at the final date, illustrative Profile A; conventions as in Figure~\ref{fig:ladder_grid}. Compared with the initial date, the informed baseline rises only at high difficulty, where the improving human skill overtakes the machine; at medium difficulty the machine remains the leader ($0.667$ against $0.625$), and the share of the stake carried by the residual inverts across the three levels.}
\label{fig:ladder_grid_late}
\end{figure}

The spacing of the rungs is a property of the skill pair, not of the ladder, and it moves as the skills evolve. Figure~\ref{fig:ladder_grid_late} redraws the ladder at the final date, after the human has improved and the machine has not. The informed baseline rises wherever the human overtakes the machine, from $0.400$ to $0.545$ at high difficulty, so rung~1 moves right and the rungs above it have less to add; and the ordering of $R/G$ across levels inverts, from $70\%$ against $12\%$ at high and low difficulty to $79\%$ at medium against $46\%$ at low and $24\%$ at high. The break-even lifts move with them: the lift of $1.15$ carries rung~3 past $b$ at low and medium difficulty, where $\gamma^{*}$ has fallen to \sqc{$1.04$} and \sqc{$1.01$}, but no longer at high difficulty, where the improved human has raised it to \sqc{$1.20$}.

\subsection{Records consumed and records produced}\label{sec:records}

Each policy states its requirements in its own \textbf{Requires} line, and those requirements are not equally hard to satisfy. The ordering is worth stating because it runs against the number of skills a policy engages. The easiest record needs no execution at all: instances with their class and signal, which grow whoever worked them and can be assembled from outside the organization. Next come the per-skill records, which each concern one skill and may be drawn from separate periods, caseloads or organizations. Hardest is the joint record, which requires the same instance to have been worked more than once and which no policy engaging a single skill produces.

The arrangements that engage least, and so cost least to operate, therefore depend on the scarcest evidence. A policy engaging one skill per instance needs a record that only double working produces; a policy engaging every skill produces that record in the course of its own operation. An organization moving to the first stops generating the evidence that made the move look worthwhile. This is what the protection argument amounts to in general. At rung~2 the second engagement insures against a commitment that was wrong; at ensembling, the weights insure against the assignment model being wrong; and in both cases what is also bought is the evidence to tell. This is the selective labels problem \citep{lakkaraju2017selective}, in which a decision-maker's choices determine which instances are labeled at all; and where the logging policy is deterministic, as an assignment criterion is, examples can be constructed that guarantee sub-optimal learning however much data accumulates \citep{swaminathan2015batch}. An arrangement that commissions one skill learns nothing about the others: it cannot find out that the skill it dropped has improved, or that the classes have shifted, and it cannot recognize its own commitment as wrong however wrong it becomes. Appendix~\ref{app:records} sets out the records, and the \textbf{Requires} and adjudication lines of each algorithm state which of them the policy consumes and which it produces.

The choice is not all or nothing. An assignment policy can execute the other skills on a fraction of instances drawn at random, keep their responses for the record, and deliver the assigned skill's response in every case.

\begin{proposition}[Assignment with audit]\label{prop:audit}
Let an assignment policy also engage, on each instance, every other available skill with probability $\epsilon(x) \geq \underline{\epsilon} > 0$, where $x$ is what its criterion reads before execution and the draw is independent of the performances given $x$, and let it deliver the assigned skill's response in every case. Then: (a) it attains what the assignment attains; (b) its expected execution cost exceeds the assignment's by $\E[\epsilon(x) \sum_{s \neq \hat{s}} c_s]$; (c) it adds to the joint record $\E[\epsilon(x)]$ paired instances per instance worked; and (d) for any function $g$ of the jointly executed performances, $\frac{1}{N}\sum_{q=1}^{N} A_q\, g_q / \epsilon(x_q)$, with $A_q$ the indicator that instance $q$ was audited, is an unbiased estimator of $\E[g]$, so that $\E[\Mx]$, $G$, $R$ and the attainment of any other policy remain estimable from the audited instances.
\end{proposition}

\begin{proof}
(a) The delivered response is the assigned skill's on every instance. (b) and (c) follow from $\E[A_q \mid x_q] = \epsilon(x_q)$. (d) $\E[A_q g_q / \epsilon(x_q) \mid x_q, g_q] = g_q$ by the conditional independence of the draw, and the law of iterated expectations gives $\E[g]$; this is the estimator of \citet{horvitz1952generalization}.
\end{proof}

The audit prices the evidence. A fraction $\epsilon$ of the second execution's cost buys a joint record growing at rate $\epsilon$, and with it the ability to detect that a dropped skill has improved, that the classes have shifted, or that the assignment model has gone wrong. Auditing more where the criterion is least certain concentrates the record where refitting needs it, and the weights $1/\epsilon(x)$ keep the estimates unbiased. They do not keep them precise: the variance of the estimator in (d) grows with $1/\epsilon(x)$, so that $\epsilon$ must stay bounded away from zero precisely where the criterion is most confident and an audit seems least needed. The problem is that of off-policy evaluation, which values a policy from instances logged under another with known probabilities, as \citet{li2010contextual} do from uniformly random logging; doubly robust estimators, which add a model of the performances to the inverse-propensity weights, tend to reduce the variance where that model is reasonable \citep{dudik2011doubly}. Algorithm~\ref{alg:audit} writes the arrangement, and Tables~\ref{tab:spectrum} and~\ref{tab:costs} give its attainment and its cost profile.

\section{Climbing the ladder: a procedure}\label{sec:procedure}

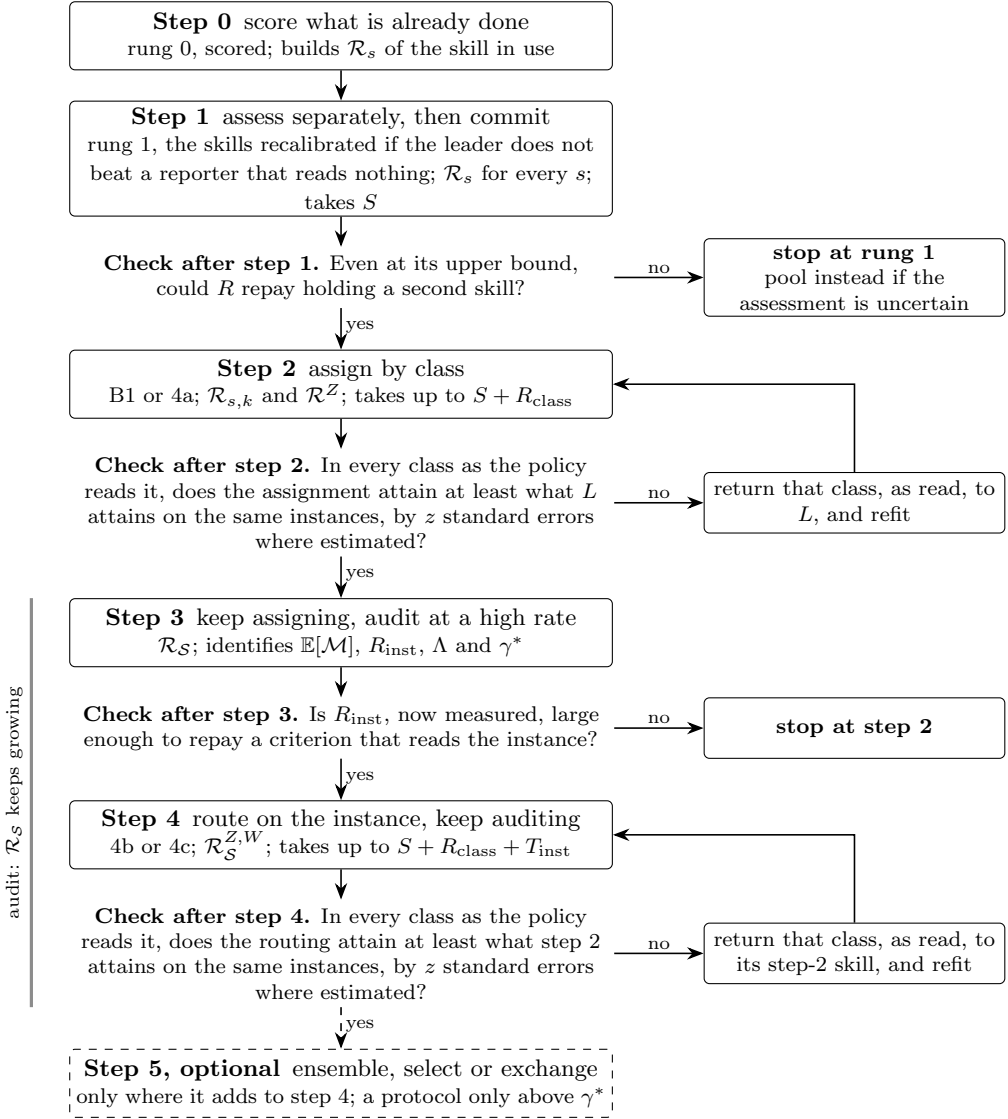
\begin{figure}[!htb]
\centering
\begin{tikzpicture}[
  font=\small, >={Stealth[length=2.2mm]},
  box/.style={draw, rounded corners=2pt, align=flush center, text width=70mm, minimum height=9mm, inner sep=2.5pt},
  gate/.style={align=flush center, text width=70mm, inner sep=3pt, font=\footnotesize},
  exit/.style={draw, rounded corners=2pt, align=flush center, text width=38mm, minimum height=8mm, inner sep=2.5pt, font=\footnotesize},
  flow/.style={->, semithick},
  lab/.style={font=\scriptsize, inner sep=1.5pt},
]
\node[box] (s0) at (0,0) {\textbf{Step 0}\enspace score what is already done\\[-1pt] {\footnotesize rung~0, scored; builds $\mathcal{R}_s$ of the skill in use}};
\node[box, below=4mm of s0] (s1) {\textbf{Step 1}\enspace assess separately, then commit\\[-1pt] {\footnotesize rung~1, the skills recalibrated if the leader does not beat a reporter that reads nothing; $\mathcal{R}_s$ for every $s$; takes $S$}};
\node[gate, below=4mm of s1] (c1) {\textbf{Check after step 1.} Even at its upper bound, could $R$ repay holding a second skill?};
\node[box, below=5.5mm of c1] (s2) {\textbf{Step 2}\enspace assign by class\\[-1pt] {\footnotesize B1 or 4a; $\mathcal{R}_{s,k}$ and $\mathcal{R}^{Z}$; takes up to $S + \Rclass$}};
\node[gate, below=4mm of s2] (c2) {\textbf{Check after step 2.} In every class as the policy reads it, does the assignment attain at least what $L$ attains on the same instances, by $z$ standard errors where estimated?};
\node[box, below=5.5mm of c2] (s3) {\textbf{Step 3}\enspace keep assigning, audit at a high rate\\[-1pt] {\footnotesize $\mathcal{R}_{\mathcal{S}}$; identifies $\E[\Mx]$, $\Rinst$, \sqc{$\Lambda$} and $\gamma^{*}$}};
\node[gate, below=4mm of s3] (c3) {\textbf{Check after step 3.} Is $\Rinst$, now measured, large enough to repay a criterion that reads the instance?};
\node[box, below=5.5mm of c3] (s4) {\textbf{Step 4}\enspace route on the instance, keep auditing\\[-1pt] {\footnotesize 4b or 4c; $\mathcal{R}^{Z,W}_{\mathcal{S}}$; takes up to $S + \Rclass + T_{\mathrm{inst}}$}};
\node[gate, below=4mm of s4] (c4) {\textbf{Check after step 4.} In every class as the policy reads it, does the routing attain at least what step~2 attains on the same instances, by $z$ standard errors where estimated?};
\node[box, dashed, below=5.5mm of c4] (s5) {\textbf{Step 5, optional}\enspace ensemble, select or exchange\\[-1pt] {\footnotesize only where it adds to step~4; a protocol only above $\gamma^{*}$}};
\foreach \a/\b in {s0/s1, s1/c1, s2/c2, s3/c3, s4/c4} \draw[flow] (\a) -- (\b);
\draw[flow] (c1) -- node[lab, right, pos=0.45] {yes} (s2);
\draw[flow] (c2) -- node[lab, right, pos=0.45] {yes} (s3);
\draw[flow] (c3) -- node[lab, right, pos=0.45] {yes} (s4);
\draw[flow, dashed] (c4) -- node[lab, right, pos=0.45] {yes} (s5);
\coordinate (ex) at ($(s0.east)+(12mm,0)$);
\node[exit, anchor=west] (x1) at (ex |- c1) {\textbf{stop at rung~1}\\ pool instead if the assessment is uncertain};
\node[exit, anchor=west] (x2) at (ex |- c2) {return that class, as read, to $L$, and refit};
\node[exit, anchor=west] (x3) at (ex |- c3) {\textbf{stop at step~2}};
\node[exit, anchor=west] (x4) at (ex |- c4) {return that class, as read, to its step-2 skill, and refit};
\draw[flow] (c1) -- node[lab, above] {no} (x1);
\draw[flow] (c2) -- node[lab, above] {no} (x2);
\draw[flow] (c3) -- node[lab, above] {no} (x3);
\draw[flow] (c4) -- node[lab, above] {no} (x4);
\draw[flow] (x2.north) |- (s2.east);
\draw[flow] (x4.north) |- (s4.east);
\coordinate (au) at ($(s3.west)+(-5mm,0)$);
\draw[very thick, black!45] (au |- s3.north) -- (au |- c4.south);
\node[lab, rotate=90, anchor=south, align=center] at ($(au |- s3.north)!0.5!(au |- c4.south)+(-0.6mm,0)$) {audit: $\mathcal{R}_{\mathcal{S}}$ keeps growing};
\end{tikzpicture}
\caption{The procedure as a flow. Each step deploys an arrangement and builds the record the next step needs; each check either lets the climb continue, stops it where the next component is not worth what pursuing it costs, or returns the classes that fall back to what the previous step assigned them while the criterion is refitted. The checks after steps~2 and~4 are made on each class as the policy reads it, observed at intake or predicted, against what the previous step attains on the same instances, by a margin of $z$ standard errors where the attainments are estimated (Proposition~\ref{prop:margin}); how the take splits between the components is read afterwards, on the true classes (Section~\ref{sec:routing}). From step~3 the audit keeps the paired record growing; step~5 grows it without one. The stated takes are those of a step carried out without error}
\label{fig:procedure}
\end{figure}

The decomposition and the ladder are of use to an organization when they are read as a sequence of steps. Each step deploys an arrangement of Section~\ref{sec:ladder}, builds the record the next step needs, and identifies the component of the stake the next step would pursue, so that the decision to take it rests on evidence rather than on expectation (Section~\ref{sec:identification}). The sequence starts at rung~0, the position of an organization that has a way of doing the work and has not examined it, and stops where the evidence says that climbing further does not pay. Table~\ref{tab:procedure} summarizes it, and Figure~\ref{fig:procedure} draws it with the checks that close each step; the paragraphs below state each step and the check that closes it, and explain why the checks must be made class by class.

\begin{table}[!htb]
\centering\footnotesize
\caption{A procedure for climbing the ladder. Each step deploys an arrangement, builds the record the next step requires, and identifies what the next step would pursue. The last column is the take when the step is carried out without error; errors in reading the class or the instance make it smaller, and split it between the components only by convention (Section~\ref{sec:routing})}
\label{tab:procedure}
\setlength{\tabcolsep}{4pt}
\begin{tabular}{c >{\raggedright\arraybackslash}p{3.6cm} >{\raggedright\arraybackslash}p{2.5cm} >{\raggedright\arraybackslash}p{3.2cm} >{\raggedright\arraybackslash}p{2.6cm}}
\toprule
Step & Arrangement & Record it builds & What it identifies & Take without error \\
\midrule
0 & rung~0, its executions scored (Alg.~\ref{alg:0}) & $\mathcal{R}_s$ of the skill in use & $\E[\theta_s]$ of that skill & $\E[\theta_s] - \E[m]$ \\[3pt]
1 & rung~1, after separate assessments (Alg.~\ref{alg:1}) & $\mathcal{R}_s$ for every $s$ & whether the leader beats a reporter that reads nothing; $b$ and $S$; an upper bound on $R$ & $S$ \\[3pt]
2 & assignment on the class: B1 where the class is observed at intake, 4a where it is predicted (Algs.~\ref{alg:B1}, \ref{alg:4a}) & $\mathcal{R}_{s,k}$ and $\mathcal{R}^{Z}$ & the incumbents and $\Rclass$ & $S + \Rclass$ \\[3pt]
3 & the same, with audit at a high rate (Alg.~\ref{alg:audit}, on the class assignment) & $\mathcal{R}^{Z,W}_{\mathcal{S}}$ & $\E[\Mx]$, $\Rinst$, \sqc{$\Lambda$} and $\gamma^{*}$ & $S + \Rclass$ \\[3pt]
4 & assignment on the instance, with audit (Algs.~\ref{alg:4b}, \ref{alg:4c}, \ref{alg:audit}) & $\mathcal{R}^{Z,W}_{\mathcal{S}}$ & the share of $\Rinst$ the criterion reaches & $S + \Rclass + T_{\mathrm{inst}}$ \\[3pt]
5 & optionally, pooling, ensembling, selection or a protocol (Algs.~\ref{alg:2}, \ref{alg:4d}, \ref{alg:4e}, \ref{alg:3}, \ref{alg:5}) & $\mathcal{R}^{Z,W,\hat{r}}_{\mathcal{S}}$ and the independent responses & the dividend; the lift $\gamma_\pi$ & adds the dividend or the lift \\
\bottomrule
\end{tabular}
\end{table}

\paragraph{Step 0: score what is already done.} An organization at rung~0 already runs a policy; what it lacks is the record of how well it runs. Adjudicating a sample of what the skill in use executes, the parenthetical steps of Algorithm~\ref{alg:0}, changes nothing in operation and is the precondition of every later step: an organization that does not score its responses can execute any policy and compare none.

\paragraph{Step 1: assess separately, then commit.} Each candidate skill is assessed on instances of the organization's own task, which need not be the same instances for every skill and may come from separate periods or pilots. The assessments identify $b$ and $S$ (Proposition~\ref{prop:S}), and the organization commits to the skill that leads on average, rung~1, gaining $2S$ over rung~0 if the skill in use was the weaker of a pair. The same records bound the residual from above, provided they keep the score of each instance and not only each skill's mean: their marginal distributions cap $R$ at what the most complementary coupling would give (Proposition~\ref{prop:anchor}), and the means alone cap it only trivially. \emph{Checks}: under a proper rule, the leader must beat a reporter that reads nothing (Section~\ref{par:rule}), which executes at no cost and which the same records supply by stating the share of positive instances in each class. A leader that falls short of it, $\sum_k p_k (\mathrm{Res}_{L,k} - \mathrm{Mis}_{L,k}) < 0$ by Equation~\eqref{eq:murphy}, is not usable as a probability as it reports; where it ranks the instances well, its resolution is there and what it lacks is calibration, which a recalibration fitted on its own record restores, and the procedure continues with the skills so recalibrated. Under the linear rule the corresponding reporter states the more frequent outcome of each class, and a leader that falls short of it while ranking the instances well signals that the rule, not the skill, is unfit for the task. And if even the cap on $R$ is small against what holding a second skill would cost, the climb stops here; if it is not, the residual may still be small, and the next steps establish how large it is. If the assessment itself is uncertain, pooling insures against a wrong commitment at the premium of Proposition~\ref{prop:insurance}.

\paragraph{Step 2: split the assessments by class, and assign by class.} The same records, sorted by class, identify the incumbent of each class and $\Rclass$, still without any instance worked twice (Proposition~\ref{prop:class}). Where the class is observed at intake, sending each class to its incumbent attains $b + \Rclass$ and takes $S + \Rclass$ (B1, Proposition~\ref{prop:nested}). Where the class must be predicted, the attainment depends on the accuracy of the prediction, and with two classes whose incumbents differ the policy beats the commitment only above the threshold accuracy $a^{*}$ of Proposition~\ref{prop:classpred}. \emph{Check}: in every class as the policy reads it, the policy attains at least what the commitment attains on the same instances, by a margin of $z$ standard errors where the attainments are estimated (see below).

\paragraph{Step 3: build the paired record while still delivering.} The instance level needs instances worked by both skills, and they can be obtained without giving up what step~2 secured. The organization keeps delivering the class assignment and has the other skill execute a share of the same instances, its responses kept for the record and not delivered: assignment with audit at a high rate, which attains what the assignment attains (Proposition~\ref{prop:audit}, Algorithm~\ref{alg:audit}). Pooling would produce the same record at a cost in performance, since it attains \sqc{$b - S + \Lambda$}, below $b$ unless the responses disagree by more than $S$. The paired record identifies $\E[\Mx]$, and with it $\Rinst$, the disagreement \sqc{$\Lambda$} and the break-even $\gamma^{*}$ (Section~\ref{sec:identification}). \emph{Check}: if $\Rinst$ is small against what a criterion reading the instance would cost to build and run, the climb stops at step~2.

\paragraph{Step 4: route on the instance, and keep auditing.} A criterion reading the class and instance signals is fitted on the paired record, evaluated on instances it was not fitted on (Section~\ref{sec:identification}), and deployed only where it adds to the class assignment it replaces (Algorithms~\ref{alg:4b} and~\ref{alg:4c}). Deployed, it would stop the paired record from growing (Section~\ref{sec:records}), so the audit continues, at a rate that can be concentrated where the criterion is least certain. The audited instances keep every population quantity estimable, and let the organization detect that a dropped skill has improved, that the classes have shifted or that the criterion has degraded (Proposition~\ref{prop:audit}). \emph{Check}: as at step~2, in every class as read, against what the class assignment attains on the same instances, with the same margin.

\paragraph{Step 5, optional: combine, or exchange.} Ensembling and selection execute every skill on every instance, and so keep the paired record growing without an audit. Ensembling pays where the disagreement it earns exceeds what it gives up in mean (Proposition~\ref{prop:ensembling}). Its simplest form is pooling, rung~2, with equal weights and nothing to fit, and the paired record of step~3 already prices it: it attains more than the commitment exactly when the responses disagree by more than $S$, \sqc{$\Lambda > S$} (Proposition~\ref{prop:insurance}). Selection pays where reading the responses tells which skill will do better on the instance more reliably than the signals do. A protocol is a separate investment: its lift is measured against the independent responses retained before the exchange, and it pays only above $\gamma^{*}$ (Proposition~\ref{prop:breakeven}).

\paragraph{The take does not climb smoothly.} The last column of Table~\ref{tab:procedure} is what each step takes when it is carried out without error, and only then does the take rise by whole components. A step carried out with error does not (Section~\ref{sec:routing}). A class prediction of accuracy $a$ takes, at the level of the class, $(2a - 1)(S + \Rclass)$: a line from $\E[m]$ at chance to $b + \Rclass$ at no error, which crosses $b$ only at $a^{*}$, so that a poorly predicted class leaves the organization below the commitment it replaced (Figure~\ref{fig:attribution_sweep}). A criterion that reads the instance can collect within some classes and give back at the level of others, and end where the commitment would have left it, as the router of the worked example does (Figure~\ref{fig:attribution}). How far the take moves at each step, and how much of the move each component accounts for, therefore depends on the policy and on how well it discriminates the instances: by class, where the incumbent is known ex ante, or one by one, where it is not.

\paragraph{The check, class by class.} The check that closes steps~2 and~4 is therefore made not on the total attainment, which can rise while part of the caseload falls back, but class by class, on the classes as the policy reads them: observed at intake, or predicted. These are the only groups of instances a policy can treat differently, and so the only ones a check can act on. For each class $\hat{k}$ as read, the new policy's attainment on the instances it places there is set against what the previous step attains on the same instances, which the class-wise means and the record $\mathcal{R}^{Z}$ give without any instance worked twice: $\sum_k \Pr(k \mid \hat{k})\, \mu_{L,k}$ after step~1, and $\sum_k \Pr(k \mid \hat{k})\, \mu_{\bar{s}(\hat{k}),k}$ after step~2. A class as read in which the new policy falls short is one the previous step served better, and the remedy is local: return it to the previous step's assignment while the criterion is refitted, which is gating in the sense of Section~\ref{sec:signals} at its coarsest grain. Each step is then taken only where it adds, and, with the class-wise means known, the climb never gives back what an earlier step held (Proposition~\ref{prop:checks}, Appendix~\ref{app:many}). On a finite record the class means are estimated, and each check compares two estimates that can err as the assessment behind a commitment can (Corollary~\ref{cor:insurance}): a switch that loses in the population can pass it. The check therefore adopts a switch on a class only where its cross-fitted gain exceeds zero by a margin, a multiple $z$ of its standard error; Proposition~\ref{prop:margin} states what the margin saves and what it forgoes. With two classes and a prediction of accuracy $a$, the check after step~2 is the threshold of Proposition~\ref{prop:classpred}: the class predicted for the skill other than $L$ is kept if and only if $a > a^{*}$. Read on the true classes, which an audited record reveals after the fact, the same take is credited as in Corollary~\ref{cor:attribution}. Classes at or above $b_k$ are resolved and their take is exact component by component; classes between $\mu_{L,k}$ and $b_k$ are unresolved, their take exact only in total (Section~\ref{sec:routing}). A true class can fall below $\mu_{L,k}$ even where the step pays, as the class led by $L$ does under any error in predicting it; that is a matter of crediting rather than a reason to step back, since no policy can treat apart the instances of a class it does not see.

\paragraph{What the procedure leaves open.} Each check weighs a component of the stake against what the step that pursues it costs, in skills held available and in executions (Table~\ref{tab:costs}), and the paper reports that cost profile without pricing it. Because the standing cost of holding a second skill is paid once, steps~2 to~4 should be judged together as well as one by one: a class assignment that does not pay on its own may pay as the first part of a route to the instance level. How the two would be priced is a matter of the setting. For illustration only, if each unit of performance were worth $\lambda$ on each of $V$ instances, a step from $\pi$ to $\pi'$ would pay when
\[
V \lambda \big(\E[\theta_{\pi'}] - \E[\theta_{\pi}]\big) \;>\; (\kappa_{\pi'} - \kappa_{\pi}) + V \big(\E[c_{\pi'}] - \E[c_{\pi}]\big),
\]
the standing cost entering once and the execution cost with the volume; the linearity is an assumption of the illustration, since the consequences of a poor execution are seldom proportional to its shortfall in performance. How large the audit should be, and when an estimate should be distrusted rather than refitted, are left open as well (Section~\ref{sec:limits}). And the procedure presumes that the organization survives the failures it incurs while it learns, since every check reads a record of them: where a single failure can cost without limit, a bound on what it may cost comes before the first step (Section~\ref{sec:contest}).

\section{Field arrangements read through the theory}\label{sec:field}

Three bodies of field evidence show how the abstractions meet deployment: the CXR experiment of \citet{agarwal2023combining}, which implemented several rungs on the same readers and films, and whose data, released as Collab-CXR \citep{moehring2025dataset}, allow the stake and the take to be measured; the double-reading arrangements of mammography screening, which name the combination rules and measure what each returns; and a field experiment in patent drafting \citep{autor2026does}, which measures what three months of an assisted workflow did to the human skill.

\subsection{The chest-radiograph experiment, read on the ladder}\label{sec:radiology}

\paragraph{Unaided reading and the model alone.} Each radiologist reading unaided is a rung-0 or rung-1 arrangement with the human skill; the model reading alone is the same arrangement with the machine skill. The model's area under the ROC curve exceeded that of $78\%$ of the participants; that measure ranks films rather than scoring each of them, and scored film by film the model outperforms about half of the participants under the squared rule and few of them under the linear rule (Section~\ref{sec:radiologydata}), so which skill supplies the informed baseline $b$ of rung~1 depends on the rule under which the skills are assessed. The diagnostic standard is binary, a case being positive when the mean of five experts' probabilities exceeds one half, so the endpoint property holds, and the deviation $|p - \omega|$ the authors report is the linear rule \citep{agarwal2023combining}, the special case of this paper's results in which the disagreement between responses earns nothing.

\paragraph{The arm with the model's prediction in view.} Radiologists read each film with the model's prediction in front of them, and nothing prescribed what to do with it. By Definition~\ref{def:synergy} this is an exchange: the returned response is a reading fixed after sight of the other skill's response, not a function of the reading the radiologist would have produced alone. It sits at rung~3 by structure, an exchange without a designed protocol. Its outcome was indistinguishable, on average, from unaided reading. The authors trace the shortfall to two errors of updating: radiologists underweighted the model's prediction, and treated it as independent of their own reading although both read the same film. The average also conceals heterogeneity, since $48\%$ of the radiologists improved with the prediction in view on the top-level findings for which the model reports a probability: the lift of the exchange varied across readers as well as across films, and averaged below one. Measured against the mechanical baseline of Equation~\eqref{eq:gamma}, the lift of this exchange was below one under the squared rule; measured against the informed baseline, the arm was level with it on average (Section~\ref{sec:radiologydata}). The within-reader design is what made this measurable at all: the same readers read the same films unaided, which supplies the independent responses that Algorithm~\ref{alg:3} retains before the exchange.

\paragraph{The arm with clinical history.} Radiologists received clinical history the model could not see. Nothing in the arrangement changed; what changed was a skill. Information the model does not hold alters $\theta_H$ itself, and with it $b$, $S$ and $R$; measured, it moved the stake from $S$ to $R$ and left its size unchanged (Section~\ref{sec:radiologydata}). The arm improved performance because it changed a skill, not because it climbed the ladder, which is the distinction between stake and take seen from the other side: enriching one skill's inputs was worth doing on its own, whereas placing the other skill's response in front of it, with no protocol for what to do with it, was not.

\paragraph{The combination rule computed afterwards.} The benchmark the authors compute is a logistic regression of the diagnostic standard on the radiologist's unaided report and the model's prediction, evaluated by the area under the ROC curve, on which it would rank above $86\%$ of the radiologists. It involves no exchange, since it operates on reports produced independently, so it is not synergy. It is not a rung-2 combination either: a logistic transform of a weighted sum of log-odds can return a probability outside the interval spanned by the two reports, and the hull bound does not apply to it (Section~\ref{sec:twochannels}); and a ranking measure does not score each film, so the comparison with the model's percentile is not a comparison of attainment in the sense of this paper. The placement needs nothing more: the rule gives no film to either skill, but combines both responses on every film, and the example of Section~\ref{sec:twochannels} shows how such a rule can attain more than routing omniscience. In the ladder of \citet{zanardelli2025navigating}, which this paper extends, the rule sat at rung~3 and the arm with the model's prediction in view at rung~2; with rung~3 defined here as synergy, which requires an exchange (Definition~\ref{def:synergy}), the arm moves up to it, and the rule, which exchanges nothing, moves outside the ladder.

\paragraph{The delegation policy.} The policy the authors derive sends each film to the model alone, to the radiologist alone, or to the radiologist with the model's prediction in view, according to the model's predicted probability, a quantity available before the outcome is known. It is a choice among arrangements of the kind Section~\ref{sec:signals} calls gating, and between its first two branches it is assignment on the standing, rung~4b (Algorithm~\ref{alg:4b}), reading a signal about one skill's standing on the instance. It could be fitted only because the within-reader design had produced a paired record of both skills on the same films. Deployed, it would stop producing that record on the films it sends to one reader (Section~\ref{sec:rung4}), unless it audited a fraction of them, as Algorithm~\ref{alg:audit} does (Proposition~\ref{prop:audit}).

Read as a whole, the CXR experiment documents an organization whose default arrangement sat at rung~3 by structure and no higher than rung~1 by outcome, while showing that a fixed rule and a rung-4 policy were within reach, and that enriching one skill's inputs moved the outcome without moving the rung.

\paragraph{An outcome performance alone cannot represent.}\label{par:floor} A theory whose arrangements act on performances, as the archetypes of Equation~\eqref{eq:archetypes} do, has a floor: the least any of them returns is what the mean policy $HM$ returns, $\E[\theta_{\bar{w}}] = \E[m] + \Lambda$ under the squared rule and $\E[m]$ under the linear one, above the weaker skill whenever the two differ. Under the squared rule the assisted arm lies below that floor, by $0.0085$ in the arm with clinical history, and below it in every variant measured, on subsets of the findings and after recalibration (Section~\ref{sec:radiologydata}). On performances it cannot be produced, because an interaction function that takes the two performances and a multiplier $\gamma \geq 1$ as given has no option that weights the skills in favor of the weaker one or scales an anchor downward. On protocols it is an ordinary case: a synergy protocol whose measured lift is below one. It is also recognizable in routing terms. Reader by reader, the arm returned about the mean of the reader's and the model's performances, $\E[m] = b - S$, what assigning each film to one of the two skills at random returns; for the readers the model outperforms, that is halfway to the $b - 2S$ of a commitment to the wrong skill (Proposition~\ref{prop:S}), with both skills engaged on every film. Rung~2 would have avoided it mechanically, since averaging the same two responses earns their disagreement above the mean anchor.

The case is more severe than any arrangement that merely fails to design a combination: in the terms of \citet{hemmer2025complementarity} it is a negative total complementarity effect, the team falling below its better member. It is the region where automation bias and miscalibrated trust operate \citep{parasuraman2010complacency, romeo2026exploring}, in either direction, and it strengthens the prescription of Section~\ref{sec:identification}: exposing one skill to another's response with nothing prescribing what to do with it can return less than combining the two responses mechanically, and the only way to know is to have retained the independent responses and measured the lift.

\subsection{The chest-radiograph experiment, measured}\label{sec:radiologydata}
 
\begin{figure}[!htb]
\centering
\includegraphics[width=\linewidth]{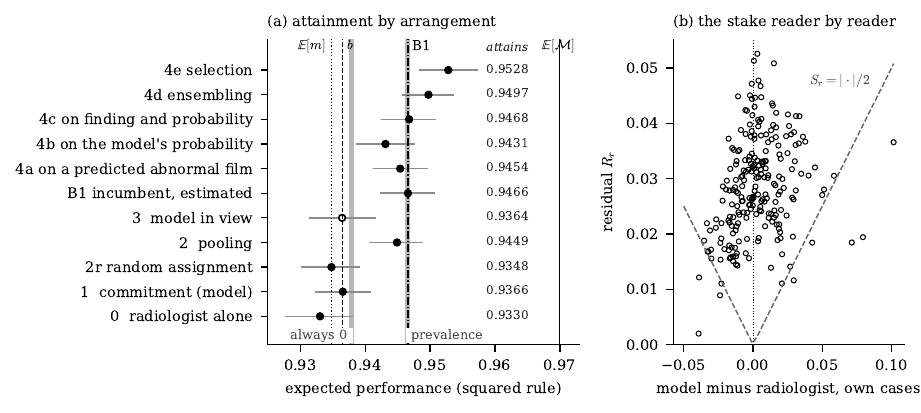}
\caption{The CXR experiment, arm with clinical history, squared rule. (a) What each arrangement attains, with 95\% intervals over resamples of films and the value in the column on the right; the vertical lines are the mean anchor, the informed baseline (the model's), the class ceiling B1 and the routed anchor, and the gray bands the two reporters that read nothing, each finding's prevalence, within $0.0001$ of B1, and always zero. The intervals are wide because the films differ in difficulty for every arrangement alike; differences between arrangements are estimated far more tightly (text). The open marker is the arm with the model's prediction in view; rung~4a is the one built in Appendix~\ref{app:cxrfilm}. (b) Each of the 227 radiologists, on the films he or she read: the model's lead over the reader, and the residual of the pair; the dashed line is the systematic component $S_r$, half the absolute lead.}
\label{fig:cxr}
\end{figure}

The Collab-CXR data hold what the theory asks of a paired record. The 227 participating radiologists, each reading a subset of 324 historical chest films, reported for each film a probability for every finding of a hierarchical list, unaided or with the model's prediction in view, with and without clinical history; the prediction of a single model, CheXpert, is recorded for every film, whether or not it was shown; and five board-certified chest radiologists, reading every film, supply the diagnostic standard. Appendix~\ref{app:cxrdata} sets out the experiment's designs and the data as they are used here. Table~\ref{tab:cxrterms} sets out the units of the data and their place in the theory. We take as the task the assessment of one finding on one film, over the 15 findings for which the model reports a probability. An instance is a read, and its class is the finding, which an organization names before anyone executes: the diagnostic question, how probable it is that pneumothorax, say, is present on this film, and not its answer, which is the truth $r$ that the panel supplies afterwards; since the question is fixed before anyone reads the film, the class is observed at intake. The truth, the model's response and so the signal $W$ are the same for every read of a film-finding, and only the radiologist's response varies: a policy that decides on $Z$ and $W$ gives all the reads of a film-finding to the same skill. Assigning each finding to its incumbent is therefore the arrangement B1, with its incumbents estimated, and not rung~4a, which assigns on a class it must predict and can misclassify; a rung~4a arises only where the class includes something not known at intake, such as whether the film is abnormal (Appendix~\ref{app:cxrfilm}). The human skill $H$ is a participant drawn at random, reading unaided; the machine skill $M$ is the model. The truth is the binary standard, positive when the mean of the five experts' probabilities exceeds one half, so that it is an endpoint; $6.2\%$ of the 4{,}860 film-findings are positive. Each film-finding weighs the same, and so does each of its reads, of which there are between 7 and 30 in each unaided arm. The findings form a hierarchy: a finding whose parent the radiologist ruled out was never shown, and its probability is recorded as zero, as it is for $58\%$ of the unaided reads, all of them below the four top-level findings. The panel read through the same hierarchy; a standard built only from the panellists who were shown each finding leaves the leader, the incumbents, the weight of the residual and the arm's place below pooling as they are (Appendix~\ref{app:cxrvar}). Every unaided read is paired with the model's response on the same instance, so each quantity of Section~\ref{sec:bounds} is identified rather than bounded. Intervals are 95\% percentiles over 1{,}000 resamples of films; resampling radiologists instead gives narrower ones, and resampling films and radiologists together wider ones, which change no conclusion (Table~\ref{tab:cxrtwoway}); the policies of rung~4 are cross-fitted, each fitted on four fifths of the films and scored on the fifth it did not see, other splits of the films moving their attainment by at most $0.0005$, and their criterion reads bins of the probabilities, a deliberately simple one. Table~\ref{tab:cxr} reports the two unaided arms, without and with clinical history, under both rules; Figure~\ref{fig:cxr} draws the arm with history under the squared rule, on which the text dwells; Appendix~\ref{app:cxr} reports each finding.

\begin{table}[!htb]
\centering
\footnotesize
\caption{The units of the Collab-CXR data, and their place in the theory. Counts are for one arm; the columns of the file are those of Table~\ref{tab:cxrcols}}
\label{tab:cxrterms}
\setlength{\tabcolsep}{4pt}
\begin{tabularx}{\textwidth}{@{}l >{\raggedright\arraybackslash}X >{\raggedright\arraybackslash}p{2.5cm} >{\raggedright\arraybackslash}p{2.1cm} >{\raggedright\arraybackslash}X@{}}
\toprule
Unit & What it is & In the file & How many & In the theory \\
\midrule
Film & one patient's chest radiograph, which comes with a short clinical history & \texttt{patient\_id} & 324 & the unit on which folds and resamples are drawn \\ \addlinespace[2pt]
Finding & a condition every film is asked about, such as pneumothorax, whether or not it turns out to be present: the diagnostic question & \texttt{pathology} & 15 of 104 & the class $k$, named at intake, so that the signal $Z$ gives it exactly \\ \addlinespace[2pt]
Film-finding & one finding on one film: whether pneumothorax is present on film 154 & \texttt{patient\_id} and \texttt{pathology} & 4{,}860 & carries the truth $r$ and the model's response $\hat{r}_M$, which is also the signal $W$; each weighs the same \\ \addlinespace[2pt]
Reading & one radiologist reading one film in one condition, answering all its findings at once & \texttt{experiment\_id}, \texttt{patient\_id} and \texttt{treatment} & 5{,}240, about 16 per film & the human skill executing the 15 instances of a film \\ \addlinespace[2pt]
Read & one radiologist's probability for one finding on one film, what a reading returns on one film-finding & a row, \texttt{probability} & 78{,}600, 7 to 30 per film-finding & an instance: $\hat{r}_H$, paired with $\hat{r}_M$ and $r$; each weighs $1/n$ of its film-finding \\
\bottomrule
\end{tabularx}
\end{table}

\begin{table}[!htb]
\centering
\footnotesize
\caption{The stake and the take in the CXR experiment, by arm and scoring rule. Instances are a film and a finding read by one radiologist, over the 15 findings the model reports; each of the 4{,}860 film-findings weighs the same. $H$ is a radiologist reading unaided, without or with clinical history, $M$ the model. Brackets: 95\% intervals over 1{,}000 resamples of films; the interval of $S$ is that of half the leader\textquoteright s margin, $(\E[\theta_L] - \E[\theta_U])/2$ with the leader held at its estimate, which can fall below zero. B1 and rungs~4b to~4e are cross-fitted over five folds of films. The two rows below the routed anchor are its analogues when a whole reading, one radiologist\textquoteright s film, or a whole film must go to one skill: the mean, over readings or over films, of the better of the two skills\textquoteright{} mean performances on its film-findings}
\label{tab:cxr}
\setlength{\tabcolsep}{4pt}
\begin{tabular}{c l c c c c}
\toprule
 & & \multicolumn{2}{c}{Squared rule} & \multicolumn{2}{c}{Linear rule} \\
\cmidrule(lr){3-4}\cmidrule(lr){5-6}
 & & X-ray only & With history & X-ray only & With history \\
\midrule
\multicolumn{6}{l}{\textit{Skills and anchors}} \\[2pt]
 & $\E[\theta_H]$, radiologist & 0.9312 & 0.9330 & 0.8908 & 0.8927 \\
 & $\E[\theta_M]$, model & 0.9366 & 0.9366 & 0.8490 & 0.8490 \\
 & $\E[m]$, mean anchor & 0.9339 & 0.9348 & 0.8699 & 0.8709 \\
 & $b$, informed baseline & 0.9366 ($M$) & 0.9366 ($M$) & 0.8908 ($H$) & 0.8927 ($H$) \\
 & B1, class ceiling & 0.9462 & 0.9466 & 0.8925 & 0.8933 \\
 & $\E[\Mx]$, routed anchor & 0.9692 & 0.9699 & 0.9320 & 0.9328 \\
 & \quad routing whole readings & 0.9558 & 0.9563 & 0.9050 & 0.9060 \\
 & \quad routing whole films & 0.9466 & 0.9473 & 0.8963 & 0.8974 \\
 & reporting each finding's prevalence & 0.9465 & 0.9465 & 0.8930 & 0.8930 \\
 & always reporting zero & 0.9379 & 0.9379 & 0.9379 & 0.9379 \\
\midrule
\multicolumn{6}{l}{\textit{The stake}} \\[2pt]
 & $G$ & 0.0353 & 0.0351 & 0.0621 & 0.0620 \\
 & & {\scriptsize[0.0331, 0.0374]} & {\scriptsize[0.0329, 0.0374]} & {\scriptsize[0.0600, 0.0643]} & {\scriptsize[0.0598, 0.0642]} \\
 & $S$ & 0.0027 & 0.0018 & 0.0209 & 0.0219 \\
 & & {\scriptsize[0.0008, 0.0047]} & {\scriptsize[$-$0.0001, 0.0038]} & {\scriptsize[0.0183, 0.0232]} & {\scriptsize[0.0194, 0.0241]} \\
 & $R$ & 0.0326 & 0.0334 & 0.0412 & 0.0401 \\
 & & {\scriptsize[0.0305, 0.0348]} & {\scriptsize[0.0312, 0.0356]} & {\scriptsize[0.0378, 0.0448]} & {\scriptsize[0.0369, 0.0436]} \\
 & \quad $\Rclass$ & 0.0097 & 0.0100 & 0.0017 & 0.0006 \\
 & & {\scriptsize[0.0080, 0.0113]} & {\scriptsize[0.0084, 0.0117]} & {\scriptsize[0.0002, 0.0036]} & {\scriptsize[0.0000, 0.0023]} \\
 & \quad $\Rinst$ & 0.0229 & 0.0233 & 0.0395 & 0.0395 \\
 & & {\scriptsize[0.0212, 0.0245]} & {\scriptsize[0.0215, 0.0251]} & {\scriptsize[0.0367, 0.0424]} & {\scriptsize[0.0365, 0.0422]} \\
 & \quad $\Rinst$, bounds from class marginals & [0.0069, 0.0477] & [0.0071, 0.0475] & [0.0177, 0.0806] & [0.0176, 0.0800] \\
 & \sqc{$\Lambda$}, disagreement & 0.0101 & 0.0101 & 0.0000 & 0.0000 \\
\midrule
\multicolumn{6}{l}{\textit{The take}} \\[2pt]
4e & selection on both responses & 0.9526 & 0.9528 & 0.9096 & 0.9101 \\
4d & ensembling on both & 0.9491 & 0.9497 & 0.8932 & 0.8941 \\
4c & assignment on both & 0.9465 & 0.9468 & 0.8932 & 0.8941 \\
4b & assignment on the model's prob. & 0.9431 & 0.9431 & 0.8910 & 0.8929 \\
4a & assignment on a predicted abnormal film & 0.9449 & 0.9454 & 0.8928 & 0.8944 \\
B1 & incumbent of each finding, estimated & 0.9462 & 0.9466 & 0.8925 & 0.8931 \\
3 & the model in view & 0.9351 & 0.9364 & 0.8899 & 0.8919 \\
 & \quad lift $\gamma$ against pooling & 0.991 & 0.991 & 1.023 & 1.024 \\
 & & {\scriptsize[0.989, 0.992]} & {\scriptsize[0.989, 0.993]} & {\scriptsize[1.021, 1.025]} & {\scriptsize[1.022, 1.026]} \\
2 & pooling & 0.9440 & 0.9449 & 0.8699 & 0.8709 \\
2r & random assignment & 0.9339 & 0.9348 & 0.8699 & 0.8709 \\
\bottomrule
\addlinespace[3pt]
\multicolumn{6}{p{0.95\textwidth}}{\footnotesize The two reporters at the foot of the anchors read nothing: one gives every film the prevalence of each finding in these films, the other always reports zero. Rung~0 with the radiologist attains $\E[\theta_H]$, with the model $\E[\theta_M]$, and rung~1 attains $b$. Row B1 is the class ceiling with its incumbents estimated, since the finding is named at intake; rung~4a predicts a class that refines the finding, whether the film is abnormal, from the model\textquoteright s probability for any abnormality (Appendix~\ref{app:cxrfilm}). The rows of the take read from rung~2r at the foot to rung~4e at the top. Rung~3 is the arm in which the radiologists read with the model\textquoteright s prediction in view, scored on the same film-findings; its lift is measured against pooling each film-finding\textquoteright s unaided reads with the model. Rungs~4b to~4e read the model\textquoteright s probability, which requires the model to execute on every instance; rungs~4d and~4e execute both skills. The bounds on $\Rinst$ use only each skill\textquoteright s distribution within each finding (Proposition~\ref{prop:classbounds}). Under the linear rule $\Lambda = 0$ exactly and pooling attains the mean anchor (Proposition~\ref{prop:hull}).}
\end{tabular}
\end{table}
 
\paragraph{A stake almost all residual.} Under the squared rule the model leads, but barely and not reliably: with clinical history $S = 0.0018$, half a lead whose interval runs from $-0.0001$ to $0.0038$, against a gross stake of $0.0351$; without it, $S = 0.0027$, from $0.0008$ to $0.0047$. The stake is small in absolute terms because positive findings are rare: a reporter that reads nothing and states the prevalence of each finding attains $0.9465$, and the two skills drawn at random fall short of a perfect score by $1 - \E[m] = 0.0652$. Against that shortfall the stake is $54\%$: routing omniscience would remove about half of it. The residual is $95\%$ of the stake, $R = 0.0334$: the regime of Proposition~\ref{prop:parity}, in which no commitment can be wrong by much on average and every commitment is wrong on the instances the other skill wins. Not all of it is within reach of a policy that can be deployed. A radiologist reads a whole film, so that a policy that sends her some of its findings has her read the film; routing whole readings, each radiologist\textquoteright s reading of a film to her or to the model, with the film and the reader known, collects at most $0.0197$ of the residual, $59\%$, and routing whole films with the reader unknown at most $0.0108$, $32\%$ (Table~\ref{tab:cxr}). And part of what remains lies between radiologists, which only a policy that knows who reads can reach (below). Which skill leads depends on the findings. On the four top-level findings, the only ones every radiologist was asked about, the radiologists lead, by $S = 0.0075$, an interval from $0.0031$ to $0.0119$; the model's lead in the aggregate comes from the findings below them, on which four reads in five are exactly zero, most of them because a parent finding was ruled out. The zeros do not create that lead: on the reads of findings the radiologists were shown, the model leads by more, $S = 0.0213$ (Appendix~\ref{app:cxrvar}). What holds across the findings is the weight of the residual: $90\%$ of the stake on the top-level findings, $73\%$ below them (Table~\ref{tab:cxrrobust}). Read reader by reader, the pair is at parity as a population. The model outperforms $53\%$ of the radiologists on the films they read, and for $96\%$ of them the residual of the pair exceeds its systematic component (Figure~\ref{fig:cxr}b). Clinical history, which the model does not see, raised the radiologists' performance by $0.0019$, an interval from $-0.0001$ to $0.0039$, and moved the stake between its components rather than changing its size: $S$ fell by $0.0009$ and $R$ rose by $0.0008$, while $G$ changed by $-0.0002$, an interval from $-0.0008$ to $0.0005$. A skill enriched toward parity with the leader transfers stake from the systematic component, which a commitment secures, to the residual, which only routing collects.
 
\paragraph{Classes, and what separate assessments establish.} The model is the incumbent of eleven findings and the radiologists of four: the call that any abnormality is present, pneumothorax, edema and the shoulder fracture (Table~\ref{tab:cxrclasses}). Four findings have at most one positive film, so that their incumbency is decided on negatives alone, by whichever skill hedges less; leaving them out leaves the shares much as they are, $S = 0.0021$ and $R = 0.0442$, the residual $95\%$ of the stake. The class residual is $0.0100$, $30\%$ of $R$; the remaining $0.0233$ lies within the findings. Readers differ as well: taking the radiologist as the class gives a class residual of $0.0052$, $15\%$ of $R$, and taking the radiologist and the finding together $0.0164$, half of $R$, an in-sample figure that the winner's curse of Proposition~\ref{prop:curse} inflates. Taking each film-finding as its own class, the finest partition in which the human skill is still a population of readers, whose incumbent only other reads of the same film can estimate, gives a class residual of $0.0219$, two thirds of $R$, or $0.0179$, half of $R$, when each film-finding's incumbent is estimated on its other readers; the rest, $0.0115$ to $0.0155$, lies between the radiologists who read the same film-finding, and only a policy that knows who reads, and what she read, can reach it (Appendix~\ref{app:cxrfinding}). On film 277, for instance, a support device is present (Table~\ref{tab:cxrrecA}). The model, the incumbent of support devices over all films, puts it at $0.45$ and scores $0.70$; the 22 radiologists who read the film unaided score $0.82$ on average, so that on this film-finding they are the incumbent, and sending it to them gains $0.12$ over sending it to the incumbent of the finding. Nineteen of them put the device between $0.47$ and $1$, and three at $0.005$ or less; on those three reads the model would have scored $0.70$ instead of at most $0.01$, and that is the $0.09$ separating the film-finding's incumbent from routing omniscience. The first gain needs to know the film; the second, who reads it. The two skills' distributions within each finding, without any pairing, identify $\Rclass$ and bound $\Rinst$ between $0.0071$ and $0.0475$ (Proposition~\ref{prop:classbounds}); the paired record places it two fifths of the way up that interval. The performances are positively coupled, with a correlation of $0.53$ across reads, and drawing the two skills independently within each finding would put $R$ at $0.0462$ rather than $0.0334$: a coupling of this strength cuts the residual by more than a quarter, in the direction Proposition~\ref{prop:coupling} states and Section~\ref{sec:limits} warns of for the illustrative profiles.
 
\paragraph{The take.} Committing to the model attains $b = 0.9366$, and assigning each instance at random $0.9348$. Both fall short of a reporter that reads nothing and gives every film the prevalence of each finding, which attains $0.9465$ (Appendix~\ref{app:cxrdata}); among the arrangements of the raw responses, assignment on the finding only matches it, and ensembling and selection exceed it. Pooling the two responses attains $0.9449$: the disagreement it earns, $\Lambda = 0.0101$, is more than five times $S$, so the premium of the insurance is negative and engaging both skills without directing work beats the informed commitment, by $0.0083$, an interval from $0.0067$ to $0.0098$ (Proposition~\ref{prop:insurance}). The arm with the model's prediction in view, rung~3 by structure, attained $0.9364$: above the mean anchor by $0.0016$, level with the commitment, and below pooling by $0.0085$, a lift of $0.991$ with an interval from $0.989$ to $0.993$; read as decisions at a threshold of one half, the arm misclassifies 17 more film-findings in every 1{,}000 than pooling, about one more error for every four films (Appendix~\ref{app:cxrdata}). The exchange gave back almost all that averaging the same two responses, neither seeing the other, would have earned. Within readers, on the 68 radiologists who read the same films unaided and with the prediction, the lift is $0.990$, and the exchange beat averaging for 11 of them. Reader by reader, the arm returned about the mean of the reader's and the model's performances. For the 36 of these radiologists whom the model outperforms on those films, it fell short of that mean by $0.0003$, an interval from $-0.0030$ to $0.0021$, after rising $0.0072$ above their unaided reading; for the 32 who outperform the model, it exceeded the mean by $0.0028$, an interval from $0.0005$ to $0.0052$, after a fall of $0.0020$ below their unaided reading, an interval from $-0.0050$ to $0.0010$. The groups are formed on the same unaided reads they are compared with, so part of each movement is regression to the mean; the pattern is that of a reading drawn partway toward the other skill's.
 
Routing collects more. Assigning each finding to its incumbent, with the incumbents learned on other films, attains $0.9466$, the class ceiling B1, and within $0.0001$ of it under every other split of the films we tried; since the finding is named at intake, this is B1 with its incumbents estimated rather than rung~4a reading a predicted class. The incumbents are stable enough to be learned, and an arrangement that executes one skill per instance collects the class residual of the responses as reported in full, $0.0100$ above the commitment. Reading the model's probability as well adds $0.0002$, an interval from $-0.0005$ to $0.0009$; reading it alone, as the first two branches of the authors' delegation policy do, attains $0.9431$, below pooling. Ensembling the two responses with weights set on the finding and the model's probability attains $0.9497$, and selecting one of the two responses once both are known, on the finding and both probabilities, $0.9528$, the most of any deployable arrangement on the ladder here and half the gross stake above the mean anchor; both execute the two skills on every instance. All of them stop far below routing omniscience, $0.9699$. Selection collects about a quarter of the instance residual. A logistic regression of the standard on the log-odds of the two probabilities, fitted finding by finding on the same folds, returns a probability outside the interval of the two responses on $27\%$ of reads and attains $0.9674$, above selection by $0.0146$, an interval from $0.0124$ to $0.0167$. Most of that margin is recalibration, which Section~\ref{sec:twochannels} credits to no channel: the model recalibrated by the same regression attains $0.9655$ alone, and the combination adds $0.0020$ to it, an interval from $0.0013$ to $0.0025$. It is the third way of combining that Section~\ref{sec:responses} leaves outside the ladder, and the routing omniscience of the responses as reported, $0.9699$, does not bound it: the combination stays below it on average, by $0.0025$, but scores above the better of the two responses on $24\%$ of reads. Its own ceiling, the score of the probability of the standard given the two responses and the finding (Proposition~\ref{prop:evidence}), is not observed, and any fitted combination estimates it from below. Part of what selection leaves, at most the $0.0064$ that knowing the radiologist adds in sample to the class residual, is the value of knowing who reads.

\paragraph{Two films, arrangement by arrangement.} Table~\ref{tab:cxrfilm} follows two films through the arrangements, each as one radiologist of the third design read it, unaided and later with the model's prediction in view, the policies fitted on the folds of films that exclude it; Appendix~\ref{app:cxrfilm} gives every finding, and every read of four of the film-findings as the data record them. Film~277 is typical: among the 2{,}855 pairs of a radiologist and a film she read both ways in the arm with clinical history, it is, of the four pairs with a finding present whose differences between the arrangements lie closest to their medians, the one with most findings present, and the only pair with a finding present on which all six differences lie between the 35th and the 65th percentile. Film~154 is not, and it is shown because it makes each mechanism visible: on it the arm with the prediction in view falls below pooling by more than on $97\%$ of these pairs, and selection rises above pooling by more than on $96\%$.
 
On film~277 three findings are present by the standard, and the records show how narrowly: two of the five experts put the airspace opacity near zero and three near one, a mean of $0.57$, and the 22 radiologists who read the film unaided gave it $0.20$ on average (Table~\ref{tab:cxrrecA}). This radiologist and the model both put any abnormality at $0.46$ and the airspace opacity at about $0.2$. The commitment sends the film to the model alone, and no radiologist sees it. Pooling has both read it and reports the average on each finding: on consolidation, absent, her $0.51$ and the model's $0.02$ become $0.26$. With the prediction in view she withdrew the consolidation, to $0.02$, and kept the support device at $0.70$, but raised any abnormality to $0.72$, a finding the panel puts at $0.07$ although it calls three of the film's findings present. Assignment by finding, B1, returns her response on the four findings she holds and the model's on the other eleven, among them the model's $0.45$ on the support device; selection, reading both probabilities, returns her $0.69$ there. The arrangements end within $0.02$ of one another, from $0.851$ for the arm the study ran to $0.869$ for selection, as they do on average.
 
On film~154 the mechanisms act at their extremes. On cardiomegaly, which the standard calls absent with a panel mean of $0.48$, the radiologist's zero was never elicited: she had put the cardiomediastinal finding, the top-level finding above it, at $0.08$, so that cardiomegaly was not shown to her (Table~\ref{tab:cxrrecB}). Two of the five experts were not shown it either, and the mean counts their zeros; the three who were asked put it at $0.70$ or more. Her zero and the model's $0.70$ pool to $0.35$; with the prediction in view she was shown the finding and answered $0.73$; assignment by finding returns the model's response, since the model holds cardiomegaly; ensembling, with a tenth of the weight on her, reports $0.63$; and selection returns her zero, because on the training films a zero from her against a high probability from the model scored better on average. Selection is right here by the standard, and wrong by every expert who was asked. With the prediction in view she also lowered any abnormality, present, from $0.83$ to $0.27$, which neither response suggested and only one other of the 17 readers who saw the prediction put below one half, and on this film the arm attains $0.781$, below her own unaided reading.
 
The last row of each block counts the findings given to the radiologist. The model runs on the film under every arrangement but the radiologist alone, and the radiologist reads it under every arrangement but the commitment, although assignment by finding gives her four of the fifteen findings, and routing on the model's probability, checked, one on film~277 and three on film~154. She must read the film to answer any of them: the policies assign film-findings, but a radiologist executes on films.
 
\begin{table}[!htb]
\centering
\footnotesize
\caption{Two films through the arrangements, each read by one radiologist of the third design, unaided ($H$) and later with the model\textquoteright s prediction in view; $M$ is the model and $r$ the standard. Each column gives the probability the arrangement reports for the finding, with the skill whose response it returns where it returns one; the arrangements are fitted on the responses as reported, on the folds of films that exclude the film. B1 assigns each finding to its incumbent, and 4c, checked, also reads the model\textquoteright s probability where the check keeps it. The film rows give the mean performance over the 15 findings, squared rule, and the findings given to the radiologist; under the hierarchy she answers only those shown to her. The model runs on the film under every arrangement but $H$ alone. Table~\ref{tab:cxrfilmfull} gives every finding, and Tables~\ref{tab:cxrrecA} and~\ref{tab:cxrrecB} every read of four of the film-findings}
\label{tab:cxrfilm}
\setlength{\tabcolsep}{2.8pt}
\begin{tabular}{l c c c c c c c c c}
\toprule
 & & \multicolumn{2}{c}{Alone} & Rung 3 & Rung 2 & B1 & 4c, checked & 4d & 4e \\
\cmidrule(lr){3-4}
Finding & $r$ & $H$ & $M$ & $M$ in view & pooling & by finding & + probability & ensembling & selection \\
\midrule
\multicolumn{10}{@{}l}{\textit{Film 277, a typical film}} \\[1pt]
Any abnormality & 0 & 0.46 & 0.46 & 0.72 & 0.46 & 0.46$_H$ & 0.46$_H$ & 0.46 & 0.46$_H$ \\
Airspace opacity & 1 & 0.17 & 0.19 & 0.19 & 0.18 & 0.19$_M$ & 0.19$_M$ & 0.18 & 0.17$_H$ \\
Atelectasis & 1 & 0.02 & 0.07 & 0.02 & 0.05 & 0.07$_M$ & 0.07$_M$ & 0.06 & 0.02$_H$ \\
Consolidation & 0 & 0.51 & 0.02 & 0.02 & 0.26 & 0.02$_M$ & 0.02$_M$ & 0.04 & 0.02$_M$ \\
Support device or hardware & 1 & 0.69 & 0.45 & 0.70 & 0.57 & 0.45$_M$ & 0.45$_M$ & 0.53 & 0.69$_H$ \\
\cmidrule(lr){1-10}
\multicolumn{2}{l}{Film, all 15 findings} & 0.852 & 0.861 & 0.851 & 0.862 & 0.863 & 0.861 & 0.866 & 0.869 \\
\multicolumn{2}{l}{Findings given to the radiologist} & 15 & 0 & 15 & 15 & 4 & 1 & 15 & 15 \\
\midrule
\multicolumn{10}{@{}l}{\textit{Film 154, a film at the extremes}} \\[1pt]
Any abnormality & 1 & 0.83 & 0.90 & 0.27 & 0.86 & 0.83$_H$ & 0.83$_H$ & 0.83 & 0.90$_M$ \\
Cardiomediastinal abnormality & 1 & 0.08 & 0.70 & 0.50 & 0.39 & 0.70$_M$ & 0.70$_M$ & 0.45 & 0.70$_M$ \\
Cardiomegaly & 0 & 0.00 & 0.70 & 0.73 & 0.35 & 0.70$_M$ & 0.70$_M$ & 0.63 & 0.00$_H$ \\
Lesion & 0 & 0.58 & 0.01 & 0.47 & 0.30 & 0.01$_M$ & 0.01$_M$ & 0.01 & 0.01$_M$ \\
\cmidrule(lr){1-10}
\multicolumn{2}{l}{Film, all 15 findings} & 0.837 & 0.886 & 0.781 & 0.887 & 0.895 & 0.895 & 0.893 & 0.929 \\
\multicolumn{2}{l}{Findings given to the radiologist} & 15 & 0 & 15 & 15 & 4 & 3 & 15 & 15 \\
\bottomrule
\end{tabular}
\end{table}
 
\paragraph{What calibration takes away.} Section~\ref{sec:insurance} warns that part of what pooling earns is the correction of each skill's own miscalibration, which a recalibration of each skill would deliver alike and which this paper credits to no channel. The data allow the correction to be made. \citet{agarwal2023combining} recalibrate the radiologists' reports to check their treatment effects, whose pattern it leaves unchanged; here both skills are recalibrated, and what the recalibration itself returns is measured. Recalibrating each skill's probability within each finding, by isotonic regression fitted on four fifths of the films and applied to the fifth, removes most of their miscalibration and keeps the order of their responses (Equation~\eqref{eq:murphy}), and raises both skills, to $0.9574$ for the radiologists and $0.9644$ for the model, and halves the stake, to $G = 0.0172$. The model's lead is $S = 0.0035$, an interval from $0.0016$ to $0.0054$, and the residual remains four fifths of the stake, $R = 0.0137$; but the class residual vanishes, to $0.0001$, so that the incumbency of the findings in the raw responses records which skill was better calibrated on each, not which was better informed. The disagreement falls to $\Lambda = 0.0049$, under half its raw value, and pooling no longer beats the commitment, $0.0014$ with an interval from $-0.0003$ to $0.0032$. The arm with the prediction in view, recalibrated alike, falls below the commitment, by $0.0037$, an interval from $-0.0067$ to $-0.0007$, to about the mean anchor, and its lift stays below one, $0.995$. Logistic recalibration gives the same picture (Table~\ref{tab:cxrrobust}). Against the routing omniscience of the recalibrated responses, $0.9781$, the logistic combination of the two probabilities collects about a fifth of the residual, $0.0030$ of $0.0137$.
 
\paragraph{The rule decides which skill leads.} Under the linear rule, the special case to which the deviation the authors report belongs, the ranking reverses. The radiologists lead by $0.044$, the model is the incumbent only of cardiomegaly and the cardiomediastinal finding, $S = 0.0219$, and almost all the residual lies within the findings. The reversal is confined to the findings below the top level: on the four top-level findings the radiologists lead under both rules. Below them, the radiologists' responses are exactly zero on four reads in five, mostly because a parent finding was ruled out, while the model reports $0.05$ at the median; the linear rule, which is not proper, rewards a response at the nearer endpoint, and the squared rule credits the evidence a hedged probability carries. What decides the leader there is the rule, acting on responses that the structure of the elicitation made extreme. On the reads of findings the radiologists were shown, the model leads under both rules, by $S = 0.0163$ under the linear rule: the radiologists\textquoteright{} lead under that rule is made of the zeros of findings they were never asked about (Appendix~\ref{app:cxrvar}). Pooling then attains the mean anchor exactly, $\Lambda = 0$ (Proposition~\ref{prop:hull}), so that the arm with the prediction in view, measured against that mechanical baseline, has a lift of $1.024$; against the commitment to the radiologists it is level, $-0.0008$ with an interval from $-0.0032$ to $0.0017$, which is the authors' finding that the prediction did not improve average performance. The model outperforms $5\%$ of the radiologists under the linear rule, $53\%$ under the squared rule, and $78\%$ on the area under the ROC curve the authors report: three measures, three answers to which skill an organization should commit to. The commitment of rung~1 is only as good as the rule under which the skills were assessed, and the decomposition, defined on performances, moves with it. Under the linear rule a reporter that always says zero attains $0.9379$, above every arrangement in Table~\ref{tab:cxr} and above routing omniscience itself.
 
\begin{figure}[!htb]
\centering
\includegraphics[width=\linewidth]{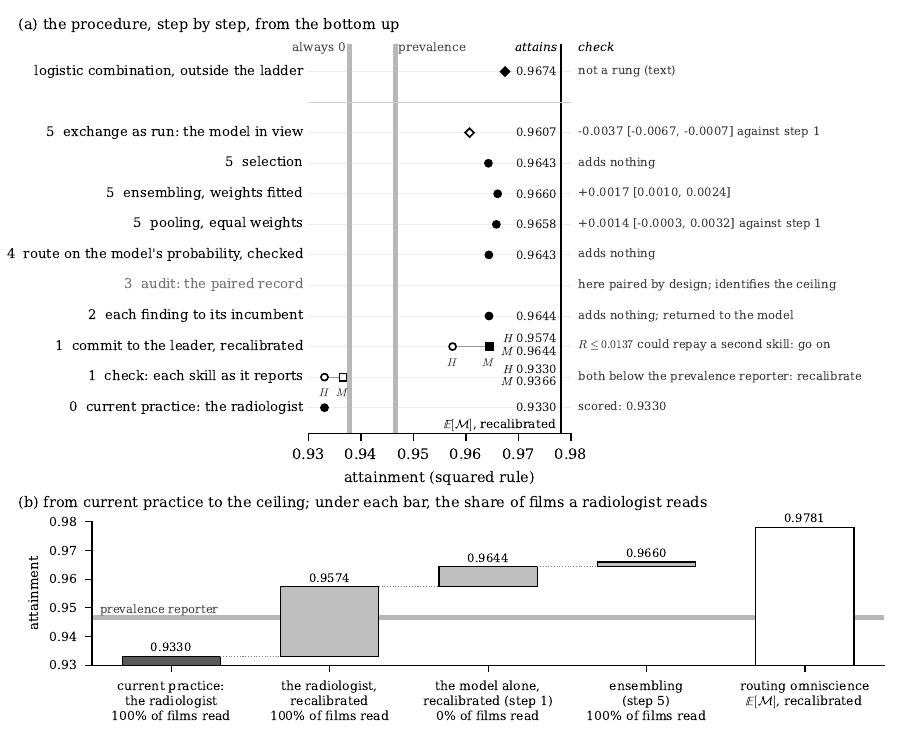}
\caption{The procedure of Section~\ref{sec:procedure} walked on the Collab-CXR data, arm with clinical history, squared rule. (a) What each step attains, fitted on four fifths of the films and scored on the fifth, with its value in the column \emph{attains}, and what its check finds, the steps read from the bottom up. The two lowest rows score the radiologists as they report; every row above them uses the two skills recalibrated within each finding, as the check of step~1 requires. Where a row shows both skills, the circle is the radiologist ($H$) and the square the model ($M$); open markers are not adopted. The gray bands are the reporters that read nothing, the black line routing omniscience between the recalibrated skills. Step~5 includes pooling, which is ensembling with equal weights, and the arm the study ran, which is an exchange; above the rule, for comparison, the logistic combination, which is not a rung. (b) The same results as a path from current practice: the radiologists recalibrated, the recalibrated model alone, and ensembling the two; under each bar, the share of films a radiologist reads}
\label{fig:cxrwalk}
\end{figure}

\paragraph{What the procedure identifies.} Run on these data, the procedure of Section~\ref{sec:procedure} reads as follows (Table~\ref{tab:cxrprocedure}, Appendix~\ref{app:cxrproc}); every step is fitted on four fifths of the films and judged on the fifth, and the check of step~4 is made finding by finding on inner folds of the training films. Step~1 opens with its check. As they report, neither skill beats the reporter that states each finding's prevalence, $0.9465$, the radiologists attaining $0.9330$ and the model $0.9366$, although both rank the film-findings well; the procedure therefore recalibrates both skills within each finding, on the training films alone and the reads with the prediction in view included, and continues with the recalibrated skills. Step~1 then commits to the model, which attains $0.9644$, with no radiologist reading. Step~2 adds nothing: it gives the radiologists pneumothorax in three folds and the shoulder fracture in four, by margins that barely hold or reverse on the films left out: it gains $0.0002$ on the first and gives back $0.0005$ on the second, and with them it has a radiologist read every film. Step~4 adds nothing either: its check keeps the model's probability only for airspace opacity, in one fold, where it gives back $0.0005$. These are the finite-sample losses that the checks, which Proposition~\ref{prop:checks} states with the class means known, do not exclude. A margin excludes them (Proposition~\ref{prop:margin}). Requiring each switch to beat the previous step by one standard error of its estimated gain, clustered by film, step~2 gives no finding to the radiologists in any fold; step~4 still keeps the model\textquoteright s probability for airspace opacity in one fold, and with a margin of $1.645$ standard errors, a one-sided test at $5\%$, in none, so that the procedure stays at the commitment, $0.9644$, with no radiologist reading. Of step~5, selection adds nothing; pooling adds $0.0014$ over the commitment, an interval from $-0.0003$ to $0.0032$, and ensembling, with its weights fitted by finding and band, $0.0017$ over step~4, an interval from $0.0010$ to $0.0024$, both with a radiologist answering every finding of every film. The arm the study ran would not have been deployed: it attains $0.9607$, below the commitment by $0.0037$, an interval from $0.0007$ to $0.0067$, and below pooling, $0.9658$. What the procedure identifies is therefore a choice between the recalibrated model alone and ensembling the two recalibrated skills, $0.0016$ higher with a radiologist on every film, a choice that rests on what a read costs, which the paper does not price. The steps that could spare films are those that add nothing, and since every finding is assessed on every film, any step that gives the radiologists even one finding has them read every film. The logistic combination of the two probabilities, outside the ladder, attains $0.9674$, above recalibrated ensembling by $0.0014$, an interval from $0.0005$ to $0.0023$. Under the linear rule the check of step~1 cannot be passed: the reporter that always says zero attains $0.9379$, above every arrangement of Table~\ref{tab:cxr} and above routing omniscience itself, and what fails is the rule. Figure~\ref{fig:cxrwalk} draws the walk, and the same results as a path from current practice. Against the radiologists' reports as they stand, $0.9330$, most of what a clinic can gain comes from recalibration: the radiologists recalibrated attain $0.9574$, and the model recalibrated $0.9644$ with no radiologist reading; ensembling adds $0.0016$ with a radiologist on every film, and showing her the prediction, as the study did, gives back $0.0037$ against the model alone.
 
\paragraph{The measurement and the original analysis.} The measurement is a secondary analysis of the released data, and its estimands are not those of \citet{agarwal2023combining}. The authors score a read by its deviation from the diagnostic standard, which is the linear rule, and the skills by the area under the ROC curve; they estimate treatment effects by regression on the reads, with errors clustered by radiologist and by patient; and they set their delegation policy against the cost of the radiologists' time. The measurement scores each read under the squared rule as well, weighs each of the 4{,}860 film-findings equally, reads each arm on its own, and prices nothing. Where the two meet, it reproduces their central finding: under the linear rule the arm with the prediction in view is level with unaided reading. It adds what their comparisons do not contain: a benchmark that needs no fitting, averaging the two responses, against which the same arm falls short; the split of the stake, mostly residual, with two thirds of the residual within films, out of reach of a rule that assigns whole films; the dependence of which skill leads on the findings, the rule and calibration, with the zeros of the elicitation hierarchy behind the radiologists' lead under the linear rule; and what recalibrating both skills returns, which their recalibration of the radiologists, a check on their treatment effects, does not measure. It qualifies two of their findings without contradicting them, since rule and scope differ: the gain from clinical history, significant in their analysis, is not established here under the squared rule; and a rule that reads only the model's probability, as the first two branches of their delegation policy do, falls below pooling in performance, a comparison their policy, which weighs the radiologists' time, does not make.
 
Read as a whole, the measurement confirms the reading of Section~\ref{sec:radiology} where it matters most. The arrangement the experiment actually ran, the model's prediction in front of its radiologists, returned less than averaging the same two responses would have in every variant, and not reliably more than committing to the better skill. This is, in performance, the setting of the safe harbor paradox of \citet{zanardelli2025navigating}: a second skill put to work with no design for what it adds returns no more than the better skill alone, and less once both are recalibrated, before the cost of the second skill, which drives the paradox in utility, is counted. The procedure, which recalibrates the skills before it commits, finds little to collect beyond the recalibrated model, which alone attains more than any deployable arrangement of the responses as reported. And part of what the arrangements of the ladder leave is within reach of a rule that recalibrates as it combines: a combination outside the ladder, which no averaging of or choice between the two responses can reproduce, attains more than any deployable arrangement on it, recalibrated or not, and still collects only about a fifth of the residual of the recalibrated responses.

\subsection{Double reading in mammography}\label{sec:mammography}

The three rules of \citet{johnston1999two} occupy three different places. Recalling a woman whenever either reader suggests it is a fixed rule that returns, for a binary decision, the more alarming of two independent responses: it returns one of its inputs unchanged, so it is routing by selection with a hand-written aggregation that reads the responses, bounded by $\E[\Mx]$. Consensus discussion is rung~3, an exchange before the joint response is fixed. Arbitration by a third reader is an exchange (Section~\ref{sec:policies}): the arbiter reads the examination after the two reports, and so is a third skill, engaged only on disagreement and shown the others\textquoteright{} work before fixing her own response. The rule is therefore routing with synergy, rung~5: routing, since whether the third skill executes depends on the first two responses, a choice between closing on agreement and referring on disagreement of the kind Section~\ref{sec:signals} calls gating; synergy, since what the arbiter returns is not a function of three independent responses. No skill bounds it, and what it adds is a lift, to be measured against what the same three readers would have returned without seeing one another\textquoteright s work, which screening records, keeping the arbiter\textquoteright s decision alone, do not retain. The evidence of \citet{taylorphillips2018double} speaks to both the stake and the take. That $8.9\%$ of detected cancers were found by the second reader alone is evidence of instances on which the underdog wins, and therefore of a positive residual (Proposition~\ref{prop:R0}); that the same two readings produce recall rates of $6.19\%$, $4.76\%$ and $4.08\%$ under different rules is the take varying with the workflow while the stake stays fixed.

The studies that place a machine among the readers change a different variable. In MASAI \citep{lang2023artificial} the model's risk score decides for each examination whether it is read once or twice, choosing per instance how many skills execute. The arm is not routing alone, since the readers also saw the model's detection marks on the examinations they read, an exchange in the sense of Definition~\ref{def:exchange}; part of the gain may therefore be synergy, and the trial does not separate the two. The rise in detection is a gain in attainment, and what the triage saves is visible in its cost profile, one reading instead of two on every examination it sends to single reading; pricing that saving is left to future work. The theory adds the evidence caveat: a criterion that sends low-risk examinations to single reading stops producing paired readings on them, so the record from which it would be refitted grows only on the examinations it already judged risky, and screening programs supply the adjudication later, through follow-up and interval cancers. Double-reading a random fraction of the examinations sent to single reading would keep the record growing on all of them, at a cost the program chooses (Algorithm~\ref{alg:audit}, Proposition~\ref{prop:audit}). The nationwide implementation of \citet{eisemann2025nationwide}, an observational study in which radiologists chose whether to use the model, shows the reconfigured arrangement at scale.

\subsection{Drafting with an assistant, and what it does to the skill}\label{sec:patent}

In the experiment of \citet{autor2026does}, patent lawyers were randomly given an AI drafting assistant for three months or left without one, and every work product was scored blind by expert attorneys on a five-dimension rubric.

\paragraph{The assisted drafts.} Drafting with the assistant is an exchange: the returned draft is fixed after the lawyer has prompted, read and edited the assistant's text, so it sits at rung~3 by structure, with a tool in place of a designed protocol. Its measured effect, $0.34$ and $0.38$ standard deviations at ten and ninety days, obtained mostly through fewer poor drafts, compares the exchange with the human skill alone, $\E[\theta_\rho] - \E[\theta_H]$, where $\theta$ stands for a draft's rating. The ratings are rubric scores standardized against the control group, not performances in the sense of Section~\ref{sec:performance}: no correct response is defined against which a draft's distance could be scored, so reading them as $\theta$ requires a mapping onto $[0,1]$ that the scale does not supply. The orderings and identification statements below do not depend on that mapping; the magnitudes do, and the lift most of all, since it is a ratio and changes with the origin of the scale, while $S$ and $R$ change only with its unit. The estimand $\E[\theta_\rho] - \E[\theta_H]$ is neither the lift of Definition~\ref{def:lift}, which is measured against the independent responses of every engaged skill, nor a comparison with the informed baseline: the assistant's own draft of the same invention was not scored, so $b$, $S$ and $\gamma$ are not identified, and whether the assisted drafts improved on a commitment to the machine is open. A drafted patent is not an endpoint truth, so the hull bound does not apply and an exchange can exceed both skills (Section~\ref{sec:responses}); whether this one did, only an arm scoring the machine alone on the same instances, on the same rubric, can say.

\paragraph{The unassisted task.} The authors measure unassisted judgment by having every lawyer mark up an AI-generated draft. Read as an arrangement, this is a sequential exchange with the machine first, and its score is that of the joint response, against a machine response whose own score the design did not record.

\paragraph{A policy that moves the skill.} The experiment's central finding concerns the skill rather than the arrangement. After three months of assisted drafting, senior lawyers marked up the draft better without the tool than untreated seniors, by $0.45$ standard deviations, while juniors showed no average gain and a wider spread, with more poor and more good markups. In the terms of Section~\ref{sec:magnitudes}, the policy changed the distribution of the human skill at the next date, and not uniformly: it raised the mean of one group and the dispersion of the other. Both reach the stake, but the second only on a condition. Each lawyer worked a single task, so the spread the experiment measures is between lawyers, not the variation of one lawyer's performance across instances, on which $\theta_H$ is defined. It enters the distribution of $\theta_H$ where the human skill is a pool: instances are assigned to lawyers without regard to who they are, and the performance an instance receives depends on which lawyer draws it. Under that reading a higher mean moves $b$ and $S$, and a wider distribution with the same mean changes $R$ (Figure~\ref{fig:gsr_equal_means}). The lawyer's identity then acts as a class in the sense of Section~\ref{sec:classes}: the machine leads on the instances of the juniors who did not learn and the human on those of the juniors who did, so part of the widened residual is $\Rclass$, collected by an organization that assesses its lawyers one by one and routes accordingly, and lost to a policy that treats the juniors as one skill. The theory takes the skills of each date as given (Section~\ref{sec:limits}); the experiment measures one link of the feedback it leaves out.

\paragraph{What the theory adds to designs of this kind.} The authors conclude that evaluating assistance requires separating the contribution of the software from the expertise of its user. The levels of evidence of Section~\ref{sec:identification} state what that separation takes, the machine alone, the human alone and the workflow on the same instances, and three additions would deliver it at small cost. Scoring the assistant's own draft of each benchmark invention, on the same rubric, would identify $b$ and $S$, and with them whether the assisted lawyers cleared the informed baseline, the criterion on which \citet{vaccaro2024combinations} find most human--AI combinations fall short; several drafts drawn from the assistant would also separate its own variability from the residual (Section~\ref{sec:identification}). Scoring the unedited draft that every lawyer redlined would turn the unassisted task into a measured lift, lawyer by lawyer, and show how many reviews left the draft worse than they found it. And reporting each lawyer's endline score against the machine's, rather than by seniority alone, would show how much of the juniors' widened spread an organization could collect by assessing them one by one, and how much a policy that treats them as one skill forgoes. The same machine score would also settle a question the shape of the drafting gains leaves open. The assisted drafts improved mostly by being poor less often, which is what an arrangement that raises the floor without adding at the top would show, as pooling does (Section~\ref{sec:insurance}), but also what an exchange that augments mainly the weaker drafts would show. Set against the machine's own draft, the two readings separate: a raised floor keeps the assisted drafts within what the two skills reach alone, pulling the weakest toward the machine's score, while augmentation places some of them above both the machine's draft and the best unassisted drafts of lawyers of the same seniority.

\section{Discussion}\label{sec:discussion}

\subsection{The decomposition against adjacent literatures}\label{sec:adjacent}

The decomposition synthesizes quantities that adjacent literatures have developed separately, under different names. The residual $R$ is the population analogue of the oracle gap in dynamic classifier selection \citep{ko2008dynamic, cruz2018dynamic}, which calls the arrangement that always picks the right classifier the Oracle, and the ceiling of learning-to-defer and algorithmic-triage methods \citep{raghu2019algorithmic, madras2018predict, mozannar2020consistent, okati2021differentiable, verma2022calibrated}, positive only when the agents err on different instances. Those methods learn the criterion of rungs~4b and~4c, deciding per instance whether the model or the human responds, and learning the best such criterion jointly with the model is computationally hard, even when both are linear \citep{mozannar2023should}. In the human--AI literature it is the performance-side mirror of the inherent complementarity potential of \citet{hemmer2025complementarity}, which measures the same per-instance selection gain on losses; their collaborative potential is in turn the analogue of the headroom synergy addresses beyond $\Mx$, and since their construction is anchored at perfect execution rather than at the mean, no counterpart of $G$ or of $S$ appears in it. The criterion that a team should beat the better of its members defines complementary team performance \citep{bansal2021does}; the result that it requires each agent to hold an advantage on some part of the domain \citep{donahue2022human} is Proposition~\ref{prop:R0} read in routing terms; and the taxonomy of \citet{rastogi2023taxonomy} distinguishes within-instance from across-instance complementarity, the qualitative counterpart of the split $R$ and $S$ make quantitative. Closest in spirit is the result of \citet{peng2025nofreelunch} that, for agents with calibrated probabilistic predictions scored by accuracy, any deterministic collaboration strategy that does not essentially always defer to the same agent will, on some instance distribution, perform worse than the least accurate agent. Theirs is worst-case and about strategies; this theory is distributional and about magnitudes. Their theorem says that no fixed rule is safe everywhere, and the decomposition says, for a given pair, what a rule is worth and which instrument collects it.

On the side of the mean anchor, Proposition~\ref{prop:hull} makes the relation to the ambiguity decomposition of \citet{krogh1994neural} exact: under the squared rule, averaging responses earns their disagreement over averaging performances, a gain over the average member that the endpoint property caps at the best one. The cap follows from the averaging principle of \citet{larrick2006intuitions} at an endpoint: inputs that do not bracket the truth, as responses to a binary truth never do, leave their average, instance by instance, no better than the better of them. The choice between committing to the better judge and averaging is the question of \citet{soll2009strategies}, whose model weighs the probability of identifying the better judge, the gap in accuracy between the judges and the rate at which their errors bracket the truth, which falls as the errors become correlated and limits what averaging gains; Proposition~\ref{prop:insurance} is its counterpart in performance, with $D$, $S$ and \sqc{$\Lambda$} in those roles, and the select crowds of \citet{mannes2014wisdom}, which average the top few judges ranked by a cue to their ability, lie between rungs~1 and~2. The non-recoverability of $R$ from the marginals is an instance of the Fr\'echet--Hoeffding bounds \citep{nelsen2006introduction}. The ensemble principle that members should be accurate but diverse \citep{breiman2001random} survives in performance space as the condition for $R > 0$, while the ensemble's harvesting mechanism, aggregation itself, survives only in part: with an endpoint truth, aggregation collects diversity against the average member and never against the best, so the intuition that averaging collects the value of diversity, correct for continuous targets, is the intuition the mean policy imports and the hull bound limits.

The residual is not a new quantity. Written as $\E[\max_s \theta_s] - \max_s \E[\theta_s]$, it is the expected value of perfect information \citep{howard1966information, raiffa1961applied}, with the skills as acts and the instance as state. In algorithm selection, the problem formalized by \citet{rice1976algorithm}, it is the gap between the virtual best solver, which picks the best algorithm for each instance, and the single best solver, and the share of that gap a selector closes is a standard measure of its performance \citep{kerschke2019automated}; and benchmarks of routing among language models measure routers against the same reference points, the best single model and an oracle that picks the best model for each query \citep{hu2024routerbench}, or score the share of the gap between a weak and a strong model that a router recovers \citep{ong2024routellm}. What is claimed is the packaging and its extensions: the additive identity $G = S + R$ pivoted at $b$, splitting the gross stake into a term recoverable from the marginals in closed form and a term they only bound; the refinement of the second into what a class-level rule and what only an instance-level rule can reach; the bridge between the performance register in which the stake is defined and the response register in which arrangements act; and the definition of the two channels, routing by how work is directed and synergy by protocol, which bounds one and measures the other. The split is what makes the pair operational for the skill policy decision, because it separates what an organization can establish from separate track records from what it must run a paired study to see. Table~\ref{tab:novelty} sets the comparison out line by line.

\begin{table}[!htb]
\centering
\footnotesize
\caption{What adjacent lines of work hold, and what the accounting adds}
\label{tab:novelty}
\begin{tabularx}{\textwidth}{@{}>{\raggedright\arraybackslash\hsize=0.7\hsize}X >{\raggedright\arraybackslash\hsize=1.15\hsize}X >{\raggedright\arraybackslash\hsize=1.15\hsize}X@{}}
\toprule
Line of work & What it holds & What this paper adds \\
\midrule
Earlier paper \citep{zanardelli2025navigating} & $G = S + R$ for a pair, with bounds; the mean, collaborative and superpower archetypes; a ladder whose upper rungs are bounded but not run; Monte Carlo analysis of utility & Any number of skills; $R$ split by class; the squared rule and the bridge between responses and performances; the two channels; every rung as an algorithm; the procedure; a measurement on field data \\ \addlinespace[2pt]
Value of information \citep{howard1966information, raiffa1961applied} & $R$ as the expected value of perfect information & The lower anchor and $S$; the split of $R$ by class; which record identifies each part \\ \addlinespace[2pt]
Algorithm selection and model routing \citep{rice1976algorithm, kerschke2019automated, hu2024routerbench, ong2024routellm} & The gap between the virtual and the single best solver, which is $R$, and the share of it a selector closes & What lies below the single best solver, $S$; the class and instance parts of $R$; combining responses and protocols, which per-instance selection does not represent; bounds on $R$ from separate assessments \\ \addlinespace[2pt]
Learning to defer, triage, dynamic selection \citep{madras2018predict, mozannar2020consistent, raghu2019algorithmic, okati2021differentiable, ko2008dynamic} & Criteria learned per instance, those of rungs~4b and~4c, with the oracle as their ceiling & The ceiling identified before any criterion is learned; the evidence each criterion consumes, and an audit that replenishes it; comparison with pooling and with protocols \\ \addlinespace[2pt]
Ensembles and forecast combination \citep{krogh1994neural, bates1969combination, soll2009strategies, mannes2014wisdom} & The ambiguity decomposition; when combining beats choosing & With an endpoint truth, combination capped at the best skill on every instance (Proposition~\ref{prop:hull}); pooling as insurance with a stated premium (Proposition~\ref{prop:insurance}) \\ \addlinespace[2pt]
Human--AI complementarity \citep{bansal2021does, donahue2022human, hemmer2025complementarity, peng2025nofreelunch} & Complementary team performance; complementarity potentials anchored at perfect execution; existence and worst-case results & Anchors at the mean and at the best skill, and so $S$ and $G$; magnitudes for a given pair rather than worst cases; synergy measured by a lift \\ \addlinespace[2pt]
Copulas and partial identification \citep{nelsen2006introduction} & The Fr\'echet--Hoeffding bounds & Their application to $R$ and $\Rinst$, class by class, from separate assessments \\
\bottomrule
\end{tabularx}
\end{table}

\subsection{Presence, protocol and evidence}\label{sec:implications}

The theory has implications that do not depend on any calibration, and they concern three decisions taken before any performance is observed: whether the skills are tested at all, whether they are tested on the same instances, and whether the workflow is treated as an object of design.

For regulation, human oversight mandates such as Article~14 of the EU AI Act\footnote{Regulation (EU) 2024/1689, Article~14 (OJ L, 2024/1689, 12.7.2024). Regulation (EU) 2026/1744 of the European Parliament and of the Council of 8 July 2026 (Digital Omnibus on AI; OJ L, 2026/1744, 24.7.2026) postponed application of the Annex~III high-risk obligations to 2~December 2027 and of Annex~I product-embedded high-risk AI to 2~August 2028, while leaving the human-oversight duty itself in place.} specify the presence of an operator. A mandate that specifies presence but not protocol selects, by default, rung~2, which forfeits \sqc{$S - \Lambda$} in exchange for an insurance, or an undesigned rung~3, whose lift can fall below one and return the outcome of a wrong commitment with both skills engaged (Section~\ref{par:floor}). Presence is not protection: what an arrangement is worth depends on the channel it operates and on what it collects, both of which can be documented and measured. A requirement to demonstrate the stake, by paired evaluation of the skills, and the take, by a third arm retaining the independent responses and reporting the lift against its break-even, would test what presence is presumed to deliver.

For organizations, the theory supports a sequence rather than a choice, which Section~\ref{sec:procedure} sets out step by step. Assess before committing, commit before routing, and route before designing an exchange, each step requiring something the one before it produces: separate assessments for rung~1, class-wise ones for assignment on the class, a paired record for instance-level routing, and a retained pre-exchange record for any claim of synergy; and each step checked, class by class, against what the one before it secured. Because the kind of interaction, and not the power of either skill, decides which rung an arrangement occupies, the design of the process by which skills interact is the lever an organization controls rather than inherits \citep{herrmann2023keeping, deranty2024artificial, vodivskar2025collaboration}. And because the arrangements that engage fewer skills consume the evidence those engaging every skill produce, the choice of rung is also a choice about what the organization will be able to learn next. The classes are part of the same choice. Refining them is, in performance, what \citet{capraro2026beyond} call reconstructing the representation of a problem rather than searching within it: it moves stake from $\Rinst$ into $\Rclass$, and can gain no more than $\Rinst$ (Proposition~\ref{prop:class}). A step's check can also change the question itself: on the Collab-CXR data, the check of step~1 turns the clinic's question from which skill should read into how each should be calibrated (Figure~\ref{fig:cxrwalk}).

\subsection{Skills of equal capability in competition}\label{sec:contest}
 
The paper treats skills that one organization holds together. The same quantities describe skills that compete: a human and a machine of comparable capability bidding for the same work, two organizations deploying the same model for the same clients, or an organization's skill set against a competitor's or a counterparty's, whose success on an instance is the organization's loss unless its own skill answers it. Parity, the case in which the question of which skill is better has no answer, is where the theory says most.
 
\paragraph{No edge from the level.} At parity $S = 0$: no commitment secures anything over the mean anchor, and the whole stake is residual, half the mean distance between the two performances (Proposition~\ref{prop:parity}). Which skill wins an instance is decided by how the two performances are spread and how they co-occur, not by their level. Two deployments of the same skill whose failures coincide have no stake between them at all: combined they add nothing to each other, and set against each other they are strong on the same instances, which, for an organization on the defensive, is the safest position parity allows (Appendix~\ref{app:contest}). Between copies of the same model, or models trained alike, the residual exists only insofar as their failures do not coincide, and language models measurably agree on the same wrong answers \citep{kim2025correlated, goel2025great}. Where competitors hold the same skill, what can still differentiate them is a second skill whose failures fall elsewhere, which is the residual the theory measures.
 
\paragraph{Who gets the work.} How the work divides between competing skills depends on the grain at which it is allocated. A commitment gives every instance to the skill better on average, so that a small change in that ordering reallocates all of it; assignment on the class gives each skill the classes it leads; assignment on the instance gives each the instances it wins. In Profile~A the human executes none of the instances under a commitment at the initial date and all of them at the final date, a third under assignment on the class at both dates, and a third rising to a half under assignment on the instance. Allocating whole tasks to the better factor, as task-based accounts of automation do at the level of the task \citep{autor2003skill, acemoglu2025simple}, turns small shifts in capability into complete reallocations; allocating at a finer grain turns them into gradual ones, but only where the paired evidence and the criterion to read the instance exist. The shares concern executions, not employment or wages, which depend on volumes and prices the paper does not model.
 
\paragraph{When outcomes accumulate.} Where the outcomes of contested instances add up, and a resolution against the organization can cost without limit, a third quantity decides the contest. Appendix~\ref{app:contest} models it: the organization's position moves against it when the opposing skill succeeds and its own fails to answer, by a consequence the organization bounds by policy, and in its favor when its skill answers. The position drifts in the organization's favor if and only if the bound lies below a threshold set by what an answer recovers and by how often the organization's skill answers where the opposing skill succeeds. At parity that frequency moves with how the two performances are coupled: it is lowest when the organization's skill fails where the opposing skill succeeds, the coupling that would make the two most valuable held together, and highest when they fail on the same instances. When the consequence of a single failure has infinite mean, no capability edge in either direction keeps the position viable without a bound (Proposition~\ref{prop:contest}). Containing the maximum damage of a single failure, which is neither a skill nor a rung, is then the instrument that remains when capability no longer separates the skills, and the commitment of Section~\ref{sec:insurance} is survivable, whichever skill it names, only under such a bound.

\section{Limitations}\label{sec:limits}

The limitations fall into three groups: questions the theory poses and does not answer, features of the setting it holds fixed, and the evidence on which its illustrations rest.

\paragraph{Nothing here says whether a skill is good enough.} Every quantity in Section~\ref{sec:bounds} is comparative: $b$ is a maximum over skills, $S$ a difference between them, $R$ a gap between an anchor and a bound. None says whether the skill that wins is adequate for the work, and adequacy cannot be answered from either side alone: a distribution is not adequate in itself, and a task does not carry a required level in itself. The theory describes the skills fully, by a distribution per class, and the task only through its classes and its scoring rule. A roster whose best member attains $0.87$ and one whose best member attains $0.55$ can produce identical decompositions, if their members differ from one another in the same way, and the arrangements are then ranked identically in both, although only one roster may be fit to do the work.

\paragraph{The take is measured, not predicted.} The theory establishes that the stake can be measured before deployment and that a workflow's arrangement can be read from its design, but neither tells an organization how much of the residual its workflow will capture, or what lift its protocol will achieve. Real workflows sit between the archetypes, and where they sit depends on mechanisms the theory does not represent: a reviewer who overrides the machine only when its confidence is low, a sequential protocol in which one party sees the other's answer before committing, a consensus procedure that pays a cost when the two disagree. Predicting the share of $R$ each recovers, or the lift each attains, requires modeling the mechanism itself rather than measuring it after the fact, which is where the human factors and computer-supported cooperative work literatures have the most to contribute to a theory of this kind. And where a workflow errs in reading the class or the instance, its take is measured exactly, but its split between $S$ and $\Rclass$ is a matter of crediting rather than of fact (Section~\ref{sec:routing}).

\paragraph{Costs are reported, not priced.} Every arrangement carries a cost profile, what it holds available and what it executes (Table~\ref{tab:costs}), and the procedure of Section~\ref{sec:procedure} weighs each component of the stake against the cost of the step that pursues it. The theory sets the two side by side without converting either into the other: it does not price a skill held available or an execution, and so does not say which rung pays, how large an audit should be, or when an organization should choose among policies instance by instance, the gating of Section~\ref{sec:signals}. Nor does it bound consequences: performance lies in $[0,1]$, so what a failure costs in performance is bounded, but what it costs the organization need not be, and where the consequence of a single failure is heavy-tailed, a bound on it can decide more than any rung (Section~\ref{sec:contest}, Appendix~\ref{app:contest}).

\paragraph{The stream is held still.} Within a date, skills do not drift and the mix of classes does not move. The illustrative profiles move between two dates, and an audit can detect that the stream has moved (Proposition~\ref{prop:audit}), but the attainments of the rungs are stated for a stream at rest. A spectrum whose rungs differ in how much they estimate is, in a moving stream, also a spectrum of how fast they go stale, and the theory gives no account of that second ordering. Nor does the stream move only of itself: a policy can move the skills it engages, as when operators confined to supervision lose the skill they no longer exercise, or when judgments formed with the model's advice converge on it and lower the residual the pair carries (Section~\ref{sec:magnitudes}), and it can do so unevenly, raising the unassisted performance of experienced operators and the dispersion of novices (Section~\ref{sec:patent}). The theory takes the skills of each date as given and does not model that feedback.

\paragraph{A policy consumes a record it may not replenish.} Every rung above assignment on the class compares the skills on the instance at hand, and that comparison is free of selection only on instances that several skills worked. The theory states what a policy attains while it is right, and says nothing about how it learns that it is no longer right: the refit blocks describe how an estimate is updated but not when it should be distrusted. Assignment with audit keeps the record growing at a chosen cost (Proposition~\ref{prop:audit}), but the theory does not say how large the audit should be. Nor does it say when the problem itself should be posed differently: the task, its classes and its scoring rule are given, and recognizing that a representation of a problem has become inadequate, and revising it, is a capacity of its own \citep{capraro2026beyond}.

\paragraph{The scoring rule is given, and the truth is an endpoint.} The theory evaluates arrangements under a rule the setting supplies and holds every response at the value the skill reported. A different monotone rule reweights the same shortfalls across instances and can reorder arrangements whose errors differ in size rather than frequency, which is why two rules of the family need not agree on which of two arrangements is better; the theory does not say which rule a setting should use, and credits recalibration to no channel, because recalibration is not an arrangement, although the procedure applies it at its first check and the measurement reports what it returns. The rule of this paper is the squared rule, under which what pooling, the insurance contract and synergy attain is priced only from paired observation; the linear rule, simpler and reported by field studies, prices them from separate assessments, and Appendix~\ref{app:linear} states the results in that special case and where it loses nothing. The ceiling on routing also needs the truth to be an endpoint of a scalar response space, as in binary classification. For multiclass responses and continuous targets a combination of responses can score above every one of them (Section~\ref{sec:responses}): the decomposition of the stake, defined on performances, still holds, and assignment and selection, which return one response, remain bounded by $\E[\Mx]$, but ensembling no longer is.

\paragraph{Evidence pooling has a ceiling, not a rung.} Combinations that read the responses as joint evidence are bounded (Proposition~\ref{prop:evidence}) but placed neither on the ladder nor in the procedure. What they require, a paired record with the responses retained and a model of the truth given them, and where their check would sit among the steps are left to future work, as is the design of the audit's exploration, on which off-policy evaluation bears \citep{dudik2011doubly}. The margin of the checks (Proposition~\ref{prop:margin}) is one rule for a finite record, not a design of the tests a procedure should run.
 
\paragraph{Human factors enter only through their consequences.} Cognitive load, alert fatigue, decision latency and the deskilling of operators confined to supervision \citep{bainbridge1983ironies, endsley1995out, parasuraman2010complacency} are not represented as mechanisms. They appear only as their consequences: a lower performance of the human skill, or a lower lift of the protocol in which it takes part.

\paragraph{Two skills, and illustrations under assumptions of convenience.} The exposition keeps a pair wherever a larger set would obscure the argument; Appendix~\ref{app:many} states what does not survive a larger set. The propositions assume nothing about the form of the distributions or about independence between the skills, but the illustrative magnitudes do. The profiles of Appendix~\ref{app:profiles} draw the two performances independently on each instance, and a positively correlated pair carries a smaller residual with the same marginals: a Gaussian coupling of $0.5$ cuts the residual of Profile~A at high difficulty by two fifths (Table~\ref{tab:gsr}). And the profiles specify performances, not responses, so the pooled response is scored with the two responses on the same side of the truth, the case in which, for given performances, they disagree least: the dividend \sqc{$\Lambda$} the profiles report is the smallest they allow.

\paragraph{Regularities, not values.} The space the theory spans is high-dimensional, and the paper establishes in it regularities rather than values: what holds from the means alone, what a bound on pairwise moments guarantees, and orderings and thresholds across the rungs (Appendix~\ref{app:factors}). How much each rung takes under realistic signals and distributions, how the lift of a protocol varies across instances, and what policies fitted on finite records attain are numerical questions, for which the appendix sets out a protocol.

\paragraph{One estimate from data.} The quantities are illustrated on synthetic profiles and measured on one public dataset, Collab-CXR (Section~\ref{sec:radiologydata}). That measurement treats the participants as one skill, a radiologist drawn at random, rather than each radiologist as a skill of its own, so that part of its residual lies between radiologists and is within reach only of a policy that knows who reads; takes the read as the instance, while a radiologist reads whole films, so that its residual overstates what a rule assigning whole films can collect; rests on a model developed on films from the hospital these films come from, whose standing may be lower elsewhere, and on radiologists\textquoteright{} responses that the hierarchy of the elicitation set to zero below every finding ruled out, which a dataset without that structure would separate from the skills; takes the findings the model reports as the classes; and fits the routing policies on bins of the probabilities, a deliberately simple criterion, so that what it reports for rung~4 is what such a criterion attains, not what the best one could. A second demonstration, on data of a different kind, such as the human labels of \citet{peterson2019human} combined with model predictions as in \citet{steyvers2022bayesian}, is left to future work.

\section{Conclusion}\label{sec:conclusion}

Putting a human and a machine on the same task is often framed as a conservative choice. This paper has argued that it is a commitment whose worth is decided by two things the decision to combine does not fix: what the skills put on the table, and what the workflow does with them.

The first is fixed by the skills. Between the mean of the skills' performances on each instance and the best of them lies a gross stake that splits, at the best single skill, into a systematic component, available in closed form from separate assessments and secured by an ex ante commitment, and a residual, the expected value of perfect information about which skill each instance favors, which factors into how often the underdog wins and by how much, and which separate assessments only bound and paired observation identifies. Where instances fall into classes the residual splits again, into what a class-level rule reaches and what only an instance-level rule can. Engaging every skill on every instance, without directing work between them, is an insurance with a stated premium and a stated coverage; what its second execution buys over a random assignment is the disagreement between the responses, lower dispersion and the paired record. And because arrangements act on responses while the stake is defined on performances, the hull bound connects the two registers: under the squared rule a combination of responses scores the combination of their performances plus their disagreement, the ambiguity decomposition of ensemble learning, and under any monotone rule on an endpoint truth it scores within the best and worst of them. The bridge is a result for binary truth, which fails for multiclass responses and continuous targets; under the linear rule, the special case in which the disagreement earns nothing, it becomes exact.

The second is decided by the workflow, and the two channels through which it reaches the stake divide on how it uses the skills. Routing alone returns a convex combination of responses produced independently, and is bounded by the best of them on every instance, however elaborately it chooses; its three implementations, assignment, ensembling and selection, differ in when they choose, in what they execute, and in how far into the residual they reach. Synergy runs a protocol in which what each skill produces enters the others' execution before a joint response is fixed, is bounded by none of the skills, and is measured as a lift against what the same skills would have returned independently, which can fall below one. That is the difference between the channels: not that synergy is better, but that it is the one the stake does not bound, in either direction. The case it makes representable, a combination returning less than averaging the responses of the skills it engages, is documented in the field. On the public data of a chest-radiograph experiment, the radiologists and the model are close to parity and most of the stake is residual; the arm that showed the radiologists the model's prediction, which the original analysis found no better on average than unaided reading, returned less than averaging the two responses; and recalibrating the model alone attains more than any deployable arrangement of the responses as reported.

Both are stated in performance. What they are worth once executions, standing commitments and the consequences of error are charged is a separate question, which future work will take up; the bounds stated here hold whatever its answer.

Between the bounds, the arrangements form a ladder whose rungs are written as algorithms, so that an organization can read off what each position requires, reads and records. Engaging a skill produces the evidence from which the arrangements engaging fewer skills are estimated, so a policy can be deployable and still unfittable, unless it audits a fraction of the instances it no longer double-executes.

Read in sequence, the ladder is a procedure. The components of the stake are reached in order, each ceiling containing the ones below it, and a step taken without error moves the take by a whole component. A step that misreads the class or the instance moves it by an amount that is exact in total but splits between the components only by convention, and can give back at the level of a class what an earlier step held; the check that keeps the climb from losing ground is therefore made class by class, on the classes as the policy reads them, against what the previous step attains on the same instances, and with a margin where the class means are estimated.

The record itself must be read with care: illustrations that draw the skills independently on each instance overstate the residual of skills whose errors coincide, and a finite record understates it.

The same quantities describe skills that compete rather than cooperate. At parity the level of capability decides nothing: the stake is all residual, every commitment is a bet without an edge, and competitors whose skills fail on the same instances have nothing left to differentiate them. Where the outcomes of contested instances accumulate and a single failure can cost without limit, what decides is how much one failure is permitted to cost, a bound that belongs to neither skill nor arrangement.
 
The theory does not say how often combinations create or destroy value, which depends on costs, error regimes and skill profiles. What it supplies is the performance accounting a study of that question requires, and an instrument an organization can apply to its own roster and caseload. The skills set the stake; the workflow decides the take; and the evidence an arrangement needs is not always the evidence it produces.

\section*{Acknowledgments}
The author declares no conflict of interest and no competing interests. AI tools were used to assist coding and editing; the author remains solely accountable for the paper's content and any errors that may persist.

\bibliographystyle{plainnat}
\bibliography{references}

@article{autor2003skill,
  title={The skill content of recent technological change: An empirical exploration},
  author={Autor, David H and Levy, Frank and Murnane, Richard J},
  journal={The Quarterly Journal of Economics},
  volume={118},
  number={4},
  pages={1279--1333},
  year={2003},
  publisher={MIT Press}
}

@techreport{agarwal2023combining,
  title={Combining Human Expertise with Artificial Intelligence: Experimental Evidence from Radiology},
  author={Agarwal, Nikhil and Moehring, Alex and Rajpurkar, Pranav and Salz, Tobias},
  year={2023},
  institution={National Bureau of Economic Research},
  type={Working Paper},
  number={31422},
  note={Revised August 2026}
}

@techreport{farboodi2021model,
  title={A Model of the Data Economy},
  author={Farboodi, Maryam and Veldkamp, Laura},
  institution={National Bureau of Economic Research},
  type={Working Paper},
  number={28427},
  year={2021}
}

@article{acemoglu2025simple,
  title={The simple macroeconomics of AI},
  author={Acemoglu, Daron},
  journal={Economic Policy},
  volume={40},
  number={121},
  pages={13--58},
  year={2025},
  publisher={Oxford University Press}
}

@article{chollet2019measure,
  title={On the measure of intelligence},
  author={Chollet, Fran{\c{c}}ois},
  journal={arXiv preprint arXiv:1911.01547},
  year={2019}
}

@article{herrmann2023keeping,
  title={Keeping the organization in the loop: a socio-technical extension of human-centered artificial intelligence},
  author={Herrmann, Thomas and Pfeiffer, Sabine},
  journal={AI \& Society},
  volume={38},
  number={4},
  pages={1523--1542},
  year={2023},
  publisher={Springer}
}

@article{deranty2024artificial,
  title={Artificial intelligence and work: a critical review of recent research from the social sciences},
  author={Deranty, Jean-Philippe and Corbin, Thomas},
  journal={AI \& Society},
  volume={39},
  number={2},
  pages={675--691},
  year={2024},
  publisher={Springer}
}

@article{simon1955behavioral,
  author    = {Simon, Herbert A.},
  title     = {A Behavioral Model of Rational Choice},
  journal   = {The Quarterly Journal of Economics},
  volume    = {69},
  number    = {1},
  pages     = {99--118},
  year      = {1955},
  publisher = {MIT Press}
}

@article{simon1956rational,
  author    = {Simon, Herbert A.},
  title     = {Rational Choice and the Structure of the Environment},
  journal   = {Psychological Review},
  volume    = {63},
  number    = {2},
  pages     = {129--138},
  year      = {1956},
  publisher = {American Psychological Association}
}

@article{schwarz2022bounded,
  author    = {Schwarz, Gavin},
  title     = {Bounded Rationality, Satisficing, Artificial Intelligence, and Decision-Making in Public Organizations: The Contributions of {Herbert Simon}},
  journal   = {Public Administration Review},
  volume    = {82},
  number    = {6},
  pages     = {1010--1019},
  year      = {2022},
  publisher = {Wiley}
}

@book{Simon1947AdministrativeBehavior,
  author    = {Herbert A. Simon},
  title     = {Administrative Behavior: A Study of Decision-Making Processes in Administrative Organization},
  publisher = {Macmillan},
  year      = {1947},
  address   = {New York}
}

@book{mitchell1993analogy,
  title={Analogy-making as perception: A computer model},
  author={Mitchell, Melanie},
  year={1993},
  publisher={MIT Press}
}

@techreport{fitts1951human,
  title={Human Engineering for an Effective Air-Navigation and Traffic-Control System},
  author={Fitts, Paul M.},
  institution={National Research Council},
  address={Washington, DC},
  year={1951}
}

@techreport{sheridan1978human,
  title={Human and Computer Control of Undersea Teleoperators},
  author={Sheridan, Thomas B and Verplank, William L},
  year={1978},
  institution={MIT Man-Machine Systems Laboratory, Cambridge, MA}
}

@article{rasmussen1983skills,
    title={Skills, Rules, and Knowledge; Signals, Signs, and Symbols, and Other Distinctions in Human Performance Models},
    author={Rasmussen, Jens},
    journal={IEEE Transactions on Systems, Man, and Cybernetics},
    volume={SMC-13}, number={3}, pages={257--266},
    year={1983}
}

@article{bainbridge1983ironies,
  title   = {Ironies of automation},
  author  = {Bainbridge, Lisanne},
  journal = {Automatica},
  volume  = {19},
  number  = {6},
  pages   = {775--779},
  year    = {1983},
  doi     = {10.1016/0005-1098(83)90046-8}
}

@article{endsley1995out,
  title={The out-of-the-loop performance problem and level of control in automation},
  author={Endsley, Mica R and Kiris, Esin O},
  journal={Human Factors},
  volume={37},
  number={2},
  pages={381--394},
  year={1995},
  publisher={SAGE}
}

@article{parasuraman1997humans,
  title={Humans and automation: Use, misuse, disuse, abuse},
  author={Parasuraman, Raja and Riley, Victor},
  journal={Human Factors},
  volume={39},
  number={2},
  pages={230--253},
  year={1997},
  publisher={SAGE}
}

@article{parasuraman2000model,
  title={A model for types and levels of human interaction with automation},
  author={Parasuraman, Raja and Sheridan, Thomas B and Wickens, Christopher D},
  journal={IEEE Transactions on systems, man, and cybernetics-Part A: Systems and Humans},
  volume={30},
  number={3},
  pages={286--297},
  year={2000},
  publisher={IEEE}
}

@article{shneiderman2020human,
  title={Human-centered artificial intelligence: Reliable, safe \& trustworthy},
  author={Shneiderman, Ben},
  journal={International Journal of Human--Computer Interaction},
  volume={36},
  number={6},
  pages={495--504},
  year={2020},
  publisher={Taylor \& Francis}
}

@article{crootof2023humans,
  title={Humans in the Loop},
  author={Crootof, Rebecca and Kaminski, Margot E. and Price, II, W. Nicholson},
  journal={Vand. L. Rev.},
  volume={76},
  pages={429},
  year={2023},
  publisher={HeinOnline}
}

@article{parasuraman2010complacency,
  title={Complacency and bias in human use of automation: An attentional integration},
  author={Parasuraman, Raja and Manzey, Dietrich H},
  journal={Human Factors},
  volume={52},
  number={3},
  pages={381--410},
  year={2010},
  publisher={SAGE}
}

@article{romeo2026exploring,
  title={Exploring automation bias in human--AI collaboration: a review and implications for explainable AI},
  author={Romeo, Giuseppe and Conti, Daniela},
  journal={AI \& Society},
  volume={41},
  number={1},
  pages={259--278},
  year={2026},
  publisher={Springer}
}

@article{green2022flaws,
  title={The flaws of policies requiring human oversight of government algorithms},
  author={Green, Ben},
  journal={Computer Law \& Security Review},
  volume={45},
  pages={105681},
  year={2022},
  publisher={Elsevier}
}

@article{vaccaro2024combinations,
  title={When combinations of humans and AI are useful: A systematic review and meta-analysis},
  author={Vaccaro, Michelle and Almaatouq, Abdullah and Malone, Thomas},
  journal={Nature Human Behaviour},
  volume={8},
  number={12},
  pages={2293--2303},
  year={2024},
  publisher={Nature Publishing Group UK London}
}

@article{dell2026navigating,
  title={Navigating the jagged technological frontier: Field experimental evidence of the effects of artificial intelligence on knowledge worker productivity and quality},
  author={Dell’Acqua, Fabrizio and McFowland III, Edward and Mollick, Ethan and Lifshitz, Hila and Kellogg, Katherine C and Rajendran, Saran and Krayer, Lisa and Candelon, Fran{\c{c}}ois and Lakhani, Karim R},
  journal={Organization Science},
  volume={37},
  number={2},
  pages={403--423},
  year={2026},
  publisher={INFORMS}
}

@article{breiman2001random,
  title   = {Random forests},
  author  = {Breiman, Leo},
  journal = {Machine Learning},
  volume  = {45},
  number  = {1},
  pages   = {5--32},
  year    = {2001}
}

@article{vodivskar2025collaboration,
  title   = {Collaboration between individuals and {AI}: fusing mental effort and {AI} for work meaningfulness},
  author  = {Vodi{\v{s}}kar, Mateja and Ruiner, Caroline},
  journal = {AI \& Society},
  volume  = {41},
  number  = {4},
  pages   = {3557--3571},
  year    = {2026},
  doi     = {10.1007/s00146-025-02772-2}
}

@techreport{randazzo2025cyborgs,
  title={Cyborgs, Centaurs and Self-Automators: The Three Modes of Human-{GenAI} Knowledge Work and Their Implications for Skilling and the Future of Expertise},
  author={Randazzo, Steven and Lifshitz-Assaf, Hila and Kellogg, Katherine and Dell'Acqua, Fabrizio and Mollick, Ethan R. and Candelon, Fran{\c{c}}ois and Lakhani, Karim R.},
  institution={Harvard Business School},
  type={Working Paper},
  number={26-036},
  year={2025}
}

@article{bates1969combination,
  title={The combination of forecasts},
  author={Bates, John M and Granger, Clive WJ},
  journal={Operational Research Quarterly},
  volume={20},
  number={4},
  pages={451--468},
  year={1969}
}

@incollection{timmermann2006forecast,
  title     = {Forecast combinations},
  author    = {Timmermann, Allan},
  booktitle = {Handbook of Economic Forecasting},
  editor    = {Elliott, Graham and Granger, Clive W. J. and Timmermann, Allan},
  volume    = {1},
  pages     = {135--196},
  publisher = {Elsevier},
  year      = {2006}
}

@article{ko2008dynamic,
  title={From dynamic classifier selection to dynamic ensemble selection},
  author={Ko, Albert HR and Sabourin, Robert and Britto Jr, Alceu Souza},
  journal={Pattern recognition},
  volume={41},
  number={5},
  pages={1718--1731},
  year={2008},
  publisher={Elsevier}
}

@article{cruz2018dynamic,
  title={Dynamic classifier selection: Recent advances and perspectives},
  author={Cruz, Rafael MO and Sabourin, Robert and Cavalcanti, George DC},
  journal={Information Fusion},
  volume={41},
  pages={195--216},
  year={2018},
  publisher={Elsevier}
}

@inproceedings{krogh1994neural,
  title     = {Neural network ensembles, cross validation, and active learning},
  author    = {Krogh, Anders and Vedelsby, Jesper},
  booktitle = {Advances in Neural Information Processing Systems},
  volume    = {7},
  pages     = {231--238},
  year      = {1994}
}

@inproceedings{madras2018predict,
  title     = {Predict Responsibly: Improving Fairness and Accuracy by Learning to Defer},
  author    = {Madras, David and Pitassi, Toniann and Zemel, Richard},
  booktitle = {Advances in Neural Information Processing Systems},
  volume    = {31},
  pages     = {6147--6157},
  year      = {2018}
}

@inproceedings{mozannar2020consistent,
  title={Consistent estimators for learning to defer to an expert},
  author={Mozannar, Hussein and Sontag, David},
  booktitle={International conference on machine learning},
  pages={7076--7087},
  year={2020},
  organization={PMLR}
}

@inproceedings{bansal2021does,
  title={Does the whole exceed its parts? the effect of ai explanations on complementary team performance},
  author={Bansal, Gagan and Wu, Tongshuang and Zhou, Joyce and Fok, Raymond and Nushi, Besmira and Kamar, Ece and Ribeiro, Marco Tulio and Weld, Daniel},
  booktitle={Proceedings of the 2021 CHI conference on human factors in computing systems},
  pages={1--16},
  year={2021}
}

@inproceedings{donahue2022human,
  title={Human-algorithm collaboration: Achieving complementarity and avoiding unfairness},
  author={Donahue, Kate and Chouldechova, Alexandra and Kenthapadi, Krishnaram},
  booktitle={Proceedings of the 2022 ACM Conference on Fairness, Accountability, and Transparency},
  pages={1639--1656},
  year={2022}
}

@inproceedings{rastogi2023taxonomy,
  title={A taxonomy of human and ML strengths in decision-making to investigate human-ML complementarity},
  author={Rastogi, Charvi and Leqi, Liu and Holstein, Kenneth and Heidari, Hoda},
  booktitle={Proceedings of the AAAI Conference on Human Computation and Crowdsourcing},
  volume={11},
  pages={127--139},
  year={2023}
}

@article{hemmer2025complementarity,
  title={Complementarity in human-AI collaboration: Concept, sources, and evidence},
  author={Hemmer, Patrick and Schemmer, Max and K{\"u}hl, Niklas and V{\"o}ssing, Michael and Satzger, Gerhard},
  journal={European Journal of Information Systems},
  volume={34},
  number={6},
  pages={979--1002},
  year={2025},
  publisher={Taylor \& Francis}
}

@article{kitagawa2018should,
  title={Who should be treated? empirical welfare maximization methods for treatment choice},
  author={Kitagawa, Toru and Tetenov, Aleksey},
  journal={Econometrica},
  volume={86},
  number={2},
  pages={591--616},
  year={2018},
  publisher={Wiley Online Library}
}

@article{roy1951some,
  title={Some thoughts on the distribution of earnings},
  author={Roy, Andrew Donald},
  journal={Oxford economic papers},
  volume={3},
  number={2},
  pages={135--146},
  year={1951},
  publisher={Oxford University Press}
}

@book{nelsen2006introduction,
  title     = {An Introduction to Copulas},
  author    = {Nelsen, Roger B.},
  edition   = {2},
  publisher = {Springer},
  year      = {2006}
}

@article{athey2021policy,
  title={Policy learning with observational data},
  author={Athey, Susan and Wager, Stefan},
  journal={Econometrica},
  volume={89},
  number={1},
  pages={133--161},
  year={2021},
  publisher={Wiley Online Library}
}

@article{howard1966information,
  title={Information value theory},
  author={Howard, Ronald A},
  journal={IEEE Transactions on systems science and cybernetics},
  volume={2},
  number={1},
  pages={22--26},
  year={1966},
  publisher={IEEE}
}

@book{raiffa1961applied,
  title     = {Applied Statistical Decision Theory},
  author    = {Raiffa, Howard and Schlaifer, Robert},
  publisher = {Division of Research, Graduate School of Business Administration, Harvard University},
  address   = {Boston},
  year      = {1961}
}

@inproceedings{peng2025nofreelunch,
  title={A No Free Lunch Theorem for Human-{AI} Collaboration},
  author={Peng, Kenny and Garg, Nikhil and Kleinberg, Jon},
  booktitle={Proceedings of the AAAI Conference on Artificial Intelligence},
  volume={39},
  pages={14369--14376},
  year={2025}
}

@inproceedings{lakkaraju2017selective,
  title={The selective labels problem: Evaluating algorithmic predictions in the presence of unobservables},
  author={Lakkaraju, Himabindu and Kleinberg, Jon and Leskovec, Jure and Ludwig, Jens and Mullainathan, Sendhil},
  booktitle={Proceedings of the 23rd ACM SIGKDD international conference on knowledge discovery and data mining},
  pages={275--284},
  year={2017}
}

@article{autor2025expertise,
  title={Expertise},
  author={Autor, David and Thompson, Neil},
  journal={Journal of the European Economic Association},
  volume={23},
  number={4},
  pages={1203--1271},
  year={2025},
  publisher={Oxford University Press}
}

@article{puranam2021human,
  author  = {Puranam, Phanish},
  title   = {Human--{AI} collaborative decision-making as an organization design problem},
  journal = {Journal of Organization Design},
  volume  = {10},
  pages   = {75--80},
  year    = {2021},
  doi     = {10.1007/s41469-021-00095-2}
}

@article{taylorphillips2018double,
  author  = {Taylor-Phillips, Sian and Jenkinson, David and Stinton, Chris and Wallis, Matthew G. and Dunn, Janet and Clarke, Aileen},
  title   = {Double Reading in Breast Cancer Screening: Cohort Evaluation in the {CO-OPS} Trial},
  journal = {Radiology},
  volume  = {287},
  number  = {3},
  pages   = {749--757},
  year    = {2018},
  doi     = {10.1148/radiol.2018171010}
}

@article{eisemann2025nationwide,
  author  = {Eisemann, Nora and Bunk, Stefan and Mukama, Trasias and Baltus, Hanna and Elsner, Susanne A. and Gomille, Thorsten and others},
  title   = {Nationwide real-world implementation of {AI} for cancer detection in population-based mammography screening},
  journal = {Nature Medicine},
  volume  = {31},
  pages   = {917--924},
  year    = {2025},
  doi     = {10.1038/s41591-024-03408-6}
}

@article{lang2023artificial,
  author  = {L{\aa}ng, Kristina and Josefsson, Viktoria and Larsson, Anna-Maria and others},
  title   = {Artificial intelligence-supported screen reading versus standard double reading in the Mammography Screening with Artificial Intelligence trial ({MASAI})},
  journal = {The Lancet Oncology},
  volume  = {24},
  number  = {8},
  pages   = {936--944},
  year    = {2023},
  doi     = {10.1016/S1470-2045(23)00298-X}
}

@article{johnston1999two,
  title={Two view mammography at incident screens: cost effectiveness analysis of policy options},
  author={Johnston, Katharine and Brown, Jackie},
  journal={BMJ},
  volume={319},
  number={7217},
  pages={1097--1102},
  year={1999},
  publisher={British Medical Journal Publishing Group}
}

@article{wood2023unified,
  author  = {Wood, Danny and Mu, Tingting and Webb, Andrew M. and Reeve, Henry W. J. and Luj{\'a}n, Mikel and Brown, Gavin},
  title   = {A Unified Theory of Diversity in Ensemble Learning},
  journal = {Journal of Machine Learning Research},
  volume  = {24},
  number  = {359},
  pages   = {1--49},
  year    = {2023}
}

@article{swaminathan2015batch,
  title   = {Batch learning from logged bandit feedback through counterfactual risk minimization},
  author  = {Swaminathan, Adith and Joachims, Thorsten},
  journal = {Journal of Machine Learning Research},
  volume  = {16},
  number  = {52},
  pages   = {1731--1755},
  year    = {2015}
}

@article{ranjan2010combining,
  title={Combining probability forecasts},
  author={Ranjan, Roopesh and Gneiting, Tilmann},
  journal={Journal of the Royal Statistical Society, Series B},
  volume={72}, number={1}, pages={71--91}, year={2010},
  doi={10.1111/j.1467-9868.2009.00726.x}}

@article{lichtendahl2022extremizing,
  title={Extremizing and antiextremizing in {B}ayesian ensembles of binary-event forecasts},
  author={Lichtendahl, Kenneth C and Grushka-Cockayne, Yael and Jose, Victor Richmond R and Winkler, Robert L},
  journal={Operations Research},
  volume={70}, number={5}, pages={2998--3014}, year={2022},
  doi={10.1287/opre.2021.2176}}

@article{kumaran2026causal,
  title   = {Causal evidence that language models use confidence to drive behaviour},
  author  = {Kumaran, Dharshan and Daw, Nathaniel and Osindero, Simon and Veli{\v{c}}kovi{\'c}, Petar and Patraucean, Viorica},
  journal = {Nature Machine Intelligence},
  volume  = {8},
  pages   = {1411--1425},
  year    = {2026},
  doi     = {10.1038/s42256-026-01293-x}
}

@article{kadavath2022know,
  title   = {Language Models (Mostly) Know What They Know},
  author  = {Kadavath, Saurav and Conerly, Tom and Askell, Amanda and Henighan, Tom and Drain, Dawn and Perez, Ethan and others},
  journal = {arXiv preprint arXiv:2207.05221}, year = {2022},
  url     = {https://arxiv.org/abs/2207.05221}
}

@article{breiman2001statistical,
  title={Statistical modeling: The two cultures (with comments and a rejoinder by the author)},
  author={Breiman, Leo},
  journal={Statistical Science},
  volume={16},
  number={3},
  pages={199--231},
  year={2001},
  doi={10.1214/ss/1009213726}
}

@incollection{scarf1958min,
  author    = {Scarf, Herbert},
  title     = {A Min-Max Solution of an Inventory Problem},
  booktitle = {Studies in the Mathematical Theory of Inventory and Production},
  editor    = {Arrow, Kenneth J. and Karlin, Samuel and Scarf, Herbert},
  publisher = {Stanford University Press},
  address   = {Stanford, CA},
  year      = {1958},
  pages     = {201--209}
}

@article{gneiting2007strictly,
  title={Strictly Proper Scoring Rules, Prediction, and Estimation},
  author={Gneiting, Tilmann and Raftery, Adrian E.},
  journal={Journal of the American Statistical Association},
  volume={102},
  number={477},
  pages={359--378},
  year={2007}
}

@article{horvitz1952generalization,
  title={A Generalization of Sampling Without Replacement from a Finite Universe},
  author={Horvitz, Daniel G. and Thompson, Donovan J.},
  journal={Journal of the American Statistical Association},
  volume={47},
  number={260},
  pages={663--685},
  year={1952}
}

@inproceedings{peterson2019human,
  title={Human Uncertainty Makes Classification More Robust},
  author={Peterson, Joshua C. and Battleday, Ruairidh M. and Griffiths, Thomas L. and Russakovsky, Olga},
  booktitle={Proceedings of the IEEE/CVF International Conference on Computer Vision},
  pages={9617--9626},
  year={2019}
}

@article{steyvers2022bayesian,
  title={Bayesian Modeling of Human--{AI} Complementarity},
  author={Steyvers, Mark and Tejeda, Heliodoro and Kerrigan, Gavin and Smyth, Padhraic},
  journal={Proceedings of the National Academy of Sciences},
  volume={119},
  number={11},
  pages={e2111547119},
  year={2022}
}

@article{zanardelli2025navigating,
  title={The safe harbor paradox: when combining human and {AI} skills creates or destroys value},
  author={Zanardelli, Riccardo},
  journal={arXiv preprint arXiv:2509.14057},
  year={2025}
}

@article{clemen1999combining,
  title={Combining Probability Distributions from Experts in Risk Analysis},
  author={Clemen, Robert T. and Winkler, Robert L.},
  journal={Risk Analysis},
  volume={19},
  number={2},
  pages={187--203},
  year={1999}
}

@article{meilijson1979convex,
  title={Convex majorization with an application to the length of critical paths},
  author={Meilijson, Isaac and N{\'a}das, Arthur},
  journal={Journal of Applied Probability}, volume={16}, number={3}, pages={671--677}, year={1979}
}

@incollection{rice1976algorithm,
  title={The algorithm selection problem}, author={Rice, John R.},
  booktitle={Advances in Computers}, volume={15}, pages={65--118}, year={1976}, publisher={Academic Press}
}

@article{kerschke2019automated,
  title={Automated algorithm selection: Survey and perspectives},
  author={Kerschke, Pascal and Hoos, Holger H. and Neumann, Frank and Trautmann, Heike},
  journal={Evolutionary Computation}, volume={27}, number={1}, pages={3--45}, year={2019}
}

@article{hu2024routerbench,
  title={{RouterBench}: A Benchmark for Multi-{LLM} Routing System},
  author={Hu, Qitian Jason and Bieker, Jacob and Li, Xiuyu and Jiang, Nan and Keigwin, Benjamin and Ranganath, Gaurav and Keutzer, Kurt and Upadhyay, Shriyash Kaustubh},
  journal={arXiv preprint arXiv:2403.12031}, year={2024}
}

@inproceedings{ong2024routellm,
  title={{RouteLLM}: Learning to Route {LLMs} with Preference Data},
  author={Ong, Isaac and Almahairi, Amjad and Wu, Vincent and Chiang, Wei-Lin and Wu, Tianhao and Gonzalez, Joseph E. and Kadous, M. Waleed and Stoica, Ion},
  booktitle={International Conference on Learning Representations (ICLR)},
  year={2025}
}

@article{larrick2006intuitions,
  title={Intuitions about combining opinions: Misappreciation of the averaging principle},
  author={Larrick, Richard P. and Soll, Jack B.},
  journal={Management Science}, volume={52}, number={1}, pages={111--127}, year={2006}
}

@article{soll2009strategies,
  title={Strategies for revising judgment: How (and how well) people use others' opinions},
  author={Soll, Jack B. and Larrick, Richard P.},
  journal={Journal of Experimental Psychology: Learning, Memory, and Cognition}, volume={35}, number={3}, pages={780--805}, year={2009}
}

@article{mannes2014wisdom,
  title={The wisdom of select crowds}, author={Mannes, Albert E. and Soll, Jack B. and Larrick, Richard P.},
  journal={Journal of Personality and Social Psychology}, volume={107}, number={2}, pages={276--299}, year={2014}
}

@article{smith2006optimizer,
  title={The optimizer's curse: Skepticism and postdecision surprise in decision analysis},
  author={Smith, James E. and Winkler, Robert L.},
  journal={Management Science}, volume={52}, number={3}, pages={311--322}, year={2006}
}

@article{fugener2021will,
  title={Will humans-in-the-loop become {B}orgs? Merits and pitfalls of working with {AI}},
  author={F{\"u}gener, Andreas and Grahl, J{\"o}rn and Gupta, Alok and Ketter, Wolfgang},
  journal={MIS Quarterly}, volume={45}, number={3}, pages={1527--1556}, year={2021}
}

@article{hernstrom2025screening,
  title   = {Screening performance and characteristics of breast cancer detected in the Mammography Screening with Artificial Intelligence trial ({MASAI}): a randomised, controlled, parallel-group, non-inferiority, single-blinded, screening accuracy study},
  author  = {Hernstr{\"o}m, Veronica and Josefsson, Viktoria and Sartor, Hanna and Schmidt, David and Larsson, Anna-Maria and Hofvind, Solveig and others},
  journal = {The Lancet Digital Health},
  volume  = {7},
  number  = {3},
  pages   = {e175--e183},
  year    = {2025},
  doi     = {10.1016/S2589-7500(24)00267-X}
}

@article{gommers2026interval,
  title={Interval cancer, sensitivity, and specificity comparing {AI}-supported mammography screening with standard double reading without {AI} in the {MASAI} study: a randomised, controlled, non-inferiority, single-blinded, population-based, screening-accuracy trial},
  author={Gommers, Jessie and others},
  journal={The Lancet}, volume={407}, number={10527}, pages={505--514}, year={2026},
  doi={10.1016/S0140-6736(25)02464-X}
}

@article{felli1998sensitivity,
  title={Sensitivity analysis and the expected value of perfect information},
  author={Felli, James C. and Hazen, Gordon B.},
  journal={Medical Decision Making}, volume={18}, number={1}, pages={95--109}, year={1998}
}

@article{blackwell1953equivalent,
  title={Equivalent comparisons of experiments}, author={Blackwell, David},
  journal={The Annals of Mathematical Statistics}, volume={24}, number={2}, pages={265--272}, year={1953}
}

@article{clark1961greatest,
  title={The greatest of a finite set of random variables}, author={Clark, Charles E.},
  journal={Operations Research}, volume={9}, number={2}, pages={145--162}, year={1961}
}

@article{heckman1997making,
  title={Making the most out of programme evaluations and social experiments: Accounting for heterogeneity in programme impacts},
  author={Heckman, James J. and Smith, Jeffrey and Clements, Nancy},
  journal={The Review of Economic Studies}, volume={64}, number={4}, pages={487--535}, year={1997}
}

@article{raghu2019algorithmic,
  title={The algorithmic automation problem: Prediction, triage, and human effort},
  author={Raghu, Maithra and Blumer, Katy and Corrado, Greg and Kleinberg, Jon and Obermeyer, Ziad and Mullainathan, Sendhil},
  journal={arXiv preprint arXiv:1903.12220}, year={2019}
}

@inproceedings{okati2021differentiable,
  title={Differentiable learning under triage}, author={Okati, Nastaran and De, Abir and Gomez-Rodriguez, Manuel},
  booktitle={Advances in Neural Information Processing Systems}, volume={34}, year={2021}
}

@inproceedings{verma2022calibrated,
  title={Calibrated learning to defer with one-vs-all classifiers}, author={Verma, Rajeev and Nalisnick, Eric},
  booktitle={Proceedings of the 39th International Conference on Machine Learning}, series={PMLR}, volume={162}, year={2022}
}

@inproceedings{mozannar2023should,
  title={Who should predict? Exact algorithms for learning to defer to humans},
  author={Mozannar, Hussein and Lang, Hunter and Wei, Dennis and Sattigeri, Prasanna and Das, Subhro and Sontag, David},
  booktitle={Proceedings of the 26th International Conference on Artificial Intelligence and Statistics}, series={PMLR}, volume={206}, year={2023}
}

@article{moehring2025dataset,
  title={A dataset for understanding radiologist--artificial intelligence collaboration},
  author={Moehring, Alex and Kutwal, Manasi and Huang, Ray and Banerjee, Oishi and Jacobi, Adam and Eber, Corey and Mendoza, Dexter and Chung, Mike and Dayan, Etan and Gupta, Yash and Bui, Tan D. T. and Truong, Steven Q. H. and Pareek, Anuj and Langlotz, Curtis P. and Lungren, Matthew P. and Agarwal, Nikhil and Rajpurkar, Pranav and Salz, Tobias},
  journal={Scientific Data},
  volume={12},
  pages={739},
  year={2025},
  doi={10.1038/s41597-025-05054-0}
}

@techreport{autor2026does,
  title={Does {AI} Assistance Enhance or Erode Expertise? Evidence from a Three-Month Field Experiment in Patent Drafting},
  author={Autor, David and Rodchenko, Tanya and Martin, Josh and Iscenko, Zanna and Strand, Scott and Pearl, David and Ferere, Melissa},
  institution={National Bureau of Economic Research},
  type={Working Paper},
  number={35720},
  year={2026}
}

@article{nemhauser1978analysis,
  title={An analysis of approximations for maximizing submodular set functions---{I}},
  author={Nemhauser, George L. and Wolsey, Laurence A. and Fisher, Marshall L.},
  journal={Mathematical Programming},
  volume={14}, number={1}, pages={265--294}, year={1978}
}

@article{neyman1934two,
  title={On the two different aspects of the representative method: the method of stratified sampling and the method of purposive selection},
  author={Neyman, Jerzy},
  journal={Journal of the Royal Statistical Society},
  volume={97}, number={4}, pages={558--625}, year={1934}
}

@article{hoeffding1963probability,
  title={Probability inequalities for sums of bounded random variables},
  author={Hoeffding, Wassily},
  journal={Journal of the American Statistical Association},
  volume={58}, number={301}, pages={13--30}, year={1963}
}

@inproceedings{kim2025correlated,
  title={Correlated errors in large language models},
  author={Kim, Elliot and Garg, Avi and Peng, Kenny and Garg, Nikhil},
  booktitle={Proceedings of the 42nd International Conference on Machine Learning},
  series={PMLR}, volume={267}, pages={30038--30066}, year={2025}
}

@inproceedings{goel2025great,
  title     = {Great Models Think Alike and This Undermines {AI} Oversight},
  author    = {Goel, Shashwat and Str{\"u}ber, Joschka and Auzina, Ilze Amanda and Chandra, Karuna K. and Kumaraguru, Ponnurangam and Kiela, Douwe and Prabhu, Ameya and Bethge, Matthias and Geiping, Jonas},
  booktitle = {Proceedings of the 42nd International Conference on Machine Learning},
  series    = {Proceedings of Machine Learning Research},
  volume    = {267},
  pages     = {19621--19678},
  year      = {2025}
}

@book{klugman2019loss,
  title={Loss Models: From Data to Decisions},
  author={Klugman, Stuart A. and Panjer, Harry H. and Willmot, Gordon E.},
  edition={5}, publisher={Wiley}, year={2019}
}

@book{asmussen2010ruin,
  title={Ruin Probabilities},
  author={Asmussen, S{\o}ren and Albrecher, Hansj{\"o}rg},
  edition={2}, publisher={World Scientific}, year={2010}
}

@article{murphy1973new,
   author  = {Murphy, Allan H.},
   title   = {A New Vector Partition of the Probability Score},
   journal = {Journal of Applied Meteorology},
   volume  = {12},
   number  = {4},
   pages   = {595--600},
   year    = {1973}
}

@misc{capraro2026beyond,
  title        = {Beyond problem solving: Adaptive reframing in human and artificial intelligence},
  author       = {Capraro, Valerio and Lieto, Antonio and Rea, Francesco and Sandini, Giulio and Sciutti, Alessandra and Gallese, Vittorio and Quattrociocchi, Walter},
  year         = {2026}
}

@inproceedings{wilder2020learning,
  title     = {Learning to Complement Humans},
  author    = {Wilder, Bryan and Horvitz, Eric and Kamar, Ece},
  booktitle = {Proceedings of the Twenty-Ninth International Joint Conference on Artificial Intelligence, {IJCAI-20}},
  pages     = {1526--1533},
  year      = {2020},
  doi       = {10.24963/ijcai.2020/212}
}

@inproceedings{kerrigan2021combining,
  title     = {Combining Human Predictions with Model Probabilities via Confusion Matrices and Calibration},
  author    = {Kerrigan, Gavin and Smyth, Padhraic and Steyvers, Mark},
  booktitle = {Advances in Neural Information Processing Systems},
  volume    = {34},
  pages     = {4421--4434},
  year      = {2021}
}

@inproceedings{cortes2016learning,
  title     = {Learning with Rejection},
  author    = {Cortes, Corinna and DeSalvo, Giulia and Mohri, Mehryar},
  booktitle = {Algorithmic Learning Theory (ALT 2016)},
  pages     = {67--82},
  year      = {2016},
  doi       = {10.1007/978-3-319-46379-7_5}
}

@article{smith2009simple,
  title   = {A Simple Explanation of the Forecast Combination Puzzle},
  author  = {Smith, Jeremy and Wallis, Kenneth F.},
  journal = {Oxford Bulletin of Economics and Statistics},
  volume  = {71},
  number  = {3},
  pages   = {331--355},
  year    = {2009}
}

@article{athey2020allocation,
  title   = {The Allocation of Decision Authority to Human and Artificial Intelligence},
  author  = {Athey, Susan C. and Bryan, Kevin A. and Gans, Joshua S.},
  journal = {AEA Papers and Proceedings},
  volume  = {110},
  pages   = {80--84},
  year    = {2020}
}

@article{agrawal2019exploring,
  title   = {Exploring the impact of artificial intelligence: Prediction versus judgment},
  author  = {Agrawal, Ajay and Gans, Joshua S. and Goldfarb, Avi},
  journal = {Information Economics and Policy},
  volume  = {47},
  pages   = {1--6},
  year    = {2019}
}

@article{acemoglu2018race,
  title   = {The Race between Man and Machine: Implications of Technology for Growth, Factor Shares, and Employment},
  author  = {Acemoglu, Daron and Restrepo, Pascual},
  journal = {American Economic Review},
  volume  = {108},
  number  = {6},
  pages   = {1488--1542},
  year    = {2018}
}

@article{acemoglu2022tasks,
  title   = {Tasks, Automation, and the Rise in {U.S.} Wage Inequality},
  author  = {Acemoglu, Daron and Restrepo, Pascual},
  journal = {Econometrica},
  volume  = {90},
  number  = {5},
  pages   = {1973--2016},
  year    = {2022}
}

@article{manski1997mixing,
  title   = {The Mixing Problem in Programme Evaluation},
  author  = {Manski, Charles F.},
  journal = {Review of Economic Studies},
  volume  = {64},
  number  = {4},
  pages   = {537--553},
  year    = {1997}
}

@article{manski2004statistical,
  title   = {Statistical Treatment Rules for Heterogeneous Populations},
  author  = {Manski, Charles F.},
  journal = {Econometrica},
  volume  = {72},
  number  = {4},
  pages   = {1221--1246},
  year    = {2004}
}

@article{fan2010sharp,
  title   = {Sharp Bounds on the Distribution of Treatment Effects and Their Statistical Inference},
  author  = {Fan, Yanqin and Park, Sang Soo},
  journal = {Econometric Theory},
  volume  = {26},
  number  = {3},
  pages   = {931--951},
  year    = {2010}
}

@article{dvijotham2023enhancing,
  title   = {Enhancing the reliability and accuracy of {AI}-enabled diagnosis via complementarity-driven deferral to clinicians},
  author  = {Dvijotham, Krishnamurthy and Winkens, Jim and Barsbey, Melih and Ghaisas, Sumedh and Stanforth, Robert and Pawlowski, Nick and Strachan, Patricia and Ahmed, Zahra and Azizi, Shekoofeh and Bachrach, Yoram and others},
  journal = {Nature Medicine},
  volume  = {29},
  number  = {7},
  pages   = {1814--1820},
  year    = {2023}
}

@inproceedings{li2010contextual,
  title     = {A Contextual-Bandit Approach to Personalized News Article Recommendation},
  author    = {Li, Lihong and Chu, Wei and Langford, John and Schapire, Robert E.},
  booktitle = {Proceedings of the 19th International Conference on World Wide Web},
  pages     = {661--670},
  year      = {2010}
}

@inproceedings{dudik2011doubly,
  title     = {Doubly Robust Policy Evaluation and Learning},
  author    = {Dud{\'\i}k, Miroslav and Langford, John and Li, Lihong},
  booktitle = {Proceedings of the 28th International Conference on Machine Learning},
  pages     = {1097--1104},
  year      = {2011}
}

@article{owen2007pigeonhole,
  title   = {The pigeonhole bootstrap},
  author  = {Owen, Art B.},
  journal = {The Annals of Applied Statistics},
  volume  = {1},
  number  = {2},
  pages   = {386--411},
  year    = {2007}
}

@inproceedings{irvin2019chexpert,
  title     = {{CheXpert}: A Large Chest Radiograph Dataset with Uncertainty Labels and Expert Comparison},
  author    = {Irvin, Jeremy and Rajpurkar, Pranav and Ko, Michael and Yu, Yifan and Ciurea-Ilcus, Silviana and Chute, Chris and Marklund, Henrik and Haghgoo, Behzad and Ball, Robyn and Shpanskaya, Katie and others},
  booktitle = {Proceedings of the AAAI Conference on Artificial Intelligence},
  volume    = {33},
  pages     = {590--597},
  year      = {2019}
}

\clearpage
\appendix
\addtocontents{toc}{\protect\vspace{0.6em}\noindent\textbf{Appendices}\protect\par\protect\vspace{0.3em}}

\section{What governs the stake and the take}\label{app:factors}

The bounds of Section~\ref{sec:bounds} and the ladder of Section~\ref{sec:ladder} depend on many features of the skills and of the policy, and the space they span is too large for a single rule. This appendix organizes it. It states how much must be known to establish each quantity (Section~\ref{app:layers}) and proves the regularities that hold across it (Section~\ref{app:regularities}).

\subsection{Four layers of knowledge}\label{app:layers}

Each quantity of the paper can be established at the first of four layers that suffices for it.

\paragraph{Layer 1: the means.} The skills' means, by class where classes apply, establish $S$, $\Rclass$ and the ceiling B1, and the attainment of rungs~0, 1 and~4a, whatever the shape of the distributions and however the skills are coupled (Propositions~\ref{prop:anchor}, \ref{prop:S} and~\ref{prop:class}). The marginal distributions, beyond their means, bound $\E[\Mx]$, $G$ and $R$ (Proposition~\ref{prop:anchor}): the most separate assessments can say about the residual.

\paragraph{Layer 2: pairwise moments.} The variances and covariances of the skills' performances within each class add a bound on the residual that holds for any distributions with those moments (Proposition~\ref{prop:twomoment} below). Taken on the distances \sqc{$\sqrt{1 - \theta_s}$} that the squared rule implies, the same moments give the disagreement \sqc{$\Lambda$} that pooling earns, and with it the attainment of rung~2 and the break-even $\gamma^{*}$.

\paragraph{Layer 3: the joint law.} $\E[\Mx]$, and with it $\Rinst$ and $G$, the attainment of every rung that reads the instance, and every attainment affected by the ceiling at one, require the joint distribution of the performances within each class: the law that says with what probability each pair of performances occurs together on an instance, with its shape and its coupling. Knowing the law is not knowing which pair occurred on which instance; it is a property of the skills on the task, from which the population quantities follow exactly.

\paragraph{Layer 4: finite records.} An organization does not know the law; it observes a finite record of instances, on which the skills may or may not have been observed together. Estimates computed from such a record scatter around the population quantities, and a policy fitted on it differs from the policy its design describes; how a protocol's lift varies across instances, too, is revealed only by an observed or simulated deployment. Knowing the pair of performances on the instance at hand, before deciding, belongs to no layer: it is the oracle of ceiling~B2, whose value the stake measures.

\begin{figure}[!htb]
\centering
\includegraphics[width=\textwidth]{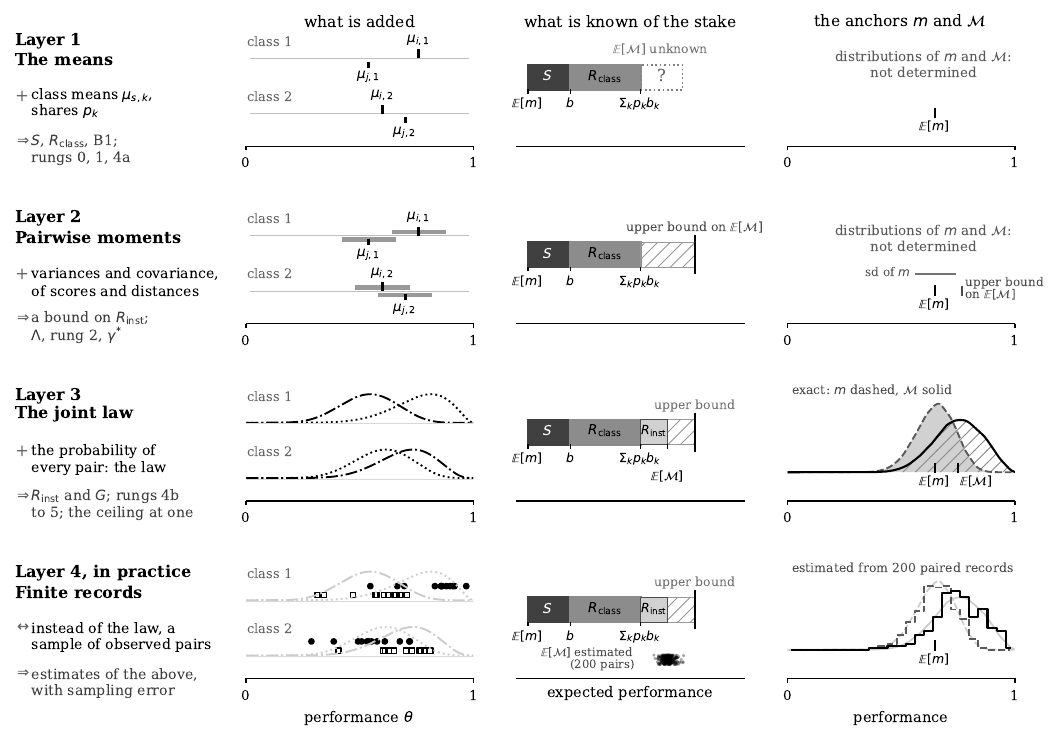}
\caption{What each layer adds, and what becomes known. The figure is drawn for illustrative Beta skills, but what each layer determines does not depend on them. Middle: what the layer adds to what is known of the skills in each class. Ticks mark the class means $\mu_{s,k}$, those of skill $i$ above the line and those of skill $j$ below it; the bars in the second row are their spreads; the dotted and dash-dotted curves are the distributions of skills $i$ and $j$ within each class (the coupling, which the law also fixes, is not drawn); and the filled dots and empty squares in the last row are observed performances of skills $i$ and $j$, a sample of pairs that in practice takes the place of the law, drawn faintly behind it because it is not known. Third column: what is then known of the anchors and of the stake. The means fix $\E[m]$, $b$ and $\sum_k p_k b_k$, and with them $S$ and $\Rclass$, but leave $\E[\Mx]$ and $\Rinst$ unknown; pairwise moments bound them from above (hatched, Proposition~\ref{prop:twomoment}) and, taken on the distances \sqc{$\sqrt{1 - \theta_s}$}, give the disagreement \sqc{$\Lambda$} that pooling earns, and with it rung~2 and $\gamma^{*}$; the joint law fixes them; and a record of 200 paired instances yields estimates of $\E[\Mx]$ scattered around its value. Right: what can be drawn of the anchors themselves. The means give $\E[m]$ alone, and pairwise moments add its spread and an upper bound on $\E[\Mx]$, but neither determines the distributions of $m$ and $\Mx$; the joint law gives them exactly ($m$ dashed and shaded, $\Mx$ solid and hatched); and a record of 200 paired instances gives histograms that estimate them, the exact curves drawn faintly behind.}
\label{fig:app-layers}
\end{figure}

Figure~\ref{fig:app-layers} summarizes the four layers and what each establishes. The layers order what must be known of the skills, and correspond to the levels of evidence of Section~\ref{sec:identification} without coinciding with them (Table~\ref{tab:evidence}). The first layer comes from separate assessments, task-wide at the first level and by class at the second; the second and third layers from the skills observed on common instances, the third level, or from a model of how their performances co-occur; and the fourth layer, how far a record of finite size takes an organization toward the joint law, qualifies every level. The fourth level, a third arm that executes the workflow itself, supplies no layer: it measures the take of a workflow, not a property of the skills.

\paragraph{The place of shape.} The shape of the distributions enters only at the third layer. The first two layers see a distribution through its means, variances and covariances alone, so that two pairs of skills with the same moments within each class have the same $S$, $\Rclass$ and B1, the same attainment for rungs~0, 1 and~4a, and the same bound on $\Rinst$, whatever their shapes. The moments do not fix the residual; they confine it. For a pair with gap $\delta_k$ between the class means and standard deviation $s_k$ of the difference, they place $R_k$ in the interval $[0,\ \frac{1}{2}(\sqrt{\delta_k^2 + s_k^2} - \delta_k)]$: the upper end is attained when the difference takes two values, and the lower end is approached when the underdog wins rarely and by small margins while the leader's advantage, where it has one, is large; normal and Beta differences fall in between. Within that interval the shape, together with the coupling, decides. With $\delta_k = 0.06$ and $s_k = 0.17$, for instance, the moments allow any residual up to $0.060$, and a normal or Beta pair yields $0.042$. The coupling enters the same way: its covariance is a second-layer quantity, but the rest of the dependence between the performances belongs to the third.

\paragraph{When the distributions of $m$ and $\Mx$ can be drawn.} The two anchors are not equally demanding. The expectation of the mean anchor, $\E[m] = \mean_s \E[\theta_s]$, is a first-layer quantity, known from the means alone, and its variance, $\frac{1}{4}(\sigma_i^2 + \sigma_j^2 + 2\,\mathrm{cov})$ within a class, a second-layer one. The routed anchor has no first- or second-layer value: the means leave $\E[\Mx]$ unknown, and pairwise moments only bound it from above, through Proposition~\ref{prop:twomoment}. The distributions of $m$ and $\Mx$, as distinct from these summaries, can be drawn only from the third layer on, because both are computed from the two performances realized on the same instance, and how they spread depends on how those performances co-occur. They can be drawn, therefore, in three cases: when the joint law is known; when the marginal distributions are known together with an assumed coupling, as with independence, under which $m$ follows by convolution and $F_{\Mx}(x) = \prod_s F_s(x)$, or with a copula of known parameter; and, in practice, when the skills have been observed together on common instances, whose values of $m$ and $\Mx$ give histograms that estimate the two distributions with sampling error. From separate assessments of each skill they cannot be drawn at all, and a picture of them drawn from such evidence displays the assumed coupling rather than the skills. Figure~\ref{fig:app-layers} shows the anchors layer by layer in its last column.

\subsection{Regularities}\label{app:regularities}

The regularities that hold across the space are mostly inequalities, monotonicities and thresholds rather than values. Those below hold for any distributions unless a family is named.

\begin{proposition}[A two-moment bound on the residual]\label{prop:twomoment}
Within class $k$, with incumbent $\bar{s}(k)$, let $\delta_{s,k} = b_k - \mu_{s,k} \geq 0$ be the gap between the class means of $\bar{s}(k)$ and of skill $s$, and $s_{s,k}^2$ the variance of $\theta_s - \theta_{\bar{s}(k)}$ in the class. Then
\begin{equation}\label{eq:twomoment}
R_k \;\leq\; \sum_{s \neq \bar{s}(k)} \tfrac{1}{2}\Big(\sqrt{\delta_{s,k}^2 + s_{s,k}^2} - \delta_{s,k}\Big), \qquad \Rinst \;=\; \textstyle\sum_k p_k R_k .
\end{equation}
For a pair the sum has a single term and the bound is tight.
\end{proposition}

\begin{proof}
$R_k = \E[(\max_{s \neq \bar{s}(k)}\theta_s - \theta_{\bar{s}(k)})^+ \mid k] \leq \sum_{s \neq \bar{s}(k)} \E[(\theta_s - \theta_{\bar{s}(k)})^+ \mid k]$, since the positive part of a maximum is at most the sum of the positive parts. For any random variable $X$ with mean $\mu$ and variance $\sigma^2$, $\E[X^+] \leq \frac{1}{2}(\mu + \sqrt{\mu^2 + \sigma^2})$, attained by a particular two-point distribution \citep{scarf1958min}; here $\mu = -\delta_{s,k}$ and $\sigma^2 = s_{s,k}^2$. For a pair the first inequality is an equality.
\end{proof}

\begin{proposition}[The value of what the criterion reads]\label{prop:information}
Let a criterion engage, on each instance, the skill with the largest expected performance given what it reads. Its attainment lies between $b$ and $\E[\Mx]$, and does not fall when it reads more: a criterion reading the class signal and the instance signal attains at least what either alone does. It attains $b$ when what it reads is independent of the performances, and $\E[\Mx]$ when it reveals the leader of every instance.
\end{proposition}

\begin{proof}
With $\mathcal{F}$ what the criterion reads, the attainment is $\E[\max_s \E[\theta_s \mid \mathcal{F}]]$. It is at least $\max_s \E[\theta_s] = b$, by Jensen's inequality applied to the convex function $\max$, and at most $\E[\max_s \theta_s] = \E[\Mx]$, by the same inequality conditionally on $\mathcal{F}$. For $\mathcal{G} \subseteq \mathcal{F}$, $\E[\max_s \E[\theta_s \mid \mathcal{G}]] = \E[\max_s \E[\E[\theta_s \mid \mathcal{F}] \mid \mathcal{G}]] \leq \E[\max_s \E[\theta_s \mid \mathcal{F}]]$, by the same inequality. The two limits follow from $\E[\theta_s \mid \mathcal{F}] = \E[\theta_s]$ and, when $\mathcal{F}$ reveals the winner $\dot{s}$ of every instance, from $\max_s \E[\theta_s \mid \mathcal{F}] = \E[\theta_{\dot{s}} \mid \mathcal{F}] = \E[\Mx \mid \mathcal{F}]$, since $\theta_{\dot{s}} \geq \theta_s$ on every instance.
\end{proof}

The proposition is the result that a decision maker never loses from a more informative experiment \citep{blackwell1953equivalent}, applied to the engagement criterion; it concerns a criterion that uses what it reads as well as it can. Assignment to the incumbent of the predicted class, as rung~4a is specified, does not: it trusts the prediction however unreliable it is, and can fall below $b$ (Proposition~\ref{prop:classpred}). A criterion that weighs the prediction by its reliability cannot.

\begin{proposition}[Ensembling and assignment]\label{prop:ensembling}
Let the truth be an endpoint. Under the squared rule, a policy whose weights depend only on what its criterion reads, $\mathcal{F}$, attains \sqc{$\E[\sum_s w_s \E[\theta_s \mid \mathcal{F}]] + \E[\sum_s w_s (\hat{r}_s - \hat{r}_w)^2]$}, and exceeds assignment to the skill with the largest expected performance given the same information if and only if the expected disagreement it earns exceeds the expected shortfall of its weighted average below the best conditional mean, $\E[\max_s \E[\theta_s \mid \mathcal{F}] - \sum_s w_s \E[\theta_s \mid \mathcal{F}]]$. Under the linear rule the disagreement term vanishes, and the policy attains at most what the assignment attains, with equality only if the weights place all their mass on such skills.
\end{proposition}

\begin{proof}
By Proposition~\ref{prop:hull}(c), \sqc{$\theta_w = \sum_s w_s \theta_s + \sum_s w_s (\hat{r}_s - \hat{r}_w)^2$}. With weights determined by $\mathcal{F}$, $\E[\sum_s w_s \theta_s \mid \mathcal{F}] = \sum_s w_s \E[\theta_s \mid \mathcal{F}]$; taking expectations gives the attainment, and the assignment attains $\E[\max_s \E[\theta_s \mid \mathcal{F}]]$ by Proposition~\ref{prop:information}, so the comparison is the difference of the two. Under the linear rule the second term is zero (Proposition~\ref{prop:hull}(a)), and $\sum_s w_s \E[\theta_s \mid \mathcal{F}] \leq \max_s \E[\theta_s \mid \mathcal{F}]$ since the weights are non-negative and sum to one.
\end{proof}

Hedging between responses therefore pays exactly where it earns more disagreement than it gives up in mean (Equation~\eqref{eq:ambiguity}). Selection can exceed assignment on the same signals only because it also reads the responses.

\begin{proposition}[Assignment on a predicted class]\label{prop:classpred}
With two skills and two classes, and a class prediction that names the true class with probability $a$ in each, assignment to the incumbent of the predicted class attains
\[
a\, B_1 + (1 - a)\, B_1', \qquad B_1 = \textstyle\sum_k p_k\, \mu_{\bar{s}(k),k}, \quad B_1' = \textstyle\sum_k p_k\, \mu_{\bar{s}(k'),k},
\]
with $k'$ the other class. If the incumbents differ across the classes, $B_1 + B_1' = 2\E[m]$, so that the attainment equals $\E[m]$ at $a = \frac{1}{2}$, reaches the ceiling B1 at $a = 1$, and exceeds $b$ if and only if $a > a^{*} = (b - B_1')/(B_1 - B_1')$. If the incumbents coincide, it equals $b$ for every $a$.
\end{proposition}

\begin{proof}
On an instance of class $k$, the policy engages $\bar{s}(k)$ with probability $a$ and $\bar{s}(k')$ otherwise; averaging over classes gives the expression. If the incumbents differ, $\bar{s}(k')$ is the other skill of class $k$, so that $\mu_{\bar{s}(k),k} + \mu_{\bar{s}(k'),k} = \mu_{i,k} + \mu_{j,k}$ and $B_1 + B_1' = 2\sum_k p_k \E[m \mid k] = 2\E[m]$. The rest is linearity in $a$. If they coincide, both terms equal $\sum_k p_k \mu_{L,k} = b$.
\end{proof}

\begin{proposition}[A normal reference]\label{prop:normalref}
For a pair whose performances in class $k$ are jointly normal, with gap $\delta_k$ between the class means and standard deviation $s_k$ of their difference,
\[
R_k \;=\; s_k\,\big[\varphi(z_k) - z_k\,\Phi(-z_k)\big], \qquad z_k = \delta_k / s_k ,
\]
which is decreasing in $z_k$ and at most $s_k/\sqrt{2\pi}$, reached when the class means coincide.
\end{proposition}

\begin{proof}
$\theta_s - \theta_{\bar{s}(k)}$, the other skill's performance less the class incumbent's, is normal with mean $-\delta_k$ and standard deviation $s_k$, and for $X \sim \mathcal{N}(\mu, \sigma^2)$, $\E[X^+] = \mu\,\Phi(\mu/\sigma) + \sigma\,\varphi(\mu/\sigma)$. The derivative of $\varphi(z) - z\Phi(-z)$ is $-\Phi(-z) < 0$, and its value at $z = 0$ is $\varphi(0) = 1/\sqrt{2\pi}$.
\end{proof}

The expression is the expected maximum of two jointly normal variables \citep{clark1961greatest}, net of the class incumbent's mean. The normal family ignores that performance lies in $[0,1]$. It is a reference against which bounded families are read: where they depart from it, the departure is the effect of shape and of the bounds of the scale.

\begin{proposition}[Synergy with binary performance]\label{prop:binarylift}
For a pair whose performances are $0$ or $1$, pooling the responses attains \sqc{$\E[m] + \frac{1}{4}\Pr(\theta_i \neq \theta_j)$} under the squared rule, and a protocol represented in reduced form by a lift \sqc{$\gamma \in [1, \frac{4}{3}]$} on the pooled response attains \sqc{$\E[m] + \frac{3\gamma - 2}{4}\Pr(\theta_i \neq \theta_j)$}.
\end{proposition}

\begin{proof}
Binary performance means each response is the truth or its opposite. When both skills are right the mean response is the truth and scores $1$; when both are wrong it scores $0$; when they disagree it is $\frac{1}{2}$ and scores \sqc{$1 - \frac{1}{4} = \frac{3}{4}$}, against a mean performance of $\frac{1}{2}$. For \sqc{$\gamma \in [1, \frac{4}{3}]$}, \sqc{$\min(1, \gamma\,\theta_{\bar{w}})$} leaves the first two cases unchanged and gives \sqc{$\frac{3\gamma}{4}$} in the third; subtracting $\E[m]$, which counts $\frac{1}{2}$ there, gives the expressions.
\end{proof}

With binary performance the lift adds only on the instances on which the skills disagree, and nothing on those on which both are already right, where the ceiling at one absorbs it. The reduced form of Proposition~\ref{prop:archetypes} is therefore a weak representation of synergy for tasks decided by hard responses.

\begin{proposition}[Capping performance at one]\label{prop:ceiling}
Let $X \in [0,1]$ be a performance the lift applies to, $\theta_{\bar{w}}$ or $\Mx$, and $\gamma \geq 1$. Then
\[
\E\big[\min(1, \gamma X)\big] \;\leq\; \min\big(1, \gamma\E[X]\big),
\]
with equality if and only if $\gamma X \leq 1$ almost surely or $\gamma X \geq 1$ almost surely. Against the unclipped mean the ceiling removes exactly $\E[(\gamma X - 1)^{+}]$, which is non-decreasing in $\gamma$ and zero if and only if $\Pr(X > 1/\gamma) = 0$.
\end{proposition}

\begin{proof}
$f(x) = \min(1, \gamma x)$ is concave, being the minimum of two affine functions, and Jensen's inequality gives the bound. Equality requires $f$ to agree almost surely with a supporting line at $\E[X]$; since $f$ is affine on each of $[0, 1/\gamma]$ and $[1/\gamma, 1]$ and no line meets it on both sides of the kink, equality holds exactly when the support of $X$ does not straddle $1/\gamma$. Pointwise, $\gamma x - \min(1, \gamma x) = (\gamma x - 1)^{+}$, non-decreasing in $\gamma$ for $x \geq 0$.
\end{proof}

Expected performance under a ceiling must be computed by clipping each instance and averaging afterwards; clipping the anchor overstates it. For the illustrative Profile A of Appendix~\ref{app:profiles} at high difficulty and $\gamma = 2$, clipping each instance gives \sqc{$0.720$} for the pooled response against \sqc{$0.751$} from clipping its mean, and $0.835$ for the routed anchor against $0.956$, with \sqc{$19.3\%$} and $44.1\%$ of instances clipped. Because the loss grows with the upper-tail mass, it is larger for $\Mx$ than for the pooled response, and the ceiling compresses $HM{+}{+}$ toward $HM{+}$ as $\gamma$ rises. A measured lift has no such artefact: responses cannot score above one, so whatever the protocol attained already lies below the ceiling.

\subsection{Crediting a take short of the climb}\label{app:crediting}
 
\begin{figure}[!htb]
\centering
\includegraphics[width=\linewidth]{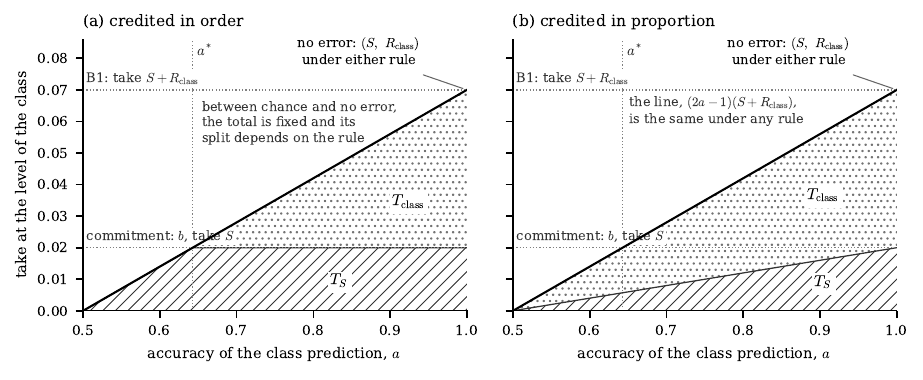}
\caption{Where the split of a take is a convention. Assignment on a predicted class, with two classes whose incumbents differ and illustrative values \sqc{$S = 0.020$} and \sqc{$\Rclass = 0.050$}, as the accuracy $a$ of the class prediction runs from chance to one. The black line is its take at the level of the class, $(2a - 1)(S + \Rclass)$, fixed by Proposition~\ref{prop:classpred} under any crediting rule; the patterns split it into $T_S$ (hatched) and $T_{\mathrm{class}}$ (dotted), (a) in order and (b) in proportion. The two agree at $a = 1$, where the policy resolves both classes and takes $(S, \Rclass)$, and at $a = \frac{1}{2}$, where it takes nothing; between them every point is a blend of right and wrong assignments, and the split is a convention. The dotted lines mark the commitment, which takes $S$, and B1, which takes $S + \Rclass$; $a^{*}$ is the accuracy at which the policy reaches $b$.}
\label{fig:attribution_sweep}
\end{figure}

Crediting in order, the convention of Corollary~\ref{cor:attribution}, has a linear alternative, crediting in proportion, which splits the take of each class whose incumbent is $L$ in a fixed ratio, a share $S / \sum_{k:\, \bar{s}(k) = L} p_k \big(b_k - \E[m \mid k]\big)$ to $S$ and the rest to $\Rclass$, and gives the whole take of every other class to $\Rclass$; for a pair it agrees with crediting in order wherever every class is resolved, and charges errors to both components alike. Assignment on a predicted class shows the difference (Figure~\ref{fig:attribution_sweep}). With two classes whose incumbents differ and accuracy $a$, its take at the level of the class is $C = (2a - 1)(S + \Rclass)$ under any rule (Proposition~\ref{prop:classpred}). At $a = 1$ it resolves both classes and takes $(S, \Rclass)$; at $a = \frac{1}{2}$ it takes nothing. In between, crediting in order gives it $\min(C, S)$ and the rest, so that $S$ fills first and is complete from the accuracy $a^{*}$ at which the policy reaches $b$; crediting in proportion gives it $(2a - 1)S$ and $(2a - 1)\Rclass$, a share of each component equal to what its accuracy delivers.
 
\section{The linear rule as a special case}\label{app:linear}

The paper scores performance by the squared rule. Its special case $\Lambda = 0$ is the linear rule, $\theta = 1 - |\hat{r} - r|$, which takes performance to be the distance from the truth itself, read in the units of the response. This appendix explains the choice, states what the results become under the linear rule, and delimits where each rule fits.

\paragraph{Why the squared rule: an example.} The two rules agree on every instance, in that both score a response closer to the truth higher, but they differ when performance is averaged over instances a skill cannot tell apart. There the linear rule can favor responses rounded toward certainty over responses that keep the strength of the evidence. The point concerns how an organization compares skills, not how the skills behave. Consider two classifiers that read the same films: half of the films are positive with probability $0.55$ and half with probability $0.95$. Classifier A is built to respond with these probabilities; classifier B is built to respond with the class it finds more likely, as hard classifiers and most human readers do, and so responds $1$ to both groups, its response no longer telling them apart.
\begin{center}
\footnotesize
\begin{tabular}{l cc}
\toprule
 & A: $\hat{r} = 0.55$ or $0.95$ & B: $\hat{r} = 1$ on every film \\
\midrule
Does the response tell the two groups apart? & yes & no \\
Mean performance under the linear rule & $0.705$ & $\mathbf{0.750}$ \\
Mean performance under the squared rule & $\mathbf{0.853}$ & $0.750$ \\
\bottomrule
\end{tabular}
\end{center}
Under the linear rule, for instance, A performs $0.55$ on a positive film of the first group and $0.45$ on a negative one, so $0.55 \times 0.55 + 0.45 \times 0.45 = 0.505$ on that group and $0.905$ on the other, $0.705$ on average, while B performs $1$ on every positive film and $0$ on every negative one, $0.750$ on average. The two classifiers act alike: A's responses, compared with one half, select the same films as B's. The linear rule nevertheless ranks B above A, penalizing in A not a worse decision but the expression of its uncertainty, although that expression is what B's response has lost; the squared rule, which penalizes an error by its square and so a confident mistake far more than a hesitant one, ranks A above B.

\paragraph{What the linear rule simplifies.} Under the linear rule the score of a combination of responses equals the combination of their scores (Proposition~\ref{prop:hull}(a)), so that $\Lambda = 0$ and every statement made on the performance anchors is also a statement about arrangements that combine responses. Table~\ref{tab:linear} lists the results that take a simpler form. The anchors, the informed baseline, the decomposition $G = S + R$ with $R = \Rclass + \Rinst$, the bounds on the stake, the ceiling on routing, the value of what a criterion reads, the effect of coupling and the identification of $R$ are the same under both rules.

\begin{table}[!htb]
\centering
\footnotesize
\caption{Results that depend on the scoring rule. The squared entries are those stated in the text; the linear entries set $\Lambda = 0$}
\label{tab:linear}
\renewcommand{\arraystretch}{1.25}
\begin{tabularx}{\textwidth}{@{}>{\raggedright\arraybackslash}p{3.6cm} >{\raggedright\arraybackslash}X >{\raggedright\arraybackslash}X@{}}
\toprule
Result & Squared rule & Linear rule \\
\midrule
Pooling, rung~2 & \sqc{$\E[m] + \Lambda = b - S + \Lambda$} & $\E[m] = b - S$ \\
Fixed combinations (Corollary~\ref{cor:fixed}) & can exceed $b$; never above $\E[\Mx]$ & never above $b$ \\
Pooling against randomizing (Proposition~\ref{prop:randomize}) & pooling higher by \sqc{$\Lambda$} & equal in expectation \\
The insurance contract (Proposition~\ref{prop:insurance}) & premium \sqc{$S - \Lambda - D$}; pooling preferable iff \sqc{$D > S - \Lambda$}; for a pair iff \sqc{$p_a < \frac{1}{2} + \frac{\Lambda}{2S}$} & premium $S - D$; iff $D > S$; for a pair iff $p_a < \frac{1}{2}$ \\
Break-even of synergy (Proposition~\ref{prop:breakeven}) & \sqc{$\gamma^{*} = \frac{b}{b - S + \Lambda}$}, \sqc{$\delta^{*} = S - \Lambda$} & $\gamma^{*} = \frac{b}{b - S}$, $\delta^{*} = S$ \\
Archetype $HM$ (Proposition~\ref{prop:archetypes}) & \sqc{$\theta_{HM} = \theta_{\bar{w}}$}; attains \sqc{$b - S + \Lambda$} & $\theta_{HM} = m$; attains $b - S$ \\
Ensembling against assignment (Proposition~\ref{prop:ensembling}) & above it if the disagreement earned exceeds the mean given up & never above assignment on the same information \\
Binary performance (Proposition~\ref{prop:binarylift}) & a disagreement pools to \sqc{$\frac{3}{4}$}; pooling attains \sqc{$\E[m] + \frac{1}{4}\Pr(\theta_i \neq \theta_j)$} & a disagreement pools to $\frac{1}{2}$; pooling attains $\E[m]$ \\
What identifies $\gamma^{*}$ (Section~\ref{sec:identification}) & paired observation, since \sqc{$\Lambda$} depends on how the responses co-occur & separate assessments \\
\bottomrule
\end{tabularx}
\end{table}

\paragraph{What it means for the decomposition.} Under the linear rule pooling sits at the mean anchor, which it then attains as random assignment does, and the take of an arrangement has only two sources, routing and synergy: combining responses earns no curvature dividend. The insurance contract then costs the whole of $S$, no fixed combination beats the commitment, and the break-even of a proposed protocol follows from the skills' separate track records, so that a protocol can be priced before the skills have ever been observed together.

\paragraph{Where each rule fits.} The two rules are not rival measures of one thing, and the territory of each can be delimited (Table~\ref{tab:territory}). They coincide on every instance when responses are hard: with a binary truth and a response of $0$ or $1$, $|\hat{r} - r|$ and $(\hat{r} - r)^2$ are both $0$ or $1$, so hard classifiers, readers who answer yes or no, and any response acted on as a decision are evaluated identically, and the simpler reading loses nothing. The linear rule is also the right valuation where a graded response is itself the action and each unit of deviation costs the same, and it has an operational advantage wherever the skills have only been assessed separately: what pooling attains, the insurance premium and the break-even of synergy then follow from the marginals. The squared rule should be preferred where a graded response is an input carrying evidence: where it is combined with other responses, read by a criterion that routes on confidence, compared with a threshold other than one half or priced with asymmetric costs, or used to compare, select or train probabilistic skills. In short, the linear rule fits a response that is final, and the squared rule a response that informs.

\begin{table}[!htb]
\centering
\footnotesize
\caption{Where each scoring rule fits}
\label{tab:territory}
\renewcommand{\arraystretch}{1.25}
\begin{tabularx}{\textwidth}{@{}>{\raggedright\arraybackslash}p{4.6cm} >{\raggedright\arraybackslash}X >{\raggedright\arraybackslash}p{3.2cm}@{}}
\toprule
Setting & Why & Rule \\
\midrule
Hard responses: classes, yes or no, decisions acted on as they stand & the two rules coincide on every instance & either; the linear reads more simply \\
Graded responses that are actions, with errors costing in proportion to their size & the linear rule is the economic valuation & linear \\
Skills assessed only separately & pooling, the insurance premium and the break-even of synergy follow from the marginals & linear, for pricing before paired data exist \\
Graded responses combined with others, mechanically or through a protocol & a mechanical combination is scored above the mean of its parts by their disagreement; a protocol is measured against that combination, and can use only the evidence the responses still carry & squared \\
Graded responses read by a criterion, or compared with a threshold other than one half & the decision depends on the probability, not on the class & squared \\
Comparing, selecting or training probabilistic skills & the linear rule can favor responses that have discarded the strength of their evidence & squared \\
\bottomrule
\end{tabularx}
\end{table}

\paragraph{Magnitudes.} For the illustrative profiles of Table~\ref{tab:gsr}, the linear rule's break-even, $b/(b - S)$, lies about two to four hundredths above the $\gamma^{*}$ the table reports, and pooling sits $0.01$ to $0.02$ lower, at $\E[m]$; the orderings of the ladder are the same under both rules. Scored by the linear rule, the responses of the worked example of Table~\ref{tab:worked} give the skills means of $0.700$ and $0.650$, so that pooling attains $\E[m] = 0.675$, below the commitment's $0.700$, and the protocol, whose joint responses then score $0.775$ on average, has a lift of $0.775/0.675 = 1.148$ against a break-even of $0.700/0.675 = 1.037$. In the CXR experiment the two rules coexist: a randomly selected subset of participants was paid a bonus by a binarized version of the squared rule, while performance is reported as the deviation from the diagnostic standard, which is the linear rule \citep{agarwal2023combining}.

\section{Proofs of the propositions on the stake}\label{app:proofs}

\begin{proof}[Proof of Proposition~\ref{prop:anchor}]
Let $n = |\mathcal{S}|$. On every instance $m = \frac{1}{n}\sum_{s \in \mathcal{S}} \theta_s$, and each $\theta_s$ takes values in $[0,1]$, so every expectation below exists. By linearity of the expectation,
\[
\E[m] \;=\; \E\Big[\frac{1}{n} \sum_{s \in \mathcal{S}} \theta_s\Big] \;=\; \frac{1}{n} \sum_{s \in \mathcal{S}} \E[\theta_s] \;=\; \mean_{s} \E[\theta_s].
\]
Linearity holds for any random variables with finite expectations, whatever their dependence, and each $\E[\theta_s]$ is determined by the distribution of $\theta_s$ alone. Hence any two joint distributions of the performances that share the same marginals give the same $\E[m]$, and $\E[m]$ can be computed from separate assessments of each skill, provided each is made on the same population of instances; where instances fall into classes, $\E[\theta_s] = \sum_k p_k \mu_{s,k}$. The same holds of $b = \max_s \E[\theta_s]$, a function of the expectations alone, and therefore of $S = b - \E[m]$.

The second statement does not follow in the same way, because the maximum is not linear: $\E[\Mx]$ does not decompose into expectations of the individual performances, and depends on how they co-occur on the same instance. With $\theta_H$ and $\theta_M$ each equal to $0$ or $1$ with probability one half, $\E[\Mx] = 0.5$ when the two always agree and $\E[\Mx] = 1$ when they never do, while both marginal distributions are identical in the two cases. Since $\E[m]$ and $b$ are fixed by the marginals, $G = \E[\Mx] - \E[m]$ and $R = \E[\Mx] - b$ move with $\E[\Mx]$, and are not determined by the marginals either. Given the marginal distributions, the smallest value $\E[\Mx]$ can take is attained under the comonotone coupling, since $\Pr(\Mx \leq x) \leq \min_s F_s(x)$ for every $x$, with equality when the performances are comonotone; for a pair, the largest is attained under the countermonotone coupling, the two being the Fr\'echet--Hoeffding bounds \citep{nelsen2006introduction}, while for any set $\E[\Mx] \leq \min_v \{v + \sum_s \E[(\theta_s - v)^{+}]\}$, since $\max_s \theta_s \leq v + \sum_s (\theta_s - v)^{+}$ for every $v$, the bound of \citet{meilijson1979convex}, which the countermonotone coupling attains for a pair.
\end{proof}

\begin{proof}[Proof of Proposition~\ref{prop:S}]
By Proposition~\ref{prop:anchor}, $\E[m] = \frac{1}{n}\sum_s \E[\theta_s]$, so $S = b - \E[m] = \frac{1}{n}\sum_s (b - \E[\theta_s]) = \frac{1}{n}\sum_s d_s$, the identity of Proposition~\ref{prop:gap} applied to the expectations, $b$ being their maximum; each $d_s \geq 0$, hence $S \geq 0$. Since $d_L = 0$, the $n - 1$ shortfalls of the other skills sum to $nS$, so their maximum is at least their mean, $\frac{n}{n-1} S$, and at most their sum, $nS$. For a pair the single non-zero shortfall is $|\E[\theta_i] - \E[\theta_j]| = 2S$, which gives both pair statements.
\end{proof}

\begin{proof}[Proof of Proposition~\ref{prop:gap}]
$\Mx - m = \Mx - \frac{1}{n}\sum_s \theta_s = \frac{1}{n}\sum_s (\Mx - \theta_s)$; for a pair, the term of the skill that leads on the instance is zero and the other is $|\theta_i - \theta_j|$. The expectations are finite because performances lie in $[0,1]$, and linearity gives the expression for $G$. For each $s$, $\E[\Mx - \theta_s] = (\E[\Mx] - b) + (b - \E[\theta_s]) = R + d_s$, and averaging over $s$ returns $G = R + S$ by Proposition~\ref{prop:S}.
\end{proof}

\begin{proof}[Proof of Proposition~\ref{prop:R}]
The maximum is convex, being a pointwise maximum of linear functions, so Jensen's inequality gives $\E[\max_s \theta_s] \geq \max_s \E[\theta_s] = b$. The decomposition follows by adding $S$ and $R$, the benchmark $b$ cancelling, and applying Proposition~\ref{prop:gap}. For the third statement, let $A = \{\theta_U > \theta_L\}$ be the event in which the underdog wins on the instance, with complement $A^c$, ties included. On $A$, $\Mx - \theta_L = \theta_U - \theta_L > 0$; on $A^c$, $\Mx - \theta_L = 0$; so $\Mx - \theta_L = (\theta_U - \theta_L)^{+}$ pathwise, and $R = \E[\Mx] - \E[\theta_L] = \E[(\theta_U - \theta_L)^{+}]$. The law of total expectation, with the positive part vanishing on $A^c$ and equal to the difference itself on $A$, gives the factorization.
\end{proof}

\begin{proof}[Proof of Proposition~\ref{prop:R0}]
Since $\Mx \geq \theta_L$ on every instance, $\E[\Mx] = \E[\theta_L]$ if and only if $\Mx = \theta_L$ almost surely, that is $\theta_L \geq \theta_U$ almost surely. The argument is symmetric in the labels and uses no property of the joint distribution.
\end{proof}

\begin{proof}[Proof of Proposition~\ref{prop:bound}]
$R \geq 0$ by Proposition~\ref{prop:R}, and $R \leq G$ because $S \geq 0$. On every instance the performances are non-negative, so $m = \frac{1}{n}\sum_s \theta_s \geq \Mx / n$, and therefore $\Delta = \Mx - m \leq \Mx (1 - 1/n) \leq 1 - 1/n$, with equality if and only if $\Mx = 1$ and every other performance is $0$; taking expectations gives $G \leq 1 - 1/n$. In the configuration described, every skill has $\E[\theta_s] = 1/n$, so $b = \E[m] = 1/n$ and $S = 0$, while $\Mx = 1$ on every instance, so $G = R = 1 - 1/n$.
\end{proof}

\begin{proof}[Proof of Proposition~\ref{prop:disp}]
Immediate from the definitions of the mean and the maximum.
\end{proof}

\section{Many skills and many classes}\label{app:many}

The pair is the unit of analysis wherever a larger set would obscure the argument, and the restriction can be relaxed at any point at the cost of heavier notation. The definitions of Sections~\ref{sec:primitives} and~\ref{sec:bounds}, the anchors, the decomposition $G = S + R$, Propositions~\ref{prop:anchor}, \ref{prop:gap}, \ref{prop:S}, \ref{prop:bound}, \ref{prop:insurance}, \ref{prop:R} (with $\theta_U$ replaced by the maximum over the skills other than the leader), \ref{prop:R0}, \ref{prop:disp}, \ref{prop:hull}, \ref{prop:class}, \ref{prop:breakeven} and~\ref{prop:archetypes}, and every policy of Table~\ref{tab:spectrum}, are stated over an available set and carry over to $|\mathcal{S}| > 2$ unchanged. What a larger set adds to the notation is that the engaged subset must itself be named, since a policy may engage some of the available skills and not others. Three properties change: the closed form for $S$ generalizes, while the symmetry of the routing range and the comparability of quantities across sets do not carry over (Section~\ref{app:change}). What a roster of several skills over several classes adds is then set out: what a skill adds to a roster, how the components move with the class mix, which errors of a class prediction cost anything and why the checks of Section~\ref{sec:procedure} never give back what an earlier step held, what separate assessments by class and a budget of paired instances establish, and when a pool should admit a skill. Throughout, $\mu_{s,k}$, $b_k$, $\bar{s}(k)$ and $L$ are as in Section~\ref{sec:classes}, and $B_1 = \sum_k p_k b_k = b + \Rclass$ is the ceiling B1.

\subsection{What a larger set changes}\label{app:change}

\paragraph{Symmetry of the routing range about the anchor.} For a pair, $\E[\min_s \theta_s] + \E[\max_s \theta_s] = 2\E[m]$, because the mean of two values is the midpoint of their minimum and maximum; a policy directing work the wrong way therefore costs what one directing it the right way earns (Proposition~\ref{prop:routing}). For larger sets the interior values enter the anchor and not the extremes, the anchor drifts away from the midpoint, and the identity fails. The per-instance identity of Proposition~\ref{prop:gap}, which measures what averaging forgoes against selection, holds for any set as a mean shortfall from the best, but its pair form, half the distance between the two scores, has no counterpart beyond two responses.

\paragraph{The closed form for $S$.} The closed form generalizes, but its pair shortcut does not. For any set, $S = \frac{1}{n}\sum_s d_s$ is the mean shortfall of the skills' expectations from the leader's (Proposition~\ref{prop:S}), recoverable from the marginals by Proposition~\ref{prop:anchor}. For a pair it reduces to the half-gap $\tfrac{1}{2}|\E[\theta_i] - \E[\theta_j]|$, a function of the two extremes alone; for larger sets it does not, because every skill's shortfall enters the mean and not only the weakest one's. With three skills whose marginals are $0.620$, $0.551$ and $0.810$, the shortfalls are $0.190$, $0.259$ and $0$, so $S = 0.150$ against a half-gap of $0.130$. The break-even \sqc{$\gamma^{*} = b/\E[\theta_{\bar{w}}]$} is correspondingly a function of the whole roster, and of how its responses co-occur.

\paragraph{Comparability of the stake across sets.} Enlarging $\mathcal{S}$ raises $\E[\Mx]$ and weakly lowers $\E[m]$ unless the skills added are no worse than the average, so $G$ tends to increase mechanically with the size of the set whether or not the added skills are any good. Its upper bound grows with the set as well, $G \leq 1 - 1/|\mathcal{S}|$, one half for a pair (Proposition~\ref{prop:bound}). A decomposition computed over a roster is not commensurable with one computed over a pair, and $G$ reads as the value at stake between the members of $\mathcal{S}$ only when $\mathcal{S}$ is held fixed. Comparison across sets requires a quantity normalized for set size, which this paper does not develop.

\subsection{The roster}\label{app:roster}

For a roster $A \subseteq \mathcal{S}$, write $\Mx_A = \max_{s \in A} \theta_s$, $b(A) = \max_{s \in A} \E[\theta_s]$ and $b_k(A) = \max_{s \in A} \mu_{s,k}$, with the value of each set to zero for the empty roster.

\begin{proposition}[What a skill adds to a roster]\label{prop:roster}
Adding skill $j$ to roster $A$ raises the informed baseline by $(\E[\theta_j] - b(A))^{+}$, the ceiling B1 by $\sum_k p_k (\mu_{j,k} - b_k(A))^{+}$, and the routed anchor by $\E[(\theta_j - \Mx_A)^{+}]$. As functions of the roster, $\E[\Mx_A]$, $b(A)$ and $\sum_k p_k b_k(A)$ are non-decreasing and submodular. Consequently, the roster of $\ell$ skills built by adding at each step the skill with the largest increment attains at least $1 - 1/e$ of the largest $\E[\Mx_A]$, or of the largest ceiling B1, over all rosters of $\ell$ skills.
\end{proposition}

\begin{proof}
The increments follow from $\max(x, y) = y + (x - y)^{+}$, applied on each instance, to the expectations, and to the class means. Each increment is non-negative, and non-increasing in $A$, since $\Mx_A$, $b(A)$ and $b_k(A)$ do not decrease as $A$ grows and $(x - y)^{+}$ does not increase in $y$; this is submodularity. The guarantee is that of greedy maximization of a non-decreasing submodular function under a cardinality constraint \citep{nemhauser1978analysis}.
\end{proof}

The ceiling B1, as a function of the roster, is a facility-location value: each class is served by the best skill held. A skill that is the incumbent of no class adds nothing to B1, and may still add to $b$ when more than one skill is held (no-incumbency bounds $\E[\theta_j]$ by $\mathrm{B1}(A)$, not by $b(A)$) and to $\E[\Mx]$; the next proposition says when it must.

\begin{proposition}[A skill that leads nowhere]\label{prop:newcomer}
Let a single skill $i$ be held, and let $F_{s,k}$ be the distribution of $\theta_s$ in class $k$. Adding skill $j$ leaves $\E[\Mx]$ unchanged under some coupling of the two performances if and only if, in every class, $\theta_i$ first-order dominates $\theta_j$, $F_{i,k} \leq F_{j,k}$. Over all couplings, the least that $j$ adds is $\sum_k p_k \E\big[(F_{j,k}^{-1}(V) - F_{i,k}^{-1}(V))^{+}\big]$, with $V$ uniform on $[0,1]$, attained when the performances are comonotone.
\end{proposition}

\begin{proof}
By Proposition~\ref{prop:roster}, $j$ adds $\E[(\theta_j - \theta_i)^{+}]$, which is zero if and only if $\theta_j \leq \theta_i$ almost surely. If such a coupling exists, $F_{i,k} \leq F_{j,k}$ in every class; conversely, if $F_{i,k} \leq F_{j,k}$, the comonotone coupling $\theta_s = F_{s,k}^{-1}(V)$ gives $\theta_j \leq \theta_i$. Since $\E[\max(\theta_i, \theta_j)]$ is smallest under the comonotone coupling (Proposition~\ref{prop:anchor}) and $\E[\theta_i]$ does not depend on the coupling, the increment is smallest there.
\end{proof}

A skill whose distribution, in some class, is not dominated by the held one's, the two quantile functions crossing, therefore adds to the routed anchor whatever the coupling, although it leads in no class.

\subsection{The class mix}\label{app:mix}

Hold the distributions of the skills within each class fixed and let the mix $p = (p_1, \dots, p_K)$ range over the simplex $\Delta_K$.

\begin{proposition}[The stake across class mixes]\label{prop:mix}
$\E[m]$, $\E[\Mx]$, $B_1$ and $\Rinst$ are linear in $p$; $b$ is convex and piecewise linear, and so is $S$; $\Rclass$ is concave and piecewise linear, and vanishes at every mix concentrated on a single class. Unless $\Rclass$ vanishes on the whole simplex, its maximum is attained at a mix at which at least two skills tie for the largest expected performance; for a pair, $S = 0$ at that mix.
\end{proposition}

\begin{proof}
$\E[m]$, $\E[\Mx]$, $B_1$ and $\Rinst = \sum_k p_k R_k$ are expectations of class-conditional quantities that do not depend on $p$, hence linear. $b(p) = \max_s \sum_k p_k \mu_{s,k}$ is a maximum of linear functions, hence convex and piecewise linear, and so is $S = b - \E[m]$; $\Rclass = B_1 - b$ is then concave and piecewise linear. At the mix concentrated on class $k$, $b = b_k = B_1$, so $\Rclass = 0$. On the region of the simplex where a given skill $s$ has the largest expected performance, $\Rclass = \sum_k p_k (b_k - \mu_{s,k})$ is linear, so its maximum over the region is attained at a vertex of the region. A vertex of the region is the intersection of $K - 1$ active constraints, each either a face of the simplex, $p_k = 0$, or a tie, $\sum_k p_k \mu_{s,k} = \sum_k p_k \mu_{t,k}$; a vertex at which no tie is active is a vertex of the simplex, where $\Rclass = 0$. A positive maximum is therefore attained where a tie is active. For a pair, $S$ is half the gap between the two expected performances (Proposition~\ref{prop:S}), zero at a tie.
\end{proof}

The class residual is therefore largest where a commitment is weakest, at the mixes where the skills are level on average, and a drift of the mix toward such a point raises $\Rclass$ with no skill changing.

\subsection{Predicted classes and the checks}\label{app:predicted}

Let a criterion predict the class $\hat{k}$ of each instance with probabilities $\Pr(\hat{k} = j \mid k)$.

\begin{proposition}[Assignment on a predicted class, many classes]\label{prop:classpredK}
Assignment to the incumbent of the predicted class attains $\sum_k p_k \sum_j \Pr(\hat{k} = j \mid k)\, \mu_{\bar{s}(j),k}$. (i) It depends on the prediction only through the probabilities $\Pr(\bar{s}(\hat{k}) = s \mid k)$: confusing two classes with the same incumbent costs nothing. (ii) If the prediction names the true class with probability $a$ and each other class with probability $(1 - a)/(K - 1)$, it attains $a B_1 + (1 - a) B_1'$, with
\[
B_1' \;=\; \sum_k p_k\, \frac{1}{K - 1} \sum_{j \neq k} \mu_{\bar{s}(j),k},
\]
and, if $B_1' < b < B_1$, exceeds $b$ if and only if $a > a^{*} = (b - B_1')/(B_1 - B_1')$.
\end{proposition}

\begin{proof}
On an instance of class $k$ the policy engages $\bar{s}(j)$ with probability $\Pr(\hat{k} = j \mid k)$; averaging over classes gives the attainment, which sums these probabilities over the classes that share an incumbent, hence (i). Under (ii) the attainment is $a \sum_k p_k \mu_{\bar{s}(k),k} + (1 - a) B_1'$, linear in $a$. For $K = 2$, $B_1'$ is the quantity of Proposition~\ref{prop:classpred}.
\end{proof}

What a class prediction must therefore separate is not the classes but the groups of classes with a common incumbent, and its accuracy over those groups is what decides whether it beats the commitment.

\begin{proposition}[Checks that never give back]\label{prop:checks}
Let $\pi_0$ and $\pi_1$ be two policies, and $\hat{k}$ a partition of the instances observable before execution. The policy that follows, on the instances of each cell of $\hat{k}$, whichever of $\pi_0$ and $\pi_1$ attains more on that cell attains at least what each of them attains. With $\pi_0$ the commitment, it attains at least $b$; with $\pi_0$ a class assignment, at least what that assignment attains.
\end{proposition}

\begin{proof}
The attainment of each policy is the sum over cells of the probability of the cell times its attainment on the cell; the combined policy takes the larger of the two terms in every cell.
\end{proof}

The checks after steps~2 and~4 of Section~\ref{sec:procedure} are this rule, with $\hat{k}$ the classes as the policy reads them and $\pi_0$ the previous step. On a predicted class $j$, the commitment attains $\sum_k \Pr(k \mid \hat{k} = j)\, \mu_{L,k}$ and the assignment $\sum_k \Pr(k \mid \hat{k} = j)\, \mu_{\bar{s}(j),k}$, both known from the class-wise means and the record $\mathcal{R}^{Z}$. The checked assignment is a restricted form of the criterion of Proposition~\ref{prop:information}, which chooses among all skills on each predicted class and attains at least as much. Both statements hold in the population; on a finite record the cell attainments are estimated, and a cell can be kept or returned in error.
 
\begin{proposition}[A margin on the checks]\label{prop:margin}
On a cell of $\hat{k}$, let $g$ be the gain of $\pi_1$ over $\pi_0$, and let its estimate be normal with mean $g$ and standard deviation $\tau$. A check that adopts $\pi_1$ on the cell if and only if the estimate exceeds $z\tau$, with $z \geq 0$, falls short of the better of the two policies on the cell, in expectation, by $|g|\,\Phi(-z - |g|/\tau)$ if $g < 0$, a give-back, and by $g\,\Phi(z - g/\tau)$ if $g > 0$, a gain forgone. The first is at most $\kappa_z\,\tau$ and the second at most $\lambda_z\,\tau$, where $\kappa_z = \max_{x \geq 0} x\,\Phi(-z - x)$ and $\lambda_z = \max_{x \geq 0} x\,\Phi(z - x)$: $\kappa_z = 0.170$, $0.033$ and $0.008$, and $\lambda_z = 0.170$, $0.507$ and $0.837$, at $z = 0$, $1$ and $1.645$.
\end{proposition}
 
\begin{proof}
The check adopts $\pi_1$ with probability $\Phi(g/\tau - z)$. If $g < 0$, adopting it costs $|g|$, with probability $\Phi(-z - |g|/\tau)$; if $g > 0$, keeping $\pi_0$ costs $g$, with probability $\Phi(z - g/\tau)$. Writing $x = |g|/\tau$ gives the two bounds, whose maxima are computed numerically; at $z = 0$ both are the $\kappa$ of Corollary~\ref{cor:insurance}.
\end{proof}
 
A margin of one standard error cuts the largest expected give-back on a cell fivefold, from $0.17\tau$ to $0.03\tau$, and triples the largest gain forgone, from $0.17\tau$ to $0.51\tau$: on a finite record it buys, approximately, the procedure's promise not to give back, at the price of a slower climb, and both fall as the record grows, since $\tau$ does. The normal form is an approximation for a cross-fitted mean over many instances, and the standard error must be estimated with the dependence of the record, by film on the Collab-CXR data, where a margin of one standard error removes what step~2 gave back and one of $1.645$ what step~4 did (Section~\ref{sec:radiologydata}).

\subsection{Evidence by class}\label{app:evidence}

The bounds of Proposition~\ref{prop:anchor} apply within each class, and the moments that limit $\Rinst$ also size the record that estimates it.

\begin{proposition}[What assessments by class establish]\label{prop:classbounds}
Given the distributions $F_{s,k}$ of the skills in each class, $\Rclass$ is identified, and
\[
\sum_k p_k \Big(\E\big[\max_s F_{s,k}^{-1}(V)\big] - b_k\Big) \;\leq\; \Rinst \;\leq\; \sum_k p_k \Big(\min_v \Big\{ v + \textstyle\sum_s \E\big[(\theta_s - v)^{+} \mid k\big] \Big\} - b_k\Big),
\]
the lower bound attained when the performances are comonotone within every class, and the upper, for a pair, when they are countermonotone.
\end{proposition}

\begin{proof}
$\Rclass = \sum_k p_k (b_k - \mu_{L,k})$ depends on the class means alone (Proposition~\ref{prop:class}). $\Rinst = \sum_k p_k R_k$, and Proposition~\ref{prop:anchor} bounds each $R_k$ within its class.
\end{proof}

A class whose upper bound is small against what a criterion reading the instance would cost there can be set aside without any instance worked twice; paired observation is needed only in the others.

\begin{proposition}[Where to pair]\label{prop:allocation}
Take the incumbents as known, and estimate $\Rinst$ by $\sum_k p_k \bar{D}_k$, with $\bar{D}_k$ the mean of $D = \Mx - \theta_{\bar{s}(k)}$ over $n_k$ paired instances of class $k$, each costing $c_k$. Under the budget $\sum_k c_k n_k = C$, the variance $\sum_k p_k^2 \sigma_k^2 / n_k$, with $\sigma_k^2$ the variance of $D$ in class $k$, is smallest for
\[
n_k \;\propto\; p_k\, \sigma_k / \sqrt{c_k},
\]
where it equals $\big(\sum_k p_k \sigma_k \sqrt{c_k}\big)^2 / C$. Since $0 \leq D \leq 1$, $\sigma_k^2 \leq R_k$, so the upper bound of Proposition~\ref{prop:classbounds} caps each $\sigma_k$ before any instance is paired.
\end{proposition}

\begin{proof}
The class means are independent, so the variance of the estimator is $\sum_k p_k^2 \sigma_k^2 / n_k$. Minimizing it subject to the budget, by a Lagrange multiplier, gives $n_k \propto p_k \sigma_k / \sqrt{c_k}$ and the stated minimum; this is the optimal allocation of stratified sampling \citep{neyman1934two}. Since $D \in [0, 1]$, $\sigma_k^2 \leq \E[D^2 \mid k] \leq \E[D \mid k] = R_k$.
\end{proof}

\begin{proposition}[The winner's curse with many skills]\label{prop:curse}
Let $\hat{b} = \max_s \hat{\E}[\theta_s]$, each mean the average of at least $N$ independent instances of its skill. Then
\[
0 \;\leq\; \E[\hat{b}] - b \;\leq\; \sqrt{\ln n \,/\, (2N)},
\]
whatever the dependence between the estimates; within each class, the estimate of $B_1$ is biased upward by at most $\sum_k p_k \sqrt{\ln n / (2 N_k)}$, with $N_k$ the smallest record of any skill in class $k$.
\end{proposition}

\begin{proof}
The lower bound is Jensen's inequality for the convex function $\max$. For the upper, $\max_s \hat{\E}[\theta_s] \leq b + \max_s (\hat{\E}[\theta_s] - \E[\theta_s])$. Each estimation error is a mean of at least $N$ independent terms bounded in an interval of length one, hence sub-Gaussian with variance proxy $1/(4N)$ \citep{hoeffding1963probability}, and the expected maximum of $n$ such variables, however dependent, is at most $\sqrt{2 \cdot \ln n / (4N)}$. The class statement applies the same bound to each $b_k$.
\end{proof}

The bound grows with the roster only as $\sqrt{\ln n}$, and is conservative: it holds for any distribution of performances in $[0,1]$, while the bias of Section~\ref{sec:identification}, computed for Beta skills, is an order of magnitude smaller.

\subsection{Combining several skills}\label{app:combining}

With more than two skills a fixed combination need not engage the whole set. Under the squared rule, pooling the responses of a subset $A$ with equal weights attains \sqc{$\mean_{s \in A} \E[\theta_s] + \Lambda_A$}, with \sqc{$\Lambda_A$} the disagreement of Equation~\eqref{eq:Lambda} over $A$. Committing ex ante to the $\ell$ skills with the largest expected performances and pooling their responses has as its ends rungs~1 and~2, $\ell = 1$ being the commitment and $\ell = n$ pooling, and for a pair there is nothing between them. In between, a smaller $\ell$ forgoes less of $S$ and earns the disagreement of fewer responses, which need not be less, so the best $\ell$ trades the shortfall of the weaker skills it admits against what their responses add: separate assessments fix the first, and only paired observation the second. These are the select crowds of \citet{mannes2014wisdom}, and the next proposition states the trade-off exactly.

\begin{proposition}[When to admit a skill to a pool]\label{prop:subset}
Let the truth be an endpoint and the rule squared. Adding skill $j$ to a pooled subset $A$ of $|A|$ skills raises the attainment if and only if
\[
\sqc{\frac{\E[\theta_j] - \mean_{s \in A} \E[\theta_s]}{|A| + 1} \;+\; \Lambda_{A \cup \{j\}} - \Lambda_A \;>\; 0 .}
\]
Within class $k$, pooling every skill attains more than sending the class to its incumbent if and only if \sqc{$\Lambda_k > S_k$}, with \sqc{$\Lambda_k$} the disagreement within the class and $S_k = b_k - \E[m \mid k]$.
\end{proposition}

\begin{proof}
The mean over $A \cup \{j\}$ exceeds the mean over $A$ by $(\E[\theta_j] - \mean_{s \in A}\E[\theta_s])/(|A| + 1)$; add the change in disagreement. Within class $k$, pooling attains \sqc{$\E[m \mid k] + \Lambda_k$} by Equation~\eqref{eq:Lambda} applied within the class, and the incumbent attains $b_k$.
\end{proof}

By the first condition, a weaker skill is admitted when the disagreement it adds outweighs its shortfall from the pool, diluted by the size of the pool. The second makes the choice between pooling and routing a choice made class by class, the coarsest form of gating (Section~\ref{sec:signals}).

\paragraph{Combining the implementations of routing.} With more than two skills the implementations can be combined: a criterion may engage a subset of the available skills, leaving the rest unengaged as assignment does, and the aggregation may then weight or select among the responses of that subset, as ensembling or selection does; with three skills, for instance, a criterion may engage the two likeliest to do well on the instance and the aggregation average their responses. With two skills the combination collapses, a subset of one being assignment and a subset of two the whole set.

\section{Skills in contest}\label{app:contest}
 
This appendix states the result of Section~\ref{sec:contest} on contests between skills. It leaves the setting of the rest of the paper, in which one organization holds every skill, for one in which the organization commits its skill $s$ to a stream of instances on each of which an opposing skill $s'$ also executes: a competitor's skill bidding for the same clients, a counterparty's, or any skill whose success on an instance is the organization's loss unless its own skill answers it.
 
\paragraph{The model.} On instance $n$ the two skills realize performances $\theta_{s,n}$ and $\theta_{s',n}$, read as probabilities of success: here $\theta$ is a skill's probability of succeeding on the instance, not the score of Section~\ref{sec:instances}, and the results depend on its distribution and coupling as the stake does on those of the score. The pairs $(\theta_{s,n}, \theta_{s',n})$ are drawn independently across instances from a joint distribution, so that the two performances on the same instance may be coupled, as everywhere else in the paper. The opposing skill succeeds with probability $\theta_{s',n}$, and the organization's skill answers a success with probability $\theta_{s,n}$, independently given the performances. A success left unanswered costs the organization a consequence $Z_n \geq 0$, drawn independently of the performances and of the resolutions, of which it bears at most $\bar{z}$: the bound the organization sets on what one unfavorable resolution may cost, by limiting the exposure of any single commitment, so that the cost is $\min(Z_n, \bar{z})$. A success answered recovers $\eta > 0$. The organization's position moves by
\begin{equation}\label{eq:contest}
Y_n \;=\; \min(Z_n, \bar{z})\,\mathbf{1}[\text{$s'$ succeeds, $s$ fails to answer}] \;-\; \eta\,\mathbf{1}[\text{$s'$ succeeds, $s$ answers}],
\end{equation}
and the organization ceases to be viable when $X_n = \sum_{j \leq n} Y_j$ first reaches a level $\bar{x} > 0$: lost clients, a withdrawn mandate, exhausted trust. Write $h(\bar{z}) = \E[\min(Z, \bar{z})]$ for the expected cost of an unanswered success, the limited expected value of the consequence \citep{klugman2019loss}; it is continuous and non-decreasing in $\bar{z}$, strictly increasing where $\Pr(Z > \bar{z}) > 0$, and tends to $\E[Z]$, possibly infinite. Write also
\[
\chi \;=\; \frac{\E[\theta_{s'}\,\theta_s]}{\E[\theta_{s'}]} \;=\; \E[\theta_s] + \frac{\operatorname{Cov}(\theta_s, \theta_{s'})}{\E[\theta_{s'}]}
\]
for the performance of the organization's skill weighted by where the opposing skill succeeds: its answer rate on the successes it faces. It depends on how the two performances are coupled, and equals $\E[\theta_s]$ when they are independent.
 
\begin{proposition}[The bound that decides a contest]\label{prop:contest}
(a) The expected step is
\[
\mu(\bar{z}) \;=\; \E[\theta_{s'}]\,\big[(1 - \chi)\,h(\bar{z}) - \eta\,\chi\big].
\]
(b) If $\E[\theta_{s'}] > 0$ and $0 < \chi < 1$, then $\mu(\bar{z}) < 0$ if and only if
\[
h(\bar{z}) \;<\; r_\eta \;=\; \eta\,\frac{\chi}{1 - \chi},
\]
and, with $Z$ unbounded above, if $r_\eta < \E[Z]$ there is a unique break-even bound $\bar{z}^{\dagger}$ with $h(\bar{z}^{\dagger}) = r_\eta$, the position drifting in the organization's favor if and only if $\bar{z} < \bar{z}^{\dagger}$; if $r_\eta \geq \E[Z]$, every bound does. (c) The threshold $r_\eta$ rises with $\chi$, which rises with the organization's own performance on every instance and with the concordance of the two performances. The opposing skill enters $\chi$ only through where it succeeds: under independence $\chi = \E[\theta_s]$, and the opposing skill sets only the rate $\E[\theta_{s'}]$ at which the position moves. At parity, $\theta_s$ and $\theta_{s'}$ sharing a distribution $F$, $\chi$ runs from $\E[F^{-1}(V)\,F^{-1}(1 - U)]/\E[\theta]$ under the countermonotone coupling, through $\E[\theta]$ under independence, to $\E[\theta^2]/\E[\theta] = \E[\theta] + \operatorname{Var}\theta/\E[\theta]$ under the comonotone one, with $V$ uniform on $[0,1]$. (d) If $\E[Z] = \infty$, then for any joint distribution with $\E[\theta_{s'}(1 - \theta_s)] > 0$, $\mu(\bar{z}) > 0$ for every large enough $\bar{z}$: without a bound, no capability edge in either direction makes the drift negative. (e) If $\Pr(Y > 0) = 0$ the position never reaches $\bar{x}$. Otherwise, if $\mu(\bar{z}) \geq 0$, it reaches $\bar{x}$ almost surely; if $\mu(\bar{z}) < 0$, it ever reaches $\bar{x}$ with probability at most $e^{-\zeta \bar{x}}$, with $\zeta > 0$ the unique root of $\E[e^{\zeta Y}] = 1$.
\end{proposition}
 
\begin{proof}
(a) Given the performances, the opposing skill succeeds and is left unanswered with probability $\theta_{s'}(1 - \theta_s)$, at expected cost $h(\bar{z})$ since the consequence is independent of the performances and of the resolutions, and succeeds and is answered with probability $\theta_{s'}\theta_s$; so $\E[Y \mid \theta_s, \theta_{s'}] = \theta_{s'}(1 - \theta_s)\,h(\bar{z}) - \eta\,\theta_{s'}\theta_s$. Taking expectations, $\E[\theta_{s'}(1 - \theta_s)] = \E[\theta_{s'}](1 - \chi)$ and $\E[\theta_{s'}\theta_s] = \E[\theta_{s'}]\,\chi$. (b) Since $\E[\theta_{s'}] > 0$, the bracket carries the sign. Existence and uniqueness of $\bar{z}^{\dagger}$ follow from $h(0) = 0 < r_\eta$, the continuity and strict increase of $h$, and the intermediate value theorem. (c) $r_\eta$ is increasing in $\chi$ on $[0,1)$. Raising $\theta_s$ on any instance raises $\E[\theta_{s'}\theta_s]$ and leaves $\E[\theta_{s'}]$ unchanged; the covariance form shows the dependence on the coupling, and vanishes under independence. For given marginals, $\E[\theta_{s'}\theta_s]$ is largest under the comonotone coupling and smallest under the countermonotone one, the Fr\'echet--Hoeffding bounds \citep{nelsen2006introduction}; with a common distribution these give the two ends stated, and independence gives $\E[\theta]^2/\E[\theta]$. (d) $\E[\theta_{s'}](1 - \chi) = \E[\theta_{s'}(1 - \theta_s)] > 0$, so in (a) the first term grows without bound with $h(\bar{z})$, and the second does not depend on $\bar{z}$. (e) The steps are independent and identically distributed, bounded below by $-\eta$ and above by $\bar{z}$. If $\mu > 0$ the position grows without bound by the strong law of large numbers, and if $\mu = 0$, a non-degenerate walk with zero mean has an unbounded upper limit \citep{asmussen2010ruin}. If $\mu < 0$ and $\Pr(Y > 0) > 0$, the moment generating function of the bounded step is finite and convex, has slope $\mu < 0$ at zero and diverges as its argument grows, so $\E[e^{\zeta Y}] = 1$ has a unique root $\zeta > 0$; $e^{\zeta X_n}$ is a martingale, and optional stopping at the first passage over $\bar{x}$ gives Lundberg's inequality \citep{asmussen2010ruin}.
\end{proof}
 
\begin{remark}
A rating-system convention credits each resolution with the strength of the skill it overcame, weighting the cost of an unanswered success by $\theta_{s,n}$ and the recovery of an answer by $\theta_{s',n}$. With independent performances the drift is then $\E[\theta_{s'}]\,\E[\theta_s(1 - \theta_s)]\,h(\bar{z}) - \eta\,\E[\theta_{s'}^2]\,\E[\theta_s]$, and at parity it is negative if and only if $h(\bar{z}) < \eta\,\E[\theta^2]/\E[\theta(1 - \theta)]$. The threshold again contains nothing that distinguishes the two skills, and part (d) holds unchanged. But the expected cost of an unanswered success, proportional to $\E[\theta_s(1 - \theta_s)]$, vanishes for a skill that never answers and is largest at $\theta_s = \frac{1}{2}$, so that under the convention exposure is not monotone in capability, and a weaker skill can bear less of it than a stronger one; the model above does without it for that reason.
\end{remark}
 
\paragraph{An illustration.} With performance uniform on $[0,1]$ for both skills, $\chi$ is $\frac{1}{3}$ under the countermonotone coupling, $\frac{1}{2}$ under independence and $\frac{2}{3}$ under the comonotone one, so that $r_\eta$ is $\eta/2$, $\eta$ and $2\eta$. With a consequence of Pareto form with tail index one and scale $z_0$, the mean is infinite, and $h(\bar{z}) = \bar{z}$ for $\bar{z} \leq z_0$ and $z_0\,[1 + \ln(\bar{z}/z_0)]$ above. With $\eta = z_0$ the break-even bound $\bar{z}^{\dagger}$ is $z_0/2$, $z_0$ and $e\,z_0 \approx 2.7\,z_0$ in the three cases: the same two skills, the same capability, and more than a fivefold difference in the exposure the organization can tolerate, decided by where its skill fails relative to the skill it faces.
 
\paragraph{The reading for skill policy.} Three features carry over to the paper's setting. The first is a mirror. When two skills are held together, failures that fall on different instances are the residual that routing collects; when they are opposed, the organization's failures that fall where the opposing skill succeeds are its exposure. The coupling that makes two skills of equal capability most valuable held together makes one of them most exposed when it faces the other, and facing a skill with a copy of itself, whose failures coincide with its own, is the safest position the marginals allow. The model is defensive: the organization loses on the opposing skill's successes and gains nothing when it fails. Were it also to gain $g$ on each instance on which the opposing skill fails and its own succeeds, the drift at parity would be $D\,\big(h(\bar{z}) + \eta - g\big) - \eta\,\E[\theta]$, with $D = \E[\theta] - \E[\theta_s\theta_{s'}]$, and the copy would remain the safest position only while $h(\bar{z}) + \eta > g$, as it does where unanswered successes are costly. The second is that the threshold is computable before any contest from the joint distribution of the two performances on the instances at stake, as the residual is, and not from their marginals alone; the bound that meets it is a property of the consequences, which must be measured separately. The third is that the bound is a third term beside the stake and the take, a lever no arrangement of the skills contains: where consequences are bounded and modest, engaging and arranging skills decide the outcome as the rest of the paper describes; where they are heavy-tailed, what decides is how much one failure is permitted to cost.

\section{Records and algorithmic notation}\label{app:records}

Everything a policy estimates comes from past instances. To know how well a skill performs on average, an organization needs instances that skill executed, with its response confronted against the correct response so that performance can be established: the record $\mathcal{R}_s$. To know how well it performs on a particular kind of instance, the same is needed sorted by class: $\mathcal{R}_{s,k}$, the part of $\mathcal{R}_s$ of class $k$. To know which of several skills performs best on an instance, none of that suffices: the comparison can only be made where the skills executed the same instance, which is collected in $\mathcal{R}_{\mathcal{S}}$, the record of instances every skill in $\mathcal{S}$ executed, with $\mathcal{R}_{\mathcal{S}} \subseteq \mathcal{R}_s$ for each $s$. To predict an instance's class no performance is needed at all, only instances whose class is known and whose signal $Z_q$ was recorded: the record $\mathcal{R}^{Z}$, which grows whoever executes the instance and can come from outside the organization. Where a record's instances must also carry a signal or retain responses, a superscript says which, as in $\mathcal{R}^{W}_{\mathcal{S}}$, $\mathcal{R}^{Z,W}_{\mathcal{S}}$ and $\mathcal{R}^{Z,W,\hat{r}}_{\mathcal{S}}$ (Table~\ref{tab:records}).

\begin{table}[H]
\small
\centering
\caption{The records. Each is a set of past instances, distinguished by which skills worked them and what was retained about them. A tick marks what an instance in that record carries; a dash marks what it need not}
\label{tab:records}
\setlength{\tabcolsep}{4pt}
\begin{tabular}{l l c c c c c >{\raggedright\arraybackslash}p{4.0cm}}
\toprule
Record & Instances it holds & $r$ & $k$ & $Z_q$ & $W_q$ & $\{\hat{r}_s\}$ & What it supports \\
\midrule
$\mathcal{R}_s$
  & worked by skill $s$
  & \checkmark & --- & --- & --- & ---
  & $\hat{\E}[\theta_s]$, hence $i^{*}$ \\[3pt]
$\mathcal{R}_{s,k}$
  & the part of $\mathcal{R}_s$ of class $k$
  & \checkmark & \checkmark & --- & --- & ---
  & $\hat{\mu}_{s,k}$, hence $\bar{s}(k)$ \\[3pt]
$\mathcal{R}^{Z}$
  & any, whether or not a skill executed them
  & --- & \checkmark & \checkmark & --- & ---
  & $\widehat{\Pr}(k \mid Z_q)$ \\[3pt]
$\mathcal{R}_{\mathcal{S}}$
  & worked by \emph{every} skill in $\mathcal{S}$
  & \checkmark & --- & --- & --- & ---
  & the winner $\dot{s}$ of each instance; with the responses retained, \sqc{$\hat{\Lambda}$} \\[3pt]
$\mathcal{R}_{\mathcal{S}}^{W}$
  & the same, carrying $W_q$
  & \checkmark & --- & --- & \checkmark & ---
  & $\hat{\E}[\theta_s \mid W_q]$ \\[3pt]
$\mathcal{R}_{\mathcal{S}}^{Z,W}$
  & the same, carrying both
  & \checkmark & --- & \checkmark & \checkmark & ---
  & $\hat{\E}[\theta_s \mid Z_q, W_q]$ \\[3pt]
$\mathcal{R}_{\mathcal{S}}^{Z,W,\hat{r}}$
  & the same, responses retained
  & \checkmark & --- & \checkmark & \checkmark & \checkmark
  & $\hat{\E}[\theta_s \mid Z_q, W_q, \{\hat{r}_s\}\!]$; the weights $\hat{w}(Z_q, W_q)$ of 4d and~5 \\
\bottomrule
\end{tabular}
\end{table}

No record is indexed by policies, and none needs to be. To know which policy would have attained most on an instance, the instance need not have been worked under more than one, which is fortunate, since it is worked under exactly one. Performance is computed rather than observed: where every skill's response is retained and the correct response is known, $\theta_\pi$ follows for every policy that forms its response by an aggregation, because its response is a function of the responses it would have engaged. The comparison among such policies is therefore drawn on $\mathcal{R}_{\mathcal{S}}^{Z,W,\hat{r}}$, the record selection after execution already requires; a synergy protocol's response is not such a function, and enters the comparison only on instances where the protocol was run. What the comparison asks beyond that record is not more instances but that every policy be standing, its criterion fitted and its estimates made, since a routing policy's attainment on a record instance is the attainment of whichever skill its criterion would have commissioned, and until the criterion exists there is no such skill. Criteria must be fitted out of fold when their policies' attainments are computed on the record, or a criterion inflates its own policy's attainment where it has seen the instance. In a setup block $Z_q$ and $W_q$ name the signal columns of the record being fitted on; in an execution block they name the values the instance at hand carries.

Table~\ref{tab:spectrum} states what each policy attains in expectation over the population of instances, which is a limit: what the policy would deliver if the world were as the profiles describe and the caseload arbitrarily large. The algorithms describe what an organization actually does, and everything they compute is drawn from a finite record: cell means are sample means and criteria are models fitted to past instances. A hat over an expectation marks that substitution, so $\hat{\E}[\theta_s]$ is the record's estimate of $\E[\theta_s]$, which appears without a hat in the table because it is intended in the long run. A hat on a value denotes a prediction about the instance at hand rather than an estimate of a population quantity: $\hat{k}$ predicts the class $k$, and $\hat{s}$ predicts the winner $\dot{s}$.

\section{Illustrative skill profiles}\label{app:profiles}

\begin{figure}[!htb]
    \centering
    \begin{subfigure}[t]{0.48\textwidth}
        \centering
        \includegraphics[width=\linewidth]{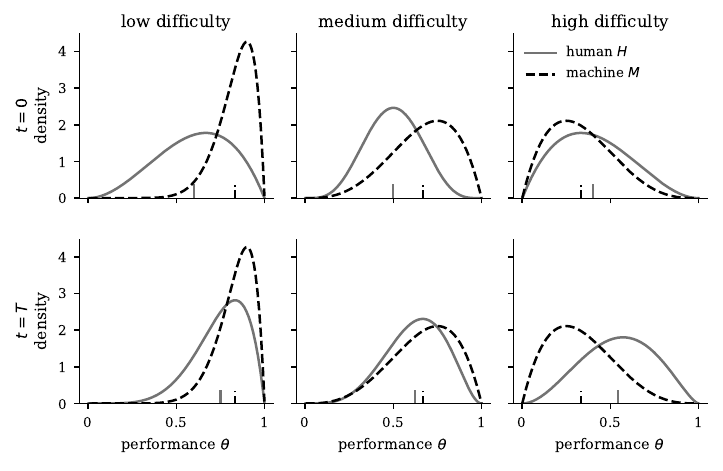}
        \subcaption{Profile A}
        \label{fig:beta_a}
    \end{subfigure}
    \hfill
    \begin{subfigure}[t]{0.48\textwidth}
        \centering
        \includegraphics[width=\linewidth]{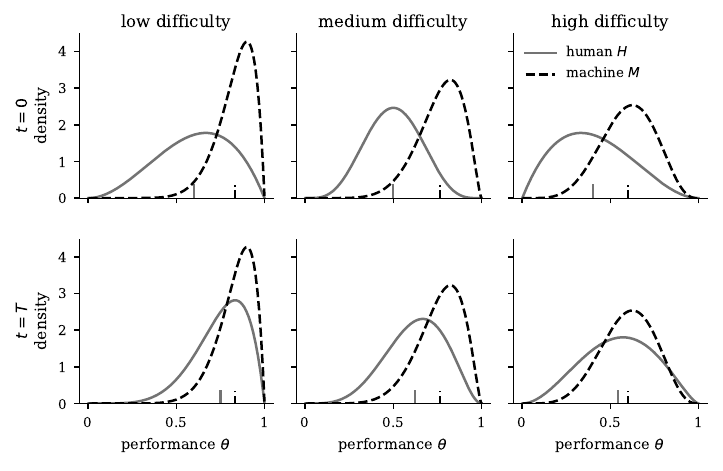}
        \subcaption{Profile B}
        \label{fig:beta_b}
    \end{subfigure}
    \caption{The illustrative performance distributions of Table~\ref{tab:beta}, for both skills and all difficulty levels, at the initial date $t = 0$ and the final date $t = T$, the human skill improving between the two and the machine skill static; densities are exact, and the ticks on the axis mark each skill's mean}
    \label{fig:beta}
\end{figure}

The figures and Table~\ref{tab:gsr} use two illustrative profiles of a human skill $H$ and a machine skill $M$ at three levels of difficulty, taken as classes, and at two dates, $t = 0$ and $t = T$, between which the human learns on the job and the machine, trained once, does not. Performance within each level is Beta-distributed, $\theta_{s,k} \sim \mathrm{Beta}(\alpha_{s,k,t}, \beta_{s,k,t})$, and the two performances are drawn independently on each instance, which fixes their coupling and makes $\E[\Mx]$ computable from the marginals by one-dimensional integration. In Profile A the machine holds a comparative advantage at low difficulty and the human at high; Profile B enhances the machine so that it leads at every level. The parameters are stylized: they illustrate the quantities of the theory and calibrate nothing (Table~\ref{tab:beta}, Figure~\ref{fig:beta}).

\begin{table}[!htb]
\centering\footnotesize
\caption{Parameters $(\alpha, \beta)$ of the illustrative performance distributions, by skill, difficulty level and date}
\label{tab:beta}
\begin{tabular}{llrrrrrrrr}
\toprule
 & & \multicolumn{4}{c}{Profile A} & \multicolumn{4}{c}{Profile B} \\
\cmidrule(lr){3-6}\cmidrule(lr){7-10}
 & & \multicolumn{2}{c}{$t = 0$} & \multicolumn{2}{c}{$t = T$} & \multicolumn{2}{c}{$t = 0$} & \multicolumn{2}{c}{$t = T$} \\
Skill & $d$ & $\alpha$ & $\beta$ & $\alpha$ & $\beta$ & $\alpha$ & $\beta$ & $\alpha$ & $\beta$ \\
\midrule
\multirow{3}{*}{$H$} & Low  & 3 & 2 & 6 & 2   & 3 & 2 & 6 & 2 \\
                     & Med  & 5 & 5 & 5 & 3   & 5 & 5 & 5 & 3 \\
                     & High & 2 & 3 & 3 & 2.5 & 2 & 3 & 3 & 2.5 \\
\midrule
\multirow{3}{*}{$M$} & Low  & 10 & 2 & 10 & 2 & 10 & 2   & 10 & 2 \\
                     & Med  & 4  & 2 & 4  & 2 & 8  & 2.5 & 8  & 2.5 \\
                     & High & 2  & 4 & 2  & 4 & 6  & 4   & 6  & 4 \\
\bottomrule
\end{tabular}
\end{table}

\section{The chest-radiograph experiment: Collab-CXR data, estimation and variations}\label{app:cxr}
 
\subsection{The experiment and the Collab-CXR release}\label{app:cxrdata}
 
The CXR experiment of \citet{agarwal2023combining}, whose data are released as Collab-CXR \citep{moehring2025dataset}, asked 227 radiologists to read historical chest films from Stanford, each with a short clinical history. For each film a radiologist reported a probability for every finding of a hierarchical list of 104, from top-level findings such as any abnormality or airspace opacity to the specific findings beneath them; a finding whose parent the radiologist ruled out was not shown. Each read took place in one of four information conditions: the film alone, the film with the clinical history, and each of these with the prediction of a single model, CheXpert, in view. Three designs allocated films and conditions to readers. In the first, 159 radiologists read 60 films each, 15 in each condition, none twice. In the second, 33 radiologists read the same 60 films in all four conditions, in sessions at least two weeks apart. In the third, 35 radiologists read 50 films, first without the model's prediction and then with it, half of them with the clinical history. The model reads the film alone, without the clinical history, and was developed on chest films from the same hospital \citep{irvin2019chexpert}, so that its standing on these films may exceed its standing on films from elsewhere. It predicts 14 labels, and its prediction is recorded on every film for the 15 findings these labels cover, shown or not, the one fracture label appearing on both fracture findings of the hierarchy, of the ribs and of the shoulder. Practice films preceded the reads, eight per radiologist in the first design and five per session in the second, drawn from the same films. Five board-certified chest radiologists at Mount Sinai, who are not among the participants, read all 324 films, and a finding is positive on a film when the mean of their five probabilities exceeds one half.
 
Table~\ref{tab:cxrdata} frames the data as they are used here, in the terms of the theory. Each film-finding is read many times, and each read is an instance: by 16 radiologists on average in each condition, and once by the model, whose response every one of those reads is paired with. The human skill is therefore a population of readers, of whom an organization deploys one per film, and the machine skill is a single model; the residual measures what routing between a radiologist drawn at random and the model could collect, not what the best of the radiologists could. A consensus of several readers would be a different, stronger human skill, and is not the one an organization deploying one reader per film holds.
 
\begin{table}[!htb]
\centering
\footnotesize
\caption{The Collab-CXR data as used here, and the study in the terms of the theory. Counts are for the arm with clinical history unless stated; the other arms have the same structure}
\label{tab:cxrdata}
\begin{tabularx}{\textwidth}{@{}>{\raggedright\arraybackslash\hsize=0.8\hsize}X >{\raggedright\arraybackslash\hsize=1.2\hsize}X@{}}
\toprule
\multicolumn{2}{@{}l}{\textit{The data}} \\[2pt]
Rows of the file & 2{,}717{,}728; 314{,}400 used ($11.6\%$): the four arms and the 15 findings, without practice or panel reads \\
Films & 324, of which 122 are positive on at least one of the 15 findings \\
Findings, the classes & 15 of the 104, those the model reports: 4 at the top of the hierarchy, 11 below \\
Film-findings & 4{,}860, of which 302 positive ($6.2\%$) \\
Radiologists & 227: 159, 33 and 35 in the three designs; the panel's reads excluded \\
Readings & 5{,}240 in each of the four arms: one radiologist reading one film in one condition, about 16 per film \\
Reads & 78{,}600 in each of the four arms; 7 to 30 per film-finding, 16.2 on average \\
Films per radiologist and arm & 15, 60 or 25, by design \\
Same reader, same film, both ways & 42{,}825 pairs, by the 68 radiologists of designs 2 and 3 \\
Responses recorded as zero & $66\%$ of the radiologists' reads, $58\%$ because the finding was not shown; $4\%$ of the model's \\
Reporters without information & squared rule: $0.947$ reporting each finding's prevalence, $0.938$ always zero; linear rule: $0.893$ and $0.938$ \\
\midrule
\multicolumn{2}{@{}l}{\textit{The study in the terms of the theory}} \\[2pt]
Task, instance & assessing one finding on one film; an instance is a read, one radiologist's probability for one film-finding (Table~\ref{tab:cxrterms}) \\
Class $k$, signal $Z$ & the finding (\texttt{pathology}): the diagnostic question, not its answer, named before anyone reads, so that $Z$ gives the class exactly and assignment on it is B1, not rung~4a; rung~4a refines the class by whether the film is abnormal, which it predicts from the model's probability for any abnormality \\
Skill $H$ & a participating radiologist drawn at random, reading unaided \\
Skill $M$ & the model, CheXpert \\
Response, truth, rule & the reported probability; the panel's binary standard; the squared rule, and the linear as the special case \\
Paired record $\mathcal{R}_{\mathcal{S}}$ & every unaided read, with the model's prediction on the same film-finding and the standard \\
Signal $W$ & the model's probability for the film-finding (\texttt{alg\_pred}), one value for all its reads, read in bins by rungs~4b to~4e; rung~4e also reads the radiologist's response, after execution \\
Rung~3 & the arm with the model's prediction in view, an exchange without a designed protocol \\
Weights & each film-finding equally, and each of its reads equally \\
\bottomrule
\end{tabularx}
\end{table}
 
\begin{figure}[!htb]
\centering
\includegraphics[width=\linewidth]{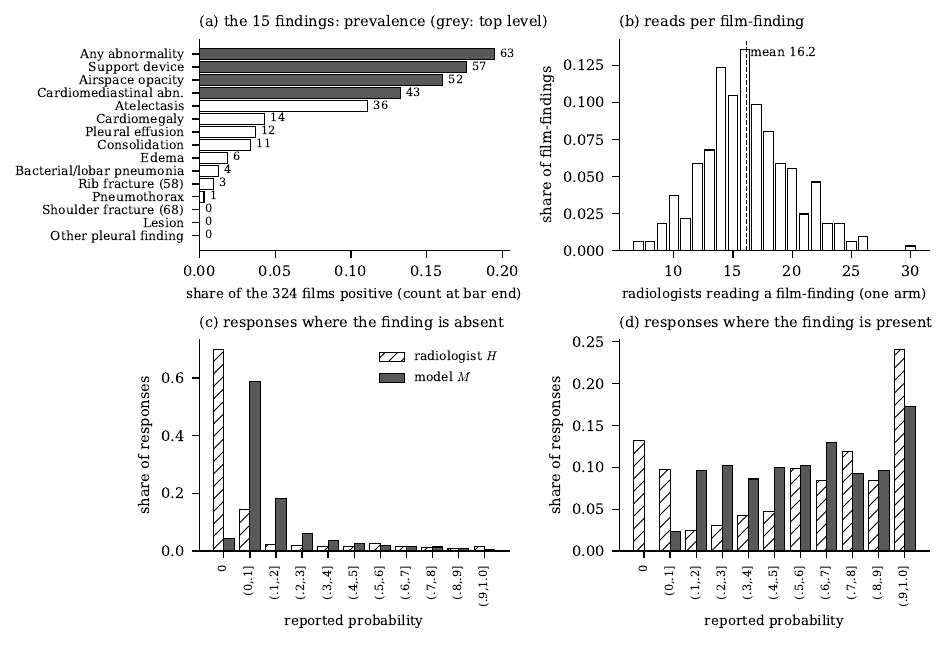}
\caption{The Collab-CXR data, arm with clinical history. (a) The 15 findings, by the share of the 324 films on which the standard is positive, with the number of positive films; the four top-level findings in gray. (b) The number of radiologists reading each film-finding. (c) and (d) The probabilities the two skills report where the standard is negative and where it is positive, the radiologists' reads weighted so that each film-finding counts once.}
\label{fig:cxrdata}
\end{figure}
 
Figure~\ref{fig:cxrdata} shows the volume and the variety behind these counts. Prevalence runs from one film in five, for any abnormality, to none of the 324 films, for three findings, and the four top-level findings are the most prevalent. A film-finding is read by 16 radiologists on average, by as few as 7 and as many as 30. And the two skills report probabilities in different ways. Where a finding is absent, the radiologists report exactly zero on $70\%$ of reads and the model on $4\%$, although it reports at most $0.1$ on $66\%$; where a finding is present, the radiologists report zero on $13\%$ of reads and more than $0.9$ on $24\%$, the model at most $0.1$ on $3\%$ and above $0.9$ on $16\%$. The radiologists commit, the model hedges.
 
\begin{figure}[!htb]
\centering
\includegraphics[width=\linewidth]{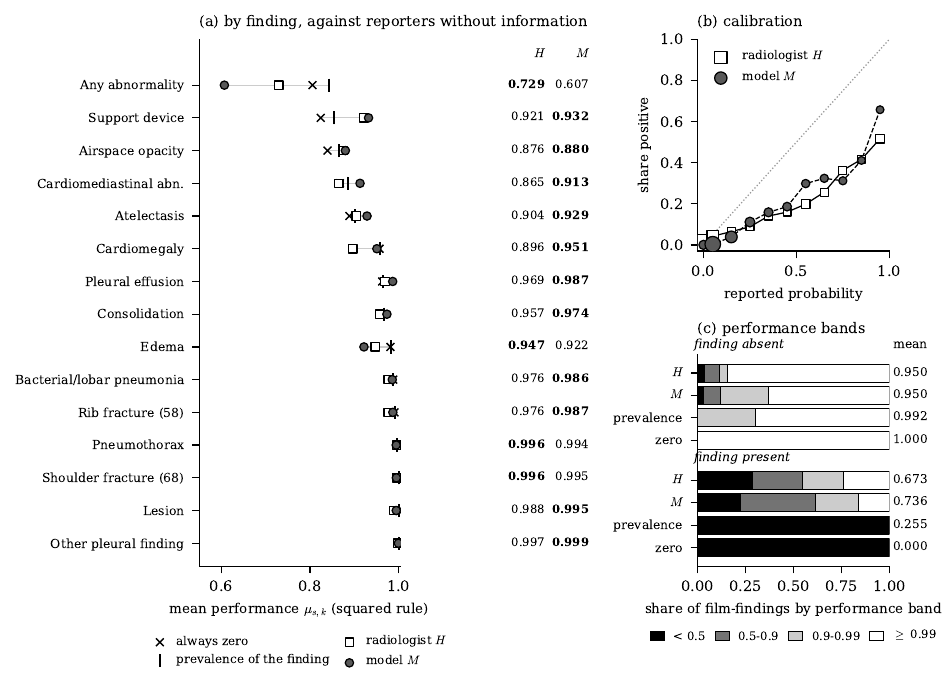}
\caption{Performance in the Collab-CXR data, arm with clinical history, squared rule, against two reporters that read nothing: one that gives every film the prevalence of each finding, and one that always reports zero. (a) The mean performance of each skill and each reporter, finding by finding, with the two skills' means beside the panel, the higher of the two, the finding's incumbent, in bold. (b) Among the responses in each band of reported probability, the share positive on the standard, marker area growing with the number of responses; the dotted line is perfect calibration. (c) The share of film-findings in each band of performance, where the finding is absent and where it is present, with the means.}
\label{fig:cxrperf}
\end{figure}
 
Performance is scored instance by instance, $\theta = 1 - (\hat{r} - r)^2$, one minus the Brier score of the probability a skill reports for one finding on one film, so that the mean performances reported here are one minus the skills' Brier scores. Under the linear rule a response that is itself a decision, zero or one, scores one when right and zero when wrong: accuracy is the linear rule on hard responses. The metrics more usual in classification are of a different kind. The areas under the ROC and precision-recall curves rank pairs of instances, and precision, recall and F1 are ratios over a set of instances once a threshold has turned probabilities into decisions; none is an average of a score each instance carries, so the decomposition of Section~\ref{sec:bounds}, which compares the skills on every instance, cannot be written in them. They are reported here only to describe the skills, and they are blind to calibration: any increasing transformation of the probabilities leaves the areas unchanged.
 
Performance is reported here against two reporters that read nothing: one that gives every film the prevalence of each finding in these films, and one that always reports zero. Because positives are rare, both are strong. Under the squared rule they attain $0.947$ and $0.938$, against $0.933$ for a radiologist drawn at random and $0.937$ for the model (Figure~\ref{fig:cxrperf}). Finding by finding, the model beats the first reporter on six findings and the radiologists on four, all among the common findings visible on the film, both on support devices, airspace opacity, atelectasis and pleural effusion and the model also on consolidation and the cardiomediastinal finding; and lose to it on the rarest, where any probability above zero costs, and on any abnormality. Where the finding is absent, both skills average $0.950$, against $0.992$ and $1$ for the two reporters; where it is present, $0.673$ and $0.736$, against $0.255$ and $0$. Both skills overstate: above $0.1$ their calibration curves lie below the diagonal, and of the responses above $0.9$ only about half of the radiologists' and two thirds of the model's fall on positive film-findings. The information is there, since ranked across all film-findings the model separates positives from negatives with an area under the ROC curve of $0.92$ and a radiologist drawn at random with $0.86$, and under the precision-recall curve with $0.44$ and $0.37$, against a prevalence of $0.062$; it is the way it is stated that the squared rule charges, and recalibrated the two skills attain $0.957$ and $0.964$, above both reporters (Section~\ref{sec:radiologydata}). In the usual form, the Brier skill score against the reporter of prevalences, one minus the ratio of the two Brier scores, which by Equation~\eqref{eq:murphy} is the resolution net of the miscalibration as a share of the uncertainty, is $-0.25$ for the radiologists and $-0.19$ for the model as reported, and $0.20$ and $0.34$ recalibrated. Read as decisions, at the threshold of one half that the squared rule\textquoteright s symmetric loss implies, a radiologist drawn at random misclassifies 98 film-findings in every 1{,}000 and the model 85, against 62 for a reporter that always says absent; pooling the two misclassifies 76, the arm with the prediction in view 93, and the better of the two on every read 41. Recalibrated, the five figures are 56, 47, 45, 54 and 30. A film carries 15 film-findings, so that the 17 errors in every 1{,}000 film-findings by which the arm exceeds pooling are about one for every four films. Much of what both skills lose is lost where the standard itself is uncertain: on the $16\%$ of film-findings on which the mean of the five experts' probabilities lies between $0.1$ and $0.7$, they lose more than half of all they lose. The panel read through the same hierarchy: $89\%$ of its probabilities on the 11 lower findings are zeros recorded because the finding was not shown, and on 78 of their film-findings the standard is negative with a panel mean between $0.3$ and $0.5$ that counts at least one such zero, as on film~154 (Appendix~\ref{app:cxrfilm}). Appendix~\ref{app:cxrvar} replaces the standard by one built from the panellists who were shown each finding. Under the linear rule the reporter that always says zero attains $0.938$, above the radiologists, $0.893$, the model, $0.849$, and even the routed anchor of the pair, $0.933$: a rule that is not proper rewards reporting the more likely outcome, which on a task of rare findings carries no information.
 
\subsection{Estimation}\label{app:cxrest}
 
The measurement of Section~\ref{sec:radiologydata} uses the primary file of the Collab-CXR release \citep{moehring2025dataset}, in its version of 16 September 2026. The file has 2{,}717{,}728 rows, one for each of the 104 findings on each film a reader read in a session. Of these, 336{,}960 are the panel's, 10 readers on 324 films, and enter only through the standard they set; 200{,}928 are the 1{,}932 practice readings; the remaining 2{,}179{,}840 are the four arms, 5{,}240 readings of a film in each. Rows are kept for the 15 findings with a model prediction, which drops the 1{,}865{,}440 rows of the other 89: the sample has 314{,}400 rows, $11.6\%$ of the file, that is 78{,}600 reads in each arm, by 227 radiologists, on the 4{,}860 film-findings, with no response, prediction or standard missing and no row repeated. A finding whose parent in the hierarchy the radiologist ruled out is recorded with probability zero, and is kept as recorded. The model's prediction and the standard are constant within a film-finding, and the standard is binary. Table~\ref{tab:cxrcols} describes every column of the file and the use made of it here.

\begin{table}[!htbp]
\centering
\footnotesize\renewcommand{\arraystretch}{0.83}
\caption{The columns of the primary file of the Collab-CXR release \citep{moehring2025dataset}, version of 16 September 2026, and their use here. Each row of the file is a read, one radiologist's response for one finding on one film in one session, 2{,}717{,}728 rows in all (Table~\ref{tab:cxrterms}). Descriptions follow the release's variable list (its Table~1), checked against the file; counts are from the file; where the list does not document a column, or documents it differently, the description says what the file shows}
\label{tab:cxrcols}
\setlength{\tabcolsep}{4pt}
\begin{tabularx}{\textwidth}{@{}>{\raggedright\arraybackslash}p{4.8cm} >{\raggedright\arraybackslash}X >{\raggedright\arraybackslash}p{3.1cm}@{}}
\toprule
Column & What it records & Use here \\
\midrule
\multicolumn{3}{@{}l}{\textit{Readers, films and findings}} \\[2pt]
\texttt{experiment\_id} & the radiologist: the 227 participants and the 10 readers of the two diagnostic panels & the reader; within-reader pairs \\
\texttt{uid\_clean} & the radiologist in one session; differs from \texttt{experiment\_id} only in design~2, whose readers have four sessions & not used \\
\texttt{experiment\_session} & the session, 0 to 3; sessions 1 to 3 exist only in design~2 & not used \\
\texttt{design} & the design, 1, 2 or 3; missing for the panel readers & the within-reader designs, 2 and 3 \\
\texttt{patient\_id} & the film, 0 to 323, which links to the image and clinical history of the Stanford release & the film; folds and resamples \\
\texttt{pathology} & the finding, one of 104 in the hierarchical list, coded by number and name & the 15 findings the model reports; the class $k$, given by $Z$ \\
\texttt{level} & the depth of the finding in the hierarchy, from 0 (10 top-level findings) to 3 & the four top-level findings among the 15 \\
\texttt{round}, \texttt{total\_num\_rounds} & the position of the film in the reader's sequence, and the number of reads in it: 68, 65 and 100 in designs 1 to 3, 324 for a panel reader & not used \\
\midrule
\multicolumn{3}{@{}l}{\textit{The condition}} \\[2pt]
\texttt{treatment} & the information condition: clinical history shown (\texttt{IN}) or not (\texttt{NIN}), the model's prediction shown (\texttt{AI}) or not (\texttt{NAI}); \texttt{warmup} for practice films & the four arms; practice reads excluded \\
\texttt{with\_ai}, \texttt{with\_ch} & indicators of the two parts of \texttt{treatment} & redundant with it \\
\midrule
\multicolumn{3}{@{}l}{\textit{The radiologist's response}} \\[2pt]
\texttt{probability} & the probability set on the slider, in $[0, 1]$; zero for a finding not shown & $\hat{r}_H$, and the read with the prediction in view \\
\texttt{visible} & whether the finding was shown when the read was submitted; a finding whose parent was ruled out is not & the zeros due to the hierarchy \\
\texttt{treat} & whether the radiologist would treat or follow up, asked on 30 findings & not used \\
\texttt{severity}, \texttt{size}, \texttt{position} & answers to follow-up questions, on 36, 19 and 24 findings & not used \\
\texttt{active\_time}, \texttt{raw\_time}, \texttt{num\_clicks} & seconds spent actively and in all on the film, and clicks on it; recorded per film and repeated across its findings & not used \\
\midrule
\multicolumn{3}{@{}l}{\textit{The reader's group}} \\[2pt]
\texttt{group\_teleradiology}, \texttt{group\_vietnam} & recruited from teleradiology firms (194 radiologists, designs 1 and 3) or from the VinMac system in Vietnam (33, design~2) & not used \\
\texttt{group\_ground\_truth} & a reader of a diagnostic panel & these rows excluded \\
\texttt{incentive\_round} & accuracy incentives, offered in designs 1 and 3: the file holds the values 1 and 1000, and is missing in design~2, which had none; the release documents an indicator & not used \\
\midrule
\multicolumn{3}{@{}l}{\textit{The model}} \\[2pt]
\texttt{alg\_pred} & the model's probability, recorded for the 15 findings it reports on every film, shown or not; its 14 labels, the fracture label on both fracture findings & $\hat{r}_M$, and $W$; on any abnormality, $Z$ of rung~4a \\
\texttt{chexbert\_label}, \texttt{chexbert\_label\_uncertainnegative} & labels extracted by CheXbert, for 13 findings; by its name, the second counts an uncertain mention as negative, and the two differ on $3\%$ of rows & not used \\
\midrule
\multicolumn{3}{@{}l}{\textit{The diagnostic standards}} \\[2pt]
\texttt{gt\_average\_us} & the mean of the five Mount Sinai panel radiologists' probabilities for the film-finding; the panel read through the same hierarchy, so that a finding not shown to one of them enters as zero & the uncertain standard (Appendix~\ref{app:cxrdata}) \\
\texttt{gt\_binary\_simple\_us} & one when that mean exceeds one half & the truth $r$ \\
\texttt{gt\_average\_vietnam}, \texttt{gt\_binary\_simple\_vietnam} & the same for a second panel of five radiologists; on the 15 findings its standard agrees with the first on $94\%$ of film-findings, and is positive on $9.0\%$ against $6.2\%$ & not used \\
\texttt{gt\_average\_all}, \texttt{gt\_binary\_simple\_all} & the same for the ten panel readers together; agrees with the first on $98\%$ & not used \\
\texttt{gt\_average\_experiment}, \texttt{gt\_binary\_simple\_experiment} & a leave-one-out standard from the participants' reads in the arm with clinical history and without the prediction; it varies from read to read and runs from $-0.16$ to $1.07$, so it is not a probability & not used \\
\texttt{gt\_average\_experimentNoCH}, \texttt{gt\_binary\_simple\_experimentNoCH} & by its name, the same from reads without clinical history; runs from $-0.17$ to $1.06$ & not used \\
\texttt{gt\_average\_active\_time\_}\emph{group} & the mean active time on the film of the readers behind each standard above & not used \\
\bottomrule
\end{tabularx}
\end{table}
 
Expectations average the reads of each film-finding and then the film-findings, with equal weights, so the classes are equally likely. B1 and the policies of rung~4 are fitted on four of five folds of films and scored on the fifth, and so is the logistic combination, which regresses the standard within each finding on the log-odds of the two probabilities, clipped to $[0.001, 0.999]$, and reports the training share where a finding has no positive on the training films: B1 estimates each finding's incumbent; rung~4b assigns on ten quantile bins of the model's probability, pooled across findings; rung~4c on five bins within each finding; rung~4d chooses, in each finding and bin, the weight on the radiologist's response from a grid of 21 values; rung~4e selects, in each finding, on four bins of each skill's probability. In each cell of an assignment or selection criterion, the skill with the larger mean performance on the training folds is chosen. The intervals for these rungs resample films with each fitted policy held fixed. The lift of rung~3 compares, film-finding by film-finding, the mean of the reads made with the model's prediction in view with the mean of the unaided reads pooled with the model; the within-reader comparison uses the 68 radiologists of the two designs in which the same reader read the same film-finding both ways. The bounds on $\Rinst$ and on each $R_k$ use the quantile functions of the two skills' performances within each finding, read on 4{,}000 points, and the bound $\min_v$ evaluated on 801 points, which can only loosen it. The residual under independence takes, within each finding, the product of the two marginals.
 
\subsection{By finding}\label{app:cxrfinding}
 
\begin{table}[!htb]
\centering
\footnotesize
\caption{The CXR experiment by finding, in the arm with clinical history under the squared rule. Prevalence is the share of films positive on the diagnostic standard; $\mu_{H,k}$ and $\mu_{M,k}$ are the class means, $\bar{s}(k)$ the incumbent, under the squared rule and, in the last column, under the linear rule; $R_k$ is the residual within the finding, with the bounds that the two class marginals alone establish (Proposition~\ref{prop:classbounds}); Corr. is the correlation of the two performances across reads}
\label{tab:cxrclasses}
\setlength{\tabcolsep}{4pt}
\begin{tabular}{l c c c c c c c c}
\toprule
Finding & Prevalence & $\mu_{H,k}$ & $\mu_{M,k}$ & $\bar{s}(k)$ & $R_k$ & Bounds on $R_k$ & Corr. & $\bar{s}(k)$, linear \\
\midrule
Any abnormality & 0.194 & 0.7293 & 0.6066 & $H$ & 0.0692 & [0.0086, 0.1959] & 0.44 & $H$ \\
Airspace opacity & 0.160 & 0.8760 & 0.8798 & $M$ & 0.0645 & [0.0336, 0.1123] & 0.45 & $H$ \\
Consolidation & 0.034 & 0.9575 & 0.9735 & $M$ & 0.0129 & [0.0030, 0.0263] & 0.50 & $H$ \\
Bacterial or lobar pneumonia & 0.012 & 0.9762 & 0.9856 & $M$ & 0.0097 & [0.0041, 0.0144] & 0.43 & $H$ \\
Edema & 0.019 & 0.9468 & 0.9217 & $H$ & 0.0290 & [0.0069, 0.0527] & 0.32 & $H$ \\
Atelectasis & 0.111 & 0.9040 & 0.9286 & $M$ & 0.0237 & [0.0079, 0.0709] & 0.76 & $H$ \\
Lesion & 0.000 & 0.9884 & 0.9946 & $M$ & 0.0046 & [0.0030, 0.0053] & 0.10 & $H$ \\
Pleural effusion & 0.037 & 0.9693 & 0.9866 & $M$ & 0.0078 & [0.0025, 0.0134] & 0.40 & $H$ \\
Other pleural finding & 0.000 & 0.9971 & 0.9993 & $M$ & 0.0006 & [0.0002, 0.0007] & 0.01 & $H$ \\
Pneumothorax & 0.003 & 0.9960 & 0.9945 & $H$ & 0.0026 & [0.0016, 0.0040] & 0.57 & $H$ \\
Cardiomediastinal abnormality & 0.133 & 0.8651 & 0.9128 & $M$ & 0.0464 & [0.0074, 0.0835] & 0.32 & $M$ \\
Cardiomegaly & 0.043 & 0.8963 & 0.9509 & $M$ & 0.0228 & [0.0055, 0.0489] & 0.40 & $M$ \\
Rib fracture (code 58) & 0.009 & 0.9764 & 0.9869 & $M$ & 0.0084 & [0.0030, 0.0131] & 0.39 & $H$ \\
Shoulder fracture (code 68) & 0.000 & 0.9961 & 0.9949 & $H$ & 0.0038 & [0.0030, 0.0039] & 0.00 & $H$ \\
Support device or hardware & 0.176 & 0.9213 & 0.9320 & $M$ & 0.0441 & [0.0157, 0.0667] & 0.35 & $H$ \\
\bottomrule
\end{tabular}
\end{table}
 
Table~\ref{tab:cxrclasses} reports each finding. The findings with the largest residual are the broad ones, any abnormality, airspace opacity, the cardiomediastinal finding and support devices, which are also the most prevalent, and on which neither skill is close to perfect. On findings that are almost never present on these films, both skills score near one and the residual is small. Under the linear rule the radiologists are the incumbent of every finding except the two cardiac findings.

\paragraph{Finer classes.} The instance of Section~\ref{sec:radiologydata} is a read: one finding on one film, read by one radiologist (Table~\ref{tab:cxrterms}). A class is any set of instances that can be named before a skill executes, and the finding is one such set: every read of, say, pneumothorax, on every film and by every radiologist. It can be cut further in two directions. Cut by radiologist, a class holds the reads of one finding by one radiologist, on the films she read. Cut by film, it holds the reads of one finding on one film by every radiologist who read that film: the film-finding, with between 7 and 30 reads. The film-finding is the finest class in which the human skill is still a population. Within it the model's response is a single number, since the model reads each film once, and only the radiologists' responses vary, from one reader to the next. Its incumbent is the model where that one response scores above the mean of the radiologists' reads, and the radiologists otherwise, which they are on $59\%$ of film-findings. Sending each film-finding to its incumbent attains $0.9585$, a class residual of $0.0219$, two thirds of $R$, which the winner's curse inflates: with the incumbent of each film-finding estimated on its other readers and scored on the read left out, it attains $0.9544$, a class residual of $0.0179$, half of $R$. What remains, between $0.0115$ and $0.0155$, lies between radiologists reading the same finding on the same film, and only a policy that knows who reads, and what she read, can collect it. No policy attains this ceiling: the incumbent of a film-finding can be estimated only from earlier reads of the same film, which a film arriving for the first time does not have. It bounds every policy that decides on the signals $Z$ and $W$ alone, since both are the same for every read of a film-finding: rungs~4b and~4c attain $0.9431$ and $0.9468$, and collect almost none of what knowing the film adds to knowing the finding.

\subsection{Variations}\label{app:cxrvar}
 
Table~\ref{tab:cxrrobust} repeats the headline quantities, for the arm with clinical history under the squared rule, on subsets of the findings and after recalibration. The top-level findings are any abnormality, airspace opacity, the cardiomediastinal finding and support devices; the four findings with at most one positive film are lesion, other pleural finding, pneumothorax and the shoulder fracture. Isotonic recalibration fits, within each finding, the map that does not decrease in the probability and best matches the share of positives on the training films, an estimate of $\bar{r}_{s,k}$ in Equation~\eqref{eq:murphy}; logistic recalibration fits a logistic regression on the log-odds of each probability, clipped to $[0.001, 0.999]$, within each finding. The row of the panellists shown replaces the standard by the mean of the probabilities of the panellists who were shown the finding, zero where none was. On it 224 film-findings turn positive, all below the top level and a third of them cardiomegaly, so that the prevalence rises from $6.2\%$ to $10.8\%$; the leader, the incumbents of every finding, the weight of the residual, $96\%$ of the stake, and the arm's place below pooling stay as they are. The two standards bracket the panel's view: a panellist who was not shown a finding had ruled out the finding above it, evidence of absence that the first standard counts as a zero and the second ignores. Table~\ref{tab:cxrtwoway} sets resampling radiologists, and films and radiologists together, beside resampling films. Four further seeds for the folds, and ten folds instead of five, move the attainments of B1 and rungs~4b to~4e by at most $0.0005$. Taking the radiologist as the class gives $\Rclass = 0.0052$, and the radiologist and the finding together $0.0164$, both in sample.
 
\begin{table}[!htb]
\centering
\footnotesize
\caption{The CXR experiment under variations, arm with clinical history, squared rule. The first row is the measurement of Table~\ref{tab:cxr}. Recalibration maps each skill\textquoteright s probability within each finding, fitted on four fifths of the films and applied to the fifth; the reads with the model\textquoteright s prediction in view are recalibrated on their own. The row \emph{Standard, panellists shown} uses a standard built from the panellists who were shown each finding; the last two keep only the reads of findings the radiologist was shown, with the model\textquoteright s paired reads, on which the arm with the prediction in view is not compared, since the findings shown differ between the arms (text). The last three columns are pooling and the arm with the prediction in view against the informed baseline, and the lift of the arm against pooling}
\label{tab:cxrrobust}
\setlength{\tabcolsep}{3pt}
\begin{tabular}{l c c c c c c c c c}
\toprule
Variant & $\E[\theta_H]$ & $\E[\theta_M]$ & $S$ & $R$ & $\Rclass$ & $\Lambda$ & Pool $-\,b$ & Arm $-\,b$ & $\gamma$ \\
\midrule
All 15 findings & 0.9330 & 0.9366 & 0.0018 & 0.0334 & 0.0100 & 0.0101 & 0.0083 & $-$0.0001 & 0.991 \\
Top level (4) & 0.8479 & 0.8328 & 0.0075 & 0.0716 & 0.0156 & 0.0209 & 0.0133 & 0.0040 & 0.989 \\
Below top level (11) & 0.9640 & 0.9743 & 0.0051 & 0.0140 & 0.0025 & 0.0062 & 0.0010 & $-$0.0071 & 0.992 \\
Two or more positives (11) & 0.9107 & 0.9150 & 0.0021 & 0.0442 & 0.0134 & 0.0131 & 0.0110 & $-$0.0002 & 0.988 \\
Isotonic recalibration & 0.9574 & 0.9644 & 0.0035 & 0.0137 & 0.0001 & 0.0049 & 0.0014 & $-$0.0037 & 0.995 \\
Logistic recalibration & 0.9563 & 0.9655 & 0.0046 & 0.0120 & 0.0000 & 0.0045 & $-$0.0001 & $-$0.0063 & 0.994 \\
Standard, panellists shown & 0.9078 & 0.9111 & 0.0017 & 0.0375 & 0.0091 & 0.0101 & 0.0085 & 0.0015 & 0.992 \\
Reads elicited only & 0.8728 & 0.9153 & 0.0213 & 0.0412 & 0.0110 & 0.0188 & $-$0.0025 & --- & --- \\
\quad below top level (11) & 0.8868 & 0.9618 & 0.0375 & 0.0156 & 0.0000 & 0.0176 & $-$0.0199 & --- & --- \\
\bottomrule
\end{tabular}
\end{table}
 
\paragraph{Reads elicited only.} Of the radiologists\textquoteright{} unaided reads, $42\%$ are of findings they were shown: all those of the four top-level findings, and $21\%$ of those below them. Keeping only these, each with the model\textquoteright s paired read, leaves 33{,}342 reads on 3{,}598 film-findings, $13.4\%$ of them reads of positive film-findings against $6.1\%$ of all reads. On them the model leads under both rules, by $S = 0.0213$ under the squared rule and by $0.0163$ under the linear, where on all reads the radiologists lead under the linear rule by $0.0219$; below the top level its lead is $S = 0.0375$. The residual is $66\%$ of the stake, and pooling falls below the commitment, by $0.0025$, since the two skills are no longer close. Under the squared rule, then, the zeros of the hierarchy narrow the model\textquoteright s lead rather than create it, and under the linear rule they create the radiologists\textquoteright. The selection is the radiologists\textquoteright{} own: a finding was shown when the reader judged its parent present, so that the reads kept are those of films on which she found something, and the model is compared with her on films she chose. The arm with the prediction in view is not compared on them, since the findings shown differ between the arms. Imputing the findings not shown, as the probability of the parent times a probability of the finding given its parent, would need the second, which the study did not elicit.
 
\paragraph{Two-way sampling.} Films and radiologists are both sampled: the films from a hospital\textquoteright s archive, the radiologists from those who took part. Resampling films holds the radiologists fixed, and resampling radiologists the films; resampling both, independently, with each read weighing the product of its film\textquoteright s and its radiologist\textquoteright s multiplicities, the pigeonhole bootstrap of \citet{owen2007pigeonhole}, accounts for both sources at once. Table~\ref{tab:cxrtwoway} sets the three side by side. The films carry most of the variation; resampling both widens the intervals of the text by about a fifth for the differences of attainment and by half for the lift, and changes no conclusion: the leader is not established as reported, pooling beats the commitment as reported and not reliably after recalibration, and the lift stays below one.
 
\begin{table}[!htb]
\centering
\footnotesize
\caption{Three ways of resampling, arm with clinical history, squared rule: 95\% intervals over 1{,}000 resamples of films, as in the text, of radiologists, and of both, drawn independently, each read weighing the product of its film\textquoteright s and its radiologist\textquoteright s multiplicities. $S$ is half the leader\textquoteright s margin, which can fall below zero; Pool $-\,b$ is pooling against the commitment, and $\gamma$ the lift of the arm with the prediction in view against pooling. Recalibrated: as in Table~\ref{tab:cxrrobust}}
\label{tab:cxrtwoway}
\setlength{\tabcolsep}{5pt}
\begin{tabular}{l c c c c}
\toprule
 & Estimate & Films & Radiologists & Both \\
\midrule
\multicolumn{5}{l}{\textit{As reported}} \\[2pt]
$S$ & 0.0018 & [$-$0.0001, 0.0038] & [0.0005, 0.0030] & [$-$0.0004, 0.0042] \\
$R$ & 0.0334 & [0.0312, 0.0356] & [0.0323, 0.0344] & [0.0309, 0.0357] \\
Pool $-\,b$ & 0.0083 & [0.0067, 0.0098] & [0.0073, 0.0094] & [0.0063, 0.0100] \\
$\gamma$ & 0.991 & [0.989, 0.993] & [0.989, 0.993] & [0.988, 0.994] \\
\midrule
\multicolumn{5}{l}{\textit{Recalibrated}} \\[2pt]
$S$ & 0.0035 & [0.0016, 0.0054] & [0.0030, 0.0039] & [0.0015, 0.0055] \\
$R$ & 0.0137 & [0.0115, 0.0159] & [0.0133, 0.0140] & [0.0114, 0.0160] \\
Pool $-\,b$ & 0.0014 & [$-$0.0003, 0.0032] & [0.0010, 0.0017] & [$-$0.0004, 0.0031] \\
$\gamma$ & 0.995 & [0.993, 0.996] & [0.994, 0.996] & [0.993, 0.996] \\
\bottomrule
\end{tabular}
\end{table}
 
\subsection{The procedure on the data}\label{app:cxrproc}
 
Table~\ref{tab:cxrprocedure} runs the procedure of Section~\ref{sec:procedure} on the arm with clinical history under the squared rule. Each step is fitted on four of five folds of films and evaluated on the fifth. The check of step~1 compares each skill, as it reports, with the reporter of each finding's prevalence; both fall short, and the procedure continues with each skill's probabilities, and the reads made with the model's prediction in view, recalibrated within each finding by isotonic regression fitted on the training films alone, so that no film is recalibrated with a map that saw it. Step~1 commits to the skill with the higher mean on the training films. Step~2 sends each finding to its incumbent on the training films. Step~4 fits the criterion of rung~4c, five bins of the model's probability within each finding, and keeps it in a finding only where, on four inner folds of the training films, it beat step~2 in that finding; elsewhere the finding stays with its incumbent. Step~5 fits ensembling and selection as rungs~4d and~4e are fitted in Section~\ref{app:cxrest}; pooling fits nothing, and the exchange is the arm the study ran. The table counts a radiologist's work in two ways, by the film-findings given to her and by the films she must read to answer them; the second is the whole of her films as soon as any one finding is given to her. The intervals quoted in Section~\ref{sec:radiologydata} resample films with the fitted policies held fixed. The variant with a margin (Proposition~\ref{prop:margin}) estimates the standard error of each check\textquoteright s gain on the training films, clustered by film, and adopts a switch only where the estimated gain exceeds it $z$ times; with $z = 0$ it is the procedure of the table. Under the linear rule the check of step~1 cannot be passed, and the procedure is not run.
 
\begin{table}[H]
\centering
\footnotesize
\caption{The procedure of Section~\ref{sec:procedure} run on the Collab-CXR data, arm with clinical history, squared rule: what each step attains, and how much of a radiologist's work it needs. As the check of step~1 requires, both skills are recalibrated within each finding on the training films; every step is fitted on four fifths of the films and evaluated on the fifth, and the check of step~4 is made finding by finding on inner folds of the training films. The steps read from the bottom up; above them, what no step reaches, and below them, the two reporters that read nothing}
\label{tab:cxrprocedure}
\setlength{\tabcolsep}{5pt}
\begin{tabular}{c l c c c}
\toprule
& & & \multicolumn{2}{c}{Radiologist\textquoteright s share} \\
\cmidrule(lr){4-5}
Step & Policy & Attains & findings given & films read \\
\midrule
 & routing omniscience, $\E[\Mx]$ & 0.9781 & --- & --- \\
 & logistic combination of the two, outside the ladder & 0.9674 & 100\% & 100\% \\
\midrule
5 & exchange: the model in view, rung~3 & 0.9607 & 100\% & 100\% \\
5 & selection & 0.9643 & 100\% & 100\% \\
5 & ensembling & 0.9660 & 100\% & 100\% \\
5 & pooling, rung~2 & 0.9658 & 100\% & 100\% \\
4 & routing on finding and probability, checked & 0.9643 & 10\% & 100\% \\
2 & incumbent of each finding & 0.9644 & 9\% & 100\% \\
1 & commitment, to the model & 0.9644 & 0\% & 0\% \\
\midrule
 & each finding\textquoteright s prevalence & 0.9465 & 0\% & 0\% \\
 & always reporting zero & 0.9379 & 0\% & 0\% \\
\bottomrule
\addlinespace[3pt]
\multicolumn{5}{p{0.8\textwidth}}{\footnotesize Findings given is the share of film-findings given to a radiologist, and films read the share of a radiologist\textquoteright s films that she must read to answer them; the model is taken to run on every film. Every finding is assessed on every film, so a policy that gives the radiologists any finding has them read every film. Step~3 builds the paired record and changes no attainment; the record of this study is paired by design. The logistic combination regresses the standard, finding by finding, on the log-odds of the two probabilities as reported, and so recalibrates as it combines; the two reporters read nothing.}
\end{tabular}
\end{table}
 
\subsection{The arrangements on a film}\label{app:cxrfilm}
 
Table~\ref{tab:cxrladder} reads each arrangement of Section~\ref{sec:ladder} in the CXR setting: how it would run on an arriving film, the share of a radiologist's films she would read and of the film-findings given to her, and what it attains in the arm with clinical history under the squared rule, on the responses as reported. The two counts differ because the policies assign film-findings while a radiologist reads films, and a policy that gives her any finding on a film has her read it. Assignment on the standing, rung~4b, is the one fitted policy that spares her some films, an eighth of them, and it attains less than pooling.
 
\begin{table}[!htb]
\centering
\footnotesize
\caption{The arrangements of Section~\ref{sec:ladder} in the CXR setting: how each would run on an arriving film, the share of a radiologist\textquoteright s films she reads and of the film-findings given to her, and what it attains in the arm with clinical history under the squared rule, on the responses as reported, rung~0 at the foot. Fitted policies are cross-fitted over five folds of films}
\label{tab:cxrladder}
\setlength{\tabcolsep}{4pt}
\begin{tabularx}{\textwidth}{@{}l >{\raggedright\arraybackslash}X c c c@{}}
\toprule
 & & \multicolumn{2}{c}{Radiologist\textquoteright s share} & \\
\cmidrule(lr){3-4}
Rung & On an arriving film & films read & findings given & Attains \\
\midrule
B3 & both report every finding, and the oracle returns the better probability & 100\% & 100\% & 0.9699 \\
B2 & an oracle names, for each finding, the skill that will score better, and that skill alone reports it & 100\% & 81\% & 0.9699 \\
5 & assembled from the study\textquoteright s arms, not run: a radiologist reads it with the prediction in view, and her read is averaged with the model\textquoteright s under the weights of rung~4d & 100\% & 100\% & 0.9475 \\
4e & both report every finding, and one of the two probabilities is returned, chosen on the finding and the bands of both & 100\% & 100\% & 0.9528 \\
4d & both report every finding, and the report is a weighted average, the weight set by the finding and the band of the model\textquoteright s probability & 100\% & 100\% & 0.9497 \\
4c+a & rung~4c, the radiologist also reporting, on a share of the films drawn at random, the findings given to the model, for the record only & 100\% & above 19\% & as 4c \\
4c, checked & rung~4c where its check keeps it, the incumbent elsewhere & 100\% & 16\% & 0.9470 \\
4c & the same, in five bands within each finding & 100\% & 19\% & 0.9468 \\
4b & the model runs first, and its probability, in ten bands pooled across findings, decides who reports each finding & 88\% & 12\% & 0.9431 \\
4a & each finding goes to the incumbent of the finding and of whether the film is abnormal, the latter predicted from the model\textquoteright s probability for any abnormality & 100\% & 31\% & 0.9454 \\
B1 & each finding goes to its incumbent: the radiologist reports any abnormality, edema, pneumothorax and the shoulder fracture, the model the other eleven & 100\% & 27\% & 0.9466 \\
3 & a radiologist reads it with the model\textquoteright s prediction in view, nothing prescribing its use, and her read is the report & 100\% & 100\% & 0.9364 \\
2 & both report every finding, and the report is the average of the two probabilities & 100\% & 100\% & 0.9449 \\
2r & a coin decides who reports; tossed for each film, a radiologist reads half of them, and tossed for each finding, nearly all & 50\% & 50\% & 0.9348 \\
1 & the skill that scored higher on films already read reports every finding: the model, under the squared rule & 0\% & 0\% & 0.9366 \\
0, $M$ & the model alone reports every finding & 0\% & 0\% & 0.9366 \\
0, $H$ & a radiologist reads it and reports every finding; the model is not used & 100\% & 100\% & 0.9330 \\
\bottomrule
\end{tabularx}
\end{table}
 
Five entries need a word. Rung~4a needs a class that is not known at intake, and the finding is known; in the procedure its place is taken by B1 with estimated incumbents. A rung~4a can be built by refining the class: the finding together with whether the film is abnormal by the standard, which 63 of the 324 films are. Sent to the incumbents of their true refined classes, the film-findings attain $0.9479$, above B1 by $0.0013$, an interval from $0.0005$ to $0.0022$: the refinement holds a little class residual. Rung~4a must predict whether the film is abnormal, and the only signal at hand is the model's probability for any abnormality. Above a threshold fitted on the training films it classifies $84\%$ of films correctly, against $81\%$ for calling every film normal, and the assignment attains $0.9454$, below B1 by $0.0012$, an interval from $0.0002$ to $0.0022$. Across thresholds the attainment rises with the accuracy of the prediction, as in Proposition~\ref{prop:classpred}, but at no threshold does it reach B1: the highest accuracy reached, $87\%$, falls short of what the refinement needs. The row 4a of Table~\ref{tab:cxrladder} is this policy. Random assignment attains the mean anchor however the coin is tossed, but what it needs of a radiologist depends on the toss: tossed for each film, half her films; tossed for each finding, nearly all of them. Rung~4c with audit was not run, since the study's record is paired by design. Under rung~4c a radiologist already reads every film, so an audit would not ask her for a second read; it would ask her to answer the findings given to the model, which the hierarchy hides whenever she rules out a parent, so that auditing changes the instrument, and the questions are not free: the reading times recorded in the data rise with the number of questions shown, by about $3.5\%$ a question on the same film net of the reader, an association that her judgment of the film confounds. Rung~5 was not run either; it is assembled from the study's arms: the read made with the prediction in view, an exchange without a designed protocol, is combined with the model's probability under the cross-fitted weights of rung~4d, fitted, as Algorithm~\ref{alg:5} fits them, on the independent responses. It attains $0.9475$, above pooling and the model in view, and below rung~4d on the independent responses by $0.0023$, an interval from $0.0013$ to $0.0033$, a lift of $0.998$: here too the exchange gave back part of what the weights alone collect. The oracle of B2 gives a radiologist $81\%$ of film-findings, more than the model, because on the absent findings that make up most of them her zero scores a little above the model's small probability; the residual is collected on the fewer film-findings on which the model wins by much.
 
Table~\ref{tab:cxrfilmfull} gives every finding of the two films of Table~\ref{tab:cxrfilm}, and Tables~\ref{tab:cxrrecA} and~\ref{tab:cxrrecB} every read of two film-findings of each in the arm with clinical history, as the primary file records them. The films were chosen among the 2{,}855 pairs of a radiologist and a film read both ways in that arm, on six differences between arrangements on the film: the arm with the prediction in view against pooling and against the unaided read, selection against pooling and against ensembling, pooling against the model, and the model against the radiologist. Film~277, read by \texttt{advtele\_rk8wlicy}, is, of the four pairs with a finding present whose six differences lie closest to their medians, the one with most findings present, and the only pair with a finding present on which all six lie between the 35th and the 65th percentile. Film~154, read by \texttt{advtele\_m7xbjpys}, lies in the tails: the arm against pooling at the 3rd percentile, selection against pooling and against ensembling at the 96th. Across all 2{,}855, selection attains the most of the eight columns on $22\%$ of pairs and the arm the least on $21\%$; the arm falls below the unaided read on $51\%$ and below pooling on $48\%$, so that on about half of the pairs the exchange did better than pooling. The ordering of the averages is rare on a single pair: with at least two findings present, 19 pairs show it, on 14 films, both of these among them. The records show what a film-finding is in these data: one response of the model, between 7 and 30 reads by different radiologists in each arm, which often disagree widely, and five probabilities of the panel, often far apart and elicited through the same hierarchy, whose mean sets the standard.
 
Film~77, read by \texttt{advtele\_0xr5ncan} and given in Table~\ref{tab:cxrfilmtwo}, was chosen for the contrast. Four of its findings are present. The radiologist put any abnormality at $0.08$ and the airspace opacity at $0.02$, so that the findings beneath the airspace opacity were not shown to her and are recorded as zero, two of the four among them. The model alone attains the most on this film, $0.853$, and every arrangement that returns her response on some finding attains less: assignment by finding returns her $0.08$ on any abnormality, and selection her $0.02$ on airspace opacity and her zero on consolidation, in the bands in which her zeros are usually right. With the prediction in view she raised any abnormality to $0.73$, and the arm attains $0.828$, more than any routing arrangement on this film. Nothing that reads the finding and the two probabilities tells this film from the many on which her zero is right; it is an instance of the part of the residual that these signals leave uncollected.
 
\begin{table}[p]
\centering
\footnotesize
\caption{The two films of Table~\ref{tab:cxrfilm}, every finding. Columns as there}
\label{tab:cxrfilmfull}
\setlength{\tabcolsep}{2.8pt}
\begin{tabular}{l c c c c c c c c c}
\toprule
 & & \multicolumn{2}{c}{Alone} & Rung 3 & Rung 2 & B1 & 4c, checked & 4d & 4e \\
\cmidrule(lr){3-4}
Finding & $r$ & $H$ & $M$ & $M$ in view & pooling & by finding & + probability & ensembling & selection \\
\midrule
\multicolumn{10}{@{}l}{\textit{Film 277, a typical film}} \\[1pt]
Any abnormality & 0 & 0.46 & 0.46 & 0.72 & 0.46 & 0.46$_H$ & 0.46$_H$ & 0.46 & 0.46$_H$ \\
Airspace opacity & 1 & 0.17 & 0.19 & 0.19 & 0.18 & 0.19$_M$ & 0.19$_M$ & 0.18 & 0.17$_H$ \\
Consolidation & 0 & 0.51 & 0.02 & 0.02 & 0.26 & 0.02$_M$ & 0.02$_M$ & 0.04 & 0.02$_M$ \\
Bacterial or lobar pneumonia & 0 & 0.01 & 0.02 & 0.00 & 0.02 & 0.02$_M$ & 0.02$_M$ & 0.02 & 0.02$_M$ \\
Edema & 0 & 0.02 & 0.19 & 0.01 & 0.11 & 0.02$_H$ & 0.19$_M$ & 0.13 & 0.02$_H$ \\
Atelectasis & 1 & 0.02 & 0.07 & 0.02 & 0.05 & 0.07$_M$ & 0.07$_M$ & 0.06 & 0.02$_H$ \\
Lesion & 0 & 0.02 & 0.02 & 0.02 & 0.02 & 0.02$_M$ & 0.02$_M$ & 0.02 & 0.02$_M$ \\
Pleural effusion & 0 & 0.00 & 0.01 & 0.00 & 0.01 & 0.01$_M$ & 0.01$_M$ & 0.01 & 0.01$_M$ \\
Other pleural finding & 0 & 0.00 & 0.00 & 0.00 & 0.00 & 0.00$_M$ & 0.00$_M$ & 0.00 & 0.00$_M$ \\
Pneumothorax & 0 & 0.00 & 0.02 & 0.00 & 0.01 & 0.00$_H$ & 0.02$_M$ & 0.02 & 0.02$_M$ \\
Cardiomediastinal abnormality & 0 & 0.01 & 0.09 & 0.02 & 0.05 & 0.09$_M$ & 0.09$_M$ & 0.08 & 0.01$_H$ \\
Cardiomegaly & 0 & 0.00 & 0.07 & 0.00 & 0.04 & 0.07$_M$ & 0.07$_M$ & 0.07 & 0.00$_H$ \\
Rib fracture (code 58) & 0 & 0.00 & 0.03 & 0.00 & 0.01 & 0.03$_M$ & 0.03$_M$ & 0.03 & 0.03$_M$ \\
Shoulder fracture (code 68) & 0 & 0.00 & 0.03 & 0.00 & 0.01 & 0.00$_H$ & 0.03$_M$ & 0.03 & 0.03$_M$ \\
Support device or hardware & 1 & 0.69 & 0.45 & 0.70 & 0.57 & 0.45$_M$ & 0.45$_M$ & 0.53 & 0.69$_H$ \\
\cmidrule(lr){1-10}
\multicolumn{2}{l}{Film, all 15 findings} & 0.852 & 0.861 & 0.851 & 0.862 & 0.863 & 0.861 & 0.866 & 0.869 \\
\multicolumn{2}{l}{Findings given to the radiologist} & 15 & 0 & 15 & 15 & 4 & 1 & 15 & 15 \\
\midrule
\multicolumn{10}{@{}l}{\textit{Film 154, a film at the extremes}} \\[1pt]
Any abnormality & 1 & 0.83 & 0.90 & 0.27 & 0.86 & 0.83$_H$ & 0.83$_H$ & 0.83 & 0.90$_M$ \\
Airspace opacity & 1 & 0.59 & 0.40 & 0.49 & 0.49 & 0.40$_M$ & 0.40$_M$ & 0.51 & 0.40$_M$ \\
Consolidation & 0 & 0.20 & 0.03 & 0.02 & 0.11 & 0.03$_M$ & 0.03$_M$ & 0.04 & 0.03$_M$ \\
Bacterial or lobar pneumonia & 0 & 0.07 & 0.03 & 0.00 & 0.05 & 0.03$_M$ & 0.03$_M$ & 0.03 & 0.03$_M$ \\
Edema & 0 & 0.05 & 0.40 & 0.14 & 0.23 & 0.05$_H$ & 0.05$_H$ & 0.11 & 0.05$_H$ \\
Atelectasis & 1 & 0.07 & 0.23 & 0.01 & 0.15 & 0.23$_M$ & 0.23$_M$ & 0.23 & 0.23$_M$ \\
Lesion & 0 & 0.58 & 0.01 & 0.47 & 0.30 & 0.01$_M$ & 0.01$_M$ & 0.01 & 0.01$_M$ \\
Pleural effusion & 0 & 0.00 & 0.06 & 0.00 & 0.03 & 0.06$_M$ & 0.06$_M$ & 0.06 & 0.06$_M$ \\
Other pleural finding & 0 & 0.00 & 0.00 & 0.00 & 0.00 & 0.00$_M$ & 0.00$_M$ & 0.00 & 0.00$_M$ \\
Pneumothorax & 0 & 0.00 & 0.01 & 0.00 & 0.01 & 0.00$_H$ & 0.01$_M$ & 0.01 & 0.01$_M$ \\
Cardiomediastinal abnormality & 1 & 0.08 & 0.70 & 0.50 & 0.39 & 0.70$_M$ & 0.70$_M$ & 0.45 & 0.70$_M$ \\
Cardiomegaly & 0 & 0.00 & 0.70 & 0.73 & 0.35 & 0.70$_M$ & 0.70$_M$ & 0.63 & 0.00$_H$ \\
Rib fracture (code 58) & 0 & 0.00 & 0.01 & 0.00 & 0.01 & 0.01$_M$ & 0.01$_M$ & 0.01 & 0.01$_M$ \\
Shoulder fracture (code 68) & 0 & 0.00 & 0.01 & 0.00 & 0.01 & 0.00$_H$ & 0.00$_H$ & 0.00 & 0.00$_H$ \\
Support device or hardware & 1 & 0.61 & 0.92 & 0.30 & 0.76 & 0.92$_M$ & 0.92$_M$ & 0.81 & 0.92$_M$ \\
\cmidrule(lr){1-10}
\multicolumn{2}{l}{Film, all 15 findings} & 0.837 & 0.886 & 0.781 & 0.887 & 0.895 & 0.895 & 0.893 & 0.929 \\
\multicolumn{2}{l}{Findings given to the radiologist} & 15 & 0 & 15 & 15 & 4 & 3 & 15 & 15 \\
\bottomrule
\end{tabular}
\end{table}
 
\begin{table}[p]
\centering
\footnotesize
\caption{Every read of two film-findings of film 277, in the arm with clinical history, as recorded in the primary file (Table~\ref{tab:cxrcols}): one row per radiologist who read the film in that arm, unaided, with the prediction in view, or both, the reader of Table~\ref{tab:cxrfilm} in bold; a dagger marks a finding not shown to the reader, participant or panel, and recorded as zero. The measurement uses every unaided read, each weighing equally within the film-finding, so that the mean of the reads is what a radiologist drawn at random returns on it. Below, the model\textquoteright s response and the five probabilities of the panel that set the standard. The film was also read by 29 radiologists without history and 30 without history with the prediction in view}
\label{tab:cxrrecA}
\setlength{\tabcolsep}{4pt}
\begin{tabular}{l c c c c c}
\toprule
 & & \multicolumn{2}{c}{Airspace opacity} & \multicolumn{2}{c}{Support device or hardware} \\
\cmidrule(lr){3-4}\cmidrule(lr){5-6}
\texttt{experiment\_id} & \texttt{design} & unaided & in view & unaided & in view \\
\midrule
\texttt{advtele\_22d1lv15} & 1 & --- & 0.07 & --- & 1.00 \\
\texttt{advtele\_3pmlgwb2} & 1 & --- & 0.04 & --- & 0.96 \\
\texttt{advtele\_8v77oqem} & 1 & --- & 0.04 & --- & 0.46 \\
\texttt{advtele\_aq9gyefs} & 1 & --- & 0.03 & --- & 0.79 \\
\texttt{advtele\_av5u00hh} & 1 & --- & 0.15 & --- & 0.81 \\
\texttt{advtele\_exe6i77m} & 1 & 0.00 & --- & 0.62 & --- \\
\texttt{advtele\_huh2mrme} & 1 & --- & 0.03 & --- & 0.82 \\
\texttt{advtele\_nk82u5gn} & 1 & 0.04 & --- & 0.78 & --- \\
\texttt{advtele\_nnq1kak3} & 1 & --- & 0.00 & --- & 0.00 \\
\texttt{advtele\_olklvaiw} & 1 & --- & 0.03 & --- & 0.92 \\
\texttt{advtele\_ptpldk0t} & 1 & 0.00 & --- & 0.97 & --- \\
\texttt{advtele\_u0ygno79} & 1 & 0.00 & --- & 0.73 & --- \\
\texttt{advtele\_ut6mnxa5} & 1 & 0.00 & --- & 0.88 & --- \\
\texttt{advtele\_wtuvezgv} & 1 & --- & 0.00 & --- & 0.47 \\
\texttt{advtele\_xy7vcz1d} & 1 & 0.03 & --- & 0.47 & --- \\
\texttt{usarad\_n6j9lbfd} & 1 & --- & 0.00 & --- & 0.95 \\
\texttt{vestarad\_n9kzj8kw} & 1 & 0.00 & --- & 0.98 & --- \\
\texttt{vn\_exp\_10822160} & 2 & 0.21 & 0.21 & 0.93 & 0.94 \\
\texttt{vn\_exp\_15842209} & 2 & 0.07 & 0.09 & 0.81 & 0.23 \\
\texttt{vn\_exp\_18071879} & 2 & 0.96 & 0.00 & 0.00 & 0.00 \\
\texttt{vn\_exp\_23492291} & 2 & 0.04 & 0.00 & 0.96 & 0.94 \\
\texttt{vn\_exp\_2798213} & 2 & 0.58 & 0.36 & 0.76 & 0.66 \\
\texttt{vn\_exp\_54949707} & 2 & 0.20 & 0.24 & 0.92 & 0.98 \\
\texttt{vn\_exp\_69495711} & 2 & 0.57 & 0.30 & 0.72 & 0.92 \\
\texttt{vn\_exp\_70635338} & 2 & 0.20 & 0.19 & 0.74 & 0.77 \\
\texttt{vn\_exp\_7363976} & 2 & 0.00 & 0.00 & 0.00 & 0.00 \\
\texttt{vn\_exp\_75517995} & 2 & 0.03 & 0.00 & 0.94 & 0.95 \\
\texttt{vn\_exp\_75911833} & 2 & 0.00 & 0.00 & 0.66 & 0.43 \\
\texttt{vn\_exp\_75946937} & 2 & 0.43 & 0.24 & 0.81 & 0.76 \\
\texttt{vn\_exp\_76798922} & 2 & 0.90 & 0.82 & 1.00 & 1.00 \\
\texttt{advtele\_7gfuz9pg} & 3 & 0.00 & 0.00 & 0.00 & 0.00 \\
\textbf{\texttt{advtele\_rk8wlicy}} & 3 & \textbf{0.17} & \textbf{0.19} & \textbf{0.69} & \textbf{0.70} \\
\midrule
mean of the reads (number) & & 0.20 (22) & 0.12 (25) & 0.70 (22) & 0.66 (25) \\
model, \texttt{alg\_pred} & & \multicolumn{2}{c}{0.19} & \multicolumn{2}{c}{0.45} \\
\midrule
\texttt{gt\_us\_1} & & \multicolumn{2}{c}{0.90} & \multicolumn{2}{c}{0.00} \\
\texttt{gt\_us\_2} & & \multicolumn{2}{c}{0.00} & \multicolumn{2}{c}{0.98} \\
\texttt{gt\_us\_3} & & \multicolumn{2}{c}{0.00} & \multicolumn{2}{c}{1.00} \\
\texttt{gt\_us\_4} & & \multicolumn{2}{c}{0.94} & \multicolumn{2}{c}{0.96} \\
\texttt{gt\_us\_5} & & \multicolumn{2}{c}{1.00} & \multicolumn{2}{c}{1.00} \\
panel mean, \texttt{gt\_average\_us} & & \multicolumn{2}{c}{0.57} & \multicolumn{2}{c}{0.79} \\
standard $r$, \texttt{gt\_binary\_simple\_us} & & \multicolumn{2}{c}{1} & \multicolumn{2}{c}{1} \\
second panel\textquoteright s mean, \texttt{gt\_average\_vietnam} & & \multicolumn{2}{c}{0.62} & \multicolumn{2}{c}{0.47} \\
\bottomrule
\end{tabular}
\end{table}
 
\begin{table}[p]
\centering
\footnotesize
\caption{Every read of two film-findings of film 154, in the arm with clinical history, as recorded in the primary file (Table~\ref{tab:cxrcols}): one row per radiologist who read the film in that arm, unaided, with the prediction in view, or both, the reader of Table~\ref{tab:cxrfilm} in bold; a dagger marks a finding not shown to the reader, participant or panel, and recorded as zero. The measurement uses every unaided read, each weighing equally within the film-finding, so that the mean of the reads is what a radiologist drawn at random returns on it. Below, the model\textquoteright s response and the five probabilities of the panel that set the standard. The film was also read by 13 radiologists without history and 11 without history with the prediction in view}
\label{tab:cxrrecB}
\setlength{\tabcolsep}{4pt}
\begin{tabular}{l c c c c c}
\toprule
 & & \multicolumn{2}{c}{Cardiomegaly} & \multicolumn{2}{c}{Any abnormality} \\
\cmidrule(lr){3-4}\cmidrule(lr){5-6}
\texttt{experiment\_id} & \texttt{design} & unaided & in view & unaided & in view \\
\midrule
\texttt{advtele\_2aoiqxz7} & 1 & --- & 0.96 & --- & 1.00 \\
\texttt{advtele\_413wwwzp} & 1 & 0.77 & --- & 0.74 & --- \\
\texttt{advtele\_59dyemkk} & 1 & 0.70 & --- & 0.60 & --- \\
\texttt{advtele\_6gqvd9r0} & 1 & --- & 0.57 & --- & 0.73 \\
\texttt{advtele\_9mkc2cw0} & 1 & --- & 0.66 & --- & 0.84 \\
\texttt{advtele\_9n3zlpnz} & 1 & --- & 0.54 & --- & 0.71 \\
\texttt{advtele\_bi8258mz} & 1 & 0.65 & --- & 0.91 & --- \\
\texttt{advtele\_exnka7n7} & 1 & 0.83 & --- & 0.66 & --- \\
\texttt{advtele\_ggrn51uf} & 1 & --- & 0.78 & --- & 0.61 \\
\texttt{advtele\_i1pm6pyg} & 1 & 0.59 & --- & 0.62 & --- \\
\texttt{advtele\_js9fea1z} & 1 & --- & 0.80 & --- & 0.60 \\
\texttt{advtele\_oonetvp4} & 1 & 0.00$^{\dagger}$ & --- & 0.56 & --- \\
\texttt{advtele\_raqnks8q} & 1 & --- & 0.62 & --- & 0.70 \\
\texttt{advtele\_rt36epwg} & 1 & 1.00 & --- & 1.00 & --- \\
\texttt{advtele\_yoe8g3eg} & 1 & --- & 0.80 & --- & 0.21 \\
\texttt{advtele\_yonikm6g} & 1 & 0.95 & --- & 0.79 & --- \\
\texttt{vestarad\_8uz26kge} & 1 & 0.00$^{\dagger}$ & --- & 0.40 & --- \\
\texttt{vestarad\_xhkgqffz} & 1 & --- & 0.98 & --- & 1.00 \\
\texttt{vn\_exp\_34564272} & 2 & 0.66 & 0.65 & 0.97 & 0.66 \\
\texttt{vn\_exp\_67023857} & 2 & 0.52 & 0.73 & 0.92 & 0.98 \\
\texttt{vn\_exp\_67496434} & 2 & 0.53 & 0.59 & 0.54 & 0.98 \\
\texttt{vn\_exp\_84282381} & 2 & 0.60 & 0.78 & 0.96 & 0.95 \\
\texttt{advtele\_bnqzo2w1} & 3 & 0.79 & 0.79 & 0.80 & 0.94 \\
\textbf{\texttt{advtele\_m7xbjpys}} & 3 & \textbf{0.00$^{\dagger}$} & \textbf{0.73} & \textbf{0.83} & \textbf{0.27} \\
\texttt{usarad\_btebmoot} & 3 & 0.00$^{\dagger}$ & 0.00$^{\dagger}$ & 0.80 & 0.96 \\
\texttt{usarad\_ts60v01e} & 3 & 0.00$^{\dagger}$ & 0.00$^{\dagger}$ & 0.90 & 0.95 \\
\midrule
mean of the reads (number) & & 0.51 (17) & 0.65 (17) & 0.76 (17) & 0.77 (17) \\
model, \texttt{alg\_pred} & & \multicolumn{2}{c}{0.70} & \multicolumn{2}{c}{0.90} \\
\midrule
\texttt{gt\_us\_1} & & \multicolumn{2}{c}{0.00$^{\dagger}$} & \multicolumn{2}{c}{0.50} \\
\texttt{gt\_us\_2} & & \multicolumn{2}{c}{0.80} & \multicolumn{2}{c}{0.83} \\
\texttt{gt\_us\_3} & & \multicolumn{2}{c}{0.91} & \multicolumn{2}{c}{0.38} \\
\texttt{gt\_us\_4} & & \multicolumn{2}{c}{0.00$^{\dagger}$} & \multicolumn{2}{c}{0.21} \\
\texttt{gt\_us\_5} & & \multicolumn{2}{c}{0.70} & \multicolumn{2}{c}{0.88} \\
panel mean, \texttt{gt\_average\_us} & & \multicolumn{2}{c}{0.48} & \multicolumn{2}{c}{0.56} \\
standard $r$, \texttt{gt\_binary\_simple\_us} & & \multicolumn{2}{c}{0} & \multicolumn{2}{c}{1} \\
second panel\textquoteright s mean, \texttt{gt\_average\_vietnam} & & \multicolumn{2}{c}{0.45} & \multicolumn{2}{c}{0.56} \\
\bottomrule
\end{tabular}
\end{table}
 
\begin{table}[p]
\centering
\footnotesize
\caption{Film 77, on which four findings are present; the radiologist put any abnormality at $0.08$ and the airspace opacity at $0.02$, so that the findings beneath the airspace opacity, two of the four among them, were not shown to her and are recorded as zero. Another radiologist of the third design; columns as in Table~\ref{tab:cxrfilm}}
\label{tab:cxrfilmtwo}
\setlength{\tabcolsep}{2.8pt}
\begin{tabular}{l c c c c c c c c c}
\toprule
 & & \multicolumn{2}{c}{Alone} & Rung 3 & Rung 2 & B1 & 4c, checked & 4d & 4e \\
\cmidrule(lr){3-4}
Finding & $r$ & $H$ & $M$ & $M$ in view & pooling & by finding & + probability & ensembling & selection \\
\midrule
Any abnormality & 1 & 0.08 & 0.72 & 0.73 & 0.40 & 0.08$_H$ & 0.08$_H$ & 0.08 & 0.08$_H$ \\
Airspace opacity & 1 & 0.02 & 0.27 & 0.36 & 0.14 & 0.27$_M$ & 0.27$_M$ & 0.12 & 0.02$_H$ \\
Consolidation & 1 & 0.00 & 0.13 & 0.01 & 0.07 & 0.13$_M$ & 0.13$_M$ & 0.08 & 0.00$_H$ \\
Bacterial or lobar pneumonia & 1 & 0.00 & 0.13 & 0.00 & 0.07 & 0.13$_M$ & 0.13$_M$ & 0.08 & 0.13$_M$ \\
Edema & 0 & 0.00 & 0.15 & 0.35 & 0.07 & 0.00$_H$ & 0.15$_M$ & 0.10 & 0.00$_H$ \\
Atelectasis & 0 & 0.00 & 0.04 & 0.01 & 0.02 & 0.04$_M$ & 0.04$_M$ & 0.03 & 0.00$_H$ \\
Lesion & 0 & 0.00 & 0.11 & 0.01 & 0.06 & 0.11$_M$ & 0.11$_M$ & 0.06 & 0.00$_H$ \\
Pleural effusion & 0 & 0.00 & 0.01 & 0.00 & 0.01 & 0.01$_M$ & 0.01$_M$ & 0.01 & 0.01$_M$ \\
Other pleural finding & 0 & 0.00 & 0.01 & 0.00 & 0.01 & 0.01$_M$ & 0.01$_M$ & 0.01 & 0.01$_M$ \\
Pneumothorax & 0 & 0.00 & 0.06 & 0.00 & 0.03 & 0.00$_H$ & 0.06$_M$ & 0.03 & 0.00$_H$ \\
Cardiomediastinal abnormality & 0 & 0.02 & 0.14 & 0.01 & 0.08 & 0.14$_M$ & 0.14$_M$ & 0.12 & 0.02$_H$ \\
Cardiomegaly & 0 & 0.00 & 0.14 & 0.00 & 0.07 & 0.14$_M$ & 0.14$_M$ & 0.14 & 0.00$_H$ \\
Rib fracture (code 58) & 0 & 0.00 & 0.03 & 0.00 & 0.01 & 0.03$_M$ & 0.03$_M$ & 0.03 & 0.03$_M$ \\
Shoulder fracture (code 68) & 0 & 0.00 & 0.03 & 0.00 & 0.01 & 0.00$_H$ & 0.03$_M$ & 0.03 & 0.03$_M$ \\
Support device or hardware & 0 & 0.04 & 0.07 & 0.05 & 0.05 & 0.07$_M$ & 0.07$_M$ & 0.07 & 0.04$_H$ \\
\cmidrule(lr){1-10}
\multicolumn{2}{l}{Film, all 15 findings} & 0.746 & 0.853 & 0.828 & 0.809 & 0.803 & 0.801 & 0.776 & 0.762 \\
\multicolumn{2}{l}{Findings given to the radiologist} & 15 & 0 & 15 & 15 & 4 & 1 & 15 & 15 \\
\bottomrule
\end{tabular}
\end{table}

\clearpage
 
\section{Review map: the formulations in sequence}\label{app:map}

\newcommand{\src}[1]{\tag*{\footnotesize[#1]}}
\newcommand{\mapstep}[1]{\par\smallskip\noindent{\small\textbf{#1}}}
\newcommand{\interp}[1]{\par\smallskip\noindent\fbox{\begin{minipage}{\dimexpr\linewidth-2\fboxsep-2\fboxrule\relax}\small\textit{Interpretation.} #1\end{minipage}}\par\medskip}

\mapstep{1. Instance, response, performance}
\begin{align*}
& k \in \mathcal{K},\quad \Pr(k) = p_k; \qquad r \in \{0,1\}; \qquad \hat{r}_s \in [0,1] \src{\S\ref{sec:instances}}\\
& \theta(\hat{r}, r) = 1 - \phi\big(|\hat{r} - r|\big), \quad \phi \text{ non-decreasing},\ \phi(0) = 0,\ \phi(1) \leq 1 \src{Def.~\ref{def:endpoint}}\\
& \text{squared (the paper's rule): } \phi(x) = x^2, \qquad \text{linear (special case): } \phi(x) = x \\
& \text{squared: } \mu_{s,k} = 1 - \bar{r}_k(1 - \bar{r}_k) + \mathrm{Res}_{s,k} - \mathrm{Mis}_{s,k}, \quad \text{calibrated} \iff \bar{r}_{s,k}(p) = \Pr(r = 1 \mid \hat{r}_s = p, k) = p \src{Eq.~\ref{eq:murphy}}\\
& \theta_s = \theta(\hat{r}_s, r), \qquad \mu_{s,k} = \E[\theta_s \mid k], \qquad \E[\theta_s] = \textstyle\sum_k p_k\, \mu_{s,k} \src{\S\ref{sec:skills}}\\
& \theta_\pi = \theta(\hat{r}_\pi, r)
\end{align*}

\interp{Every instance has a correct response, which becomes known only later. A skill's response is scored by how far it lies from that response, and its performance is that score, between zero and one. The squared rule, the paper's, credits a response for the strength of the evidence it carries, which is what the skill knows, and charges it for miscalibration, the gap between what it says and what its responses turn out to mean. Everything that follows is an average of these scores, over all instances or over the instances of one class.}

\mapstep{1b. Skill policies and the two channels}
\begin{align*}
& \mathcal{S} \text{ available: standing cost } \kappa_\pi = \textstyle\sum_{s \in \mathcal{S}} \kappa_s; \qquad \mathcal{E}(x) \subseteq \mathcal{S} \text{ engaged on the instance: } c_\pi = \textstyle\sum_{s \in \mathcal{E}(x)} c_s \src{Defs.~\ref{def:available}, \ref{def:engaged}}\\
& \pi = \big(\mathcal{S},\ \mathcal{E}(\cdot),\ A \text{ or } \rho\big), \qquad \mathcal{E}: \text{what is read before execution} \mapsto \text{skills engaged} \src{Defs.~\ref{def:policy}, \ref{def:criterion}}\\
& \text{aggregation: } \hat{r}_\pi = A\big(\{\hat{r}_s\}_{s \in \mathcal{E}}, \cdot\big), \ \hat{r}_s \text{ independent}; \qquad \text{protocol: } \hat{r}_\pi = \hat{r}_\rho \src{Defs.~\ref{def:aggregation}, \ref{def:protocol}}\\
& \text{exchange: a skill's response is fixed after sight of another's} \src{Def.~\ref{def:exchange}}\\
& \text{who forms } \hat{r}_\pi: \ \text{reads the responses only} \Rightarrow \text{aggregation} \src{\S\ref{sec:policies}}\\
& \quad \text{executes the task after seeing them} \Rightarrow \text{a skill, in an exchange (a third reader who arbitrates: rung 5)} \\
& \text{augmentation} \iff \text{routing or synergy}: \ \text{can attain more than any fixed use (commitment, fixed pooling)} \src{Defs.~\ref{def:augmentation}, \ref{def:routing}, \ref{def:synergy}}\\
& \text{assignment: } |\mathcal{E}| = 1,\ \hat{r}_\pi = \hat{r}_{\hat{s}}; \quad \text{ensembling: } |\mathcal{E}| \geq 2,\ \hat{r}_\pi = \textstyle\sum_s \hat{w}_s \hat{r}_s; \quad \text{selection: } |\mathcal{E}| \geq 2,\ \hat{r}_\pi = \hat{r}_{\hat{s}} \src{Defs.~\ref{def:assignment}, \ref{def:ensembling}, \ref{def:selection}}
\end{align*}

\interp{A skill policy states which skills the organization keeps available, which of them it engages on each instance on the strength of what it can read beforehand, and how the delivered response is formed: by combining responses produced independently, or by letting the skills see each other's work first. Holding a skill costs whatever the volume; engaging it costs per instance. Doing better than any fixed use of the skills, augmentation, comes only from directing work, routing, or from interaction, synergy. Whoever forms the response decides which: a rule that only reads the responses combines them, while a person who examines the instance again after seeing them is a further skill in an exchange, as a third reader settling a disagreement is.}

\mapstep{2. Anchors and the stake}
\begin{align*}
& m = \mean_s \theta_s, \qquad \Mx = \max_s \theta_s, \qquad \Delta = \Mx - m \geq 0, \qquad n = |\mathcal{S}| \src{\S\ref{sec:anchors}}\\
& b = \max_s \E[\theta_s] = \E[\theta_L], \qquad \theta_U = \max_{s \neq L} \theta_s \ \text{(a pair: the other skill)} \\
& \E[m] \;\leq\; b \;\leq\; \E[\Mx] \\
& G = \E[\Mx] - \E[m], \qquad S = b - \E[m], \qquad R = \E[\Mx] - b \src{Defs.~\ref{def:G}, \ref{def:S}, \ref{def:R}}\\
& G = \E[\Delta] = \underbrace{b - \E[m]}_{S} + \underbrace{\E[\Mx] - b}_{R} \src{Eq.~\ref{eq:gsr}}\\
& \E[m] = \mean_s \E[\theta_s],\ b,\ S \ \text{from the marginals}; \qquad \E[\Mx],\ G,\ R \ \text{depend on the coupling} \src{Prop.~\ref{prop:anchor}}\\
& \Delta = \tfrac{1}{n} \textstyle\sum_s (\Mx - \theta_s) \qquad \text{(a pair: } \tfrac{1}{2}|\theta_i - \theta_j|) \src{Prop.~\ref{prop:gap}}\\
& S = \tfrac{1}{n} \textstyle\sum_s d_s, \quad d_s = b - \E[\theta_s]; \qquad \max_s d_s \in \big[\tfrac{n}{n-1} S,\ nS\big] \src{Prop.~\ref{prop:S}}\\
& \quad \text{(a pair: } S = \tfrac{1}{2}|\E[\theta_i] - \E[\theta_j]|,\ \max_s d_s = 2S) \\
& G = \tfrac{1}{n} \textstyle\sum_s \E[\Mx - \theta_s], \qquad \E[\Mx - \theta_s] = R + d_s \qquad \text{(a pair: } \tfrac{1}{2}\E|\theta_i - \theta_j|) \src{Prop.~\ref{prop:gap}}\\
& R = \E\big[(\theta_U - \theta_L)^{+}\big] = \Pr(\theta_U > \theta_L)\; \E\big[\theta_U - \theta_L \mid \theta_U > \theta_L\big] \geq 0 \src{Prop.~\ref{prop:R}}\\
& R = 0 \iff \Pr(\theta_U > \theta_L) = 0 \src{Prop.~\ref{prop:R0}}\\
& 0 \leq R \leq G \leq 1 - \tfrac{1}{n} \qquad \text{(a pair: } \tfrac{1}{2}) \src{Prop.~\ref{prop:bound}}\\
& \min_s \theta_s \leq m \leq \Mx, \qquad \Mx \geq \theta_s\ \ \forall s \src{Prop.~\ref{prop:disp}}
\end{align*}

\interp{On each instance the skills' performances have a mean and a best. Over the instances, the mean is what handing work out blindly attains and the best is what perfect allocation would attain; the gap between them, the gross stake, is all that knowing the skills could be worth. It starts from the mean, not from the worst performance, because the mean is what handing work out with no information attains and is readable from separate track records, while reaching the worst would take as much information as reaching the best. It splits at the best single skill. The systematic component is what choosing the better skill once secures, readable from separate track records. The residual is what remains for allocating instance by instance; it exists only because the better skill changes from one instance to another, and only observing both skills on the same instances reveals it.}

\mapstep{3. Engaging every skill as insurance}
\begin{align*}
& \text{commitment: } b \text{ w.p. } p_a, \quad b - d_s \text{ w.p. } q_s \ (s \neq L), \quad \textstyle\sum_{s \neq L} q_s = 1 - p_a \src{Prop.~\ref{prop:insurance}}\\
& \quad\Rightarrow\ \E = b - D, \quad D = \textstyle\sum_{s \neq L} q_s d_s, \qquad \text{spread } \max_s d_s \\
&\text{pooling: } \E[\theta_{\bar{w}}] = b - S + \Lambda \text{ whatever the assessment}, \quad \Lambda = \E\big[\tfrac{1}{n}\textstyle\sum_s (\hat{r}_s - \hat{r}_{\bar{w}})^2\big] \src{Eq.~\ref{eq:Lambda}}\\
&\text{premium} = (b - D) - (b - S + \Lambda) = S - \Lambda - D, \qquad \text{pooling} \succ \text{commitment} \iff D > S - \Lambda \\
&\text{uniform errors: } D = (1 - p_a)\tfrac{n}{n-1} S, \quad \text{pooling} \succ \text{commitment} \iff p_a < \tfrac{1}{n} + \tfrac{n-1}{n}\tfrac{\Lambda}{S} \\
&\text{a pair: } \E = b - 2S(1 - p_a), \quad \text{spread } 2S, \quad \text{premium } S(2p_a - 1) - \Lambda, \quad \iff p_a < \tfrac{1}{2} + \tfrac{\Lambda}{2S} \\
&\text{under sampling, a pair: } 1 - p_a = \Phi(-2S/\tau), \quad D = 2S\,\Phi(-2S/\tau) \leq \min\{S,\ \kappa\tau\} \src{Cor.~\ref{cor:insurance}}\\
&\quad \kappa = \max_{x \geq 0} x\,\Phi(-x) \approx 0.170; \quad \tau = \sigma/\sqrt{N}: \ \text{pooling} \succ \text{commitment only if } S - \Lambda < \kappa\sigma/\sqrt{N} \\
&\quad \text{distribution-free: } D \leq 2S\,e^{-2NS^2} \leq e^{-1/2}/\sqrt{N} \\
&\text{randomizing (one skill at random): } \E = \E[m] \text{ with one execution; pooling } \E[m] + \Lambda \text{ with } n \src{Prop.~\ref{prop:randomize}}\\
& \quad \mathrm{Var}(\theta_{\mathrm{rand}}) - \mathrm{Var}(m) = \E\big[\tfrac{1}{n}\textstyle\sum_s (\theta_s - m)^2\big] \\
&\quad \text{(a pair: } \tfrac{1}{4}\E[(\theta_i - \theta_j)^2]); \qquad \text{linear: } \Lambda = 0,\ \text{pooling} = \text{randomizing in expectation} \\
&\text{parity, a pair } (\E[\theta_H] = \E[\theta_M]):\ \ S = 0, \quad R = G = \tfrac{1}{2}\E|\theta_H - \theta_M|, \quad \text{premium } -\Lambda \leq 0 \src{Prop.~\ref{prop:parity}}
\end{align*}

\interp{Engaging every skill without directing work does not augment; it insures. A commitment pays the baseline when the assessment is right and less when it is wrong; pooling pays the same whatever the assessment and, under the squared rule, earns the disagreement between the responses. The premium is what pooling gives up in expectation when the assessment is usually right; random assignment attains the mean anchor and buys neither. At parity no skill is wrong on average, every commitment is still wrong on the instances the other skill wins, and the insurance against it is free. Against an assessment made on instances like those the commitment will serve, the insurance is worth little once the record is long, since a wrong assessment is likely only where the skills are close, and there it costs little; it is worth more where the assessment may be stale or made elsewhere.}

\mapstep{4. From responses to performances}
\begin{align*}
& \hat{r}_w = \textstyle\sum_s w_s \hat{r}_s, \quad w_s \geq 0,\ \textstyle\sum_s w_s = 1; \qquad \theta_w = \theta(\hat{r}_w, r) \src{Prop.~\ref{prop:hull}}\\
& |\hat{r}_w - r| = \textstyle\sum_s w_s\, |\hat{r}_s - r| \quad\text{(endpoint truth)} \\
& \Rightarrow\ \min_s \theta_s \;\leq\; \theta_w \;\leq\; \max_s \theta_s \\
& \text{linear: } \theta_w = \textstyle\sum_s w_s \theta_s; \qquad \phi \text{ convex: } \theta_w \geq \textstyle\sum_s w_s \theta_s; \qquad \phi \text{ concave: } \theta_w \leq \textstyle\sum_s w_s \theta_s \\
&\text{squared: } \theta_w = \textstyle\sum_s w_s \theta_s + \textstyle\sum_s w_s (\hat{r}_s - \hat{r}_w)^2 \src{Eq.~\ref{eq:ambiguity}}\\
&w = \bar{w}:\quad \text{linear } \E[\theta_{\bar{w}}] = \E[m]; \qquad \text{squared } \E[\theta_{\bar{w}}] = \E[m] + \E\big[\tfrac{1}{n}\textstyle\sum_s (\hat{r}_s - \hat{r}_{\bar{w}})^2\big] \ \text{(a pair: } \tfrac{1}{4}\E[(\hat{r}_i - \hat{r}_j)^2]) \\
&w \text{ fixed:}\quad \text{linear } \textstyle\sum_s w_s \E[\theta_s] \leq b; \qquad \text{squared } \textstyle\sum_s w_s \E[\theta_s] + \E\big[\textstyle\sum_s w_s (\hat{r}_s - \hat{r}_w)^2\big] \lesseqgtr b,\ \leq \E[\Mx] \src{Cor.~\ref{cor:fixed}}\\
& \hat{r}_{\mathcal{E}} = \Pr(r = 1 \mid Z, W, \{\hat{r}_s\}), \ \ \hat{r} \text{ any function of } (Z, W, \{\hat{r}_s\}): \ \ \E[\theta(\hat{r}, r)] = \E[\theta_{\mathcal{E}}] - \E\big[(\hat{r} - \hat{r}_{\mathcal{E}})^2\big] \leq \E[\theta_{\mathcal{E}}] \src{Prop.~\ref{prop:evidence}}\\
& \quad \E[\theta_{\mathcal{E}}] = 1 - \E\big[\hat{r}_{\mathcal{E}}(1 - \hat{r}_{\mathcal{E}})\big]; \quad \text{routing on what is observable} \leq \min\{\E[\Mx], \E[\theta_{\mathcal{E}}]\}; \quad \text{evidence pooling} \leq \E[\theta_{\mathcal{E}}] \\
& \quad \text{not ordered: } \E[\theta_{\mathcal{E}}] = 0.9571 > \E[\Mx] = 0.9509 \ \text{(Table~\ref{tab:outside})}; \quad 0.75 < 1 \ \text{(opposite certainties)} \\
& \text{scope: fails for a continuous truth and for multiclass responses} \src{\S\ref{sec:responses}}\\
& \quad r = (1,0,0):\ \|(0,1,0) - r\|^2 = \|(0,0,1) - r\|^2 = 2 \;>\; \|(0,\tfrac12,\tfrac12) - r\|^2 = 1.5 \\
&\text{linear rule not proper: } \argmax_{\hat{r}} \E\big[1 - |\hat{r} - r|\big] = \mathbb{1}\{\Pr(r = 1) > \tfrac{1}{2}\}; \ \text{squared rule proper} \src{\S\ref{par:rule}}\\
& \text{rule-free: the stake, the hull bound, routing range, value of information, identification} \\
&\text{the paper's rule, squared: pooling} = \E[m] + \Lambda,\ \gamma^{*} = b / (b - S + \Lambda),\ \text{ensembling can exceed assignment} \\
&\text{linear } (\Lambda = 0): \ \E[\theta_{\bar{w}}] = \E[m],\ \gamma^{*} = b/(b - S); \ \text{Cor.~\ref{cor:fixed} (second part), Props.~\ref{prop:ensembling} and~\ref{prop:randomize}}
\end{align*}

\interp{Where the truth is binary, any blend of responses scores between the worst and the best of the skills blended: averaging never beats the best skill on an instance. Under the squared rule it does beat the average of the scores, by the disagreement between the responses, which is what combining earns without any information. The cap fails where the truth is continuous or has several classes, the territory of forecast combination. A rule that reads the responses as joint evidence, rather than blending them, can beat the best skill on an instance; it is bounded instead by what the responses and the signals reveal together about the truth, a ceiling that can lie above or below perfect routing.}

\mapstep{5. Classes: the residual refined}
\begin{align*}
& b_k = \max_s \mu_{s,k}, \qquad \bar{s}(k) = \argmax_s \mu_{s,k}, \qquad b = \textstyle\sum_k p_k\, \mu_{L,k} \src{\S\ref{sec:classes}}\\
& R = \Rclass + \Rinst, \qquad \Rclass = \textstyle\sum_k p_k b_k - b = \textstyle\sum_k p_k (b_k - \mu_{L,k}) \geq 0, \qquad \Rinst = \textstyle\sum_k p_k R_k \geq 0 \src{Prop.~\ref{prop:class}}\\
& \Rclass = 0 \iff \mu_{L,k} = b_k \ \ \forall k: p_k > 0 \\
& \textstyle\sum_k p_k G_k = G, \qquad \textstyle\sum_k p_k S_k = S + \Rclass \\
& \Rightarrow\quad \E[m] \;\xrightarrow{\;+S\;}\; b \;\xrightarrow{\;+\Rclass\;}\; \textstyle\sum_k p_k b_k \;\xrightarrow{\;+\Rinst\;}\; \E[\Mx]
\end{align*}

\interp{Where instances fall into kinds, the residual splits. Part of it is collected by a standing rule sending each kind to the skill best at it, and identified from track records sorted by kind; the rest is collected only by reading each instance. The stake is climbed in this order: mean anchor, best skill, best skill per class, best skill per instance.}

\mapstep{6. Channel one: routing}
\begin{align*}
& \hat{r}_\pi = \textstyle\sum_s w_s \hat{r}_s, \quad \hat{r}_s \text{ independent}; \quad \text{assignment, ensembling, selection} \src{Def.~\ref{def:routing}}\\
& \Rightarrow\ \min_s \theta_s \leq \theta_\pi \leq \max_s \theta_s, \qquad \E[\theta_\pi] \in \big[\E[\min_s \theta_s],\ \E[\Mx]\big] \src{Prop.~\ref{prop:routing}}\\
& \text{pair: } \E[\min_s \theta_s] = 2\E[m] - \E[\Mx] = \E[m] - G \ \Rightarrow\ \E[\theta_\pi] \in \big[\E[m] - G,\ \E[m] + G\big] \\
& \text{routing reading nothing} \to \E[m]; \quad \text{the marginals} \to b; \quad \text{the class} \to\ \leq b + \Rclass; \quad \text{the instance} \to\ \leq \E[\Mx] \src{Prop.~\ref{prop:nested}}\\
& \text{criterion maximizing } \E[\theta_s \mid \mathcal{F}]:\quad b \;\leq\; \E\big[\max_s \E[\theta_s \mid \mathcal{F}]\big] \;\leq\; \E[\Mx] \src{Prop.~\ref{prop:information}}\\
& \quad \text{non-decreasing as } \mathcal{F} \text{ grows} \\
& \text{weights measurable in } \mathcal{F}:\quad \E[\theta_w] = \E\big[\textstyle\sum_s w_s \E[\theta_s \mid \mathcal{F}]\big] + \E\big[\textstyle\sum_s w_s (\hat{r}_s - \hat{r}_w)^2\big]; \ \text{linear: } \leq \E\big[\max_s \E[\theta_s \mid \mathcal{F}]\big] \src{Prop.~\ref{prop:ensembling}}\\
& \text{4a, two classes, accuracy } a:\quad a B_1 + (1 - a) B_1', \qquad > b \iff a > a^{*} = \frac{b - B_1'}{B_1 - B_1'} \src{Prop.~\ref{prop:classpred}}\\
& \quad B_1 + B_1' = 2\E[m] \text{ if the incumbents differ across classes} \\
& \tilde{\theta}_\pi = \textstyle\sum_s w_s \theta_s,\ a_k = \E[\tilde{\theta}_\pi \mid k]:\quad \E[\theta_\pi] - \E[m] = T_S + T_{\mathrm{class}} + T_{\mathrm{inst}} + \sqc{\E[\theta_w - \tilde{\theta}_\pi]} + \E[\theta_\pi - \theta_w] \src{Cor.~\ref{cor:attribution}}\\
& \quad C = \textstyle\sum_k p_k (\min(a_k, b_k) - \E[m \mid k]),\ \ T_S = \min(C, S) \leq S \\
& \quad T_{\mathrm{class}} = (C - S)^{+} \in [0, \Rclass],\ \ T_{\mathrm{inst}} = \textstyle\sum_k p_k (a_k - b_k)^{+} \in [0, \Rinst]
\end{align*}

\interp{Routing directs work but cannot make any skill better than it is: it attains between always picking the worse and always picking the better skill on each instance. What it reads decides where in that range it lands, and reading more never hurts a criterion that uses what it reads as well as it can, while a criterion that trusts an unreliable class prediction can fall below the commitment. What a policy takes can be read component by component: exactly in the classes where it attains what the commitment or the incumbent attains there, by a crediting convention in the others, in order in the main text and in proportion in Appendix~\ref{app:crediting}.}

\mapstep{7. Channel two: synergy}
\begin{align*}
& \hat{r}_\rho = \rho\big(\{\hat{r}_s\}, \ldots\big) \ \text{over at least two engaged skills, not a function of the independent } \hat{r}_s \src{Defs.~\ref{def:protocol}, \ref{def:synergy}}\\
& \gamma = \frac{\theta(\hat{r}_\rho, r)}{\theta_w}, \qquad \gamma_\pi = \frac{\E[\theta(\hat{r}_\rho, r)]}{\E[\theta_w]}, \qquad \Pr(\gamma < 1) \src{Def.~\ref{def:lift}}\\
& \E[\theta(\hat{r}_\rho, r)] = \gamma_\pi\, \E[\theta_w] \;\geq\; b \iff \gamma_\pi \geq \gamma^{*} = \frac{b}{\E[\theta_w]} \src{Prop.~\ref{prop:breakeven}}\\
&w = \bar{w}:\quad \gamma^{*} = \frac{b}{b - S + \Lambda}; \qquad \text{linear: } \frac{b}{b - S} \\
&\delta = \theta(\hat{r}_\rho, r) - \theta_w, \qquad \delta_\pi = \E[\theta(\hat{r}_\rho, r)] - \E[\theta_w] \;\geq\; \delta^{*} = b - \E[\theta_w] \quad (w = \bar{w}:\ \delta^{*} = S - \Lambda) \src{Def.~\ref{def:lift}, Prop.~\ref{prop:breakeven}}\\
&\text{binary performance: pooling } \E[m] + \tfrac{1}{4}\Pr(\theta_i \neq \theta_j); \ \gamma \in [1,\tfrac{4}{3}]:\ \E[\min(1, \gamma \theta_{\bar{w}})] = \E[m] + \tfrac{3\gamma - 2}{4} \Pr(\theta_i \neq \theta_j) \src{Prop.~\ref{prop:binarylift}}
\end{align*}

\interp{A protocol lets the skills see each other's work before the response is fixed, so its result is not a blend of what they would have done alone and is bounded by none of them. Its contribution, the lift, can only be measured, against the independent responses kept before the exchange. It matches the commitment only above a break-even lift, which rises with the systematic component and falls with the disagreement that pooling already earns.}

\mapstep{8. Reduced form: the archetypes}
\begin{align*}
&\theta_{HM} = \theta_{\bar{w}}, \qquad \theta_{HM{+}} = \min(1, \gamma \theta_{\bar{w}}), \qquad \theta_{HM{+}{+}} = \min(1, \gamma \Mx) \src{Prop.~\ref{prop:archetypes}, Eq.~\ref{eq:archetypes}}\\
&\text{unclipped, relative to } b:\quad -(S - \Lambda), \qquad (\gamma - 1)\,b - \gamma (S - \Lambda), \qquad (\gamma - 1)\,b + \gamma R \\
& \E\big[\min(1, \gamma X)\big] \leq \min\big(1, \gamma \E[X]\big), \qquad \gamma\E[X] - \E\big[\min(1, \gamma X)\big] = \E\big[(\gamma X - 1)^{+}\big] \src{Prop.~\ref{prop:ceiling}}
\end{align*}

\interp{The archetypes of the earlier paper are the corner settings of the two channels: pooling without information, a protocol on the pooled response, and a protocol on perfect routing. Because performance cannot exceed one, a uniform lift delivers less than it promises wherever it would push performance past the ceiling.}

\mapstep{9. Ceilings of the routing families}
\begin{align*}
& \text{B1 } i,j \mid \mathrm{class}: \quad \textstyle\sum_k p_k\, \mu_{\bar{s}(k),k} = b + \Rclass; \qquad \kappa_\pi = \textstyle\sum_{s \in \bar{s}(\mathcal{K})} \kappa_s, \quad \E[c_\pi] = \textstyle\sum_k p_k\, c_{\bar{s}(k)} \src{\S\ref{sec:ceilings}}\\
& \text{B2 } i,j \mid \mathrm{inst}: \quad \E[\Mx] = b + R, \quad \dot{s} = \argmax_s \theta_s; \qquad \kappa_\pi = \textstyle\sum_s \kappa_s, \quad \E[c_\pi] = \E[c_{\dot{s}}] \\
& \text{B3 } ij(w \mid \mathrm{inst}): \quad \argmax_w \theta\big(\textstyle\sum_s w_s \hat{r}_s, r\big) = e_{\dot{s}} \ \Rightarrow\ \E[\Mx]; \qquad \kappa_\pi = \textstyle\sum_s \kappa_s, \quad \E[c_\pi] = \textstyle\sum_s c_s \\
& \text{B3 and B2: same response on every instance; execution costs differ by } \textstyle\sum_s c_s - \E[c_{\dot{s}}]
\end{align*}

\interp{Each routing family has a ceiling, what it would attain with perfect information: each class sent to its incumbent, each instance to its winner, or the responses weighted on each instance. The last two attain the same performance and differ only in what they execute, one skill per instance against every skill.}

\mapstep{10. The ladder: attains}
\begin{equation*}
\renewcommand{\arraystretch}{1.35}
\begin{array}{@{}l l l@{}}
 & \text{policy} & \text{attains} \\
5 & ij(w \mid Z_q, W_q){+} & \gamma_\pi\, \E[\theta_{\hat{w}}] \\
\text{4e} & i,j \mid Z_q, W_q, \{\hat{r}_s\} & \E[\theta_{\hat{s}}] \;\leq\; \E[\Mx] \\
\text{4d} & ij(w \mid Z_q, W_q) & \E[\theta_{\hat{w}}] \;\leq\; \E[\Mx] \\
\text{4c+a} & i,j \mid Z_q, W_q,\ \text{audit } \epsilon & \text{as 4c} \\
\text{4b, 4c} & i,j \mid W_q \ \text{(and } Z_q) & \E[\theta_{\hat{s}}] \;\leq\; \E[\Mx] \\
\text{4a} & i,j \mid Z_q & \sum_k p_k \sum_{\hat{k}} \Pr(\hat{k} \mid k)\, \mu_{\bar{s}(\hat{k}),k} \;\leq\; b + \Rclass \\
3 & ij{+} & \gamma_\pi\, \E[\theta_{\bar{w}}] \\
\text{2r} & i,j \mid \varnothing \ (\text{random assignment}) & \E[m] = b - S \\
2 & ij \ (\text{pooling}) &\E[\theta_{\bar{w}}] = b - S + \Lambda \ \ (\text{linear: } b - S) \\
1 & i^{*} = \argmax_s \hat{\E}[\theta_s] & b \quad (\E[\theta_{i^{*}}] < b \text{ if the assessment is wrong; } b - 2S \text{ for a pair}) \\
0 & i & \E[\theta_i]
\end{array}
\src{Tab.~\ref{tab:spectrum}}
\end{equation*}

\interp{The ladder orders the arrangements by what they need and what they can attain: one skill unexamined, the assessed commitment, pooling and random assignment, a protocol, routing on the class and on the instance, and both channels together. Each rung's attainment has a closed form or is bounded by its family's ceiling.}

\mapstep{11. The ladder: cost profile}
\begin{align*}
& \kappa_\pi = \textstyle\sum_{s \in \mathcal{S}} \kappa_s \ \text{(standing, held available)}, \qquad c_\pi = \textstyle\sum_{s \text{ engaged}} c_s \;(+\, c_\rho) \ \text{(execution, per instance)} \src{\S\ref{sec:costs}}
\end{align*}
\begin{equation*}
\renewcommand{\arraystretch}{1.35}
\begin{array}{@{}l l l l@{}}
 & \text{policy} & \kappa_\pi & \E[c_\pi] \\
5 & ij(w \mid Z_q, W_q){+} & \sum_s \kappa_s & \sum_s c_s + c_\rho \\
\text{4e} & i,j \mid Z_q, W_q, \{\hat{r}_s\} & \sum_s \kappa_s & \sum_s c_s \\
\text{4d} & ij(w \mid Z_q, W_q) & \sum_s \kappa_s & \sum_s c_s \\
\text{4c+a} & i,j \mid Z_q, W_q,\ \text{audit } \epsilon & \sum_s \kappa_s & \E[c_{\hat{s}}] + \E[\epsilon \textstyle\sum_{s \neq \hat{s}} c_s] \\
\text{4b, 4c} & i,j \mid W_q \ \text{(and } Z_q) & \sum_s \kappa_s & \E[c_{\hat{s}}] \\
\text{4a} & i,j \mid Z_q & \sum_{s \in \bar{s}(\mathcal{K})} \kappa_s & \sum_k p_k \sum_{\hat{k}} \Pr(\hat{k} \mid k)\, c_{\bar{s}(\hat{k})} \\
3 & ij{+} & \sum_s \kappa_s & \sum_s c_s + c_\rho \\
\text{2r} & i,j \mid \varnothing \ (\text{random assignment}) & \sum_s \kappa_s & \tfrac{1}{n}\sum_s c_s \\
2 & ij \ (\text{pooling}) & \sum_s \kappa_s & \sum_s c_s \\
1 & i^{*} & \kappa_{i^{*}} & c_{i^{*}} \\
0 & i & \kappa_i & c_i
\end{array}
\src{Tab.~\ref{tab:costs}}
\end{equation*}

\interp{Every arrangement carries two costs that follow from its design: a standing cost for each skill it keeps available, paid whatever the volume, and an execution cost for each skill it engages on an instance. Rungs that engage every skill pay for every execution, assignment for one, and an audit adds a share of the others.}

\mapstep{12. What identifies what}
\begin{equation*}
\renewcommand{\arraystretch}{1.3}
\begin{array}{@{}l c l@{}}
\{\E[\theta_s]\} \ \text{(separate assessments)} & \Rightarrow & b,\ S,\ i^{*} \ (\text{and } \gamma^{*} \text{ under the linear rule}) \\
\{\mu_{s,k}\} \ \text{(separate assessments by class)} & \Rightarrow & \bar{s}(k),\ \sum_k p_k b_k,\ \Rclass \\
\text{joint law of } (\theta_s)_s \ \text{(paired, same instances)} & \Rightarrow & \E[\Mx],\ G,\ R,\ \Rinst,\ \Lambda,\ \gamma^{*} \\
\text{third arm, independent } \hat{r}_s \text{ retained} & \Rightarrow & \E[\theta_\pi],\ \gamma_\pi
\end{array}
\src{\S\ref{sec:identification}}
\end{equation*}
\begin{align*}
& \text{pairwise moments: } R_k \leq \textstyle\sum_{s \neq \bar{s}(k)} \tfrac{1}{2}\Big(\sqrt{\delta_{s,k}^2 + s_{s,k}^2} - \delta_{s,k}\Big) \src{Prop.~\ref{prop:twomoment}, Eq.~\ref{eq:twomoment}}\\
& \quad s_{s,k}^2 = \mathrm{Var}(\theta_s - \theta_{\bar{s}(k)} \mid k); \ \text{tight for a pair over all laws with those moments} \\
& \text{more concordant coupling, fixed marginals: } \E[m], b, S, \Rclass \text{ unchanged; } \E[\Mx], G, R, \Rinst \text{ do not rise} \src{Prop.~\ref{prop:coupling}}\\
& \text{finite record: } \hat{b} = \max_s \hat{\E}[\theta_s] \text{ biased up} \ \Rightarrow\ \hat{S} \text{ up}, \ \hat{R} = \hat{\E}[\Mx] - \hat{b} \text{ down} \src{\S\ref{sec:identification}}\\
& \quad \mathrm{se}(\hat{R}) \approx \mathrm{sd}(\Mx - \theta_L)/\sqrt{N} \\
& \text{audit at rate } \epsilon(x) > 0:\ \text{attains as the assignment; paired record grows at rate } \E[\epsilon] \src{Prop.~\ref{prop:audit}}\\
& \quad \tfrac{1}{N}\textstyle\sum_q A_q\, g_q / \epsilon(x_q) \ \text{unbiased for } \E[g], \ \text{with variance growing in } 1/\epsilon(x) \ \text{(inverse propensity)} \\
& \text{anchors: } \E[m] \text{ from the means; } \mathrm{Var}(m) = \tfrac{1}{4}(\sigma_i^2 + \sigma_j^2 + 2\,\mathrm{cov}) \text{ from moments} \src{App.~\ref{app:factors}}\\
& \quad \text{laws of } m \text{ and } \Mx \text{ only from the joint law} \\
& \quad \text{(independent skills: } m \text{ by convolution, } F_{\Mx}(x) = \textstyle\prod_s F_s(x))
\end{align*}

\interp{Each quantity needs its own evidence. Separate track records give the baseline and the systematic component; records sorted by class give the incumbents and the class residual; only the skills observed on the same instances give the instance residual, the disagreement and the break-even lift; only an arm running the workflow itself gives its take. Without paired data the residual is only bounded, and errors that coincide shrink it. A finite record overstates the baseline and understates the residual, and an audit keeps the paired record growing at a known cost, with estimates as precise as its rate is high where the criterion is confident.}

\mapstep{13. Regularities across the space}
\begin{align*}
& \text{normal pair in class } k:\quad R_k = s_k\big[\varphi(z_k) - z_k \Phi(-z_k)\big], \quad z_k = \delta_k / s_k, \quad R_k \leq s_k/\sqrt{2\pi} \src{Prop.~\ref{prop:normalref}}\\
& \text{layers: means} \to \text{pairwise moments} \to \text{joint law} \to \text{finite records} \src{App.~\ref{app:layers}}
\end{align*}

\interp{For normally distributed skills, the residual of a class depends only on the gap between the two means relative to the spread of their difference, and is largest when the means coincide. What can be established about the stake grows with what is known of the skills: their means, then their pairwise moments, then their joint law, each read through a record of finite size.}

\mapstep{14. Many skills and many classes}
\begin{align*}
& \E[m],\ \E[\Mx],\ B_1,\ \Rinst \text{ linear in the mix } p; \quad b,\ S \text{ convex} \\
& \quad \Rclass \text{ concave, largest where skills tie on average} \src{Prop.~\ref{prop:mix}}\\
& \text{adding } j \text{ to } A:\ \ \Delta b = (\E[\theta_j] - b(A))^{+}, \quad \Delta B_1 = \textstyle\sum_k p_k (\mu_{j,k} - b_k(A))^{+}, \quad \Delta \E[\Mx] = \E[(\theta_j - \Mx_A)^{+}] \src{Prop.~\ref{prop:roster}}\\
& \quad \text{submodular in the roster; greedy roster} \geq (1 - 1/e) \times \text{best} \\
& j \text{ adds nothing under some coupling} \iff F_{i,k} \leq F_{j,k} \ \ \forall k \src{Prop.~\ref{prop:newcomer}}\\
& \text{4a, } K \text{ classes: } \textstyle\sum_k p_k \sum_j \Pr(\hat{k} = j \mid k)\, \mu_{\bar{s}(j),k}; \\
& \quad \text{symmetric errors: } a B_1 + (1 - a) B_1', \ \ a^{*} = \frac{b - B_1'}{B_1 - B_1'} \src{Prop.~\ref{prop:classpredK}}\\
& \text{the better of } \pi_0, \pi_1 \text{ on each cell of } \hat{k} \text{ attains at least each of them} \src{Prop.~\ref{prop:checks}}\\
& \hat{g} \sim N(g, \tau^2), \ \text{adopt } \pi_1 \iff \hat{g} > z\tau: \ \ \text{give-back} \leq \kappa_z \tau, \ \ \text{gain forgone} \leq \lambda_z \tau \src{Prop.~\ref{prop:margin}}\\
& \quad (\kappa_z, \lambda_z) = (0.170, 0.170),\ (0.033, 0.507),\ (0.008, 0.837) \ \text{at } z = 0,\ 1,\ 1.645 \\
& \textstyle\sum_k p_k \big(\E[\max_s F_{s,k}^{-1}(V)] - b_k\big) \;\leq\; \Rinst \\
& \quad \;\leq\; \textstyle\sum_k p_k \big(\min_v \{v + \sum_s \E[(\theta_s - v)^{+} \mid k]\} - b_k\big) \src{Prop.~\ref{prop:classbounds}}\\
& n_k \propto p_k \sigma_k / \sqrt{c_k}, \qquad \sigma_k^2 \leq R_k \src{Prop.~\ref{prop:allocation}}\\
& 0 \leq \E[\hat{b}] - b \leq \sqrt{\ln n / (2N)} \src{Prop.~\ref{prop:curse}}\\
& \text{admit } j \text{ to pool } A \iff \tfrac{1}{|A| + 1}\big(\E[\theta_j] - \mean_{s \in A} \E[\theta_s]\big) + \Lambda_{A \cup \{j\}} - \Lambda_A > 0 \\
& \quad \text{pool class } k \iff \Lambda_k > S_k \src{Prop.~\ref{prop:subset}}
\end{align*}
\interp{With several skills and classes, the class residual is largest where the skills are level on average. A skill is worth to a roster what it adds where it beats everything already held, with diminishing returns as the roster grows, and a skill that leads nowhere can still be worth holding if its failures fall elsewhere. A class prediction needs to separate only the classes with different incumbents. A check that keeps, class by class, the better of two policies never gives back what the previous step held when the class means are known; on a finite record, a margin of one standard error cuts what it can give back fivefold, at the price of small gains forgone. Paired evidence is best spent on frequent classes with a variable residual, and the winner's curse grows only slowly with the number of skills compared.}
 
\mapstep{15. Skills in contest}
\begin{align*}
& \text{parity: } S = 0, \qquad R = G = \tfrac{1}{2}\E|\theta_H - \theta_M| \src{Prop.~\ref{prop:parity}}\\
& Y_n = \min(Z_n, \bar{z})\, \mathbf{1}[s' \text{ succeeds, } s \text{ fails to answer}] - \eta\, \mathbf{1}[s' \text{ succeeds, } s \text{ answers}] \src{Eq.~\ref{eq:contest}}\\
& \chi = \E[\theta_{s'}\theta_s] / \E[\theta_{s'}] = \E[\theta_s] + \operatorname{Cov}(\theta_s, \theta_{s'}) / \E[\theta_{s'}], \qquad h(\bar{z}) = \E[\min(Z, \bar{z})] \\
& \mu(\bar{z}) = \E[\theta_{s'}]\,\big[(1 - \chi)\, h(\bar{z}) - \eta\chi\big]; \qquad \mu < 0 \iff h(\bar{z}) < r_\eta = \eta\chi/(1 - \chi) \src{Prop.~\ref{prop:contest}}\\
& \text{parity: } \chi \text{ from } \E[F^{-1}(V)F^{-1}(1 - U)]/\E[\theta] \text{ (countermonotone)} \\
& \quad \text{through } \E[\theta] \text{ (independent) to } \E[\theta^2]/\E[\theta] \text{ (comonotone)} \\
& \E[Z] = \infty \Rightarrow \mu > 0 \text{ for large } \bar{z}, \text{ whatever the edge}; \qquad \mu < 0 \Rightarrow \Pr(\text{the position ever reaches } \bar{x}) \leq e^{-\zeta \bar{x}}
\end{align*}

\interp{When two skills compete, the opposing skill's capability sets how often the organization is tested, and its own skill's performance on the instances where the opposing skill succeeds sets which way its position drifts. At parity that performance turns on how the two are coupled: failing where the other succeeds, the coupling that makes two skills most valuable held together, is the most exposed position. When outcomes accumulate and a single loss can be very costly, the contest is decided by how much one unfavorable outcome is allowed to cost, a bound set by policy that belongs to neither skill nor arrangement and, under infinite-mean consequences, is the only instrument that keeps the position viable.}

\section{Interpretations of the propositions}\label{app:interpretations}
 
Each proposition and corollary, in the order in which they appear, read in plain words.
 
\begin{description}[leftmargin=0pt, itemsep=4pt, parsep=0pt, font=\normalfont\bfseries, style=unboxed]
\item[Proposition~\ref{prop:S} (Closed form of the systematic component).] What committing to the skill best on average secures over handing work out blindly is the mean of the skills' expected shortfalls from that skill; committing to the weakest skill instead falls short of it by between $n/(n-1)$ and $n$ times as much, twice as much for a pair.
\item[Proposition~\ref{prop:gap} (The gap and the gross stake).] On every instance, the distance between the mean and the best performance is the mean shortfall of the skills from the best of them. In expectation, each skill's shortfall from the best on each instance splits into a part common to all skills, the residual, and the skill's own shortfall from the skill best on average.
\item[Proposition~\ref{prop:R} (Decomposition and factorization of the stake).] The gross stake is the systematic component plus the residual, and the residual is how often a skill other than the one best on average does better on an instance, times by how much the best of them does better when it does.
\item[Proposition~\ref{prop:R0} (When the residual vanishes).] There is something to gain from allocating instance by instance if and only if the better skill is not always the same one.
\item[Proposition~\ref{prop:bound} (Bounds on the stake).] The stake is at most one minus one over the number of skills, one half for a pair, reached only when on every instance one skill is perfect and the others fail, each skill in turn.
\item[Proposition~\ref{prop:disp} (Instance-wise ordering of the anchors).] On every instance, the best performance is at least that of any single skill, and the mean lies strictly between the worst and the best whenever they differ.
\item[Proposition~\ref{prop:class} (Class decomposition of the residual).] Where instances fall into kinds, the residual splits into what a standing rule sending each kind to its best skill collects and what only reading each instance collects; the first is zero exactly when the skill best on average is best in every kind.
\item[Proposition~\ref{prop:hull} (The hull bound).] When the truth is binary, a blend of the skills' responses, however its weights are chosen, scores between the worst and the best of them on every instance; under the squared rule it scores the blend of their scores plus the disagreement between the responses.
\item[Corollary~\ref{cor:fixed} (Fixed combinations against the informed baseline).] Blending the responses with fixed weights can beat the commitment under the squared rule, thanks to the disagreement it earns, but never the best skill on every instance; under the linear rule it never beats the commitment.
\item[Proposition~\ref{prop:evidence} (The evidence ceiling).] A combination that reads the responses as joint evidence can beat the best of them on an instance, but not the best that the responses and signals together reveal about the truth; that ceiling and routing omniscience are not ordered, and every deployable routing policy lies at or below both.
\item[Proposition~\ref{prop:insurance} (The insurance contract).] Engaging every skill on every instance trades the expected return of a commitment, which is high when the assessment is right and low when it is wrong, for a fixed return; it pays in expectation only when the assessment is wrong often enough, or the responses disagree enough. \item[Corollary~\ref{cor:insurance} (The insurance contract under sampling).] An assessment made on instances like those the commitment will serve errs mostly where the skills are close and a wrong commitment costs little, so the insurance is worth at most a fraction of the assessment's standard error, falling with the square root of the record; it is worth more where the assessment may be stale or made elsewhere.
\item[Proposition~\ref{prop:parity} (Parity).] When two skills are equally good on average, no commitment is wrong on average, yet each is wrong on the instances the other wins; all of the stake is residual, half the mean distance between the two performances, and pooling costs nothing in expectation.
\item[Proposition~\ref{prop:randomize} (Pooling against randomizing).] Picking one skill at random for each instance attains the mean anchor with one execution; pooling attains it plus the disagreement with every execution, and is also less dispersed.
\item[Proposition~\ref{prop:routing} (The routing range).] Directing work among skills that respond independently attains between always picking the worse skill and always picking the better one on each instance, whatever information it uses; for a pair, as far below the mean anchor as perfect routing is above it.
\item[Proposition~\ref{prop:nested} (The components are reached in order).] Sending each class to its best skill collects the class residual only by using the best skill of every class, and then attains in every class at least what the commitment attains there; an arrangement collects the whole stake only by matching the best performance on every instance.
\item[Corollary~\ref{cor:attribution} (What a policy takes, component by component).] What a policy attains above the mean anchor splits into what it takes of each component of the stake, what combining responses earns, and what a protocol adds or subtracts; the split among the components is exact in the classes where the policy attains exactly what the commitment or the incumbent attains there, and a crediting convention in the others.
\item[Proposition~\ref{prop:breakeven} (Break-even of synergy).] A protocol pays only if its lift exceeds a break-even that rises with the systematic component and falls with the disagreement pooling already earns.
\item[Proposition~\ref{prop:archetypes} (The archetypes under the squared rule).] The three archetypes of the earlier paper, pooling, a protocol on the pooled response and a protocol on perfect routing, attain closed forms that the squared rule shifts by the disagreement.
\item[Proposition~\ref{prop:coupling} (Coupling).] When the skills' errors coincide more, the systematic component and the class residual stay where they are, and the instance residual can only shrink.
\item[Proposition~\ref{prop:anchor} (What the marginals determine).] Separate track records fix the mean anchor, the baseline and the systematic component, but only bound the residual: between what it would be if the skills failed together and what it would be if they took turns.
\item[Proposition~\ref{prop:audit} (Assignment with audit).] Having the other skills also execute a random share of instances, for the record only, keeps the paired record growing without changing what the policy delivers, at a known execution cost, and lets every paired quantity be estimated without bias.
\item[Proposition~\ref{prop:twomoment} (A two-moment bound on the residual).] The residual of a class is capped by the gap between the skills' means and the spread of their difference, knowledge that falls short of the full joint law.
\item[Proposition~\ref{prop:information} (The value of what the criterion reads).] A criterion that sends each instance to the skill expected to do best, given what it reads, attains at least the commitment and at most perfect routing, and reading more never hurts it.
\item[Proposition~\ref{prop:ensembling} (Ensembling and assignment).] Weighting the responses beats sending the instance to the single best-expected skill only when the disagreement it earns exceeds what it gives up by not picking that skill; under the linear rule it never does.
\item[Proposition~\ref{prop:classpred} (Assignment on a predicted class).] When the two classes have different best skills, trusting a class prediction of accuracy $a$ moves attainment in a straight line from the mean anchor at chance to the class ceiling at no error, so it beats the commitment only above a threshold accuracy; when they have the same best skill, the prediction does not matter.
\item[Proposition~\ref{prop:normalref} (A normal reference).] For normally distributed skills, the residual of a class depends only on how far apart the means are relative to the spread of the difference, and is largest when the means coincide.
\item[Proposition~\ref{prop:binarylift} (Synergy with binary performance).] When performance is all or nothing, pooling earns a quarter of the probability that the skills disagree, and a protocol's lift raises the coefficient on that probability.
\item[Proposition~\ref{prop:ceiling} (Capping performance at one).] Because performance cannot exceed one, a uniform lift delivers less than it promises, by exactly what it would have pushed past the ceiling.
\item[Proposition~\ref{prop:roster} (What a skill adds to a roster).] A skill is worth to a roster what it adds where it beats everything already held; its worth falls as the roster grows, and building the roster greedily, the largest addition first, attains at least $1 - 1/e$ of what the best roster of the same size attains.
\item[Proposition~\ref{prop:newcomer} (A skill that leads nowhere).] A skill that is never the best on average in any class can still raise what perfect routing attains, and must, whatever the coupling, unless the held skill's distribution dominates its own in every class.
\item[Proposition~\ref{prop:mix} (The stake across class mixes).] As the mix of classes shifts, the class residual is largest at a mix where the skills are level on average, which for a pair is where a commitment secures nothing.
\item[Proposition~\ref{prop:classpredK} (Assignment on a predicted class, many classes).] A class prediction only needs to tell apart classes with different best skills; mistakes between classes with the same best skill cost nothing.
\item[Proposition~\ref{prop:checks} (Checks that never give back).] Keeping, in every class as the policy reads it, the better of the old and the new policy never does worse than either, which is why the climb of Section~\ref{sec:procedure} never gives back what an earlier step held.
\item[Proposition~\ref{prop:margin} (A margin on the checks).] On a finite record a check can adopt a switch that loses; requiring the estimated gain to exceed its standard error cuts what a check can give back fivefold, at the price of forgoing small gains.
\item[Proposition~\ref{prop:classbounds} (What assessments by class establish).] Track records sorted by class identify the class residual exactly and bound the instance residual class by class, so that classes where it cannot be large are set aside before any paired study.
\item[Proposition~\ref{prop:allocation} (Where to pair).] A budget of paired instances is best spent in proportion to how frequent a class is and how variable its residual, and less where pairing is costly; the class bounds cap that variability before any pairing.
\item[Proposition~\ref{prop:curse} (The winner's curse with many skills).] Picking the best of several estimated skills overstates the baseline, but only by a margin that grows slowly with the number of skills and shrinks with the size of each record.
\item[Proposition~\ref{prop:subset} (When to admit a skill to a pool).] A weaker skill belongs in a pool when the disagreement it adds outweighs its shortfall from the pool, diluted by the pool's size; within a class, pooling beats the class's best skill when the disagreement exceeds that class's systematic component.
\item[Proposition~\ref{prop:contest} (The bound that decides a contest).] When an organization's skill faces an opposing skill and outcomes accumulate, its position gains ground if and only if what one unanswered success may cost stays below a threshold set by how often its skill answers where the opposing skill succeeds; at parity that depends on how the two performances are coupled, and when single losses can be unboundedly large no capability edge suffices without a bound.
\end{description}

\end{spacing}
\end{document}